\documentclass[printer,longauth,usenatbib]{aa}
\usepackage{graphicx}
\usepackage{natbib}
\usepackage{times}
\usepackage[T1]{fontenc}
\usepackage{multirow} 
\usepackage{tabularx}
\usepackage[skip=1ex]{caption}
\usepackage{amsmath}
\usepackage{longtable}
\usepackage{pdflscape}
\usepackage{adjustbox}
\usepackage{amsmath, systeme}
\usepackage{hyperref}

\PassOptionsToPackage{table}{xcolor}
\usepackage{xcolor}       

\usepackage{soul}

\usepackage[utf8]{inputenc}
\usepackage{tabularx}
\usepackage{supertabular}

\definecolor{highlight}{rgb}{0, 0 ,0}

\usepackage[normalem]{ulem}

\title{Asteroid sizes determined with thermophysical model \break and stellar occultations}
\author{A. Choukroun \inst{\ref{inst:UAM}}
  \and A. Marciniak \inst{\ref{inst:UAM}}
  \and J. \v{D}urech \inst{\ref{inst:Prague}}
  \and J.~Per{\l}a \inst{\ref{inst:UAM}}
  \and W.~Og{\l}oza \inst{\ref{inst:Suh}}
  \and R.~Szak{\'a}ts \inst{\ref{inst:Konkoly},\ref{inst:Excellence}}
  \and L.~Moln{\'a}r \inst{\ref{inst:Konkoly},\ref{inst:Excellence},\ref{inst:Lendulet},\ref{inst:ELTE}}
  \and A.~P{\'a}l \inst{\ref{inst:Konkoly},\ref{inst:Excellence},\ref{inst:ELU}} 
  \and F.~Monteiro \inst{\ref{inst:OASI}, \ref{inst:bz-grp-dyn}}
  \and I.~Mieczkowska \inst{\ref{inst:UAM}}
  \and W.~Beisker \inst{\ref{inst:IOTA-ES}}
 \and D.~Agnetti \inst{\ref{inst:IOTA-ES}}
 \and C.~Anderson \inst{\ref{inst:IOTA}}
  \and S.~Andersson \inst{\ref{inst:IOTA-ES}}
 \and D.~Antuszewicz \inst{\ref{inst:IOTA-ES}}
\and P.~Arcoverde \inst{\ref{inst:OASI}}
  \and R.-L.~Aubry \inst{\ref{inst:T60}}
 \and P.~Bacci \inst{\ref{inst:IOTA-ES}}
 \and R.~Bacci \inst{\ref{inst:IOTA-ES}}
  \and P.~Baruffetti \inst{\ref{inst:IOTA-ES}}
  \and L.~Benedyktowicz \inst{\ref{inst:IOTA-ES}}
 \and M.~Bertini  \inst{\ref{inst:IOTA-ES}}
  \and D.~B{\l}a{\.z}ewicz \inst{\ref{inst:IOTA-ES}}
\and {\color{highlight} R.~Boninsegna} \inst{\ref{inst:IOTA-ES}}
  \and Zs.~Bora \inst{\ref{inst:Konkoly}}
  \and M.~Borkowski \inst{\ref{inst:IOTA-ES}}
\and {\color{highlight} E.~Bredner} \inst{\ref{inst:IOTA}}
 \and J.~Broughton \inst{\ref{inst:IOTA}}
  \and M.~Butkiewicz - B\k{a}k \inst{\ref{inst:UAM}}
 \and N.~Carlson \inst{\ref{inst:IOTA}}
  \and G.~Casalnuovo \inst{\ref{inst:IOTA-ES}}
 \and F.~Casarramona  \inst{\ref{inst:IOTA-ES}}
  \and Y.-J.~Choi \inst{\ref{inst:KASI}}
 \and S.~Cikota \inst{\ref{inst:CalarAlto}}
 \and M.~Collins \inst{\ref{inst:IOTA}}
  \and B.~Cseh \inst{\ref{inst:Konkoly}, \ref{inst:Lendulet}, \ref{inst:MTA-ELTe}}
  \and G.~Cs\"{o}rnyei \inst{\ref{inst:Konkoly}}
 \and H.~De Groot \inst{\ref{inst:IOTA-ES}}
  \and P.~Delincak \inst{\ref{inst:IOTA-ES}}
 \and P.~Denyer \inst{\ref{inst:IOTA}}
\and {\color{highlight} R.~Dequinze} \inst{\ref{inst:IOTA}}
 \and M.~Dogramatzidis \inst{\ref{inst:IOTA-ES}}
  \and M.~Dr{\'o}{\.z}d{\.z} \inst{\ref{inst:Suh}}
  \and R.~Duffard \inst{\ref{inst:Granada}}
  \and D.~Eisfeldt \inst{\ref{inst:IOTA}}
\and M.~Eleftheriou \inst{\ref{inst:IOTA-ES}}
\and {\color{highlight} C.~Ellington} \inst{\ref{inst:Maastricht}}
  \and S.~Fauvaud \inst{\ref{inst:T60},\ref{inst:Bardon}}
  \and M.~Fauvaud \inst{\ref{inst:T60},\ref{inst:Bardon}}
  \and M.~Ferrais \inst{\ref{inst:Florida}}
  \and M.~Filipek \inst{\ref{inst:IOTA-ES}}
  \and P.~Fini \inst{\ref{inst:IOTA-ES}}
  \and M.~Frits \inst{\ref{inst:Konkoly}}
  \and B.~G\"{a}hrken \inst{\ref{inst:IOTA-ES}}
 \and G.~Galli \inst{\ref{inst:IOTA-ES}}
 \and D.~Gault \inst{\ref{inst:TTOA}}
  \and S.~Geier \inst{\ref{inst:IAC},\ref{inst:GRANTECAN}}
 \and B.~Gimple \inst{\ref{inst:IOTA}}
  \and J.~Golonka \inst{\ref{inst:Torun}}
 \and L.~Grazzini \inst{\ref{inst:IOTA-ES}}
  \and J.~Grice \inst{\ref{inst:Open}}
  \and K.~Guhl \inst{\ref{inst:IOTA-ES}}
  \and W.~Hanna \inst{\ref{inst:TTOA}}
  \and M.~Harman \inst{\ref{inst:IOTA-ES}}
  \and W.~Hasubick \inst{\ref{inst:IOTA-ES}}
 \and T.~Haymes \inst{\ref{inst:IOTA-ES}}
 \and D.~Herald \inst{\ref{inst:TTOA}}
  \and D.~Higgins \inst{\ref{inst:Hill}}
  \and R.~Hirsch \inst{\ref{inst:UAM}}
  \and J.~Horbowicz \inst{\ref{inst:UAM}}
  \and \'{A}.~Horti - D{\'a}vid \inst{\ref{inst:Konkoly}}
  \and B.~Ign{\'a}cz \inst{\ref{inst:Konkoly},\ref{inst:Excellence}}
  \and E.~Jehin \inst{\ref{inst:Liege}}
  \and A.~Jones \inst{\ref{inst:BAA}}
 \and R.~Jones \inst{\ref{inst:IOTA}}
 \and D.~Dunham \inst{\ref{inst:IOTA}}
  \and Cs.~Kalup \inst{\ref{inst:Konkoly}}
  \and K.~Kami{\'n}ski \inst{\ref{inst:UAM}}
  \and M.~K.~Kami{\'n}ska \inst{\ref{inst:UAM}}
  \and P.~Kankiewicz \inst{\ref{inst:Kielce}}
\and M. Kaplan \inst{\ref{inst:Akdeniz}}
 \and A.~Karagiannidis \inst{\ref{inst:IOTA-ES}}
  \and B.~Kattentidt \inst{\ref{inst:IOTA-ES}}
 \and S.~Kidd \inst{\ref{inst:IOTA-ES}}
 \and B.~Kirpluk \inst{\ref{inst:Torun}}
  \and D.-H.~Kim \inst{\ref{inst:KASI},\ref{inst:Chungbuk}}
  \and M.-J.~Kim \inst{\ref{inst:KASI}}
  \and I.~Konstanciak \inst{\ref{inst:UAM}}
  \and G.~Krannich \inst{\ref{inst:IOTA-ES}}
 \and M.~Kretlow \inst{\ref{inst:Granada}}
  \and J.~Kub{\'a}nek \inst{\ref{inst:IOTA-ES}, \ref{inst:cz-as}, \ref{inst:cz-obs-roky}}
  \and V.~Kudak \inst{\ref{inst:Uzhhorod}}
  \and P.~Kulczak \inst{\ref{inst:UAM}}
  \and M.~Lecossois \inst{\ref{inst:belg}}
 \and R.~Leiva \inst{\ref{inst:Granada}}
 \and M.~Libert \inst{\ref{inst:IOTA-ES}}
  \and J.~Licandro \inst{\ref{inst:IAC},\ref{inst:ULL}}
 \and P.~Lindner \inst{\ref{inst:IOTA-ES}}
 \and R.~Liu \inst{\ref{inst:IOTA}}
 \and Y.~Liu \inst{\ref{inst:IOTA}}
 \and G.~Lyzenga \inst{\ref{inst:IOTA}}
 \and M.~Maestripieri \inst{\ref{inst:IOTA-ES}}
 \and C.~Malagon \inst{\ref{inst:IOTA-ES}}
  \and P.~Maley  \inst{\ref{inst:IOTA},\ref{inst:Houston}}
  \and A.~Manna \inst{\ref{inst:IOTA-ES}}
  \and S.~Messner  \inst{\ref{inst:IOTA}}
  \and O.~Michniewicz \inst{\ref{inst:Kepler}}
\and  {\color{highlight} M.~A.~Miftah} \inst{\ref{inst:Liege}}
 \and M.~Mizutani \inst{\ref{inst:JOIN}}
  \and N.~Morales \inst{\ref{inst:Granada}}
  \and M.~Murawiecka \inst{\ref{inst:UAM}}
  \and J.~Nadolny \inst{\ref{inst:UAM}, \ref{inst:IAC},  \ref{inst:ULL}}
 \and T.~Nemoto \inst{\ref{inst:JOIN}}
  \and J.~Newman  \inst{\ref{inst:TTOA}}
 \and V.~Nikitin \inst{\ref{inst:IOTA}}
 \and P.~Nosal \inst{\ref{inst:IOTA-ES}}
 \and P.~Nosworthy \inst{\ref{inst:TTOA}}
\and {\color{highlight} M.~O'Connell} \inst{\ref{inst:IOTA-ES}}
   \and J.~Oey \inst{\ref{inst:Blue}}
\and A.~M.~Ortiz-Ochoa \inst{\ref{inst:Brussels}}
 \and A.~Ossola \inst{\ref{inst:IOTA-ES}}
  \and D.~Oszkiewicz \inst{\ref{inst:UAM}}
  \and E.~Pak\v{s}tien{\.e} \inst{\ref{inst:Vilnius}}
  \and M.~Paw{\l}owski \inst{\ref{inst:UAM}}
  \and V.~Perig \inst{\ref{inst:Uzhhorod}}
\and {\color{highlight} E.~Petrescu} \inst{\ref{inst:Liege}}
  \and F.~Pilcher \inst{\ref{inst:Organ}}
  \and E.~Podlewska-Gaca \inst{\ref{inst:UAM}}
 \and M.~Pol{\'a}\v{c}ek \inst{\ref{inst:IOTA-ES}}
  \and J.~Pol{\'a}k \inst{\ref{inst:IOTA-ES}}
  \and T.~Polakis \inst{\ref{inst:Tempe}}
  \and M.~Poli{\'n}ska \inst{\ref{inst:UAM}}
  \and A.~Popowicz \inst{\ref{inst:Gliwice}}
  \and V.~Reddy \inst{\ref{inst:Lunar}}
  \and J.-J.~Rives \inst{\ref{inst:T60}}
  \and M.~Rottenborn \inst{\ref{inst:cz-obs-roky}}
 \and N.~Ruocco \inst{\ref{inst:IOTA-ES}, \ref{inst:osservatorio}}
  \and A.~Rutkowski \inst{\ref{inst:Suh}}
  \and K.~Saci \inst{\ref{inst:IOTA-ES}}
  \and T.~Santana-Ros \inst{\ref{inst:Alicante},\ref{inst:Barcelona}}
  \and K.~S{\'a}rneczky \inst{\ref{inst:Konkoly}}
  \and O.~Schreurs \inst{\ref{inst:IOTA-ES}, \ref{inst:belg}}
 \and V.~Sempronio \inst{\ref{inst:IOTA}}
  \and B.~Skiff \inst{\ref{inst:Lowell}}
  \and J.~Skrzypek \inst{\ref{inst:UAM}}
 \and D.~Smith \inst{\ref{inst:IOTA-ES}}
  \and K.~Sobkowiak \inst{\ref{inst:UAM}}
\and E.~Sonbas \inst{\ref{inst:Adiyaman}, \ref{inst:Adiyaman2}}
  \and S.~Sposetti \inst{\ref{inst:IOTA-ES}}
 \and C.~Stewart \inst{\ref{inst:IOTA}}
 \and W.~Stewart \inst{\ref{inst:IOTA-ES}}
 \and T.~Swift \inst{\ref{inst:IOTA}}
  \and M.~Szkudlarek \inst{\ref{inst:Kepler}}
  \and K.~Szyszka \inst{\ref{inst:Torun}}
  \and N.~Tak{\'a}cs \inst{\ref{inst:Konkoly}}
  \and {\L}.~Tychoniec \inst{\ref{inst:ESO}}
 \and M.~Uno$^\dagger$
  \and S.~Urakawa  \inst{\ref{inst:Bisei}}
  \and K.~Vida \inst{\ref{inst:Konkoly}}
 \and C.~Weber \inst{\ref{inst:IOTA-ES}}
 \and N.~W\"{u}nsche \inst{\ref{inst:IOTA-ES}}
  \and H.~Yamamura  \inst{\ref{inst:JOIN}}
  \and H.~Yoshihara  \inst{\ref{inst:JOIN}}
  \and M.~Zawilski \inst{\ref{inst:IOTA-ES}}
  \and P.~Zelen{\'y} \inst{\ref{inst:ValMez},\ref{inst:IOTA-ES}}
 \and S.~Zo{\l}a\inst{\ref{inst:UJ}}
  \and M.~{\.Z}ejmo \inst{\ref{inst:Kepler}}
  \and K.~{\.Z}ukowski \inst{\ref{inst:UAM}}
}

 \institute{
   Astronomical Observatory Institute, Faculty of Physics and Astronomy, Adam Mickiewicz University,
  S{\l}oneczna 36, 60-286 Pozna{\'n}, Poland. E-mail: antoine.choukroun@amu.edu.pl \label{inst:UAM}
  \and Astronomical Institute, Faculty of Mathematics and Physics, Charles University, V Hole\v{s}ovi\v{c}k{\'a}ch 2,
  180 00 Prague 8, Czech Republic \label{inst:Prague}
  \and Mt. Suhora Observatory, University of the National Education Commission, Podchor\k{a}żych 2, 30-084 Cracow, Poland \label{inst:Suh}
  \and Konkoly Observatory, HUN-REN Research Centre for Astronomy and Earth Sciences, Konkoly Thege Mikl{\'o}s {\'u}t 15-17, H-1121 Budapest, Hungary \label{inst:Konkoly}
  \and CSFK, MTA Centre of Excellence, Budapest, Konkoly Thege Mikl{\'o}s {\'u}t 15-17, H-1121, Hungary \label{inst:Excellence}
  \and MTA CSFK Lend{\"u}let Near-Field Cosmology Research Group, Hungary \label{inst:Lendulet}
  \and ELTE E{\"o}tv{\"o}s Lor{\'a}nd University, Institute of Physics and Astronomy, 1117, P\'azm\'any P\'eter s\'et\'any 1/A, Budapest, Hungary\label{inst:ELTE}
  \and Astronomy Department, E\"otv\"os Lor\'and University, P\'azm\'any P. s. 1/A, H-1171 Budapest, Hungary \label{inst:ELU}
  \and Observat{\'o}rio Nacional, R. Gen. Jos{\'e} Cristino, 77 - S{\~a}o Crist{\'o}v{\~a}o, 20921-400, Rio de Janeiro - RJ, Brazil \label{inst:OASI}
    \and Grupo de Din{\^a}mica Orbital e Planetologia, S{\~a}o Paulo State University, UNESP, Guaratinguet{\'a}, CEP 12516-410, S{\~a}o Paulo, Brazil \label{inst:bz-grp-dyn}
  \and International Occultation Timing Association/European Section e.V. (IOTA/ES), Am Brombeerhag 13, 30459 Hannover, Germany \label{inst:IOTA-ES}
  \and International Occultation Timing Association (IOTA) PO Box 20313 Fountain Hills, AZ 85269-0313 USA \label{inst:IOTA}
  \and Association T60, Observatoire Midi-Pyr{\'e}n{\'e}es, 14, avenue Edouard Belin, 31400 Toulouse, France \label{inst:T60}
  \and Korea Astronomy and Space Science Institute, 776 Daedeok-daero, Yuseong-gu, Daejeon 34055, Korea \label{inst:KASI}
  \and Centro Astron{\'o}nomico Hispano en Andaluc{\'i}a, Observatorio de Calar Alto, Sierra de los Filabres, E-04550 G{\'e}rgal, Spain \label{inst:CalarAlto}
  \and MTA-ELTE Lend{\"u}let "Momentum" Milky Way Research Group, Hungary \label{inst:MTA-ELTe}
  \and Instituto de Astrof{\'i}sica de Andaluc{\'i}a (CSIC), Glorieta de la Astronom{\'i}a s/n, 18008 Granada, Spain \label{inst:Granada}
  \and {\color{highlight} Maastricht University, Maastricht Science Programme, Maastricht, Netherlands} \label{inst:Maastricht}
  \and Observatoire du Bois de Bardon, 16110 Taponnat, France \label{inst:Bardon}
  \and Florida Space Institute, University of Central Florida, Orlando, FL, USA \label{inst:Florida}
  \and Trans-Tasman Occultation Alliance (TTOA), Wellington, PO Box 3181, New Zealand \label{inst:TTOA}
  \and Instituto de Astrof{\'i}sica de Canarias, C/ V{\'i}a Lactea, s/n, E-38205 La Laguna, Tenerife, Spain \label{inst:IAC}
  \and Gran Telescopio Canarias (GRANTECAN), Cuesta de San Jos{\'e} s/n, E-38712, Bre{\~n}a Baja, La Palma, Spain \label{inst:GRANTECAN}
  \and Institute of Astronomy, Faculty of Physics, Astronomy and Informatics, Nicolaus Copernicus University in Toru{\'n}, 
  ul.~Grudzi\k{a}dzka~5, 87-100 Toru{\'n}, Poland \label{inst:Torun}
  \and Open University, School of Physical Sciences, The Open University, MK7 6AA, UK \label{inst:Open}
  \and Hunters Hill Observatory, 7 Mawalan Street, Ngunnawal, ACT 2913, Australia \label{inst:Hill}
  \and Space sciences, Technologies and Astrophysics Research Institute, Universit{\'e} de Li{\`e}ge, All{\'e}e du 6 Ao{\^u}t 17,
  4000 Li{\`e}ge, Belgium \label{inst:Liege}
  \and British Astronomical Association, PO Box 702, Tonbridge TN9 9TX, UK \label{inst:BAA}
  \and Institute of Physics, Jan Kochanowski University, ul. Uniwersytecka 7, 25-406 Kielce \label{inst:Kielce}
  \and Akdeniz University, Department of Space Sciences and Technologies, Antalya, T{\"u}rkiye \label{inst:Akdeniz}
  \and Chungbuk National University, 1, Chungdae-ro, Seowon-gu, Cheongju-si, Chungcheongbuk-do, Republic of Korea \label{inst:Chungbuk}
    \and Czech Astronomical Society, Fri\v{c}ova 298, 251 65, Ond\v{r}ejov, Czechia \label{inst:cz-as}
  \and Observatory in Rokycany and Pilsen, p.o., Voldu\v{s}sk{\'a} 721, 337 01 Rokycany, Czechia \label{inst:cz-obs-roky}
  \and Laboratory of Space Researches, Uzhhorod National University, Daleka st. 2a, 88000, Uzhhorod, Ukraine \label{inst:Uzhhorod}
  \and Observatoire de Nandrin, Soci{\'e}t{\'e} Astronomique de Li{\`e}ge, 17 Avenue des Platanes, 4000 Li{\`e}ge, Belgium \label{inst:belg}
  \and Departamento de Astrof{\'i}sica, Universidad de La Laguna - ULL, E-38205 Tenerife, Spain \label{inst:ULL}
  \and NASA Johnson Space Center Astronomical Society, Houston, TX USA \label{inst:Houston}
  \and Janusz Gil Institute of Astronomy, University of Zielona G{\'o}ra, Szafrana 2, 65-516 Zielona G{\'o}ra, Poland \label{inst:Kepler}
  \and Japan Occultation Information Network (JOIN), Japan \label{inst:JOIN}
  \and B-PHOT, Building F, Pleinlaan 2, 1050 Elsene, Brussels, Belgium \label{inst:Brussels}
  \and Blue Mountains Observatory, 94 Rawson Parade, Leura NSW 2780, Australia \label{inst:Blue}
  \and Institute of Theoretical Physics and Astronomy, Vilnius University, Saul{\.e}tekio al. 3, 10257 Vilnius, Lithuania \label{inst:Vilnius}
  \and Organ Mesa Observatory, 4438 Organ Mesa Loop, Las Cruces, New Mexico 88011 USA \label{inst:Organ}
  \and Command Module Observatory, 121 W. Alameda Dr., Tempe, AZ 85282 USA \label{inst:Tempe}
  \and Silesian University of Technology, Department of Electronics, Electrical Engineering and Microelectronics, Akademicka 16,
  44-100 Gliwice, Poland \label{inst:Gliwice}
  \and Lunar and Planetary Laboratory, University of Arizona, Tucson, AZ 85721, USA \label{inst:Lunar}
  \and Osservatorio Astronomico Via Nastro Verde 50, 80067, Sorrento (Na), Italy  \label{inst:osservatorio}
  \and Departamento de F{\'i}sica, Ingenier{\'i}a de Sistemas y Teor{\'i}a de la Se{\~n}al, Universidad de Alicante, Carr. de San Vicente
  del Raspeig, s/n, 03690 San Vicente del Raspeig, Alicante, Spain \label{inst:Alicante}
  \and Institut de Ci{\`e}ncies del Cosmos (ICCUB), Universitat de Barcelona (IEEC-UB), Carrer de Mart{\'i} i Franqu{\`e}s, 1, 08028, 
  Barcelona, Spain\label{inst:Barcelona}
  \and Lowell Observatory, 1400 West Mars Hill Road, Flagstaff, Arizona, 86001 USA \label{inst:Lowell}
  \and Department of Physics, Adiyaman University, 02040 Adiyaman, T{\"u}rkiye \label{inst:Adiyaman}
  \and Astrophysics Application and Research Center, Adıyaman University, Adiyaman 02040, T{\"u}rkiye \label{inst:Adiyaman2}
  \and European Southern Observatory, Karl-Schwarzschild-Strasse 2, 85748 Garching bei M{\"u}nchen, Germany \label{inst:ESO}
  \and Japan Spaceguard Association, Bisei Spaceguard Center, 1716-3, Okura, Bisei, Ibara, Okayama 714-1411 Japan \label{inst:Bisei}
  \and Observatory, Vset{\'i}nsk{\'a} 78, Vala\v{s}sk{\'e} Mezi\v{r}{\'i}\v{c}{\'i}, Czechia \label{inst:ValMez} 
  \and Astronomical Observatory, Jagiellonian University, ul. Orla 171, 30-244 Cracow, Poland \label{inst:UJ}
}

\date{February 2025}

\begin{document}

\def\nAst{15}

\date{Received 11 March 2025 / Accepted 17 April 2025}

\abstract
{
The sizes of many asteroids, especially slowly rotating, low-amplitude targets, remain poorly constrained due to selection effects. These biases limit the availability of high-quality data, leaving size estimates reliant on spherical shape assumptions. Such approximations introduce significant uncertainties propagating, for example, into density determinations or thermophysical and compositional studies, affecting our understanding of asteroid properties.}
{
 This work targets poorly studied main-belt asteroids, for most of which no shape models were previously available. Using only high-quality, dense light curves, thermal infrared observations (systematically including WISE data), and stellar occultations, we aimed to produce reliable shape models and scale them using two independent techniques, allowing for size comparison at the end. We conducted two observing campaigns to achieve this: one to obtain dense photometric light curves and another to acquire multi-chord stellar occultations by these objects.}
{
 Shape and spin models were reconstructed using light curve inversion techniques. Sizes were determined via two methods: (1) advanced thermophysical modelling using the convex inversion thermophysical model (CITPM), which optimises spin and shape models to light curve data in the visible range together with infrared data, and (2) scaling the shape models with stellar occultations.}
{
 We obtained precise sizes and shape models for \nAst\ asteroids. CITPM and occultation-derived sizes agree within 5\% for most cases, demonstrating the reliability of the modelling approach. Larger discrepancies are usually linked to incomplete occultation chord coverage. The study also provides insights into surface properties, including albedo, surface roughness and thermal inertia.} 
{
The use of high-quality data, coupled with an advanced TPM that uses both thermal and visible data while allowing the shape model to be adjusted according to both types of data, enabled us to determine sizes with precision comparable to those derived from multichord stellar occultations. We resolved substantial inconsistencies in previous size determinations for target asteroids, providing good input for future studies on asteroid densities and surface properties.}


\authorrunning{Choukroun et al.} 
\titlerunning{Asteroid sizes from thermophysical model and stellar occultations}

\maketitle

\section{Introduction}
\label{sec:intro}

Asteroids, large ones in particular, as remnants of the early Solar System, play a crucial role in our understanding of its formation processes \citep{2009Icar..204..558M}. Consequently, determining their physical parameters such as size, density, or thermal inertia is in high demand. Many areas of study are directly dependent on these parameters. For example, asteroid surface properties are linked to how the Sun heats them. This heating is correlated with the orientation of the asteroid spin axis and the rotation period. An asteroid with a very slow rotation or a low pole inclination has part of its surface exposed for long durations (\citealp{delbo2015}; \citealp{capek_thermal_2010}). These differences in exposure times directly influence how the surface evolves. Thermal inertia also plays a crucial role and can provide insights into the past of asteroids. Young asteroid surfaces tend to be composed mainly of coarse regolith grains, which have higher thermal inertia, while older surfaces consist of finer regolith, resulting in lower thermal inertia \citep{delbo_thermal_2009}. {\color{highlight} However, recent studies \citep{Cambioni2021} have shown that some young, rocky surfaces with high porosity may also exhibit low thermal inertia despite having minimal regolith coverage. So the presence or lack of surface regolith inferred from thermal inertia is not always a good indicator of age.}

The size of an asteroid is a parameter that remains difficult to determine without thermophysical modelling,  stellar occultations, radar observations, or adaptive optics. Since all these techniques have their own limitations, for most asteroids this parameter is poorly constrained. In many cases, the size is estimated with an assumed geometric albedo  \citep{harris_revision_1997} leading to errors and inconsistencies on size estimation reaching sometimes 30\% or more (\citealp{tanga_mp3c_2013}, see also Table~\ref{tab:diam_lit}), which can propagate to uncertainties on density estimation of more than 90\%. The reasons for such large errors can vary, but they are often linked to the quality of the data used as well as the methods employed. Without advanced observing techniques like radar or adaptive optics, extracting precise diameter (and other physical parameters) often relies on thermophysical modelling (TPM, see \citealp{delbo2015} for the review). Such modelling techniques require an input shape model, while the majority of asteroids lack such models. Consequently, to extract these parameters, the use of ellipsoidal shape models \citep{ellipsoide} or spherical shape assumption becomes an undesirable necessity, as in the most commonly used technique called near-earth thermal model (NEATM, \citealt{neatm}). 

To obtain a representative shape model of an asteroid, several methods exist. One can consider the relatively recent SAGE (shaping asteroid using genetic evolution) modelling approach by \citet{sage}, for example. However, the most widely used technique is the light curve inversion technique producing convex shape representations (\citealp{kaasalainen_optimization_2001}, \citealp{kaasalainen_optimization_2001-1}). Asteroid models generated from this technique are stored in the DAMIT database (database of asteroid models from inversion techniques, \citealp{damit})\footnote{\url{https://astro.troja.mff.cuni.cz/projects/damit/}}. Although these models are mostly scale-free due to unknown albedo \citep{2020arXiv200509947K}, they are still representative of the general shape of asteroids, making them necessary inputs for thermophysical modelling \citep{delbo2015}. 

However, certain problems remain affecting the accuracy and completeness of available models. It has been shown that there are a few selection biases favouring targets with high light curve amplitudes and fast rotators \citep{M2015}. With the increasing number of surveys, space missions, and ground-based light curve observations, these biases are gradually diminishing. Nevertheless, the available spin and shape models are often based on sparse data \citep{durech2009_sparse_data}, good for spin state determination, but resulting in limited applicability of shape models \citep{Hanu__2013}. Moreover, popular thermophysical modelling techniques (see ATPM from \citealp{ATPM} or TPM from \citealp{lagerros_thermal_1996, lagerros_thermal_1996-1, lagerros_thermal_1997, lagerros_thermal_1998})  use the input shape model as it is, without modification, merely optimising model thermal flux. As a result, shape model errors propagate into the final thermophysical parameters \citep{maclennan2019}. Even when a shape model is available, it is possible to end up with relatively large uncertainties in size as well as in thermal inertia \citep{hanus2015}. The solution to this problem is a simultaneous shape optimisation to fit both visible light curves and thermal data as proposed by the convex inversion thermophysical model (CITPM, \citealp{durech_asteroid_2017}), the method used here.

\vspace{0.2cm}

In parallel to that area of asteroid studies, advances in technology and observing techniques now allow us to observe stellar occultations by asteroids with high precision (\citealp{rommel2020}, \citealp{2022A&A...658A..73F}). With the developments in techniques capable of annotating precise times obtained via GPS to the images, it has become possible to obtain practically direct measurements of asteroid sizes with great accuracy (\citealp{herald2020}), sometimes even sufficient to extract topological information when shape features are large enough compared to measurement uncertainties (see e.g. \citealp{Dias-Oliveira2017}, \citealp{Leiva2017}, or \citealp{rommel2023}).
Although this type of observation is much more feasible than radar or adaptive optics observations, since it can be done with amateur equipment, obtaining multiple chords (segments on asteroid projection defined by star reappearance and disappearance) remains a complex task. Indeed, there are three constraints for this type of observation:

\begin{itemize}
    \item Temporal constraint: although occultations are relatively frequent events, those involving a particular target and a sufficiently bright star, within the reach of most observers' equipment, occur rarely. 
    
    \item Geographical constraint: just like a solar eclipse is not observable everywhere on the Earth's surface, asteroid stellar occultation shadow paths are very narrow, as the shadow path width is comparable to the size of the object.

    \item Human resource constraint: to overcome the previous geographical constraint and to obtain usable observations, multiple observers are needed. Occultations resulting in only one or two positive chords provide limited constraints on size. Multichord (with at least three chords) stellar occultation observations are therefore usually only the result of coordinated observing campaigns.
\end{itemize}

\noindent Due to these constraints, the number of asteroids with available multichord occultations remains relatively limited. However, it is important to note that, thanks to significant collaborative efforts and the participation of amateur astronomers, more and more asteroids are being observed in occultations with such results \citep{herald2020}. 

\begin{table*}
\noindent
\centering
\caption{{\small Target asteroid diameters from the literature.
 }}
 \begin{tabular}{lccccc}
\hline
 Asteroid         & D$_{\text{min}}$    & D$_{\text{min}}$   & D$_{\text{max}}$    & D$_{\text{max}}$ & {\color{highlight}$
\overline{\mathrm{D}}
$} \\
                  &      [km]           &   reference        &      [km]           &   reference      &  {\color{highlight}[km]}     \\
\hline
 (215) Oenone    & 35.21 $\pm$  0.4 & {\color{highlight}[1]} & 43.69 $\pm$  4.370 & {\color{highlight}[2]} &  {\color{highlight} 41.82 $\pm$ 0.10}\\
 (279) Thule    & 113.04 $\pm$  3.11 & {\color{highlight}[3]}   & 136.78 $\pm$  7.105 & {\color{highlight}[4]} & {\color{highlight} 122.62 $\pm$ 2.05} \\
 (357) Ninina    & 92.54 $\pm$  3.015 &  {\color{highlight}[4]}   & 124.11 $\pm$  0.86 &  {\color{highlight}[5]} & {\color{highlight} 113.77 $\pm$ 0.61}\\
 (366) Vincentina    & 83.84 $\pm$  8.38 &  {\color{highlight}[2]}   & 98.25 $\pm$  4.638 &  {\color{highlight}[4]} & {\color{highlight} 86.65 $\pm$ 0.30}\\
  (373) Melusina    & 84.55 $\pm$  8.45 &  {\color{highlight}[2]}   & 107.74 $\pm$  5.815 &  {\color{highlight}[4]} & {\color{highlight} 95.94 $\pm$ 0.61}\\
  (395) Delia    & 44.19 $\pm$  0.45 &  {\color{highlight}[5]}   & 61.49 $\pm$  0.7 &  {\color{highlight}[5]} & {\color{highlight} 50.16 $\pm$ 0.34 }\\
  (429) Lotis    & 54.2 $\pm$  4.56 & {\color{highlight}[6]}    & 89.69 $\pm$  38.25 &  {\color{highlight}[5]} & {\color{highlight} 69.79 $\pm$ 0.68}\\
  (527) Euryanthe     & 48.55 $\pm$  13.62 &  {\color{highlight}[7]}   & 58.56 $\pm$  0.62 &  {\color{highlight}[5]} & {\color{highlight} 53.99 $\pm$ 0.27}\\
  (541) Deborah      & 49.04 $\pm$  17.985 &  {\color{highlight}[8]}    & 65.60 $\pm$  3.801 &  {\color{highlight}[4]} & {\color{highlight} 55.42 $\pm$ 0.31}\\
  (672) Astarte     & 27.49 $\pm$ 2.75 &  {\color{highlight}[2]}  & 35.58 $\pm$ 0.495 & {\color{highlight}[9]} & {\color{highlight} 33.37 $\pm$ 0.33}\\ 
  (814) Tauris      & 98.77 $\pm$  33.9 & {\color{highlight}[6]}    & 122.26 $\pm$  1.819 &  {\color{highlight}[4]} & {\color{highlight} 111.9 $\pm$ 0.97}\\
  (859) Bouzareah       & 65.21 $\pm$  2.758 &  {\color{highlight}[4]}   & 86.02 $\pm$  1 &  {\color{highlight}[5]} & {\color{highlight} 70.28 $\pm$ 0.39}\\
  (907) Rhoda       & 62.73 $\pm$  1.7 &  {\color{highlight}[10]}    & 98.01 $\pm$  32.58 &  {\color{highlight}[11]} & {\color{highlight} 80.91 $\pm$ 0.30}\\
  (931) Whittemora       & 40.62 $\pm$  2.02 &  {\color{highlight}[5]}   & 63.51 $\pm$  11.434 &  {\color{highlight}[4]} & {\color{highlight} 50.1 $\pm$ 0.42}\\
  (1062) Ljuba       & 51.02 $\pm$  0.887 & {\color{highlight}[9]}    & 63.16 $\pm$  12.63 &  {\color{highlight}[2]} & {\color{highlight} 54.91 $\pm$ 0.52}\\
 
\hline
\end{tabular}
    \tablefoot{Data extracted from the MP3C database (\url{https://mp3c.oca.eu/}). D$_{\text{min}}$ and D$_{\text{max}}$ are the minimum and the maximum published value, respectively, {\color{highlight}and $\overline{\mathrm{D}}$ is the error-weighted mean diameter from all the literature values computed as follows: \(
\overline{\mathrm{D}}
 = \sum (\frac{D_i}{\sigma^2_i}) / \sum(\frac{1}{\sigma^2_i})\). The uncertainties were evaluated with: \(\sigma_{
\overline{\mathrm{D}}
} = \sqrt{1 / \sum(\frac{1}{\sigma^2_i})}\) }. References: [1] \cite{Masiero2014}, [2] \cite{alilagoa2018}, [3] \cite{AKARI}, [4] \cite{ryan2010}, [5] \cite{masiero2012}, [6] \cite{nugent2016}, [7] \cite{masiero2020}, [8] \cite{masiero2017}, [9] \cite{masiero2011}, [10] \cite{tedesco2002}, [11] \cite{masiero2021}}
\label{tab:diam_lit}
\end{table*}

In this work, we use a methodology to consecutively combine two modelling techniques (the light curve inversion, followed by the CITPM) to obtain precise shape models and spin parameters, with the main aim of determining the asteroid sizes. In parallel, we also organised occultation observing campaigns to determine target asteroid sizes through an independent method. In total, we modelled and precisely scaled 15 main-belt asteroids. The targets we chose were relatively understudied until now, being slow rotators with small light curve amplitudes \citep{M2015}. This procedure allowed us first to provide precise spin parameters and detailed shapes for the target asteroids, and second, to demonstrate that using only flux measurement techniques, it is possible to obtain highly precise size values (by ‘size’ we mean the diameter of the equivalent volume sphere, see Section \ref{sec:Meth}). In the end, we compare diameters for the same targets obtained with both methods to check their agreement and identify trends and biases. The current paper is the continuation of our work presented in \citet{marciniak_properties_2021} and \citet{M2023}, now combining the two approaches from these works. Section 2 is dedicated to explaining the methodology we employed and detailing the used techniques. Section 3 presents the asteroids we selected and the current state of knowledge existing on them.  Results are presented in Section 4, with comparisons to a broad picture based on data from the literature. Section 5 contains conclusions and prospects for future work.

\section{Methodology}
\label{sec:Meth} 

Modern methods of thermophysical modelling take an asteroid shape model as an input and optimise physical parameters to best fit thermal data. The results are therefore very dependent on input shape models, and it is not uncommon to obtain the results with a high value of reduced $\chi^2_{\mathrm{IR}}$ indicating an imperfect fit between the physical model and the data (\citealp{hanus2015}, \citealp{Hung2022}). In this paper, we use a methodology to obtain precise physical parameters of asteroids (such as spin axis orientation, 3D shape approximation, size, albedo, and thermal inertia) based on high-quality data in both the visible and infrared ranges. We applied this methodology to characterise around a dozen main-belt asteroids.

\vspace{0.2cm}

- First, we apply the convex inversion method which allows us to construct three-dimensional shape models from photometric light curves, as long as data from five independent apparitions were gathered in the form of dense light curves. The resulting asteroid models describe the shape and rotational state of our targets. Since the studied asteroids are located in the main belt, this technique often results in two possible solutions for the spin axis orientation due to the mirror pole ambiguity \citep{Kaasalainen_2006}. However, these shape models are scale-free, preventing the direct determination of object physical dimensions. Complementary techniques are therefore required to precisely derive asteroid sizes. 

\vspace{0.2cm}

\begin{figure}
    \centering
    \includegraphics[width=1\linewidth]{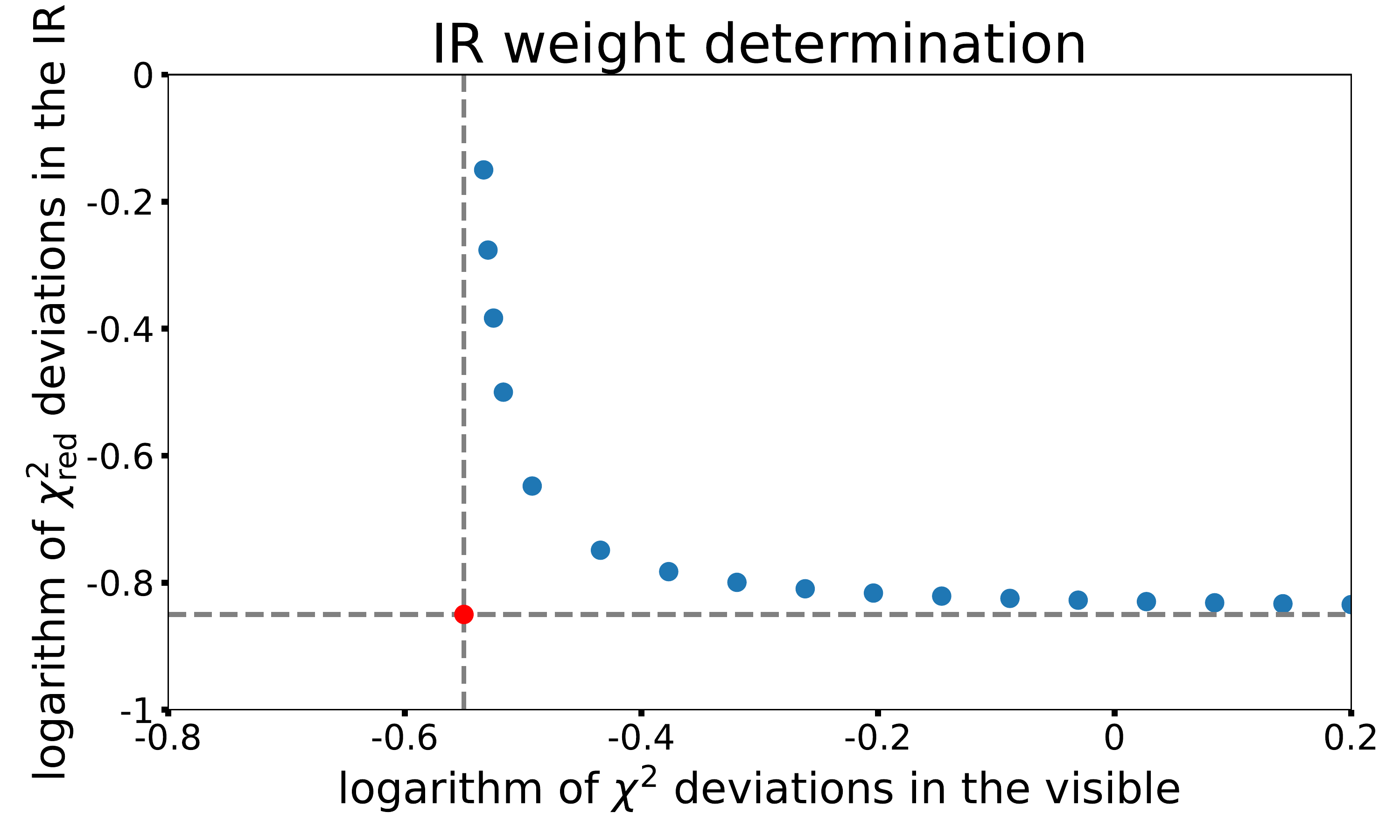}
    \caption{\color{highlight}{Ideal} evolution of $\chi^2_{\mathrm{vis}}$ and reduced $\chi^2_{\mathrm{IR}}$ as a function of $\omega_{\mathrm{IR}}$. Each blue point corresponds to a different IR weight. {\color{highlight}This plot is constrained by two asymptotes along the x axis and the y axis. These two asymptotes are used to determine a new origin (red point). The best $\omega_{\mathrm{IR}}$ value corresponds to the point closest to the new origin.}}
    \label{fig:ir_weight_determination}
\end{figure}

- The second step involves utilising these spin solutions as a necessary input to the CITPM. This method combines convex inversion with the thermophysical model (‘model’ refers here the modelling method) in the approach of Lagerros, following the equation:

\begin{equation}
    \chi^2 = \chi^2_{\mathrm{vis}} + \omega_{\mathrm{IR}} \times \chi^2_{\mathrm{IR}}, 
\end{equation}

\noindent where $\omega_{\mathrm{IR}}$ is a parameter dedicated to weighing infrared data \citep{Kaasalainen2011}, $\chi^2_{\mathrm{vis}}$ evaluates how well the model from the CITPM method (‘model’ here and throughout the rest of this text meaning the shape representation with accompanying set of physical parameters) fits the visible data, and $\chi^2_{\mathrm{IR}}$ evaluates the model fit to the thermal data. An important step to ensure good results is to well estimate $\omega_{\mathrm{IR}}$. A weight value that is too small will result in models that overfit visible data, yielding a low $\chi^2_{\mathrm{vis}}$ but a high $\chi^2_{\mathrm{IR}}$. On the other hand, a weight value that is too high will cause models to overfit thermal data. {\color{highlight}Determining analytically the best $\omega_{\mathrm{IR}}$ remains challenging due to the highly non-linear nature of the CITPM. We investigated the influence of $\omega_{IR}$ on the diameter and found it to be non-trivial. Models react differently, with the diameter varying by a few percent when starting near a good solution—and more so when starting from a poor one. The main effect of $\omega_{IR}$ is on convergence. Moreover focusing solely on diameter changes does not fully assess model quality. Since CITPM also refines the shape, $\omega_{IR}$ influences also shape and spin axis orientation, of the order of several percent too. However, a good $\omega_{\mathrm{IR}}$ can be obtained by following the procedure described by \citet{Kaasalainen2012} and shown in Figure \ref{fig:ir_weight_determination}. The idea is to run the CITPM multiple times with different values of $\omega_{\mathrm{IR}} \in [0.001, 0.05]$ (blue points) and to determine a vertical and a horizontal asymptote defined by the logarithm of the minimum $\chi^2_{\mathrm{vis}}$ and the logarithm  of the minimum $\chi^2_{\mathrm{IR}}$ obtained. The intersection between these two asymptotes defines a new origin. The best $\omega_{\mathrm{IR}}$ value is the one which minimises the distance between this new origin and the points $P_{\omega_{\mathrm{IR}}}(\log(\chi^2_{\mathrm{vis}}), \log(\chi^2_{\mathrm{IR}}))$.} However, plotting  \textcolor{highlight}{the logarithm of} $\chi^2_{\mathrm{IR}}$ vs \textcolor{highlight}{the logarithm of }$\chi^2_{\mathrm{vis}}$ does not always produce results as clean as the one shown in Figure \ref{fig:ir_weight_determination}. Some noise or instabilities may be present. For this reason, the obtained $\omega_{\mathrm{IR}}$ is not perfect but remains {\color{highlight}close. We also examined how the amount of visible and infrared data affect $\omega_{IR}$, but found no clear trend. Data quality remains the primary source of $\omega_{IR}$ variability, assuming their sufficient amount and diversity.} This combined approach adjusts the shape model to both types of data, providing significant advantages. Traditional TPM methods treat the input shape model as absolute truth, which can pose problems when the shape model is of poor resolution or, in extreme cases, when no shape model exists and a simple spherical model is used as input. However, the CITPM allows us to improve the spin and shape models, scale them in size, and determine other thermophysical parameters such as thermal inertia, hemispherical albedo — derived using Hapke parameters, which differs from the geometric albedo that assumes a spherical shape for asteroids — and an estimate of surface roughness. A few studies using the TPM also took into account the uncertainties introduced by the input shape (see e.g. \citealp{hanus2015}), but they did not optimise the shape model to both data types at the same time. The CITPM optimise Hapke parameters (\citealp{hapke1981, hapke1984, hapke1986}) to derive the hemispherical albedo of the model, providing a more accurate approach, and simultaneously optimises parameters for the facets (surface elements), allowing a precise determination of global model parameters like the asteroid size, thermal inertia $\Gamma$ and the surface roughness.

\vspace{0.2cm}

- Finally, the third step involves using stellar occultations to scale the initial shape models obtained with convex inversion, independently from the second step. To perform this, the occultation chords are projected onto the fundamental plane (plane perpendicular to the star-asteroid direction and goes through the Earth) following the procedure described in \citet{DURECH2011}, using the equations:

\begin{equation}
    \begin{aligned}
        \hat{\vec{s}}_\xi &= (-\sin\delta\cos\alpha,\ -\sin\delta\sin\alpha,\  \cos\delta), \\
        \hat{\vec{s}}_\eta &= (\sin\alpha,\  -\cos\alpha, \ 0),
    \end{aligned}
\end{equation}
\noindent and
\begin{equation}
    \begin{pmatrix}
        \xi \\
        \eta
    \end{pmatrix} 
    =  
    \begin{bmatrix}
         \hat{\vec{s}}_\xi \cdot \left( \vec{x} + \Delta \vec{v} \Delta t + \frac{1}{2} \Delta \dot{\vec{v}} (\Delta t)^2 \right)
        \\
        \hat{\vec{s}}_\eta \cdot \left( \vec{x} + \Delta \vec{v} \Delta t + \frac{1}{2} \Delta \dot{\vec{v}} (\Delta t)^2 \right)
     \end{bmatrix},
\end{equation}

\noindent where $\hat{\vec{s}}_\xi$ and $\hat{\vec{s}}_\eta$ correspond to unit vectors on the fundamental plane, and $\alpha$ and $\delta$ are the right ascension and declination of the occulted star. Next, $\xi$ and $\eta$ correspond to the geocentric co-ordinates of an observer's position projected on the fundamental plane, $\vec{x}$ is the observer's position on the Earth in the sidereal equatorial frame, $\Delta \vec{v}$ is the relative velocity of the Earth and the asteroid, and $\Delta \vec{t}$ corresponds to the time between the star disappearance or reappearance and an arbitrary epoch. The sidereal equatorial frame is used here for two main reasons. The first is that the geographic co-ordinate system (longitude, latitude) does not account for the Earth's orientation at the moment of observation. The second is connected to the sidereal equatorial system definition: the z axis points towards the celestial pole, the x axis points towards the vernal equinox, and the y axis is chosen such that it forms a right-handed orthonormal reference frame. This co-ordinate system is expressed on the same basis as the equatorial system, used for the co-ordinates of the occulted star. This significantly simplifies calculations, as transitioning between the two systems does not require a change of basis. The corresponding silhouettes of the shape model (for both pole solutions) constructed during the first step are also projected onto the fundamental plane. The centroid position and the size of the asteroid silhouette are then optimised using both positive and negative chords. At this stage, shape models are scaled based on occultation observations, and the final size corresponds to the diameter of a sphere with the same volume as the scaled shape model. It is often possible to verify the obtained dimensions and resolve ambiguities in pole solutions (see e.g. \citealt{M2023}). 

After this stage we compare the results for diameters from the second and third steps, providing cross-validation between the two approaches. 
The diameters from thermophysical modelling are defined for a sphere of the same surface area as our shape model, so we recalculate them for a sphere of the same volume as our model, and then thermally derived diameters can be directly compared to diameters from occultation fitting. If the sizes obtained by thermophysical modelling and by occultation fitting are in agreement, we could reasonably assume that for any target modelled following CITPM methodology, obtained size values are accurate without requiring a double-check through occultations.

\section{Targets and data}

The targets we selected are main-belt asteroids that have been poorly studied before. The reason for this was the fact that these targets are challenging to observe. They are all slow rotators (P > 12h) with low maximum light curve amplitude (below 0.3 mag). Available photometric data on these targets were mostly sparse data from sky surveys, and not many dense light curves that we needed were available. Even in cases where the Nyquist-Shannon theorem was satisfied \citep{Shannon1949}, there were often too few data points to cover a complete light curve. Furthermore, the photometric quality of sparse data from ground-based surveys is not always sufficient, with photometric scatter being of an order of 0.1 mag at best \citep{hanus2011}. On the other hand, the low-scatter data from the Gaia mission is not abundant enough to enable unique spin and shape models of our targets.

Also, to obtain the size values as precisely as possible, our input data must be abundant and of high quality. To achieve these objectives, we used only dense light curves with low photometric scatter from a large photometric campaign described in \citet{M2015}. Thanks to the wide collaboration and a long-time campaign, we managed to cover enough apparitions for all our targets to obtain unique spin and shape models using the convex inversion method. Also, for each light curve, we carefully eliminated any problematic data points (e.g., star passages, outlying points, too noisy fragments, etc.). 

Regarding infrared data, it is important to note that the transition between reflected light and thermal emission is continuous, and the transition phase where the flux is a mix of both, for main-belt asteroids, is usually located around 3 - 5 $\mu$m \citep{harrisetlagerros}. Therefore, our thermal dataset does not include any points observed at wavelengths shorter than 9 $\mu$m. Moreover, intending to obtain the most accurate parameters possible, we focused on targets that were observed by WISE spacecraft in W3 (11.1 $\mu$m) and W4 - (22.64 $\mu$m) bands (\citealp{wise}, \citealp{neowise}), taking care of rejecting all of the partially saturated points. We also included thermal data from IRAS \citep{IRAS} and AKARI missions \citep{AKARI}, provided that WISE data were available. All thermal data have been downloaded from the infrared database ({\color{highlight}IRDB}, \citealp{Szakats})\footnote{\url{https://ird.konkoly.hu/}}. {\color{highlight} According to IRDB documentation, for most instruments and filters, the colour correction was derived from the relative response profiles of the specific filters while assuming an estimated effective temperature for the target, calculated as described in \citet{lang1999}.} The number of visible light curves and IR data points is presented in Table~\ref{tab:data_summary}. Since the determination of Hapke model parameters in CITPM requires calibrated data in the visible range, we added the V-band data from USNO (US naval observatory) archive\footnote{\url{https://newton.spacedys.com/astdys2/index.php?pc=3.0}} in this step. 

For occultation observations, in October 2020 we initiated another observing campaign as outlined in \citet{M2023}. Information regarding both the photometric and occultation observing campaigns can be found \href{https://zenodo.org/records/15274451}{here}. The ultimate goal was to achieve sufficiently well-covered occultation events to accurately determine the sizes of our targets using an independent approach. For earlier events, we utilised archival data from the asteroid occultations v4.0 archive of the NASA planetary data system \citep{Herald2024}, and from the software ‘Occult v4.2024’\footnote{\url{http://lunar-occultations.com/iota/occult4.htm}}. 

\begin{table}[h]
    \centering
    \caption{Data summary.}
    \small 
    \setlength{\tabcolsep}{4pt}
    \begin{tabular}{c|c|c|c|c|c|c}
        Target  & Taxon. & N\textsubscript{lc} & N\textsubscript{app} &  N\textsubscript{IR} & N\textsubscript{WISE} & N\textsubscript{occ} (N\textsubscript{pos}) \\
        \hline
        (215) Oenone     & S & $78$  & $8$  & $53$ & $22$ &  $1 (2)$\\  
        (279) Thule      & X & $107$ & $13$ & $33$ & $15$ & $3 (5, 2, 4)$\\
        (357) Ninina     & CX & $89$  & $7$  & $38$ & $10$ & $4 (7, 2, 4, 7)$ \\
        (366) Vincentina & Ch & $69$  & $9$  & $67$ & $21$ & $3 (6, 6, 2)$\\
        (373) Melusina   & C & $31$  & $6$  & $40$ & $11$ & $1 (2)$ \\
        (395) Delia      & Ch & $68$  & $7$  & $20$ & $11$ & $3 (3, 2, 2)$\\
        (429) Lotis      & C & $79$  & $8$  & $41$ & $11$ & $1 (2)$ \\
        (527) Euryanthe  & Cb & $90$  & $6$  & $35$ & $19$ & $2 (6, 5)$ \\
        (541) Deborah    & B & $44$  & $6$  & $48$ & $30$ & ${\color{highlight}3} (2,4,{\color{highlight}6})$ \\
        (672) Astarte    & S & $62$  & $8$  & $24$ & $16$ & $1 (4)$ \\
        (814) Tauris     & C & $101$ & $11$ & $19$ & $8$  & $1 (5)$ \\
        (859) Bouzareah  & X & $49$  & $7$  & $39$ & $12$ & $3 (2, 3, 4)$ \\
        (907) Rhoda      & Xk & $92$  & $9$  & $46$ & $22$ & $1 (2)$ \\
        (931) Whittemora & M & $60$  & $9$  & $49$ & $24$ & $3 (3, 2, 2)$ \\
        (1062) Ljuba     & C & $132$ & $9$  & $27$ & $14$ & $2 (2,2)$   \\
    \hline
    \end{tabular}
    \tablefoot{‘Taxon.’ refers to the taxonomic type. N\textsubscript{lc} is the number of dense light curves for each target registered at N\textsubscript{app} apparitions, N\textsubscript{IR} is the number of all infrared data points, and N\textsubscript{WISE} is the number of data points obtained by WISE spacecraft. N\textsubscript{occ} represents the number of stellar occultation events, with the number of positive chords for each event shown in parentheses. Taxonomic types come from \cite{tholen1984}, \cite{bus2002} and from \cite{Vere2015}.}
    
    \label{tab:data_summary}
\end{table}

\vspace{0.2cm}

Since our research focused on poorly studied targets, for the majority of them, neither spin solutions nor shape models existed. Only for five of them, spin and shape models derived from sparse data were available in the DAMIT database. In addition, for (279) Thule, a spin solution was listed in the LCDB\footnote{light curve database \citep{LCDB}, \url{https://minplanobs.org/mpinfo/php/lcdb.php}} \citep{2014Pilcher}, although no shape model was provided. The spin axis solutions for (279) Thule were determined there using an approximative method based on the amplitudes of light curves. Information on all prior spin/shape models is gathered in Table~\ref{tab:prior_spin_shape}.

\begin{table}[h]
    \centering
    \caption{Prior spin/shape solutions available for our targets.}
    \resizebox{0.5\textwidth}{!}{
    \begin{tabular}{c|c|c|c|c}
         Target  & $\lambda$ & $\beta$ & Period & Reference \\
                   &   [°]     &   [°]   &  [hours]       \\
         \hline
         
         (215) Oenone & $227 \pm 4$ & $66 \pm 5$ & $27.9081$ & [1] \\

         (279) Thule & $70 \pm 8$ & $0$ & $15.94$ & [2]\\ 
          & $250 \pm 8$ & $0$ & $15.94$ & [2] \\

          (357) Ninina & $49 \pm 10$ & $0 \pm 10$ & $35.984$ & [3]\\ 
                        & $230 \pm 10$ & $36 \pm 10$ & $35.984$ & [3] \\

          (373) Melusina & $19 \pm 5$ & $ -48 \pm 5$ & $12.98629$ & [4]\\
           & $147 \pm 5$ & $-61 \pm 5$ & $12.98632$ & [4]\\

          (395) Delia & $15 \pm 3$ & $-50 \pm 4$ & $19.6802$ & [1] \\
           & $196 \pm 4$ & $-53 \pm 4$ & $19.6803$ & [1]\\

           (672) Astarte & - & $59 \pm 18$ & $22.5799$ & [1] \\

           \hline

    \end{tabular}
    }
    \label{tab:prior_spin_shape}
    \tablefoot{Ecliptic J2000 co-ordinates $\lambda$ and $\beta$ correspond to the spin axis orientation in space. References: [1] \cite{durech2020}, [2] \cite{2014Pilcher}, [3] \cite{franco2024}, [4] \cite{durech2018} }
\end{table}

\section{Results}

\subsection{Finding the best model in CITPM}

As described in Section 2, the first step involves obtaining the shape models and spin solutions of each of our targets. This step is crucial for accurately determining the solutions for the spin axis and the sideral rotation period, where a sub-second precision is required to achieve reliable results with the CITPM. Also, a good shape model is important to determine the size via stellar occultation fitting (see Sections \ref{sec:occ_fitting}, \ref{sec:detailed_targets}, and Table~\ref{tab:occ_diam}). 

To constrain the thermophysical parameters of the models derived in the first step, it is essential to determine the weight of the infrared data in the first place. Following the procedure described in Section 2 and illustrated in Figure 1, our approach involves running CITPM multiple times for one of the two spin solutions while certain parameters remain unchanged, such as thermal inertia and surface roughness. The only parameter that varies between each run is the infrared data weight. These successive runs allow us to plot the curve like the one shown in Figure 1, thereby identifying the weight that provides the best balance between thermal and visible data.

\begin{figure}
    \centering
    \includegraphics[width=1\linewidth]{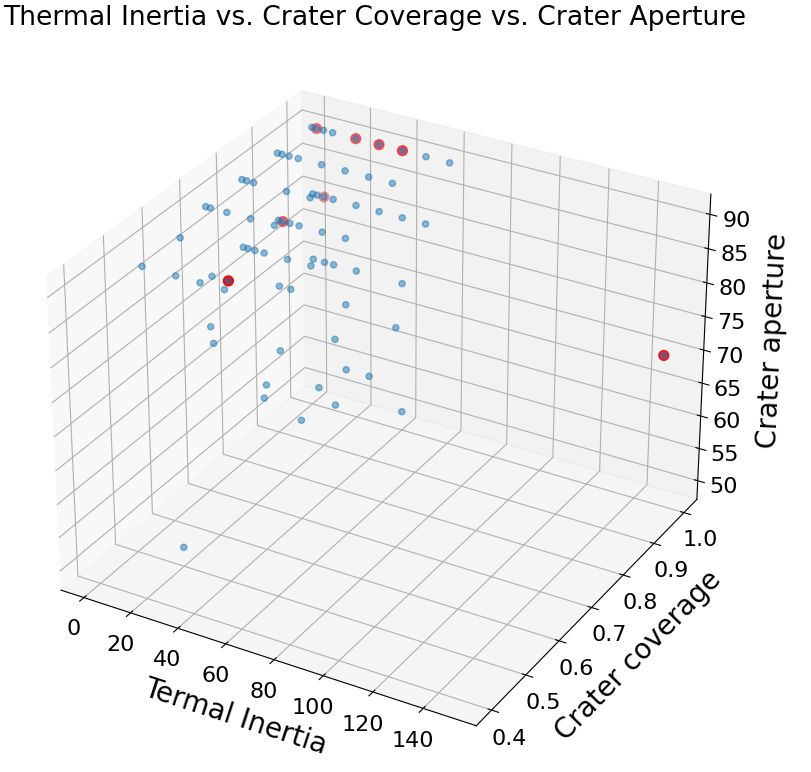}
    \caption{Crater coverage $\rho_c$ against crater aperture $\gamma_c$ against thermal inertia $\Gamma$ in $\text{J}\,\text{m}^{-2}\,\text{s}^{-\frac{1}{2}}\,\text{K}^{-1}$ of the best solutions for asteroid (814) Tauris. All blue points correspond to acceptable solutions according to equations \ref{eq:treshold_condition}. The red points represent the 10 best solutions among all acceptable ones. All points are grouped, except for the red point with $\Gamma > 140$, which is isolated. This point was not used to determine final parameter uncertainties. }
    \label{fig:ti_ca_cc}
\end{figure}

The next step involves running the code multiple times (in a loop as an iterative way) for a given thermal inertia value while varying the surface roughness parameters: the crater aperture that is changed from 10° to 90° with a step of 10°, and crater coverage, changed from 0.1 to 1 with a step of 0.1. These parameters correspond to the hemispherical crater opening angle and the extent to which the surface of the target is covered by craters, and they are not optimised during the run and remain fixed. This process is repeated for a range of thermal inertia values. Although this step is computationally intensive, it is critical because it ensures a good representation of the model within the parameter space. A solution is considered acceptable if:  

\begin{equation}
\begin{cases}
    \chi^2_{\text{vis}} < \min(\chi^2_{\text{vis}}) \cdot 1.1 \\
    \chi^2_{\text{IR}} < 1 + \sqrt{\frac{2}{\nu}},
\end{cases}%
\label{eq:treshold_condition}
\end{equation}%
where $\nu$ is the number of thermal measurements minus the number of degrees of freedom. The number of degrees of freedom is set to four (size, thermal inertia and two parameters for surface roughness). The parameters of the best of all acceptable solutions from the loop over surface roughness are used as input for a final CITPM run. The other acceptable solutions are used to determine uncertainty ranges. To not overestimate uncertainties, we also produce a 3D plot of crater coverage vs crater apertures vs thermal inertia of all acceptable solutions. If an isolated point is found out of the general cluster, this solution is not taken into account to calculate uncertainties (see Figure \ref{fig:ti_ca_cc}). 

All the results obtained with this procedure are presented in Table~\ref{tab:citpm}, where each target entry is split into two lines corresponding to the solutions for each pole. Columns 2 to 5 describe the spin axis parameters, as these are also optimised by the CITPM. The last four columns provide the diameter of the sphere with equivalent volume, {\color{highlight}geometric} albedo derived from Hapke parameters {\color{highlight}following equation 65 of \citet{hapke1984}}, thermal inertia in SI units ($\text{J}\,\text{m}^{-2}\,\text{s}^{-\frac{1}{2}}\,\text{K}^{-1}$), and the mean surface slope according to \citet{hapke1984}, \citet{lagerros_thermal_1996}. To achieve these results, we automatised a large part of the procedure. As shown in Table~\ref{tab:citpm}, some parameters are well constrained, such as spin parameters, diameter, and albedo. However, surface roughness and thermal inertia are not always well constrained because of the entanglement existing between them (see Figures \ref{fig:ti_215} to \ref{fig:ti_1062}). Although, for some of our targets, everything proceeded smoothly without setbacks, for others, direct decisions had to be made to ensure that the obtained results were robust. The reasons for this can be multiple and varied: insufficient data, insufficient diversity in the observation geometry, convex assumption of the model, inadequate spectral diversity in the thermal data, etc. The CITPM remains a multivariable optimisation model that is highly sensitive to the input data. The number of local minima can, therefore be large.

Model fits to data in the visible range are shown in Figures \ref{215lc_fit} - \ref{1062lc_fit}. We present only a small part of the whole light curve dataset here, and only for targets where data from Kepler or TESS spacecraft were available. The fit to all of the light curve data can be found in the DAMIT database. Only the uninterrupted observations from space show substantial coverage of rotation phases for our slow rotators. Model fits to thermal data, on the other hand, are shown in Figures \ref{215irW3} to \ref{1062irW4}. These plots present fits to data from WISE spacecraft because only these observations were rich enough to reveal thermal light curves. Both kinds of fit are presented for the pole solution preferred by occultations, or pole 1 in case of no preference.

With thermophysical modelling, the most common issue we encountered was the difficulty in determining a suitable value for \(\omega_{\mathrm{IR}}\). Sometimes, the procedure described in Section~\ref{sec:Meth} did not yield a plot as clear as the one shown in Figure \ref{fig:ir_weight_determination}; instead, some plots of both $\chi^2$ values tended to be chaotic, showing no discernible trend. In such cases, our approach was to produce two additional plots. As previously explained, these plots were generated by running the CITPM code multiple times, with the only difference between runs being the \(\omega_{\mathrm{IR}}\) weight value. This means that other parameters remained the same for each run of a given target. We also used the same parameters across different targets (except for \(\lambda\), \(\beta\), and the period, which are specific to each target), to facilitate direct comparability of the results. Originally, we used crater opening angle \(\gamma_c\) of 50° and crater coverage \(\rho_c\) of 0.5, as the parameters of the surface roughness. However, due to the difficulties we faced, we modified the procedure to produce three plots for different surface crater combinations: \(\gamma_c = 20, \rho_c = 0.2\); \(\gamma_c = 50, \rho_c = 0.5\); and \(\gamma_c = 80, \rho_c = 0.8\). This updated method had two advantages. First, comparing each of these plots helped us identify a \(\omega_{\mathrm{IR}}\) weight value that better reflected the surface roughness. Second, in situations where the plots appeared chaotic, it was often the case that at least one of them contained a less chaotic region that allowed us to estimate \(\omega_{\mathrm{IR}}\). Although we could not be certain this value was optimal, it provided a sufficiently reliable approximation for \(\omega_{\mathrm{IR}}\). The targets affected by this issue were: (373)~Melusina, (814)~Tauris, (859)~Bouzareah, and (1062)~Ljuba.

\vspace{0.2cm}
\subsection{Size determination via occultation fitting}
\label{sec:occ_fitting}

CITPM is one way to determine asteroid sizes. In parallel to this, we also determined their sizes using stellar occultations. Table \ref{tab:occ_diam} presents the results of scaling the models from light curve inversion with the chords from stellar occultations. The asteroid models we used here were based on dense light curves in the visible 
range only (first step, see Section \ref{sec:Meth}). As such, their shapes are less smooth (we found that adding thermal data tends to smooth the shape models out), 
however, they had poorly constrained stretch along the spin axis in some cases.
The ‘D RMS’ in Table~\ref{tab:occ_diam} are formal diameter uncertainties from the occultation fitting.
On the other hand, the ‘+/-’ diameter uncertainties in the

\onecolumn

\hfill
\renewcommand{\arraystretch}{1.5}
\begin{longtable}{|c|c|c|c|c|c|c|c|c|c}
    \caption{Results obtained through modelling with the Convex Inversion Thermophysical Model. } \\ 
    
    \hline
    Target name & Pole & $\lambda$ & $\beta$ & Period & D & {\color{highlight}p\textsubscript{V}}& $\Gamma$ & $\bar{\theta}$ \\ 
                &       &   [°]    &   [°]   &  [h]   & [km] &     &   [$\text{J}\,\text{m}^{-2}\,\text{s}^{-\frac{1}{2}}\,\text{K}^{-1}$]   &   [°]         \\
    \endfirsthead
    \caption*{(Next)} \\ 
    \hline
    \endhead
    \hline \multicolumn{3}{r}{\textit{(Continued on next page...)}} \\
    \hline
    \endfoot
    \hline
    \endlastfoot
    \label{tab:citpm}
    
    (215) Oenone & 1 & $45_{-7}^{+13}$ & $85_{-5}^{+5}$ & $27.90773\pm 0.00005$ & $37_{-2}^{+2}$ & $0.227_{-0.038}^{+0.006}$ & $63_{-62}^{+86}$ & $68$ \\
     & 2 & $233_{-6}^{+14}$ & $85_{-9}^{+5}$ & $27.90774 \pm 0.00006$ & $37_{-2}^{+2}$ &  
     $0.22_{-0.03}^{+0.01}$ & $62_{-61}^{+88}$ & $42$\\
    
    \hline
    
    (279) Thule & 1 & $58_{-4}^{+24}$ & $2_{-7}^{+9}$ & $23.8964 \pm 0.0003$ & $116_{-6}^{+9}$ & $0.039_{-0.005}^{+0.005}$ & $76_{-75}^{+44}$ & $68$\\
     & 2 & $236_{-1}^{+28}$ & $3_{-8}^{+5}$ & $23.8963 \pm 0.0004$ & $117_{-8}^{+10}$ & $0.038_{-0.005}^{+0.007}$ & $69_{-68}^{+81}$ & $68$ \\

    \hline
    
    (357) Ninina &  \textbf{1} & ${\bf41_{-16}^{+2}}$ & ${\bf 7_{-4}^{+25}}$ & ${\bf 35.9845 \pm 0.0005}$ & ${\bf 104_{-8}^{+9}}$ & ${\bf 0.08_{-0.02}^{+0.03}}$ & ${\bf 40_{-39}^{+208}}$ & ${\bf 59}$ \\
     & 2 & $224_{-21}^{+7}$ & $42_{-6}^{+13}$ & $35.9845 \pm 0.0004$ & $101_{-2}^{+12}$ & $0.08_{-0.03}^{+0.01}$ & $6_{-5}^{+194}$ & $34$ \\

     \hline
     
    (366) Vincentina & 1 & $51_{-1}^{+2}$ & $19_{-1}^{+9}$ & $17.3484 \pm 0.0002$ & $87_{-4}^{+5}$ & $0.081_{-0.009}^{+0.037}$ & $48_{-47}^{+151}$ & $25$ \\
     & 2 & $234_{-1}^{+2}$ & $-0_{-2}^{+5}$ & $17.3484 \pm 0.0003$ & $87_{-5}^{+5}$ & $0.080_{-0.009}^{+0.043}$ & $41_{-40}^{+109}$ & $24$ \\

     \hline

    (373) Melusina & 1 & $20_{-1}^{+24}$ & $-68_{-2}^{+1}$ & $12.986274 \pm 0.00003$ & $99_{-4}^{+13}$ & $0.040_{-0.007}^{+0.006}$  & $13_{-12}^{+87}$ & $19$  \\
     & \textbf{2} & ${\bf 132_{-1}^{+14}}$ & ${\bf -54_{-4}^{+4}}$ & ${\bf 12.986283 \pm0.00002}$ & ${\bf 101_{-10}^{+1}}$ & ${\bf 0.038_{-0.002}^{+0.009}}$ & ${\bf 5_{-4}^{+95}}$ & ${\bf 13}$  \\

     \hline

    (395) Delia & 1 & $5_{-2}^{+24}$ & $-49_{-15}^{+5}$ & $19.68097 \pm 0.00005$ & $52_{-4}^{+7}$ & $0.039_{-0.007}^{+0.007}$ & $26_{-25}^{+94}$ & $45$ \\
     & 2 & $207_{-23}^{+14}$ & $-56_{-20}^{+5}$ & $19.68103 \pm 0.00008$ & $53_{-5}^{+11}$ & $0.038_{-0.01}^{+0.008}$ & $39_{-38}^{+61}$ & $59$ \\

     \hline

    (429) Lotis & 1 & $133_{-6}^{+3}$ & $-16_{-10}^{+5}$ & $13.58081 \pm 0.00002$ & $67_{-4}^{+1}$ & $0.036_{-0.001}^{+0.06}$ & $183_{-182}^{+317}$ & $21$ \\
     & 2 & $312_{-4}^{+4}$ & $5_{-6}^{+5}$ & $13.58081 \pm 0.00003$ & $68_{-7}^{+1}$ & $0.036_{-0.001}^{+0.059}$ & $128_{-127}^{+372}$ & $21$ \\

     \hline

    (527) Euryanthe & 1 & $147_{-6}^{+7}$ & $-59_{-4}^{+11}$ & $42.8379 \pm 0.0004$ & $53_{-2}^{+1}$ & $0.039_{-0.003}^{+0.005}$ & $31_{-30}^{+68}$ & $3$ \\
     & \textbf{2} & ${\bf 284_{-6}^{+9}}$ & ${\bf -58_{-6}^{+9}}$ & ${\bf 42.8378 \pm 0.0002}$ & ${\bf 51_{-1}^{+3}}$ & ${\bf 0.039_{-0.003}^{+0.005}}$ & ${\bf 31_{-30}^{+68}}$ & ${\bf 28}$ \\

    \hline

    (541) Deborah & 1 & $155_{-18}^{+11}$ & $63_{-20}^{+9}$ & $29.4223\pm0.0001$ & $57_{-4}^{+1}$ & $0.0488_{-0.009}^{+0.0001}$ & $2_{-1}^{+60}$ & $38$ \\
             &\textbf{2} & ${\bf 310_{-2}^{+19}}$ &${\bf 63_{-5}^{+16}}$ & ${\bf 29.422425\pm0.0002}$ & ${\bf 54_{-1}^{+4}}$ & ${\bf 0.043_{-0.002}^{+0.003}}$ &${\bf 2_{-1}^{+28}}$ & ${\bf 17}$ \\

    \hline

    (672) Astarte & 1 & $184_{-6}^{+6}$ & $65_{-26}^{+1}$ & $22.57993\pm0.00004$ & $30_{-2}^{+1}$ & $0.047_{-0.004}^{0.01}$ & $4_{-3}^{+446}$ & $55$\\
                  & 2 & $307_{-8}^{+9}$ & $65_{-11}^{+3}$ & $22.57990\pm0.00005$ & $31_{-1}^{+3}$ & $0.048_{-0.008}^{+0.003}$ & $37_{-36}^{+263}$ & $ 59 $ \\

    \hline
     
    (814) Tauris & 1 & $169_{-2}^{+2}$ & $-43_{-7}^{+10}$ & $36.0581\pm0.0001$ & $101_{-5}^{+4}$ & $0.043_{-0.003}^{+0.011}$ & $4_{-3}^{+75}$ & $38$ \\
     & 2 & $298_{-11}^{+26}$ & $-84_{-1}^{+6}$ & $36.05858 \pm 0.0002$ & $110_{-11}^{+4}$ & $0.040_{-0.004}^{+0.015}$ & $2_{-1}^{+148}$ & $10$ \\

     \hline

    (859) Bouzareah & 1 & $29_{-12}^{+24}$ & $-45_{-17}^{+20}$ & $24.9287\pm0.0003$ & $68_{-4}^{+9}$ & $0.041_{-0.006}^{+0.009}$ & $6_{-5}^{+198}$ & $38$ \\
     & 2 & $200_{-21}^{+24}$ & $-32_{-30}^{+4}$ & $24.9286\pm0.0002$ & $67_{-3}^{+13}$ & $0.041_{-0.006}^{+0.017}$ & $6_{-5}^{+198}$ & $32$ \\

     \hline

    (907) Rhoda & 1 & $94_{-11}^{+1}$ & $30_{-4}^{+14}$ & $22.4511\pm0.0001$ & $71_{-3}^{+5}$ & $0.045_{-0.016}^{+0.009}$ & $8_{-7}^{+242}$ & $16$ \\
    & 2 & $279_{-2}^{+2}$ & $-7_{-2}^{+14}$ & $22.45117\pm0.00007$ & $72_{-3}^{+5}$ & $0.038_{-0.008}^{+0.012}$ & $4_{-3}^{+146}$ & $13$ \\

     \hline

    (931) Whittemora & 1 & $32_{-14}^{+6}$ & $46_{-9}^{+12}$ & $19.17588\pm0.00006$ & $50_{-2}^{+4}$ & $0.13_{-0.02}^{+0.01}$ & $86_{-85}^{+114}$ & $45$ \\
     & \textbf{2} &  ${\bf 220_{-23}^{+11}}$ & ${\bf 70_{-8}^{+5}}$ & ${\bf 19.17589\pm0.00006}$ & ${\bf 51_{-2}^{+2}}$ & ${\bf 0.12_{-0.01}^{+0.02}}$ & ${\bf 40_{-39}^{+160}}$ & ${\bf 21}$ \\

     \hline

    (1062) Ljuba & 1 & $122_{-4}^{+1}$ & $-13_{-4}^{+7}$ & $33.7907\pm0.0006$ & $50_{-2}^{+4}$ & $0.077_{-0.023}^{+0.005}$ & $29_{-28}^{+220}$ & $42$ \\
     & \textbf{2} &  ${\bf 305_{-3}^{+3}}$ & ${\bf -20_{-8}^{+4}}$ & ${\bf 33.7909\pm0.0005}$ & ${\bf 51_{-2}^{+3}}$ & ${\bf 0.063_{-0.009}^{+0.015}}$ & ${\bf 29_{-28}^{+220}}$ & ${\bf 55}$ \\

\end{longtable}

\tablefoot{The columns present parameters derived from the CITPM for the pole 1 and pole 2 solutions of each target: J2000 ecliptic pole longitude ($\lambda$) and latitude ($\beta$), sidereal rotation period, diameter (D), geometric albedo {\color{highlight}p\textsubscript{V}}, thermal inertia ($\Gamma$), and Hapke's mean surface slope ($\bar{\theta}$).
    The diameter corresponds to the diameter of a sphere with a volume equivalent to that of the model. Pole solutions preferred by occultations are marked with boldface.}

\twocolumn
\renewcommand{\arraystretch}{1}

preceding column also take model uncertainties into account if these were larger than the formal uncertainties. 
To determine the influence of the shape model uncertainty on diameter, we created ten versions of each shape 
model using various levels of shape regularisation. Such a procedure, described in more detail in \citet{M2023}, resulted in spin and shape 
models similarly well-fitting light curves but with different levels of stretch along the rotation axis 
(‘flatter’ and ‘rounder’ shapes), since that shape dimension is usually worse determined by light curve 
inversion. Each version of the shape was then fitted to all available occultations with at least two well-spaced chords, resulting in a range 
of plausible diameters. 
Some variations in the shape misfitted the occultations by more than three sigma, and in such a case, 
their diameters were not taken into account to assess the diameter range. 
On the other hand, some regularised shapes provided a better 
fit than the original shape model did, and then they replaced it. Figures \ref{215_occ} to \ref{1062_occ} present the best-fitting 
shapes in this respect, and Table~\ref{tab:occ_diam} contains diameters for such improved models.

\begin{figure}
    \centering
    \includegraphics[scale = 0.24]{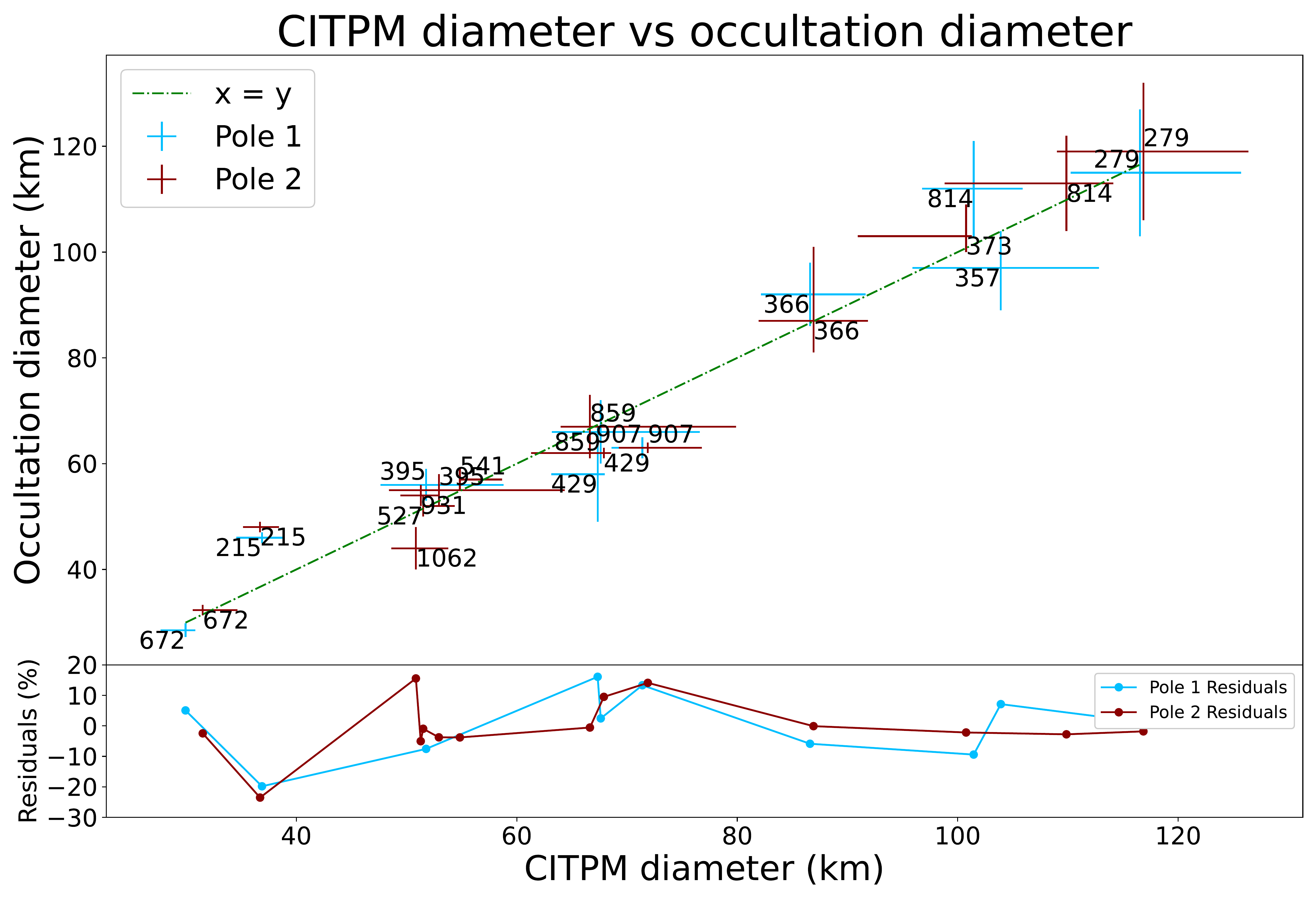}
    \caption{Comparison between thermally derived asteroid diameters and occultation-based diameters, both determined in this work (Tables \ref{tab:citpm} and \ref{tab:occ_diam}). The horizontal axis corresponds to the CITPM-derived size, while the vertical axis is the occultation-derived size. The dashed green line is for y = x. Blue points are for the pole 1 solution, while brown points are for the pole 2 solution. Relative residuals are computed as follows: $100\times(D_{\mathrm{CITPM}} - D_{\mathrm{occ}}) / D_\mathrm{occ}$}
    \label{fig:sizevssize}
\end{figure}

\begin{table}[h!]
\begin{center}
\noindent
\caption{{\small Occultation derived diameters.
 }}
 
\begin{tabular}{lccc}
\hline
 Target  & spin solution    &       D         &   D RMS  \\
         &                  &     [km]        &   [km]   \\
\hline
 (215) Oenone&  1       &  46 $\pm$ 1     &   1 \\
             &  2        &  48 $\pm$ 1     &   1 \\
 (279) Thule&  1        & 115 $\pm$ 12    &  12 \\
            &  2        & 119 $\pm$ 13    &  13 \\
 (357) Ninina& {\bf 1}  &${\bf 97^{+7}_{-8}}$ &{\bf 6} \\
             &  2       & 113 $\pm$ 11    & 11 \\
 (366) Vincentina& 1   &  92 $\pm$  6    &  5 \\ 
             & 2   &  87 $^{+14}_{-6}$& 6 \\ 
 (373) Melusina& 1     &  99 $\pm$  8    &  8 \\ 
               & {\bf 2}&${\bf 103^{+6}_{-3}}$ &{\bf 3} \\
 (395) Delia& 1        &  56 $\pm$  3    &  3 \\ 
            & 2        &  55 $\pm$  3    &  3 \\ 
 (429) Lotis& 1        &  58 $\pm$  9    &  9 \\ 
            & 2        &  62 $\pm$  1    &  1 \\
 (527) Euryanthe& 1    &  53 $\pm$  6    &  6 \\ 
                & {\bf 2}&{\bf 52 $\pm$ 2} &{\bf 2} \\
 (541) Deborah& 1      &  {\color{highlight}60} $\pm$ {\color{highlight}5}    &  {\color{highlight}5} \\
               & {\bf 2}&{\bf {\color{highlight}57} $\pm$ {\color{highlight}2}} &{\bf {\color{highlight}2}} \\ 
 (672) Astarte&  1      & 28.5 $\pm$ 1.3 & 1.3 \\
              &  2      & 32.3 $\pm$ 1.0 & 0.9 \\
 (814) Tauris&  1       & 112 $\pm$  9    &  9 \\ 
             &  2       & 113 $\pm$  9    &  9 \\ 
 (859) Bouzareah&  1    &  66 $\pm$  6    &  6 \\ 
                &  2    &  67 $\pm$  6    &  6 \\ 
 (907) Rhoda& 1        &  63 $\pm$   2    &  2 \\ 
            & 2        &  63 $\pm$   1    &  1 \\ 
 (931) Whittemora& 1   &  49 $\pm$  5     &  5 \\  
                 & {\bf 2}&{\bf 54 $\pm$ 2} &{\bf 2} \\
(1062) Ljuba& 1        &  41 $\pm$  4     &  4 \\  
            & {\bf 2}&{\bf 44 $\pm$ 4} &{\bf 4} \\
\hline
\end{tabular}

\label{tab:occ_diam}
\tablefoot{For each target and pole solution presented in Table~\ref{tab:citpm}, the diameter of the equivalent volume sphere is given, together with its best estimate of uncertainty, 
influenced by the uncertainty of both occultation timings and the shape model itself. The last column contains the formal RMS uncertainty of the occultation fit only (see text in Section \ref{sec:occ_fitting} for details). 
Solutions preferred by occultations are marked with boldface.}

\end{center}
\end{table}
 
The plots shown in Figures \ref{215_occ} to \ref{1062_occ} present the instantaneous silhouettes of the light curve inversion models on the fundamental plane, with the scale expressed in kilometres. Blue and magenta contours pertain to mirror solutions for pole 1 and pole 2, respectively, unless stated otherwise, see Table~\ref{tab:citpm} for the spin axis parameters. To be more precise, the original $\lambda$ and $\beta$ from pure light curve inversion (our step one) were a few degrees different than those from CITPM presented in Tab. \ref{tab:citpm} (our step two), as the other method optimises the spin axis orientation as well. However, the values for the pole solution from both methods were close enough to each other (within the error bars) that we decided not to present those from step one here to avoid confusion. Both versions of the shape models with their spin parameters are available in the DAMIT database. When occultation fits do not show a clear preference for one model versus the other, both models are shown as solid contours. Otherwise, the preferred model is shown with a solid contour, while the less preferred one is shown with a dashed contour.

\begin{figure}[!h]
    \centering
    \includegraphics[scale = 0.265]{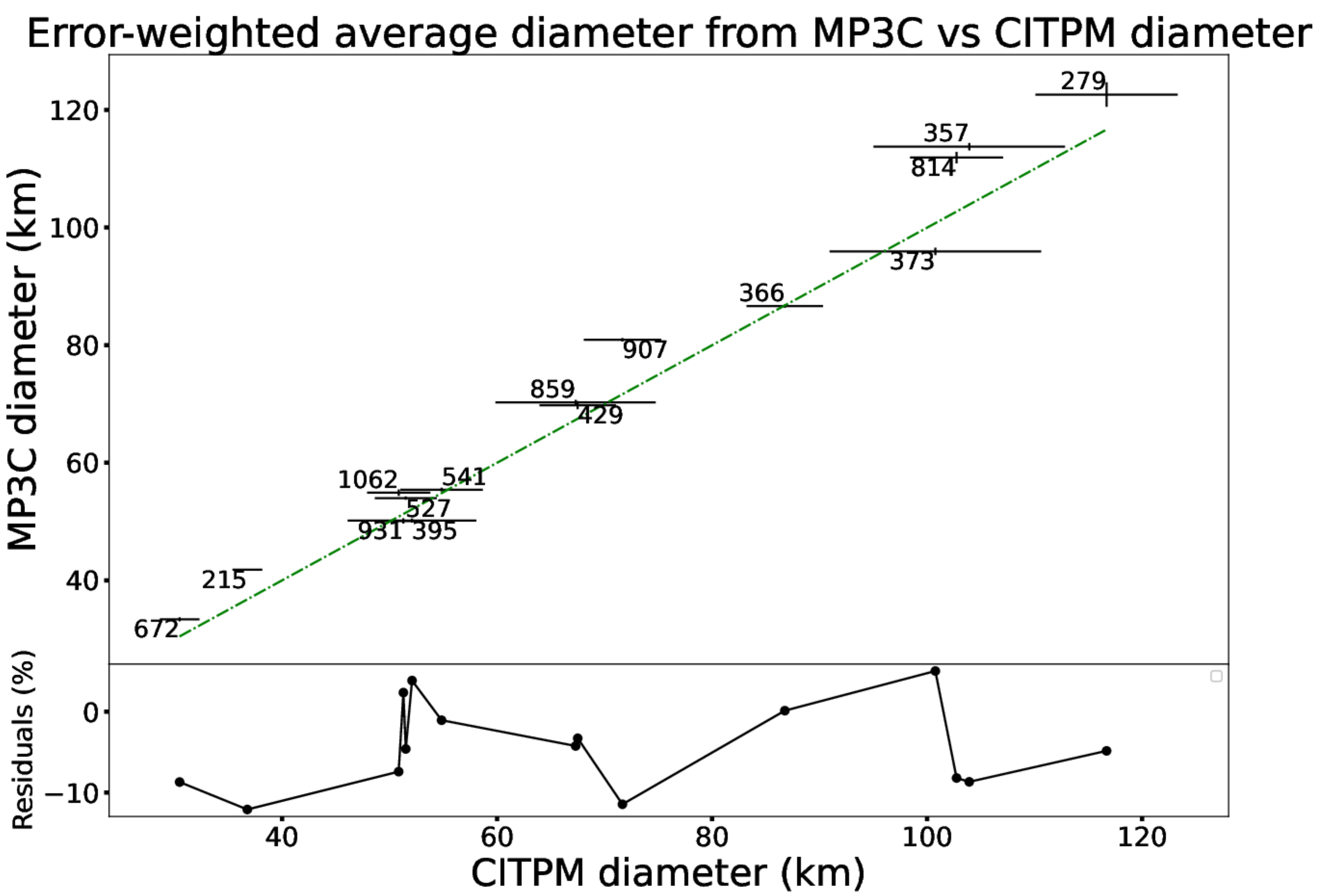}
    \caption{{\color{highlight}Comparison between the error-weighted average of thermally derived asteroid diameters from the MP3C database and CITPM diameters derived in this work (Tables \ref{tab:diam_lit} and \ref{tab:citpm}). The horizontal axis corresponds to the CITPM-derived size, while the vertical axis represents the average diameter from values listed in MP3C database. The dashed green line indicates y = x. If a certain pole solution is preferred, this solution is directly used for the CITPM value. If no solution is preferred, we apply the same error-weighted average that we applied to MP3C values (see Tab. \ref{tab:diam_lit}).}}
    \label{fig:mp3c}
\end{figure}

\subsection{Detailed description of specific targets} 
\label{sec:detailed_targets}
Below, we describe the results for a few particular targets in more detail.

\subsubsection{(279) Thule}

None of the two solutions for spin and shape fit the occultations perfectly,  
particularly the event from 2008/04/03 (see Fig. \ref{279_occ}). 
There were five chords in this event, but one of them could have been registered with a gross timing error, 
so we decided to exclude it. \citet{Sato2015} interpreted that chord as produced by a nearby, large satellite,

but we did not find any other signatures from additional large objects around Thule, neither in other occultation events nor in thermophysical analysis. Thermal flux from two 
bodies would lead to overestimated size when treated as a single object,
inconsistent with a single body of the size determined by occultations. 
On the contrary, we find a good agreement of sizes determined with both methods (see the values for D in Table~\ref{tab:citpm} and Table~\ref{tab:occ_diam}), 
thus excluding the large satellite hypothesis. The two shortest chords from that occultation outline the very edges of the real body, 
but such small details of real shapes are problematic to be reproduced by light curve inversion 
or sometimes tend to be shifted over the surface of the shape model. 
Taking this into account, the first pole (blue contour) is slightly preferred.

\subsubsection{(366) Vincentina}

Solution for pole 2 seems slightly preferred by occultations, and as such is marked in blue in Fig. \ref{366_occ}.
Although two multichord events are available, they do not constrain the size for model 1 well. 
Due to the lack of chords in the northernmost part of that model silhouette, longer shapes, up to 100 km are almost 
equally possible as the best fitting one, with its 87 km. 

\subsubsection{(373) Melusina}

In spite of only two-chord event available for this asteroid, the lengths and positions of these chords 
relative to each other enabled us to identify a preferred solution (pole 2, marked with the solid contour 
in Fig. \ref{373_occ}). Here, only the procedure of testing shapes with various regularisations enabled us to find  
a good-fitting shape. On the other hand, each version of the shape for pole 1 fitted worse than the one shown with the dashed contour.

\subsubsection{(395) Delia}

Here, similarly to the case of 279 Thule, we had to exclude one occultation observation 
(from the event on 2006/08/27), to be able to fit any shape model to it. 
Interestingly, this time that observation was negative. That negative chord was placed 8 km south of the southernmost positive chord, 50.5 km long, 
shown in the first panel of Fig. \ref{395_occ}. It was too close not to be intersected by each of 
the tested model contours. 
We investigated that observation closely and found that it was a visual observation of an event 
with a very bright, 6-magnitude star, where another observer reported ingress and egress of a certain duration. 
These facts probably point to effects complicating the observations, like for instance non-negligible angular 
extent of an occulted star.

\subsubsection{(429) Lotis}

Although for Lotis, there was only a two-chord event available, the size is quite secure: it cannot be much 
larger or smaller than 60 km due to the long chord and its placement against the short chord 
(see Fig. \ref{373_occ}). 
The shape solution connected to pole 2 (marked here in blue) is slightly preferred, but still, 
it probably has an underestimated uncertainty (Tab. \ref{tab:occ_diam}).

\subsubsection{814 Tauris}

The only multichord occultation for Tauris has been registered in our ‘SlowRotators’ campaign 
and both shape solutions fit it well (Fig. \ref{814_occ}). 
Interestingly, a few years earlier, there was another event with Tauris 
\citep{George2019}, where a single station registered a double dip in brightness, which might have been caused by 
distinct topographic features on the surface of an asteroid. 
A chord from this event, treated as a single-dip one, would fit the edge of our shape model 1 of Tauris, 
confirming this hypothesis. However, the light curve inversion method used here for shape reconstruction 
produces only convex shape models, so we cannot confirm this nonconvex topography. 
The contact binary or double asteroid hypotheses also made by \citet{George2019} are rather improbable, 
because our shape model of Tauris would have a brick-like or a cone-like appearance if that was the case (see the DAMIT database for a full view of the model).

\subsubsection{(907) Rhoda}

The fit to the two-chord event does not give reliable size determination in this case. Both chords 
could as well intersect the northern hemisphere, leading to a larger size determination 
(see Fig. \ref{814_occ}). 
That probably explains the inconsistency between the occultation ($63\pm2$ km) and the CITPM ($74^{+5}_{-3}$ km) diameters 
that we obtained. Values from the latter method are more robust in this case.

\subsubsection{(1062) Ljuba}

The size and RMS from occultation fitting are not robust here (see Fig. \ref{1062_occ}). 
Shape model uncertainties are larger 
than the timing uncertainties, and the fit is accidentally forced to chords with perfect timings, 
but the size can also be 10\% larger than the adopted 44 km. Here too the size from CITPM ($51^{+4}_{-2}$ km) is more reliable. 
Curiously, despite such a relatively poorly constrained size, a preferred pole solution (pole 2) is found: 
only this shape shows the abrupt narrowing in the southern part, necessary to explain 
the observed occultation durations (though imperfectly). 

\vspace{0.2cm}

Models for the remaining asteroids (357) Ninina, (527) Euryanthe, (541) Deborah and (931) Whittemora provided very good 
fit to occultations and showed a clear preference for one of two mirror pole solutions 
(see the Figures: \ref{357_occ}, \ref{527_occ}, \ref{541_occ} and \ref{931_occ}). 
On the other hand, both mirror-pole models for (215) Oenone (Fig. \ref{215_occ}), (672) Astarte (Fig. \ref{672_occ}) and (859) Bouzareah (Fig. \ref{859_occ}) fitted similarly.

\subsection{Size comparison}

When comparing the size values determined through modelling with the CITPM method and those determined through detailed shape model fitting to occultations (see Figure \ref{fig:sizevssize}), we observe that the two methods yield consistent results. This demonstrates that, although CITPM determines size through thermophysical modelling and occultations do so through direct observation if the input data are of high quality, the CITPM provides as accurate results as multichord occultations. Targets which lay a little bit farther from the agreement line in Figure \ref{fig:sizevssize} can find an explanation in insufficient chord number or their poor placement, leading to the conclusion that CITPM size should be considered as more reliable in these cases (see Section \ref{sec:detailed_targets}). 

\begin{figure}[!h]
    \centering
    \includegraphics[scale=0.25]{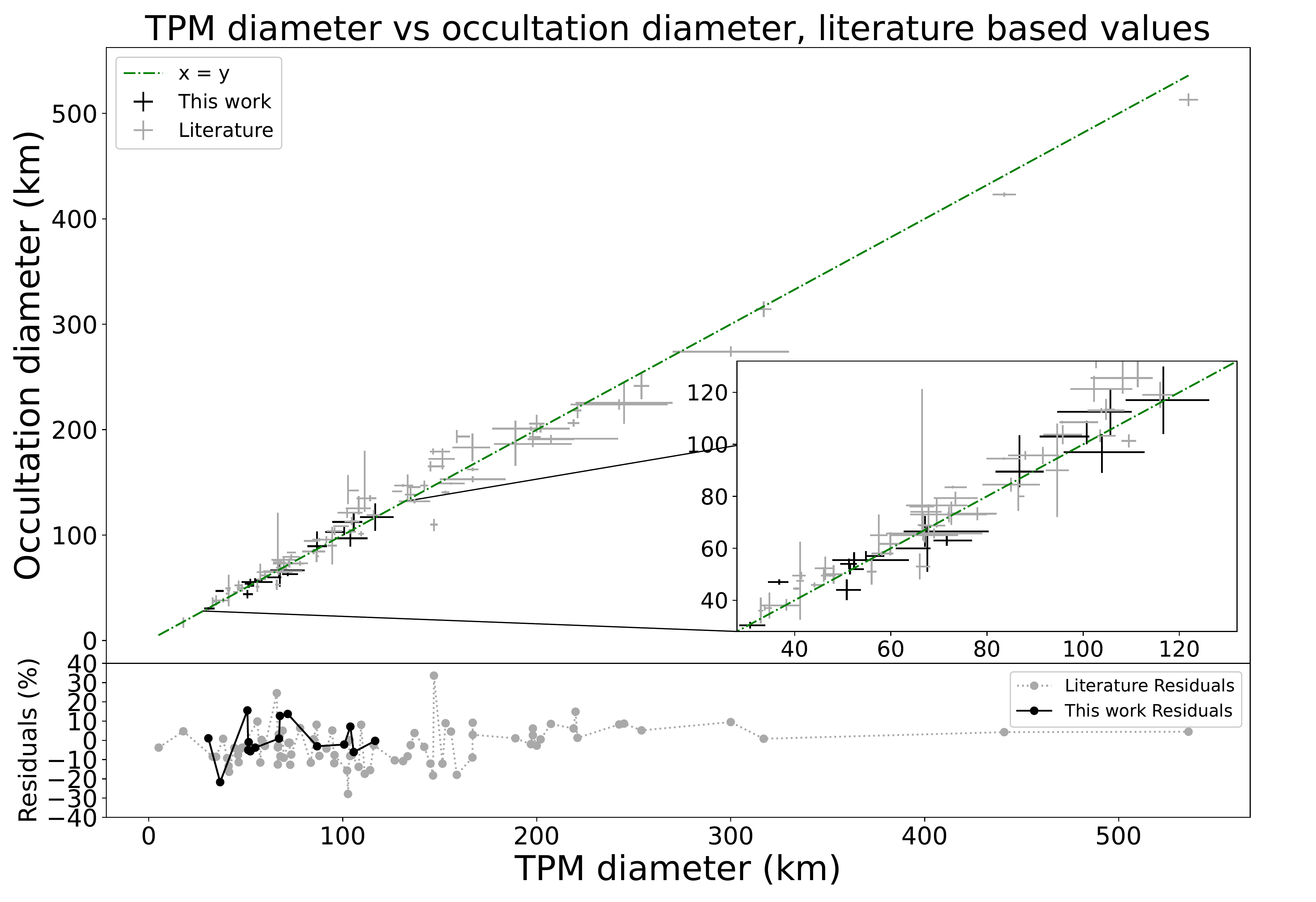}
    \caption{Comparison between thermally derived diameters and occultation-based diameters from the literature (Table~\ref{tab:literature}). The horizontal axis corresponds to TPM size while the vertical axis is the occultation size. The dashed green line is for y = x. Grey points correspond to values from the literature, while black points correspond to pole-averaged values from this work. Relative residuals are computed as follows: $100\times(D_{\mathrm{TPM}} - D_{\mathrm{occ}}) / D_\mathrm{occ}$}
    \label{fig:sizevssize_lit}
\end{figure}

In this work, we also compiled results for asteroid sizes from the literature, and present them in Table~\ref{tab:literature}. We focused on size values determined either through thermophysical modelling or via occultations. In some cases, particularly for occultations, a single value was reported in the literature, corresponding to the best-fit solution. These values are directly included in Table~\ref{tab:literature}. In other cases, both for occultations and for size values derived from TPM, the authors provided averaged size estimates. These values are also included as they were in Table~\ref{tab:literature}. For cases where the authors presented multiple values for diameters (often corresponding to mirror pole solutions), we averaged the results to ensure consistency with the rest of the dataset and to avoid having multiple entries for a single target. When occultation data did not allow us to resolve the mirror pole ambiguity, we also averaged our results between pole 1 and pole 2 to facilitate comparison with the values reported in the literature. The results are shown in Figure \ref{fig:sizevssize_lit}. The black points represent the values obtained in this work, while the grey points correspond to values from the literature. This plot demonstrates that our values are consistent with those published by other teams. Furthermore, when examining the residuals plot, it can be seen that our residuals are, on average, smaller than those based on the literature values.

 {\color{highlight}Figure~\ref{fig:mp3c} shows MP3C error-weighted average values for diameters (available in Table~\ref{tab:diam_lit}) compared to CITPM values. CITPM values are also error-weighted averages between pole 1 and pole 2 in cases where no pole solution is preferred. Values in the last column of Table~\ref{tab:diam_lit} correspond to the error-weighted mean diameters obtained by averaging results from multiple techniques listed in the MP3C database, whereas CITPM values are derived using a single, consistent modelling approach. Interestingly, when comparing the values from Table~\ref{tab:diam_lit} and Table~\ref{tab:citpm} with those in Table~\ref{tab:occ_diam}, CITPM estimates are generally slightly closer to those derived from occultations and provide a more accurate estimation of the associated uncertainties, despite relying on a single methodology. }

\subsection{Thermal inertia versus period}

\begin{figure}[!h]
    \centering
    \includegraphics[width=1\linewidth]{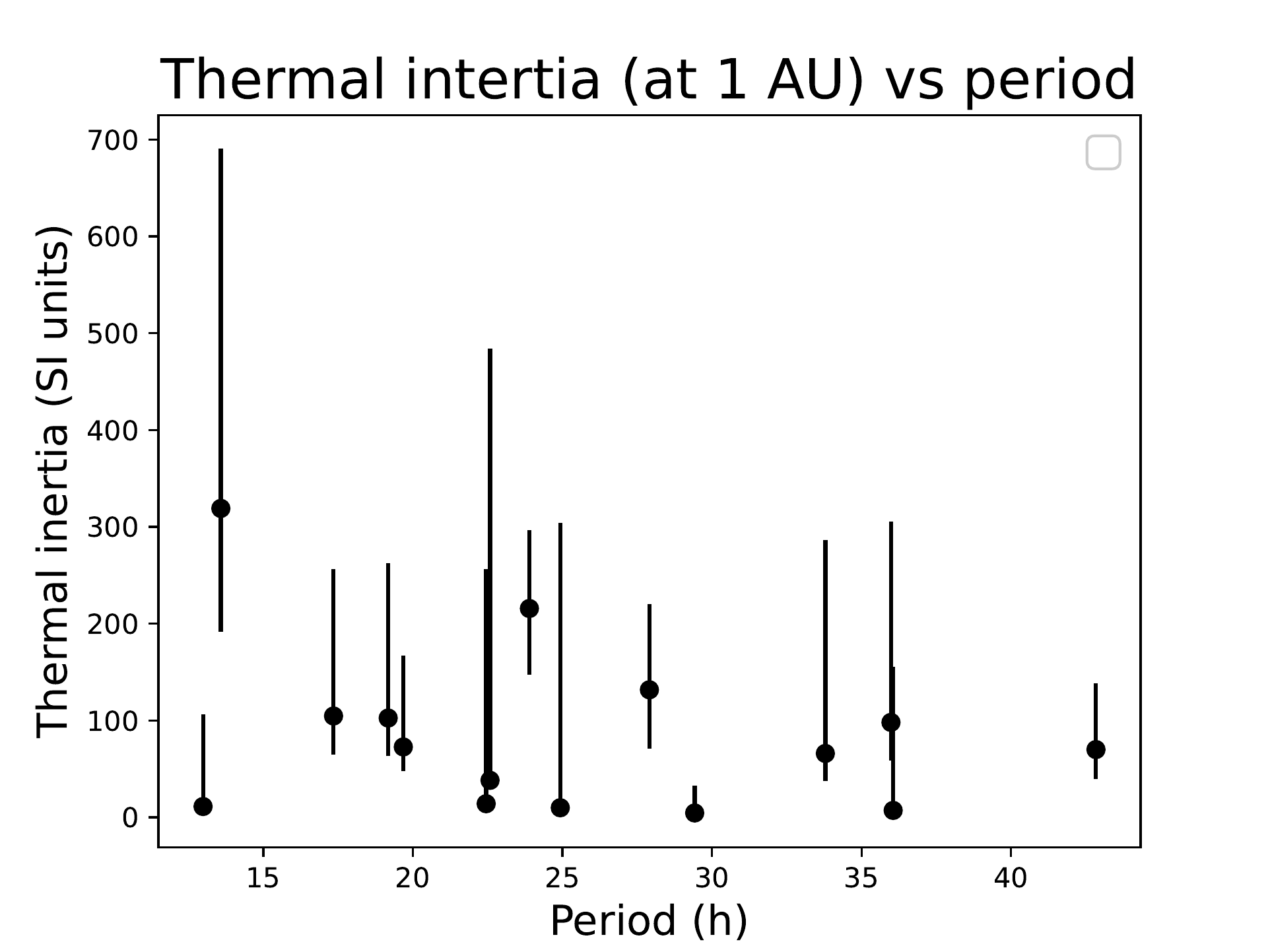}
    \caption{Thermal inertia normalised to 1 AU (in $\text{J}\,\text{m}^{-2}\,\text{s}^{-\frac{1}{2}}\,\text{K}^{-1}$, vertical axis) vs. rotation period (in hours, horizontal axis). Values for the mirror pole solutions have been averaged. No trend can be seen.}
    \label{fig:periodvsti}
\end{figure}

In this work, we primarily focused on the precise determination of asteroid sizes. However, since thermal inertia and rotation period are two other parameters simultaneously optimised in the CITPM, our results provide a good opportunity to compare these two parameters and assess a potential correlation, as suggested by \citet{harris2016}. The result can be seen in Figure \ref{fig:periodvsti}, where the x axis is the rotation period and the y axis represents the thermal inertia normalised to 1 AU following this equation \citep{mueller2010}:

\begin{equation}
    \Gamma_{\text{norm}} = \Gamma * (\frac{R_{\text{norm}}}{R_{\text{target}}})^{-\frac{3}{4}},    
\end{equation} 

\noindent where $\Gamma$ and $\Gamma_{\text{norm}}$ are, respectively, the thermal inertia and the normalised thermal inertia, $R_{\text{norm}}$ and $R_{\text{target}}$ being, respectively, the distance from the Sun at which the thermal inertia is normalised and an average distance between the target and the Sun during infrared observations.

In Figure 5 of \citet{harris2016}, an increasing trend of greater thermal inertia for longer rotation periods is seen for periods above 10 hours. Our results cover a range of periods between 10 and 40 hours, which should be sufficient to observe this correlation. However, Figure \ref{fig:periodvsti} shows no such correlation, aligning with the conclusions drawn by other studies (\citealt{M2019, marciniak_properties_2021} and \citealt{hanus2018}).

\section{Conclusions and future work}

In this study, we combined two modelling techniques to determine the shapes and sizes of \nAst\ main-belt asteroids: 
the light curve Inversion followed by CITPM. In parallel, we scaled their shape models using stellar occultation observations. Through two extensive observing campaigns, requiring significant human and observational resources, we obtained a substantial amount of dense light curves and occultation data (see the information in Zenodo repository described in Section 6). Using thermal data, systematically including observations from WISE spacecraft, coupled with an innovative thermophysical modelling approach (CITPM), allowed us to accurately determine asteroid sizes. Furthermore, the results show a good agreement with those obtained from stellar occultation fitting, confirming the robustness of the methodology. Discrepancies were generally around 5\%, and in cases of greater differences, they were often due to the insufficient number of positive occultations or poor distribution of chords, making the CITPM model-derived values more reliable. Despite these significant advances, some challenges remain. For example the dependence on convex approximation for the shape models may affect the ability to decouple thermal inertia from surface roughness (\citealp{delbo2015}, \citealp{durech_asteroid_2017}). 

Albedo values found here lay within a range for the target taxonomic types defined in the literature. However, contrary to studies suggesting a correlation between rotation period and thermal inertia \citep{harris2016}, our results did not find any significant trend, aligning with conclusions from other recent investigations. 

The results of this study resolve substantial inconsistencies in size determinations for some large and medium-sized asteroids. These are essential for refining models of the evolution of the solar system, particularly concerning asteroid internal structure, composition and collisional processes, which require precise knowledge of the density, so a well-determined mass and even better-determined diameter (\citealp{carry2012}, \citealp{hanus2017}). Future advancements, including thermal data from space missions such as JWST or NEO Surveyor, could enable the determination of more precise thermophysical parameters for a larger number of asteroids. Finally, strengthening multi-chord observation campaigns for stellar occultations, supported by international collaborations, remains a primary objective to improve asteroid size determinations. In future work, reliable sizes from multichord stellar occultations could be used as a fixed parameter in (CI)TPM, better constraining the thermal inertia and surface roughness \citep{muller2018}.

\section{Data availability}
Details of the observing campaigns: \url{https://zenodo.org/records/15274451}

\begin{acknowledgements}

This work was supported by the National Science Centre, Poland, through grant no. 2020/39/O/ST9/00713,
and by the Polish National Agency for Academic Exchange through stipend agreement no. BPN/PRE/2022/U/0001.

This project has been supported by the Lend{\"u}let grant LP2012-31 of the Hungarian Academy of Sciences and by the KKP-137523 grant of the Hungarian 	National Research, Development and Innovation Office (NKFIH).

This work was partially supported by a program of the Polish Ministry of Science under the title ‘Regional Excellence Initiative’, project no. 	RID/SP/0050/2024/1.

This work was partly supported by the grant K-138962 of the National Research, Development and Innovation Office (NKFIH, Hungary).

F.M. thanks the fellowship granted by FAPESP (2024/16260-6) and the IMPACTON team for the technical support at OASI.

This work uses data obtained from the Asteroid Light-Curve Data Exchange Format (ALCDEF) database,
which is supported by funding from NASA grant 80NSSC18K0851.

This work is based on data provided by the Minor Planet Physical Properties Catalogue (MP3C) of the Observatoire
de la C{\^o}te d'Azur.

Observations at the Mol{\.e}tai Observatory has been partially funded by Europlanet 2024.
Europlanet 2024 RI has received funding from the European Union's Horizon 2020 research and innovation programme
under grant agreement No 871149.

The Joan Or{\'o} Telescope (TJO) of the Montsec Astronomical Observatory (OAdM)
is owned by the Catalan Government and operated by the Institute for Space Studies of Catalonia (IEEC).

This article is based on observations made in the Observatorios de Canarias del IAC with the
       0.82m IAC80 telescope operated on the island of Tenerife by the Instituto de Astrof{\'i}sica de Canarias (IAC)
       in the Observatorio del Teide.
This article is based on observations made with the SARA telescopes (Southeastern Association for Research in Astronomy),
whose nodes are located at the Observatorios de Canarias del IAC on the island of La Palma in the Observatorio
del Roque de los Muchachos; Kitt Peak, AZ under the auspices of the National Optical Astronomy Observatory (NOAO);
and Cerro Tololo Inter-American Observatory (CTIO) in La Serena, Chile.

This project uses data from the SuperWASP archive. The WASP project is currently funded and operated by Warwick University
and Keele University, and was originally set up by Queen's University Belfast, the Universities of Keele, St. Andrews,
and Leicester, the Open University, the Isaac Newton Group, the Instituto de Astrofisica de Canarias,
the South African Astronomical Observatory, and by STFC.

TRAPPIST-South is a project funded by the Belgian Fonds (National) de la Recherche Scientifique (F.R.S.-FNRS) under
grant PDR T.0120.21. TRAPPIST-North is a project funded by the University of Li{\`e}ge, in collaboration with the Cadi Ayyad
University of Marrakech (Morocco).

Funding for the Kepler and K2 missions are provided by the NASA Science Mission Directorate.
The data presented in this paper were obtained from the Mikulski Archive for Space
Telescopes (MAST). STScI is operated by the Association of Universities for Research in Astronomy,
Inc., under NASA contract NAS5-26555. Support for MAST for non-HST data is provided by the NASA Office 
of Space Science via grant NNX09AF08G and by other grants and contracts.

Data from Pic du Midi Observatory have been obtained with the 0.6-m telescope, a facility operated by
Observato{\'i}re Midi Pyr{\'e}n{\'e}es and Association T60, an amateur association.

This research is partly based on observations obtained with the 60-cm Cassegrain telescope (TC60) and
90-cm Schmidt-Cassegrain telescope (TSC90) of the Institute of Astronomy of the Nicolaus Copernicus University in Toru{\'n}.

This publication makes use of data products from the Wide-field Infrared Survey Explorer,
which is a joint project of the University of California, Los Angeles, and the Jet Propulsion
Laboratory/California Institute of Technology, funded by the National Aeronautics and Space Administration.

We acknowledge the contributions of the occultation observers
who have provided the observations in the dataset. Most of 
those observers are affiliated with one or more of:
European Asteroidal Occultation Network (EAON), International Occultation Timing Association (IOTA),
IOTA European Section (IOTA/ES), and IOTA East Asia Section (IOTA/EA), Japanese Occultation Information Network (JOIN),
and Trans Tasman Occultation Alliance (TTOA).

\end{acknowledgements}

\clearpage

\newpage

\twocolumn
\bibliographystyle{aa}
\bibliography{bib}

\begin{thebibliography}{94}
\expandafter\ifx\csname natexlab\endcsname\relax\def\natexlab#1{#1}\fi

\bibitem[{{Al{\'\i}-Lagoa} {et~al.}(2020){Al{\'\i}-Lagoa}, {M{\"u}ller}, {Kiss}, {Szak{\'a}ts}, {Marton}, {Farkas-Tak{\'a}cs}, {Bartczak}, {Butkiewicz-B{\k{a}}k}, {Dudzi{\'n}ski}, {Marciniak}, {Podlewska-Gaca}, {Duffard}, {Santos-Sanz}, \& {Ortiz}}]{alilagoa2020}
{Al{\'\i}-Lagoa}, V., {M{\"u}ller}, T.~G., {Kiss}, C., {et~al.} 2020, \aap, 638, A84

\bibitem[{{Al{\'\i}-Lagoa} {et~al.}(2018){Al{\'\i}-Lagoa}, {M{\"u}ller}, {Usui}, \& {Hasegawa}}]{alilagoa2018}
{Al{\'\i}-Lagoa}, V., {M{\"u}ller}, T.~G., {Usui}, F., \& {Hasegawa}, S. 2018, \aap, 612, A85

\bibitem[{Bartczak \& Dudziński(2017)}]{sage}
Bartczak, P. \& Dudziński, G. 2017, Monthly Notices of the Royal Astronomical Society, 473, 5050

\bibitem[{{Bus} \& {Binzel}(2002)}]{bus2002}
{Bus}, S.~J. \& {Binzel}, R.~P. 2002, \icarus, 158, 146

\bibitem[{{Cambioni} {et~al.}(2021){Cambioni}, {Delbo}, {Poggiali}, {Avdellidou}, {Ryan}, {Deshapriya}, {Asphaug}, {Ballouz}, {Barucci}, {Bennett}, {Bottke}, {Brucato}, {Burke}, {Cloutis}, {DellaGiustina}, {Emery}, {Rozitis}, {Walsh}, \& {Lauretta}}]{Cambioni2021}
{Cambioni}, S., {Delbo}, M., {Poggiali}, G., {et~al.} 2021, \nat, 598, 49

\bibitem[{{\v C}apek \& Vokrouhlický(2010)}]{capek_thermal_2010}
{\v C}apek, D. \& Vokrouhlický, D. 2010, \aap, 519, A75

\bibitem[{{Carry}(2012)}]{carry2012}
{Carry}, B. 2012, \planss, 73, 98

\bibitem[{{Carry} {et~al.}(2019){Carry}, {Vachier}, {Berthier}, {Marsset}, {Vernazza}, {Grice}, {Merline}, {Lagadec}, {Fienga}, {Conrad}, {Podlewska-Gaca}, {Santana-Ros}, {Viikinkoski}, {Hanu{\v{s}}}, {Dumas}, {Drummond}, {Tamblyn}, {Chapman}, {Behrend}, {Bernasconi}, {Bartczak}, {Benkhaldoun}, {Birlan}, {Castillo-Rogez}, {Cipriani}, {Colas}, {Drouard}, {{\v{D}}urech}, {Enke}, {Fauvaud}, {Ferrais}, {Fetick}, {Fusco}, {Gillon}, {Jehin}, {Jorda}, {Kaasalainen}, {Keppler}, {Kryszczynska}, {Lamy}, {Marchis}, {Marciniak}, {Michalowski}, {Michel}, {Pajuelo}, {Tanga}, {Vigan}, {Warner}, {Witasse}, {Yang}, \& {Zurlo}}]{carry2019}
{Carry}, B., {Vachier}, F., {Berthier}, J., {et~al.} 2019, \aap, 623, A132

\bibitem[{{Carry} {et~al.}(2021){Carry}, {Vernazza}, {Vachier}, {Neveu}, {Berthier}, {Hanu{\v{s}}}, {Ferrais}, {Jorda}, {Marsset}, {Viikinkoski}, {Bartczak}, {Behrend}, {Benkhaldoun}, {Birlan}, {Castillo-Rogez}, {Cipriani}, {Colas}, {Drouard}, {Dudzi{\'n}ski}, {Desmars}, {Dumas}, {{\v{D}}urech}, {Fetick}, {Fusco}, {Grice}, {Jehin}, {Kaasalainen}, {Kryszczynska}, {Lamy}, {Marchis}, {Marciniak}, {Michalowski}, {Michel}, {Pajuelo}, {Podlewska-Gaca}, {Rambaux}, {Santana-Ros}, {Storrs}, {Tanga}, {Vigan}, {Warner}, {Wieczorek}, {Witasse}, \& {Yang}}]{carry2021}
{Carry}, B., {Vernazza}, P., {Vachier}, F., {et~al.} 2021, \aap, 650, A129

\bibitem[{{Delbo'} {et~al.}(2015){Delbo'}, {Mueller}, {Emery}, {Rozitis}, \& {Capria}}]{delbo2015}
{Delbo'}, M., {Mueller}, M., {Emery}, J.~P., {Rozitis}, B., \& {Capria}, M.~T. 2015, in Asteroids IV, ed. P.~{Michel}, F.~E. {DeMeo}, \& W.~F. {Bottke}, 107--128

\bibitem[{Delbo’ \& Tanga(2009)}]{delbo_thermal_2009}
Delbo’, M. \& Tanga, P. 2009, Planetary and Space Science, 57, 259

\bibitem[{{Dias-Oliveira} {et~al.}(2017){Dias-Oliveira}, {Sicardy}, {Ortiz}, {Braga-Ribas}, {Leiva}, {Vieira-Martins}, {Benedetti-Rossi}, {Camargo}, {Assafin}, {Gomes-J{\'u}nior}, {Baug}, {Chandrasekhar}, {Desmars}, {Duffard}, {Santos-Sanz}, {Ergang}, {Ganesh}, {Ikari}, {Irawati}, {Jain}, {Liying}, {Richichi}, {Shengbang}, {Behrend}, {Benkhaldoun}, {Brosch}, {Daassou}, {Frappa}, {Gal-Yam}, {Garcia-Lozano}, {Gillon}, {Jehin}, {Kaspi}, {Klotz}, {Lecacheux}, {Mahasena}, {Manfroid}, {Manulis}, {Maury}, {Mohan}, {Morales}, {Ofek}, {Rinner}, {Sharma}, {Sposetti}, {Tanga}, {Thirouin}, {Vachier}, {Widemann}, {Asai}, {Hayato}, {Hiroyuki}, {Owada}, {Yamamura}, {Hayamizu}, {Bradshaw}, {Kerr}, {Tomioka}, {Andersson}, {Dangl}, {Haymes}, {Naves}, \& {Wortmann}}]{Dias-Oliveira2017}
{Dias-Oliveira}, A., {Sicardy}, B., {Ortiz}, J.~L., {et~al.} 2017, \aj, 154, 22

\bibitem[{{\v D}urech {et~al.}(2017){\v D}urech, Delbo’, Carry, Hanuš, \& Alí-Lagoa}]{durech_asteroid_2017}
{\v D}urech, J., Delbo’, M., Carry, B., Hanuš, J., \& Alí-Lagoa, V. 2017, \aap, 604, A27

\bibitem[{{\v D}urech {et~al.}(2018){\v D}urech, {Hanu{\v{s}}}, \& {Al{\'\i}-Lagoa}}]{durech2018}
{\v D}urech, J., {Hanu{\v{s}}}, J., \& {Al{\'\i}-Lagoa}, V. 2018, \aap, 617, A57

\bibitem[{{\v D}urech {et~al.}(2011){\v D}urech, {Kaasalainen}, {Herald}, {Dunham}, {Timerson}, {Hanu{\v{s}}}, {Frappa}, {Talbot}, {Hayamizu}, {Warner}, {Pilcher}, \& {Gal{\'a}d}}]{DURECH2011}
{\v D}urech, J., {Kaasalainen}, M., {Herald}, D., {et~al.} 2011, \icarus, 214, 652

\bibitem[{{\v D}urech {et~al.}(2009){\v D}urech, {Kaasalainen}, {Warner}, {Fauerbach}, {Marks}, {Fauvaud}, {Fauvaud}, {Vugnon}, {Pilcher}, {Bernasconi}, \& {Behrend}}]{durech2009_sparse_data}
{\v D}urech, J., {Kaasalainen}, M., {Warner}, B.~D., {et~al.} 2009, \aap, 493, 291

\bibitem[{{\v D}urech {et~al.}(2010){\v D}urech, {Sidorin}, \& {Kaasalainen}}]{damit}
{\v D}urech, J., {Sidorin}, V., \& {Kaasalainen}, M. 2010, \aap, 513, A46

\bibitem[{{\v D}urech {et~al.}(2020){\v D}urech, {Tonry}, {Erasmus}, {Denneau}, {Heinze}, {Flewelling}, \& {Van{\v{c}}o}}]{durech2020}
{\v D}urech, J., {Tonry}, J., {Erasmus}, N., {et~al.} 2020, \aap, 643, A59

\bibitem[{{Ferrais} {et~al.}(2022){Ferrais}, {Jorda}, {Vernazza}, {Carry}, {Bro{\v{z}}}, {Rambaux}, {Hanu{\v{s}}}, {Dudzi{\'n}ski}, {Bartczak}, {Vachier}, {Aristidi}, {Beck}, {Marchis}, {Marsset}, {Viikinkoski}, {Fetick}, {Drouard}, {Fusco}, {Birlan}, {Podlewska-Gaca}, {Burbine}, {Dyar}, {Bendjoya}, {Benkhaldoun}, {Berthier}, {Castillo-Rogez}, {Cipriani}, {Colas}, {Dumas}, {{\v{D}}urech}, {Fauvaud}, {Grice}, {Jehin}, {Kaasalainen}, {Kryszczynska}, {Lamy}, {Le Coroller}, {Marciniak}, {Michalowski}, {Michel}, {Prieur}, {Reddy}, {Rivet}, {Santana-Ros}, {Scardia}, {Tanga}, {Vigan}, {Witasse}, \& {Yang}}]{ferrais2022}
{Ferrais}, M., {Jorda}, L., {Vernazza}, P., {et~al.} 2022, \aap, 662, A71

\bibitem[{{Ferrais} {et~al.}(2020){Ferrais}, {Vernazza}, {Jorda}, {Rambaux}, {Hanu{\v{s}}}, {Carry}, {Marchis}, {Marsset}, {Viikinkoski}, {Bro{\v{z}}}, {Fetick}, {Drouard}, {Fusco}, {Birlan}, {Podlewska-Gaca}, {Jehin}, {Bartczak}, {Berthier}, {Castillo-Rogez}, {Cipriani}, {Colas}, {Dudzi{\'n}ski}, {Dumas}, {{\v{D}}urech}, {Kaasalainen}, {Kryszczynska}, {Lamy}, {Le Coroller}, {Marciniak}, {Michalowski}, {Michel}, {Santana-Ros}, {Tanga}, {Vachier}, {Vigan}, {Witasse}, \& {Yang}}]{ferrais2020}
{Ferrais}, M., {Vernazza}, P., {Jorda}, L., {et~al.} 2020, \aap, 638, L15

\bibitem[{{Ferreira} {et~al.}(2022){Ferreira}, {Tanga}, {Spoto}, {Machado}, \& {Herald}}]{2022A&A...658A..73F}
{Ferreira}, J.~F., {Tanga}, P., {Spoto}, F., {Machado}, P., \& {Herald}, D. 2022, \aap, 658, A73

\bibitem[{{Franco} {et~al.}(2024){Franco}, {Pilcher}, {Oey}, {Marchini}, {Papini}, {Scarfi}, {Iozzi}, {Ruocco}, {Bacci}, {Maestripieri}, {Montigiani}, \& {Mannucci}}]{franco2024}
{Franco}, L., {Pilcher}, F., {Oey}, J., {et~al.} 2024, Minor Planet Bulletin, 51, 100

\bibitem[{{Fuksa} {et~al.}(2023){Fuksa}, {Bro{\v{z}}}, {Hanu{\v{s}}}, {Ferrais}, {Fatka}, \& {Vernazza}}]{fuksa2023}
{Fuksa}, M., {Bro{\v{z}}}, M., {Hanu{\v{s}}}, J., {et~al.} 2023, \aap, 677, A189

\bibitem[{{George} {et~al.}(2019){George}, {Garlitz}, \& {Morton}}]{George2019}
{George}, T., {Garlitz}, J., \& {Morton}, A. 2019, Journal for Occultation Astronomy, 9, 3

\bibitem[{{Hanu{\v{s}}} {et~al.}(2015){Hanu{\v{s}}}, {Delbo'}, {\v D}urech, \& {Al{\'\i}-Lagoa}}]{hanus2015}
{Hanu{\v{s}}}, J., {Delbo'}, M., {\v D}urech, J., \& {Al{\'\i}-Lagoa}, V. 2015, \icarus, 256, 101

\bibitem[{{Hanu{\v{s}}} {et~al.}(2018){Hanu{\v{s}}}, {Delbo'}, {{\v{D}}urech}, \& {Al{\'\i}-Lagoa}}]{hanus2018}
{Hanu{\v{s}}}, J., {Delbo'}, M., {{\v{D}}urech}, J., \& {Al{\'\i}-Lagoa}, V. 2018, \icarus, 309, 297

\bibitem[{{Hanu{\v{s}}} {et~al.}(2016){Hanu{\v{s}}}, {Delbo'}, {Vokrouhlick{\'y}}, {Pravec}, {Emery}, {Al{\'\i}-Lagoa}, {Bolin}, {Devog{\`e}le}, {Dyvig}, {Gal{\'a}d}, {Jedicke}, {Korno{\v{s}}}, {Ku{\v{s}}nir{\'a}k}, {Licandro}, {Reddy}, {Rivet}, {Vil{\'a}gi}, \& {Warner}}]{hanus2016}
{Hanu{\v{s}}}, J., {Delbo'}, M., {Vokrouhlick{\'y}}, D., {et~al.} 2016, \aap, 592, A34

\bibitem[{{Hanu{\v{s}}} {et~al.}(2011){Hanu{\v{s}}}, {{\v{D}}urech}, {Bro{\v{z}}}, {Warner}, {Pilcher}, {Stephens}, {Oey}, {Bernasconi}, {Casulli}, {Behrend}, {Polishook}, {Henych}, {Lehk{\'y}}, {Yoshida}, \& {Ito}}]{hanus2011}
{Hanu{\v{s}}}, J., {{\v{D}}urech}, J., {Bro{\v{z}}}, M., {et~al.} 2011, \aap, 530, A134

\bibitem[{{Hanu{\v{s}}} {et~al.}(2017){Hanu{\v{s}}}, {Viikinkoski}, {Marchis}, {{\v{D}}urech}, {Kaasalainen}, {Delbo'}, {Herald}, {Frappa}, {Hayamizu}, {Kerr}, {Preston}, {Timerson}, {Dunham}, \& {Talbot}}]{hanus2017}
{Hanu{\v{s}}}, J., {Viikinkoski}, M., {Marchis}, F., {et~al.} 2017, \aap, 601, A114

\bibitem[{Hanuš {et~al.}(2013)Hanuš, Ďurech, Brož, Marciniak, Warner, Pilcher, Stephens, Behrend, Carry, Čapek, Antonini, Audejean, Augustesen, Barbotin, Baudouin, Bayol, Bernasconi, Borczyk, Bosch, Brochard, Brunetto, Casulli, Cazenave, Charbonnel, Christophe, Colas, Coloma, Conjat, Cooney, Correira, Cotrez, Coupier, Crippa, Cristofanelli, Dalmas, Danavaro, Demeautis, Droege, Durkee, Esseiva, Esteban, Fagas, Farroni, Fauvaud, Fauvaud, Del~Freo, Garcia, Geier, Godon, Grangeon, Hamanowa, Hamanowa, Heck, Hellmich, Higgins, Hirsch, Husarik, Itkonen, Jade, Kamiński, Kankiewicz, Klotz, Koff, Kryszczyńska, Kwiatkowski, Laffont, Leroy, Lecacheux, Leonie, Leyrat, Manzini, Martin, Masi, Matter, Michałowski, Michałowski, Michałowski, Michelet, Michelsen, Morelle, Mottola, Naves, Nomen, Oey, Ogłoza, Oksanen, Oszkiewicz, Pääkkönen, Paiella, Pallares, Paulo, Pavic, Payet, Polińska, Polishook, Poncy, Revaz, Rinner, Rocca, Roche, Romeuf, Roy, Saguin, Salom, Sanchez, Santacana, Santana-Ros, Sareyan, Sobkowiak,
  Sposetti, Starkey, Stoss, Strajnic, Teng, Trégon, Vagnozzi, Velichko, Waelchli, Wagrez, \& Wücher}]{Hanu__2013}
Hanuš, J., Ďurech, J., Brož, M., {et~al.} 2013, \aap, 551, A67

\bibitem[{{Hapke}(1981)}]{hapke1981}
{Hapke}, B. 1981, \jgr, 86, 3039

\bibitem[{{Hapke}(1984)}]{hapke1984}
{Hapke}, B. 1984, \icarus, 59, 41

\bibitem[{{Hapke}(1986)}]{hapke1986}
{Hapke}, B. 1986, \icarus, 67, 264

\bibitem[{{Harris}(1998)}]{neatm}
{Harris}, A.~W. 1998, \icarus, 131, 291

\bibitem[{{Harris} \& {Drube}(2016)}]{harris2016}
{Harris}, A.~W. \& {Drube}, L. 2016, \apj, 832, 127

\bibitem[{Harris \& Harris(1997)}]{harris_revision_1997}
Harris, A.~W. \& Harris, A.~W. 1997, Icarus, 126, 450

\bibitem[{{Harris} \& {Lagerros}(2002)}]{harrisetlagerros}
{Harris}, A.~W. \& {Lagerros}, J.~S.~V. 2002, in Asteroids III, ed. W.~F. {Bottke}, Jr., A.~{Cellino}, P.~{Paolicchi}, \& R.~P. {Binzel}, 205--218

\bibitem[{{Herald} {et~al.}(2020){Herald}, {Gault}, {Anderson}, {Dunham}, {Frappa}, {Hayamizu}, {Kerr}, {Miyashita}, {Moore}, {Pavlov}, {Preston}, {Talbot}, \& {Timerson}}]{herald2020}
{Herald}, D., {Gault}, D., {Anderson}, R., {et~al.} 2020, \mnras, 499, 4570

\bibitem[{{Herald} {et~al.}(2024){Herald}, {Gault}, {Carlson}, {Guhl}, {Frappa}, {Giacchini}, {Hayamizu}, {Kerr}, \& {Moore}}]{Herald2024}
{Herald}, D., {Gault}, D., {Carlson}, N., {et~al.} 2024, {Small Bodies Occultations Bundle V4.0}, NASA Planetary Data System, urn:nasa:pds:smallbodiesoccultations::4.0

\bibitem[{{Hung} {et~al.}(2022){Hung}, {Hanu{\v{s}}}, {Masiero}, \& {Tholen}}]{Hung2022}
{Hung}, D., {Hanu{\v{s}}}, J., {Masiero}, J.~R., \& {Tholen}, D.~J. 2022, \psj, 3, 56

\bibitem[{{Jiang} {et~al.}(2020){Jiang}, {Ji}, \& {Yu}}]{jiang2020}
{Jiang}, H., {Ji}, J., \& {Yu}, L. 2020, \aj, 159, 264

\bibitem[{Kaasalainen(2011)}]{Kaasalainen2011}
Kaasalainen, M. 2011, Multimodal inverse problems: Maximum compatibility estimate and shape reconstruction

\bibitem[{{Kaasalainen} \& {\v D}urech(2020)}]{2020arXiv200509947K}
{Kaasalainen}, M. \& {\v D}urech, J. 2020, arXiv e-prints, arXiv:2005.09947

\bibitem[{Kaasalainen \& Lamberg(2006)}]{Kaasalainen_2006}
Kaasalainen, M. \& Lamberg, L. 2006, Inverse Problems, 22, 749

\bibitem[{Kaasalainen \& Torppa(2001)}]{kaasalainen_optimization_2001}
Kaasalainen, M. \& Torppa, J. 2001, \icarus, 153, 24

\bibitem[{Kaasalainen {et~al.}(2001)Kaasalainen, Torppa, \& Muinonen}]{kaasalainen_optimization_2001-1}
Kaasalainen, M., Torppa, J., \& Muinonen, K. 2001, \icarus, 153, 37

\bibitem[{{Kaasalainen} \& {Viikinkoski}(2012)}]{Kaasalainen2012}
{Kaasalainen}, M. \& {Viikinkoski}, M. 2012, \aap, 543, A97

\bibitem[{Lagerros(1996{\natexlab{a}})}]{lagerros_thermal_1996}
Lagerros, J.~S.~V. 1996{\natexlab{a}}, \aap, 310, 1011

\bibitem[{Lagerros(1996{\natexlab{b}})}]{lagerros_thermal_1996-1}
Lagerros, J.~S.~V. 1996{\natexlab{b}}, \aap, 315, 625

\bibitem[{Lagerros(1997)}]{lagerros_thermal_1997}
Lagerros, J.~S.~V. 1997, in AAS/Division for Planetary Sciences Meeting Abstracts, Vol.~29, AAS/Division for Planetary Sciences Meeting Abstracts \#29, 07.23

\bibitem[{Lagerros(1998)}]{lagerros_thermal_1998}
Lagerros, J. S.~V. 1998, \aap, 332, 1123

\bibitem[{{Lang}(1999)}]{lang1999}
{Lang}, K.~R. 1999, {Astrophysical formulae}

\bibitem[{Leiva {et~al.}(2017)Leiva, Sicardy, Camargo, Ortiz, Desmars, Bérard, Lellouch, Meza, Kervella, Snodgrass, Duffard, Morales, Gomes-Júnior, Benedetti-Rossi, Vieira-Martins, Braga-Ribas, Assafin, Morgado, Colas, De~Witt, Sickafoose, Breytenbach, Dauvergne, Schoenau, Maquet, Bath, Bode, Cool, Lade, Kerr, \& Herald}]{Leiva2017}
Leiva, R., Sicardy, B., Camargo, J. I.~B., {et~al.} 2017, The Astronomical Journal, 154, 159

\bibitem[{{Lu} {et~al.}(2013){Lu}, {Zhao}, \& {You}}]{ellipsoide}
{Lu}, X.-P., {Zhao}, H.-B., \& {You}, Z. 2013, Research in Astronomy and Astrophysics, 13, 471

\bibitem[{{MacLennan} \& {Emery}(2019)}]{maclennan2019}
{MacLennan}, E.~M. \& {Emery}, J.~P. 2019, \aj, 157, 2

\bibitem[{{MacLennan} \& {Emery}(2021)}]{maclennan2021}
{MacLennan}, E.~M. \& {Emery}, J.~P. 2021, \psj, 2, 161

\bibitem[{{Mainzer} {et~al.}(2011){Mainzer}, {Bauer}, {Grav}, {Masiero}, {Cutri}, {Dailey}, {Eisenhardt}, {McMillan}, {Wright}, {Walker}, {Jedicke}, {Spahr}, {Tholen}, {Alles}, {Beck}, {Brandenburg}, {Conrow}, {Evans}, {Fowler}, {Jarrett}, {Marsh}, {Masci}, {McCallon}, {Wheelock}, {Wittman}, {Wyatt}, {DeBaun}, {Elliott}, {Elsbury}, {Gautier}, {Gomillion}, {Leisawitz}, {Maleszewski}, {Micheli}, \& {Wilkins}}]{neowise}
{Mainzer}, A., {Bauer}, J., {Grav}, T., {et~al.} 2011, \apj, 731, 53

\bibitem[{{Marciniak} {et~al.}(2019){Marciniak}, {Al{\'\i}-Lagoa}, {M{\"u}ller}, {Szak{\'a}ts}, {Moln{\'a}r}, {P{\'a}l}, {Podlewska-Gaca}, {Parley}, {Antonini}, {Barbotin}, {Behrend}, {Bernasconi}, {Butkiewicz-B{\k{a}}k}, {Crippa}, {Duffard}, {Ditteon}, {Feuerbach}, {Fauvaud}, {Garlitz}, {Geier}, {Goncalves}, {Grice}, {Grze{\'s}kowiak}, {Hirsch}, {Horbowicz}, {Kami{\'n}ski}, {Kami{\'n}ska}, {Kim}, {Kim}, {Konstanciak}, {Kudak}, {Kulczak}, {Maestre}, {Manzini}, {Marks}, {Monteiro}, {Og{\l}oza}, {Oszkiewicz}, {Pilcher}, {Perig}, {Polakis}, {Poli{\'n}ska}, {Roy}, {Sanabria}, {Santana-Ros}, {Skiff}, {Skrzypek}, {Sobkowiak}, {Sonbas}, {Thizy}, {Trela}, {Urakawa}, {{\.Z}ejmo}, \& {{\.Z}ukowski}}]{M2019}
{Marciniak}, A., {Al{\'\i}-Lagoa}, V., {M{\"u}ller}, T.~G., {et~al.} 2019, \aap, 625, A139

\bibitem[{{Marciniak} {et~al.}(2018){Marciniak}, {Bartczak}, {M{\"u}ller}, {Sanabria}, {Al{\'{\i}}-Lagoa}, {Antonini}, {Behrend}, {Bernasconi}, {Bronikowska}, {Butkiewicz-B{\c a}k}, {Cikota}, {Crippa}, {Ditteon}, {Dudzi{\'n}ski}, {Duffard}, {Dziadura}, {Fauvaud}, {Geier}, {Hirsch}, {Horbowicz}, {Hren}, {Jerosimic}, {Kami{\'n}ski}, {Kankiewicz}, {Konstanciak}, {Korlevic}, {Kosturkiewicz}, {Kudak}, {Manzini}, {Morales}, {Murawiecka}, {Og{\l}oza}, {Oszkiewicz}, {Pilcher}, {Polakis}, {Poncy}, {Santana-Ros}, {Siwak}, {Skiff}, {Sobkowiak}, {Stoss}, {{\.Z}ejmo}, \& {{\.Z}ukowski}}]{M2018}
{Marciniak}, A., {Bartczak}, P., {M{\"u}ller}, T., {et~al.} 2018, \aap, 610, A7

\bibitem[{{Marciniak} {et~al.}(2021){Marciniak}, {\v D}urech, {Al{\'\i}-Lagoa}, {Og{\l}oza}, {Szak{\'a}ts}, {M{\"u}ller}, {Moln{\'a}r}, {P{\'a}l}, {Monteiro}, {Arcoverde}, {Behrend}, {Benkhaldoun}, {Bernasconi}, {Bosch}, {Brincat}, {Brunetto}, {Butkiewicz-B{\k{a}}k}, {Del Freo}, {Duffard}, {Evangelista-Santana}, {Farroni}, {Fauvaud}, {Fauvaud}, {Ferrais}, {Geier}, {Golonka}, {Grice}, {Hirsch}, {Horbowicz}, {Jehin}, {Julien}, {Kalup}, {Kami{\'n}ski}, {Kami{\'n}ska}, {Kankiewicz}, {Kecskem{\'e}thy}, {Kim}, {Kim}, {Konstanciak}, {Krajewski}, {Kudak}, {Kulczak}, {Kundera}, {Lazzaro}, {Manzini}, {Medeiros}, {Michimani-Garcia}, {Morales}, {Nadolny}, {Oszkiewicz}, {Pak{\v{s}}tien{\.{e}}}, {Paw{\l}owski}, {Perig}, {Pilcher}, {Pinel}, {Podlewska-Gaca}, {Polakis}, {Richard}, {Rodrigues}, {Rond{\'o}n}, {Roy}, {Sanabria}, {Santana-Ros}, {Skiff}, {Skrzypek}, {Sobkowiak}, {Sonbas}, {Stachowski}, {Strajnic}, {Trela}, {Tychoniec}, {Urakawa}, {Verebelyi}, {Wagrez}, {{\.Z}ejmo}, \& {{\.Z}ukowski}}]{marciniak_properties_2021}
{Marciniak}, A., {\v D}urech, J., {Al{\'\i}-Lagoa}, V., {et~al.} 2021, \aap, 654, A87

\bibitem[{{Marciniak} {et~al.}(2023){Marciniak}, {\v D}urech, {Choukroun}, {Hanu{\v{s}}}, {Og{\l}oza}, {Szak{\'a}ts}, {Moln{\'a}r}, {P{\'a}l}, {Monteiro}, {Frappa}, {Beisker}, {Pavlov}, {Moore}, {Adomavi{\v{c}}ien{\.{e}}}, {Aikawa}, {Andersson}, {Antonini}, {Argentin}, {Asai}, {Assoignon}, {Barton}, {Baruffetti}, {Bath}, {Behrend}, {Benedyktowicz}, {Bernasconi}, {Biguet}, {Billiani}, {B{\l}a{\.z}ewicz}, {Boninsegna}, {Borkowski}, {Bosch}, {Brazill}, {Bronikowska}, {Bruno}, {Butkiewicz-B{\k{a}}k}, {Caron}, {Casalnuovo}, {Castellani}, {Ceravolo}, {Conjat}, {Delincak}, {Delpau}, {Demeautis}, {Demirkol}, {Dr{\'o}{\.z}d{\.z}}, {Duffard}, {Durandet}, {Eisfeldt}, {Evangelista}, {Fauvaud}, {Fauvaud}, {Ferrais}, {Filipek}, {Fini}, {Fukui}, {G{\"a}hrken}, {Geier}, {George}, {Goffin}, {Golonka}, {Goto}, {Grice}, {Guhl}, {Hal{\'\i}{\v{r}}}, {Hanna}, {Harman}, {Hashimoto}, {Hasubick}, {Higgins}, {Higuchi}, {Hirose}, {Hirsch}, {Hofschulz}, {Horaguchi}, {Horbowicz}, {Ida}, {Ign{\'a}cz}, {Ishida}, {Isobe}, {Jehin},
  {Joachimczyk}, {Jones}, {Juan}, {Kami{\'n}ski}, {Kami{\'n}ska}, {Kankiewicz}, {Kasebe}, {Kattentidt}, {Kim}, {Kim}, {Kitazaki}, {Klotz}, {Komraus}, {Konstanciak}, {K{\"o}nyves-T{\'o}th}, {Kouno}, {Kowald}, {Krajewski}, {Krannich}, {Kreutzer}, {Kryszczy{\'n}ska}, {Kub{\'a}nek}, {Kudak}, {Kugel}, {Kukita}, {Kulczak}, {Lazzaro}, {Licandro}, {Livet}, {Maley}, {Manago}, {M{\'a}nek}, {Manna}, {Matsushita}, {Meister}, {Mesquita}, {Messner}, {Michelet}, {Michimani}, {Mieczkowska}, {Morales}, {Motyli{\'n}ski}, {Murawiecka}, {Newman}, {Nikitin}, {Nishimura}, {Oey}, {Oszkiewicz}, {Owada}, {Pak{\v{s}}tien{\.{e}}}, {Paw{\l}owski}, {Pereira}, {Perig}, {Per{\l}a}, {Pilcher}, {Podlewska-Gaca}, {Pol{\'a}k}, {Polakis}, {Poli{\'n}ska}, {Popowicz}, {Richard}, {Rives}, {Rodrigues}, {Rogi{\'n}ski}, {Rond{\'o}n}, {Rottenborn}, {Sch{\"a}fer}, {Schnabel}, {Schreurs}, {Selva}, {Simon}, {Skiff}, {Skrutskie}, {Skrzypek}, {Sobkowiak}, {Sonbas}, {Sposetti}, {Stuart}, {Szyszka}, {Terakubo}, {Thomas}, {Trela}, {Uchiyama}, {Urbanik},
  {Vaudescal}, {Venable}, {Watanabe}, {Watanabe}, {Winiarski}, {Wr{\'o}blewski}, {Yamamura}, {Yamashita}, {Yoshihara}, {Zawilski}, {Zelen{\'y}}, {{\.Z}ejmo}, {{\.Z}ukowski}, \& {{\.Z}ywica}}]{M2023}
{Marciniak}, A., {\v D}urech, J., {Choukroun}, A., {et~al.} 2023, \aap, 679, A60

\bibitem[{{Marciniak} {et~al.}(2015){Marciniak}, {Pilcher}, {Oszkiewicz}, {Santana-Ros}, {Urakawa}, {Fauvaud}, {Kankiewicz}, {Tychoniec}, {Fauvaud}, {Hirsch}, {Horbowicz}, {Kami{\'n}ski}, {Konstanciak}, {Kosturkiewicz}, {Murawiecka}, {Nadolny}, {Nishiyama}, {Okumura}, {Poli{\'n}ska}, {Richard}, {Sakamoto}, {Sobkowiak}, {Stachowski}, \& {Trela}}]{M2015}
{Marciniak}, A., {Pilcher}, F., {Oszkiewicz}, D., {et~al.} 2015, \planss, 118, 256

\bibitem[{{Marsset} {et~al.}(2020){Marsset}, {Bro{\v{z}}}, {Vernazza}, {Drouard}, {Castillo-Rogez}, {Hanu{\v{s}}}, {Viikinkoski}, {Rambaux}, {Carry}, {Jorda}, {{\v{S}}eve{\v{c}}ek}, {Birlan}, {Marchis}, {Podlewska-Gaca}, {Asphaug}, {Bartczak}, {Berthier}, {Cipriani}, {Colas}, {Dudzi{\'n}ski}, {Dumas}, {Durech}, {Ferrais}, {F{\'e}tick}, {Fusco}, {Jehin}, {Kaasalainen}, {Kryszczynska}, {Lamy}, {Le Coroller}, {Marciniak}, {Michalowski}, {Michel}, {Richardson}, {Santana-Ros}, {Tanga}, {Vachier}, {Vigan}, {Witasse}, \& {Yang}}]{marsset2020}
{Marsset}, M., {Bro{\v{z}}}, M., {Vernazza}, P., {et~al.} 2020, Nature Astronomy, 4, 569

\bibitem[{{Marsset} {et~al.}(2017){Marsset}, {Carry}, {Dumas}, {Hanu{\v{s}}}, {Viikinkoski}, {Vernazza}, {M{\"u}ller}, {Delbo}, {Jehin}, {Gillon}, {Grice}, {Yang}, {Fusco}, {Berthier}, {Sonnett}, {Kugel}, {Caron}, \& {Behrend}}]{marsset2017}
{Marsset}, M., {Carry}, B., {Dumas}, C., {et~al.} 2017, \aap, 604, A64

\bibitem[{{Masiero} {et~al.}(2014){Masiero}, {Grav}, {Mainzer}, {Nugent}, {Bauer}, {Stevenson}, \& {Sonnett}}]{Masiero2014}
{Masiero}, J.~R., {Grav}, T., {Mainzer}, A.~K., {et~al.} 2014, \apj, 791, 121

\bibitem[{{Masiero} {et~al.}(2020){Masiero}, {Mainzer}, {Bauer}, {Cutri}, {Grav}, {Kramer}, {Pittichov{\'a}}, {Sonnett}, \& {Wright}}]{masiero2020}
{Masiero}, J.~R., {Mainzer}, A.~K., {Bauer}, J.~M., {et~al.} 2020, \psj, 1, 5

\bibitem[{{Masiero} {et~al.}(2021){Masiero}, {Mainzer}, {Bauer}, {Cutri}, {Grav}, {Kramer}, {Pittichov{\'a}}, \& {Wright}}]{masiero2021}
{Masiero}, J.~R., {Mainzer}, A.~K., {Bauer}, J.~M., {et~al.} 2021, \psj, 2, 162

\bibitem[{{Masiero} {et~al.}(2011){Masiero}, {Mainzer}, {Grav}, {Bauer}, {Cutri}, {Dailey}, {Eisenhardt}, {McMillan}, {Spahr}, {Skrutskie}, {Tholen}, {Walker}, {Wright}, {DeBaun}, {Elsbury}, {Gautier}, {Gomillion}, \& {Wilkins}}]{masiero2011}
{Masiero}, J.~R., {Mainzer}, A.~K., {Grav}, T., {et~al.} 2011, \apj, 741, 68

\bibitem[{{Masiero} {et~al.}(2012){Masiero}, {Mainzer}, {Grav}, {Bauer}, {Cutri}, {Nugent}, \& {Cabrera}}]{masiero2012}
{Masiero}, J.~R., {Mainzer}, A.~K., {Grav}, T., {et~al.} 2012, \apjl, 759, L8

\bibitem[{{Masiero} {et~al.}(2017){Masiero}, {Nugent}, {Mainzer}, {Wright}, {Bauer}, {Cutri}, {Grav}, {Kramer}, \& {Sonnett}}]{masiero2017}
{Masiero}, J.~R., {Nugent}, C., {Mainzer}, A.~K., {et~al.} 2017, \aj, 154, 168

\bibitem[{{Matter} {et~al.}(2013){Matter}, {Delbo}, {Carry}, \& {Ligori}}]{matter2013}
{Matter}, A., {Delbo}, M., {Carry}, B., \& {Ligori}, S. 2013, \icarus, 226, 419

\bibitem[{{Morbidelli} {et~al.}(2009){Morbidelli}, {Bottke}, {Nesvorn{\'y}}, \& {Levison}}]{2009Icar..204..558M}
{Morbidelli}, A., {Bottke}, W.~F., {Nesvorn{\'y}}, D., \& {Levison}, H.~F. 2009, \icarus, 204, 558

\bibitem[{{Mueller} {et~al.}(2010){Mueller}, {Marchis}, {Emery}, {Harris}, {Mottola}, {Hestroffer}, {Berthier}, \& {di Martino}}]{mueller2010}
{Mueller}, M., {Marchis}, F., {Emery}, J.~P., {et~al.} 2010, \icarus, 205, 505

\bibitem[{Müller {et~al.}(2018)Müller, Marciniak, Kiss, Duffard, Alí-Lagoa, Bartczak, Butkiewicz-Bąk, Dudziński, Fernández-Valenzuela, Marton, Morales, Ortiz, Oszkiewicz, Santana-Ros, Szakáts, Santos-Sanz, {Takácsné Farkas}, \& Varga-Verebélyi}]{muller2018}
Müller, T., Marciniak, A., Kiss, C., {et~al.} 2018, Advances in Space Research, 62, 2326, past, Present and Future of Small Body Science and Exploration

\bibitem[{{Neugebauer} {et~al.}(1984){Neugebauer}, {Habing}, {van Duinen}, {Aumann}, {Baud}, {Beichman}, {Beintema}, {Boggess}, {Clegg}, {de Jong}, {Emerson}, {Gautier}, {Gillett}, {Harris}, {Hauser}, {Houck}, {Jennings}, {Low}, {Marsden}, {Miley}, {Olnon}, {Pottasch}, {Raimond}, {Rowan-Robinson}, {Soifer}, {Walker}, {Wesselius}, \& {Young}}]{IRAS}
{Neugebauer}, G., {Habing}, H.~J., {van Duinen}, R., {et~al.} 1984, \apjl, 278, L1

\bibitem[{{Nugent} {et~al.}(2016){Nugent}, {Mainzer}, {Bauer}, {Cutri}, {Kramer}, {Grav}, {Masiero}, {Sonnett}, \& {Wright}}]{nugent2016}
{Nugent}, C.~R., {Mainzer}, A., {Bauer}, J., {et~al.} 2016, \aj, 152, 63

\bibitem[{{Pilcher}(2014)}]{2014Pilcher}
{Pilcher}, F. 2014, Minor Planet Bulletin, 41, 73

\bibitem[{{Podlewska-Gaca} {et~al.}(2020){Podlewska-Gaca}, {Marciniak}, {Al{\'\i}-Lagoa}, {Bartczak}, {M{\"u}ller}, {Szak{\'a}ts}, {Duffard}, {Moln{\'a}r}, {P{\'a}l}, {Butkiewicz-B{\k{a}}k}, {Dudzi{\'n}ski}, {Dziadura}, {Antonini}, {Asenjo}, {Audejean}, {Benkhaldoun}, {Behrend}, {Bernasconi}, {Bosch}, {Chapman}, {Dintinjana}, {Farkas}, {Ferrais}, {Geier}, {Grice}, {Hirsh}, {Jacquinot}, {Jehin}, {Jones}, {Molina}, {Morales}, {Parley}, {Poncy}, {Roy}, {Santana-Ros}, {Seli}, {Sobkowiak}, {Vereb{\'e}lyi}, \& {{\.Z}ukowski}}]{edyta2020}
{Podlewska-Gaca}, E., {Marciniak}, A., {Al{\'\i}-Lagoa}, V., {et~al.} 2020, \aap, 638, A11

\bibitem[{{Rommel} {et~al.}(2020){Rommel}, {Braga-Ribas}, {Desmars}, {Camargo}, {Ortiz}, {Sicardy}, {Vieira-Martins}, {Assafin}, {Santos-Sanz}, {Duffard}, {Fern{\'a}ndez-Valenzuela}, {Lecacheux}, {Morgado}, {Benedetti-Rossi}, {Gomes-J{\'u}nior}, {Pereira}, {Herald}, {Hanna}, {Bradshaw}, {Morales}, {Brimacombe}, {Burtovoi}, {Carruthers}, {de Barros}, {Fiori}, {Gilmore}, {Hooper}, {Hornoch}, {Jacques}, {Janik}, {Kerr}, {Kilmartin}, {Winkel}, {Naletto}, {Nardiello}, {Nascimbeni}, {Newman}, {Ossola}, {P{\'a}l}, {Pimentel}, {Pravec}, {Sposetti}, {Stechina}, {Szak{\'a}ts}, {Ueno}, {Zampieri}, {Broughton}, {Dunham}, {Dunham}, {Gault}, {Hayamizu}, {Hosoi}, {Jehin}, {Jones}, {Kitazaki}, {Kom{\v{z}}{\'\i}k}, {Marciniak}, {Maury}, {Miku{\v{z}}}, {Nosworthy}, {F{\'a}brega Polleri}, {Rahvar}, {Sfair}, {Siqueira}, {Snodgrass}, {Sogorb}, {Tomioka}, {Tregloan-Reed}, \& {Winter}}]{rommel2020}
{Rommel}, F.~L., {Braga-Ribas}, F., {Desmars}, J., {et~al.} 2020, \aap, 644, A40

\bibitem[{Rommel {et~al.}(2023)Rommel, {Braga-Ribas, F.}, {Ortiz, J. L.}, {Sicardy, B.}, {Santos-Sanz, P.}, {Desmars, J.}, {Camargo, J. I. B.}, {Vieira-Martins, R.}, {Assafin, M.}, {Morgado, B. E.}, {Boufleur, R. C.}, {Benedetti-Rossi, G.}, {Gomes-Júnior, A. R.}, {Fernández-Valenzuela, E.}, {Holler, B. J.}, {Souami, D.}, {Duffard, R.}, {Margoti, G.}, {Vara-Lubiano, M.}, {Lecacheux, J.}, {Plouvier, J. L.}, {Morales, N.}, {Maury, A.}, {Fabrega, J.}, {Ceravolo, P.}, {Jehin, E.}, {Albanese, D.}, {Mariey, H.}, {Cikota, S.}, {Ruždjak, D.}, {Cikota, A.}, {Szakáts, R.}, {Baba Aissa, D.}, {Gringahcene, Z.}, {Kashuba, V.}, {Koshkin, N.}, {Zhukov, V.}, {Fişek, S.}, {Çakir, O.}, {Özer, S.}, {Schnabel, C.}, {Schnabel, M.}, {Signoret, F.}, {Morrone, L.}, {Santana-Ros, T.}, {Pereira, C. L.}, {Emilio, M.}, {Burdanov, A. Y.}, {de Wit, J.}, {Barkaoui, K.}, {Gillon, M.}, {Leto, G.}, {Frasca, A.}, {Catanzaro, G.}, {Sanchez, R. Zanmar}, {Tagliaferri, U.}, {Di Sora, M.}, {Isopi, G.}, {Krugly, Y.}, {Slyusarev, I.}, {Chiorny,
  V.}, {Mikuž, H.}, {Bacci, P.}, {Maestripieri, M.}, {Grazia, M. D.}, {de la Cueva, I.}, {Yuste-Moreno, M.}, {Ciabattari, F.}, {Kozhukhov, O. M.}, {Serra-Ricart, M.}, {Alarcon, M. R.}, {Licandro, J.}, {Masi, G.}, {Bacci, R.}, {Bosch, J. M.}, {Behem, R.}, {Prost, J.-P.}, {Renner, S.}, {Conjat, M.}, {Bachini, M.}, {Succi, G.}, {Stoian, L.}, {Juravle, A.}, {Carosati, D.}, {Gowe, B.}, {Carrillo, J.}, {Zheleznyak, A. P.}, {Montigiani, N.}, {Foster, C. R.}, {Mannucci, M.}, {Ruocco, N.}, {Cuevas, F.}, {Di Marcantonio, P.}, {Coretti, I.}, {Iafrate, G.}, {Baldini, V.}, {Collins, M.}, {Pál, A.}, {Csák, B.}, {Fernández-Garcia, E.}, {Castro-Tirado, A. J.}, {Hudin, L.}, {Madiedo, J. M.}, {Anghel, R. M.}, {Calvo-Fernández, J. F.}, {Valvasori, A.}, {Guido, E.}, {Gherase, R. M.}, {Kamoun, S.}, {Fafet, R.}, {Sánchez-González, M.}, {Curelaru, L.}, {Vîntdevară, C. D.}, {Danescu, C. A.}, {Gout, J.-F.}, {Schmitz, C. J.}, {Sota, A.}, {Belskaya, I.}, {Rodríguez-Marco, M.}, {Kilic, Y.}, {Frappa, E.}, {Klotz, A.},
  {Lavayssière, M.}, {Oliveira, J. Marques}, {Popescu, M.}, {Mammana, L. A.}, {Fernández-Lajús, E.}, {Schmidt, M.}, {Hopp, U.}, {Komžík, R.}, {Pribulla, T.}, {Tomko, D.}, {Husárik, M.}, {Erece, O.}, {Eryilmaz, S.}, {Buzzi, L.}, {Gährken, B.}, {Nardiello, D.}, {Hornoch, K.}, {Sonbas, E.}, {Er, H.}, {Burwitz, V.}, {Sybilski, P. Waldemar}, {Bykowski, W.}, {Müller, T. G.}, {Ogloza, W.}, {Gonçalves, R.}, {Ferreira, J. F.}, {Ferreira, M.}, {Bento, M.}, {Meister, S.}, {Bagiran, M. N.}, {Tekeş, M.}, {Marciniak, A.}, {Moravec, Z.}, {Delinčák, P.}, {Gianni, G.}, {Casalnuovo, G. B.}, {Boutet, M.}, {Sanchez, J.}, {Klemt, B.}, {Wuensche, N.}, {Burzynski, W.}, {Borkowski, M.}, {Serrau, M.}, {Dangl, G.}, {Klös, O.}, {Weber, C.}, {Urbaník, M.}, {Rousselot, L.}, {Kubánek, J.}, {André, P.}, {Colazo, C.}, {Spagnotto, J.}, {Sickafoose, A. A.}, {Hueso, R.}, {Sánchez-Lavega, A.}, {Fisher, R. S.}, {Rengstorf, A. W.}, {Perelló, C.}, {Dascalu, M.}, {Altan, M.}, {Gazeas, K.}, {de Santana, T.}, {Sfair, R.}, {Winter,
  O. C.}, {Kalkan, S.}, {Canales-Moreno, O.}, {Trigo-Rodríguez, J. M.}, {Tsamis, V.}, {Tigani, K.}, {Sioulas, N.}, {Lekkas, G.}, {Bertesteanu, D. N.}, {Dumitrescu, V.}, {Wilberger, A. J.}, {Barnes, J. W.}, {Fieber-Beyer, S. K.}, {Swaney, R. L.}, {Fuentes, C.}, {Mendez, R. A.}, {Dumitru, B. D.}, {Flynn, R. L.}, \& {Wake, D. A.}}]{rommel2023}
Rommel, F.~L., {Braga-Ribas, F.}, {Ortiz, J. L.}, {et~al.} 2023, \aap, 678, A167

\bibitem[{Rozitis \& Green(2011)}]{ATPM}
Rozitis, B. \& Green, S.~F. 2011, Monthly Notices of the Royal Astronomical Society, 415, 2042

\bibitem[{{Ryan} \& {Woodward}(2010)}]{ryan2010}
{Ryan}, E.~L. \& {Woodward}, C.~E. 2010, \aj, 140, 933

\bibitem[{{Sat{\={o}}} {et~al.}(2015){Sat{\={o}}}, {Hamanowa}, {Tomioka}, \& {Uehara}}]{Sato2015}
{Sat{\={o}}}, I., {Hamanowa}, H., {Tomioka}, H., \& {Uehara}, S. 2015, International Journal of Astronomy and Astrophysics, 5, 193

\bibitem[{Shannon(1949)}]{Shannon1949}
Shannon, C. 1949, Proceedings of the {IRE}, 37, 10

\bibitem[{{Szak{\'a}ts} {et~al.}(2020){Szak{\'a}ts}, {M{\"u}ller}, {Al{\'\i}-Lagoa}, {Marton}, {Farkas-Tak{\'a}cs}, {B{\'a}nyai}, \& {Kiss}}]{Szakats}
{Szak{\'a}ts}, R., {M{\"u}ller}, T., {Al{\'\i}-Lagoa}, V., {et~al.} 2020, \aap, 635, A54

\bibitem[{{Tanga} {et~al.}(2015){Tanga}, {Carry}, {Colas}, {Delbo}, {Matter}, {Hanu{\v{s}}}, {Al{\'\i} Lagoa}, {Andrei}, {Assafin}, {Audejean}, {Behrend}, {Camargo}, {Carbognani}, {Cedr{\'e}s Reyes}, {Conjat}, {Cornero}, {Coward}, {Crippa}, {de Ferra Fantin}, {Devog{\'e}le}, {Dubos}, {Frappa}, {Gillon}, {Hamanowa}, {Jehin}, {Klotz}, {Kryszczy{\'n}ska}, {Lecacheux}, {Leroy}, {Manfroid}, {Manzini}, {Maquet}, {Morelle}, {Mottola}, {Poli{\'n}ska}, {Roy}, {Todd}, {Vachier}, {Vera Hern{\'a}ndez}, \& {Wiggins}}]{tanga2015}
{Tanga}, P., {Carry}, B., {Colas}, F., {et~al.} 2015, \mnras, 448, 3382

\bibitem[{Tanga {et~al.}(2013)Tanga, Delbo', \& Gerakis}]{tanga_mp3c_2013}
Tanga, P., Delbo', M., \& Gerakis, J. 2013, in {AAS}/{Division} for {Planetary} {Sciences} {Meeting} {Abstracts}, Vol.~45, {AAS}/{Division} for {Planetary} {Sciences} {Meeting} {Abstracts} \#45, 208.29

\bibitem[{{Tedesco} {et~al.}(2002){Tedesco}, {Noah}, {Noah}, \& {Price}}]{tedesco2002}
{Tedesco}, E.~F., {Noah}, P.~V., {Noah}, M., \& {Price}, S.~D. 2002, \aj, 123, 1056

\bibitem[{{Tholen}(1984)}]{tholen1984}
{Tholen}, D.~J. 1984, PhD thesis, University of Arizona

\bibitem[{{Usui} {et~al.}(2011){Usui}, {Kuroda}, {M{\"u}ller}, {Hasegawa}, {Ishiguro}, {Ootsubo}, {Ishihara}, {Kataza}, {Takita}, {Oyabu}, {Ueno}, {Matsuhara}, \& {Onaka}}]{AKARI}
{Usui}, F., {Kuroda}, D., {M{\"u}ller}, T.~G., {et~al.} 2011, \pasj, 63, 1117

\bibitem[{{Vere{\v{s}}} {et~al.}(2015){Vere{\v{s}}}, {Jedicke}, {Fitzsimmons}, {Denneau}, {Granvik}, {Bolin}, {Chastel}, {Wainscoat}, {Burgett}, {Chambers}, {Flewelling}, {Kaiser}, {Magnier}, {Morgan}, {Price}, {Tonry}, \& {Waters}}]{Vere2015}
{Vere{\v{s}}}, P., {Jedicke}, R., {Fitzsimmons}, A., {et~al.} 2015, \icarus, 261, 34

\bibitem[{{Warner} {et~al.}(2009){Warner}, {Harris}, \& {Pravec}}]{LCDB}
{Warner}, B.~D., {Harris}, A.~W., \& {Pravec}, P. 2009, \icarus, 202, 134

\bibitem[{{Wright} {et~al.}(2010){Wright}, {Eisenhardt}, {Mainzer}, {Ressler}, {Cutri}, {Jarrett}, {Kirkpatrick}, {Padgett}, {McMillan}, {Skrutskie}, {Stanford}, {Cohen}, {Walker}, {Mather}, {Leisawitz}, {Gautier}, {McLean}, {Benford}, {Lonsdale}, {Blain}, {Mendez}, {Irace}, {Duval}, {Liu}, {Royer}, {Heinrichsen}, {Howard}, {Shannon}, {Kendall}, {Walsh}, {Larsen}, {Cardon}, {Schick}, {Schwalm}, {Abid}, {Fabinsky}, {Naes}, \& {Tsai}}]{wise}
{Wright}, E.~L., {Eisenhardt}, P. R.~M., {Mainzer}, A.~K., {et~al.} 2010, \aj, 140, 1868

\bibitem[{{Yu} {et~al.}(2017){Yu}, {Yang}, {Ji}, \& {Ip}}]{yu2017}
{Yu}, L.~L., {Yang}, B., {Ji}, J., \& {Ip}, W.-H. 2017, \mnras, 472, 2388

\end{thebibliography}
\onecolumn

\begin{appendix}

\section{Published TPM and Occultation values}

\renewcommand{\arraystretch}{0.8}
\begin{longtable}{|c|c|c|c|c|c}
    
    \caption{Diameters from the literature obtained through TPM and those determined with the use of stellar occultations.} \\ 
    \hline
    Target & D\textsubscript{tpm} & references\textsubscript{TPM} & D\textsubscript{occ} & references\textsubscript{occ}\\ 
    \hline
    \endfirsthead
    \caption*{Table~\ref{tab:literature} continued} \\ 
    \hline
    Target & D\textsubscript{tpm} & references\textsubscript{TPM} & D\textsubscript{occ} & references\textsubscript{occ}\\ 
    \hline
    \endhead
    \hline \multicolumn{3}{r}{\textit{(Continued on next page...)}} \\
    \hline
    \endfoot
    \hline
    \endlastfoot
    \label{tab:literature}
    
(2) Pallas & $536.0 \pm 5.0$ & \cite{alilagoa2020} & $513.0 \pm 6.0$ & \cite{marsset2020} \\ 
(3) Juno & $254.0 \pm 4.0$ & \cite{alilagoa2020} & $241.5 \pm 12.7$ & \cite{herald2020} \\ 
(5) Astraea & $103.8_{-2.9}^{+4.8}$ & \cite{Hung2022} & $113.0 \pm 1.0$ & \cite{herald2020} \\ 
(6) Hebe & $198.0^{+4.0}_{-2.0}$ & \cite{marsset2017} & $193.0 \pm 6.0$ & \cite{marsset2017} \\ 
(8) Flora & $142.0 \pm 2.0$ & \cite{alilagoa2020} & $147.0 \pm 4.7$ & \cite{herald2020} \\ 
(10) Hygeia & $441.0 \pm 6.0$ & \cite{alilagoa2020} & $423.0 \pm 2.0$ & \cite{herald2020} \\ 
(14) Irene & $133.45_{-0.0}^{+6.6}$ & \cite{Hung2022} & $145.5 \pm 12.0$ & \cite{edyta2020} \\ 
(16) Psyche & $242.5 \pm 25.0$ & \cite{matter2013} & $224.0 \pm 5.0$ & \cite{ferrais2020} \\ 
(17) Thetis & $72.6_{-0.0}^{+1.6}$ & \cite{Hung2022} & $73.5 \pm 4.5$ & \cite{herald2020} \\ 
(18) Melpomene & $135.0 \pm 3.0$ & \cite{alilagoa2020} & $138.4 \pm 6.3$ & \cite{herald2020} \\ 
(19) Fortuna & $219.0 \pm 3.0$ & \cite{alilagoa2020} & $206.3 \pm 3.6$ & \cite{herald2020} \\ 
(20) Massalia & $147.0 \pm 2.0$ & \cite{alilagoa2020} & $110.0_{-6.35}^{+5.5}$ & \cite{edyta2020} \\ 
(21) Lutetia & $95.76 \pm 4.0$ & \cite{alilagoa2020} & $103.7 \pm 3.7$ & \cite{herald2020} \\ 
(22) Kalliope & $167.0 \pm 17.0$ &  \cite{delbo2015} & $153.0 \pm 3.0$ & \cite{ferrais2022} \\ 
(27) Euterpe & $95.6_{-0.5}^{+7.5}$ & \cite{Hung2022} & $108.5 \pm 0.5$ & \cite{herald2020} \\ 
(29) Amphitrite & $202.0 \pm 3.0$ & \cite{alilagoa2020} & $201.0 \pm 3.8$ & \cite{herald2020} \\ 
(41) Daphne & $189.0 \pm 1.0$ &  \cite{delbo2015} & $187.0 \pm 21.5$ & \cite{carry2019} \\ 
(43) Ariadne & $59.9_{-2.6}^{+2.0}$ & \cite{Hung2022} & $61.7 \pm 4.4$ & \cite{herald2020} \\ 
(45) Eugenia & $198.0 \pm 20.0$ &  \cite{delbo2015} & $186.5 \pm 3.2$ & \cite{herald2020} \\ 
(52) Europa & $317.0 \pm 4.0$ & \cite{alilagoa2020} & $314.4 \pm 7.5$ & \cite{herald2020} \\ 
(54) Alexandra & $153.0 \pm 2.0$ & \cite{alilagoa2020} & $140.5 \pm 1.5$ & \cite{herald2020} \\ 
(56) Melete & $104.75_{-0.3}^{+1.8}$ & \cite{Hung2022} & $113.4 \pm 4.0$ & \cite{herald2020} \\ 
(63) Ausonia & $94.6 \pm 2.4$ & \cite{jiang2020} & $90.0 \pm 18.0$ & \cite{DURECH2011} \\ 
(64) Angelina & $46.05_{-0.1}^{+2.1}$ & \cite{Hung2022} & $49.8_{-2.65}^{+2.95}$ & \cite{edyta2020} \\ 
(68) Leto & $102.7_{-0.3}^{+5.6}$ & \cite{Hung2022} & $142.4_{-13.15}^{+14.6}$ & \cite{edyta2020} \\ 
(71) Niobe & $72.9_{-1.7}^{+2.9}$ & \cite{Hung2022} & $83.5 \pm 0.5$ & \cite{herald2020} \\ 
(76) Freia & $151.4_{-7.3}^{+6.3}$ & \cite{Hung2022} & $172.3 \pm 10.1$ & \cite{herald2020} \\ 
(80) Sappho & $66.7_{-0.8}^{+4.6}$ & \cite{Hung2022} & $68.7 \pm 5.7$ & \cite{herald2020} \\ 
(85) Io & $145.2_{-1.3}^{+7.2}$ & \cite{Hung2022} & $165.2 \pm 4.8$ & \cite{herald2020} \\ 
(87) Sylvia & $300.0 \pm 30.0$ &  \cite{delbo2015} & $274.0 \pm 5.0$ & \cite{carry2021} \\ 
(88) Thisbe & $221.0 \pm 2.0$ & \cite{alilagoa2020} & $218.2 \pm 7.4$ & \cite{herald2020} \\ 
(93) Minerva & $167.0 \pm 3.0$ & \cite{alilagoa2020} & $162.3 \pm 1.7$ & \cite{herald2020} \\ 
(94) Aurora & $158.8_{-0.0}^{+6.7}$ & \cite{Hung2022} & $193.5 \pm 6.2$ & \cite{herald2020} \\ 
(95) Arethusa & $131.1_{-4.5}^{+5.2}$ & \cite{Hung2022} & $147.0 \pm 1.4$ & \cite{herald2020} \\ 
(99) Dike & $66.5 \pm 0.9$ & \cite{hanus2018} & $69.0 \pm 3.0$ & \cite{herald2020} \\ 
(107) Camilla & $245.0 \pm 25.0$ &  \cite{delbo2015} & $225.4 \pm 19.8$ & \cite{herald2020} \\ 
(109) Felicitas & $85.0 \pm 6.0$ & \cite{maclennan2019} & $84.5 \pm 2.7$ & \cite{herald2020} \\ 
(121) Hermione & $220.0 \pm 22.0$ &  \cite{delbo2015} & $191.5 \pm 0.5$ & \cite{herald2020} \\ 
(130) Elektra & $197.0 \pm 20.0$ &  \cite{delbo2015} & $201.0 \pm 2.0$ & \cite{fuksa2023} \\ 
(135) Hertha & $73.45_{-4.5}^{+4.6}$ & \cite{Hung2022} & $79.3 \pm 2.5$ & \cite{herald2020} \\ 
(152) Atala & $57.5 \pm 1.8$ & \cite{hanus2018} & $65.0 \pm 8.0$ & \cite{DURECH2011} \\ 
(158) Koronis & $34.75_{-1.9}^{+4.4}$ & \cite{Hung2022} & $38.0 \pm 5.0$ & \cite{DURECH2011} \\ 
(159) Aemilia & $137.0 \pm 8.0$ & \cite{M2018} & $132.0 \pm 2.0$ & \cite{herald2020} \\ 
(166) Rhodope & $48.1_{-2.3}^{+1.9}$ & \cite{Hung2022} & $50.0 \pm 3.6$ & \cite{herald2020} \\ 
(167) Urda & $41.15 \pm 0.8$ & \cite{hanus2018} & $47.5 \pm 15.0$ & \cite{DURECH2011} \\ 
(187) Lamberta & $126.8_{-1.4}^{+3.8}$ & \cite{Hung2022} & $141.5 \pm 0.5$ & \cite{herald2020} \\ 
(192) Nausikaa & $87.95 \pm 3.6$ & \cite{Hung2022} & $95.7 \pm 1.7$ & \cite{herald2020} \\ 
(199) Byblis & $66.0_{-0.8}^{+2.2}$ & \cite{Hung2022} & $53.0 \pm 5.0$ & \cite{herald2020} \\ 
(208) Lacrimosa & $40.4 \pm 0.7$ & \cite{maclennan2019} & $44.5 \pm 0.5$ & \cite{herald2020} \\ 
(234) Barbara & $44.1_{-0.7}^{+1.9}$ & \cite{Hung2022} & $45.9 \pm 1.0$ & \cite{tanga2015} \\ 
(276) Adelheid & $102.25_{-4.9}^{+8.0}$ & \cite{Hung2022} & $121.3 \pm 5.0$ & \cite{herald2020} \\ 
(306) Unitas & $56.0 \pm 1.0$ &  \cite{delbo2015} & $51.0 \pm 5.0$ & \cite{DURECH2011} \\ 
(328) Gudrun & $103.55_{-0.6}^{+3.2}$ & \cite{Hung2022} & $103.3 \pm 2.5$ & \cite{herald2020} \\ 
(329) Svea & $78.0 \pm 4.0$ & \cite{M2018} & $73.3 \pm 2.5$ & \cite{herald2020} \\ 
(334) Chicago & $146.55_{-1.6}^{+8.8}$ & \cite{Hung2022} & $179.3 \pm 2.9$ & \cite{herald2020} \\ 
(347) Pariana & $46.35_{-2.2}^{+1.7}$ & \cite{Hung2022} & $52.3 \pm 4.5$ & \cite{herald2020} \\ 
(349) Dembowska & $155.8 \pm 7.1$ & \cite{yu2017} & $149.0 \pm 1.0$ & \cite{herald2020} \\ 
(350) Ornamenta & $116.0_{-3.7}^{+3.1}$ & \cite{Hung2022} & $119.0 \pm 5.0$ & \cite{herald2020} \\ 
(360) Carlova & $114.1_{-0.7}^{+3.3}$ & \cite{Hung2022} & $135.0 \pm 3.0$ & \cite{hanus2017} \\ 
(365) Corduba & $83.5_{-3.6}^{+3.5}$ & \cite{Hung2022} & $94.5 \pm 0.5$ & \cite{herald2020} \\ 
(372) Palma & $166.85_{-10.3}^{+9.2}$ & \cite{Hung2022} & $183.2 \pm 13.1$ & \cite{herald2020} \\ 
(380) Fiducia & $72.0 \pm 8.0$ & \cite{M2019} & $73.0 \pm 1.0$ & \cite{herald2020} \\ 
(381) Myrrha & $111.35_{-4.2}^{+4.9}$ & \cite{Hung2022} & $134.8_{-12.8}^{+45.3}$ & \cite{edyta2020} \\ 
(404) Arsinoe & $91.6_{-2.3}^{+3.4}$ & \cite{Hung2022} & $95.7 \pm 3.3$ & \cite{herald2020} \\ 
(423) Diotima & $200.0 \pm 4.0$ & \cite{alilagoa2020} & $205.7 \pm 8.4$ & \cite{herald2020} \\ 
(426) Hippo & $108.2_{-6.7}^{+6.3}$ & \cite{Hung2022} & $125.5_{-6}^{+11.5}$ & \cite{M2023} \\ 
(433) Eros & $17.8 \pm 0.5$ &  \cite{delbo2015} & $17.0 \pm 5.0$ & \cite{herald2020} \\ 
(439) Ohio & $67.85_{-3.8}^{+2.7}$ & \cite{Hung2022} & $74.0_{-8}^{+3}$ & \cite{M2023} \\ 
(441) Bathilde & $66.5_{-2.6}^{+2.5}$ & \cite{Hung2022} & $76.05_{-10.0}^{+45.25}$ & \cite{edyta2020} \\ 
(458) Hercynia & $33.8_{-0.0}^{+1.5}$ & \cite{Hung2022} & $37.0 \pm 1.0$ & \cite{herald2020} \\ 
(464) Megaira & $69.56 \pm 6.38$ & \cite{maclennan2021} & $76.5_{-8.5}^{+2.5}$ & \cite{M2023} \\ 
(468) Lina & $69.0 \pm 10.0$ & \cite{M2019} & $65.7 \pm 1.9$ & \cite{herald2020} \\ 
(532) Herculina & $207.3 \pm 12.0$ &  \cite{delbo2015} & $191.0 \pm 4.0$ & \cite{hanus2017} \\ 
(580) Selene & $47.35_{-1.9}^{+1.2}$ & \cite{Hung2022} & $49.5 \pm 0.5$ & \cite{herald2020} \\ 
(657) Gunlod & $38.3_{-2.7}^{+2.8}$ & \cite{Hung2022} & $38.0_{-2}^{+2.5}$ & \cite{M2023} \\ 
(694) Ekard & $109.5 \pm 1.5$ &  \cite{delbo2015} & $101.3 \pm 2.5$ & \cite{herald2020} \\ 
(695) Bella & $41.4_{-1.9}^{+0.9}$ & \cite{Hung2022} & $49.5 \pm 1.5$ & \cite{herald2020} \\ 
(757) Portlandia & $32.95 \pm 0.5$ & \cite{hanus2018} & $36.0 \pm 5.0$ & \cite{herald2020} \\ 
(791) Ani & $86.5_{-0.0}^{+1.3}$ & \cite{Hung2022} & $80.0 \pm 5.6$ & \cite{herald2020} \\ 
(834) Burnhamia & $67.0 \pm 7.0$ & \cite{M2019} & $65.0 \pm 2.8$ & \cite{herald2020} \\ 
(849) Ara & $72.1_{-2.3}^{+1.7}$ & \cite{Hung2022} & $73.0 \pm 3.0$ & \cite{hanus2017} \\ 
(925) Alphonsina & $58.15_{-2.2}^{+2.4}$ & \cite{Hung2022} & $58.0 \pm 0.8$ & \cite{herald2020} \\ 
(3200) Phaethon & $5.1 \pm 0.2$ & \cite{hanus2016} & $5.3 \pm 0.4$ & \cite{herald2020} \\

\end{longtable}

\tablefoot{ If one value was provided or a specific pole was preferred, the value is included directly in this table. In cases where solutions for two poles were reported in the literature without distinction, their respective sizes are averaged here, and the maximum uncertainty is adopted.}

\newpage

\section{Chi-squared plots vs. thermal inertia}
Plots of $\chi^2_\text{red}$ versus thermal inertia for various levels of surface roughness described by $\bar{\theta}$ (symbol coded) and optimised diameter (colour coded). Plots present the results for one pole solution since the results for the mirror-pole solution were similar.

\begin{figure}[h!]
    \centering
    \begin{minipage}{0.5\textwidth}
        \centering
        \includegraphics[width=\linewidth]{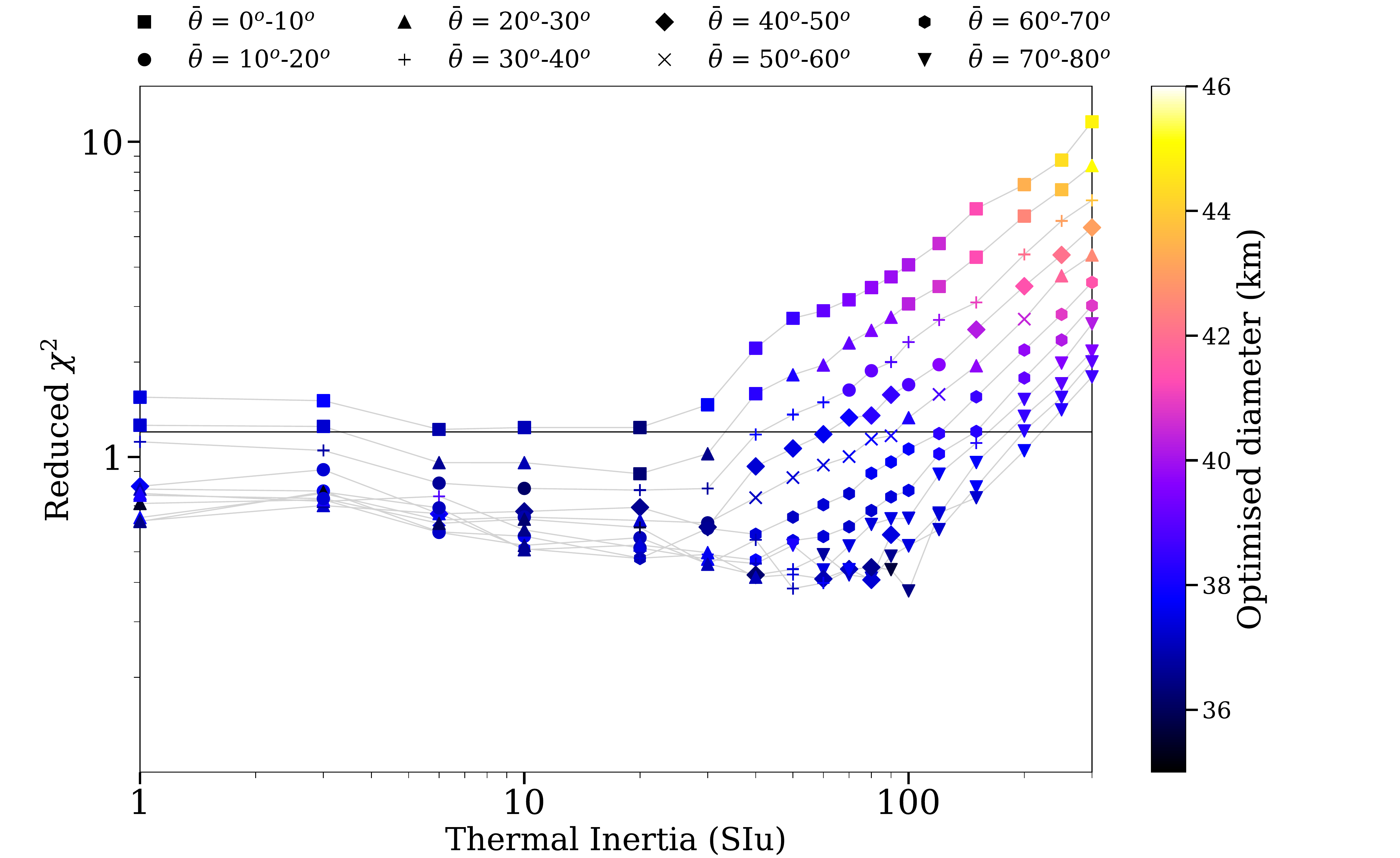} 
        \caption{(215) Oenone: reduced $\chi^2$ vs. $\Gamma$. The horizontal line represents the threshold for accepting a solution based on thermal data. The plot corresponds to pole solution 1.}
        \label{fig:ti_215}
    \end{minipage}\hfill
    \begin{minipage}{0.5\textwidth}
        \centering
        \includegraphics[width=\linewidth]{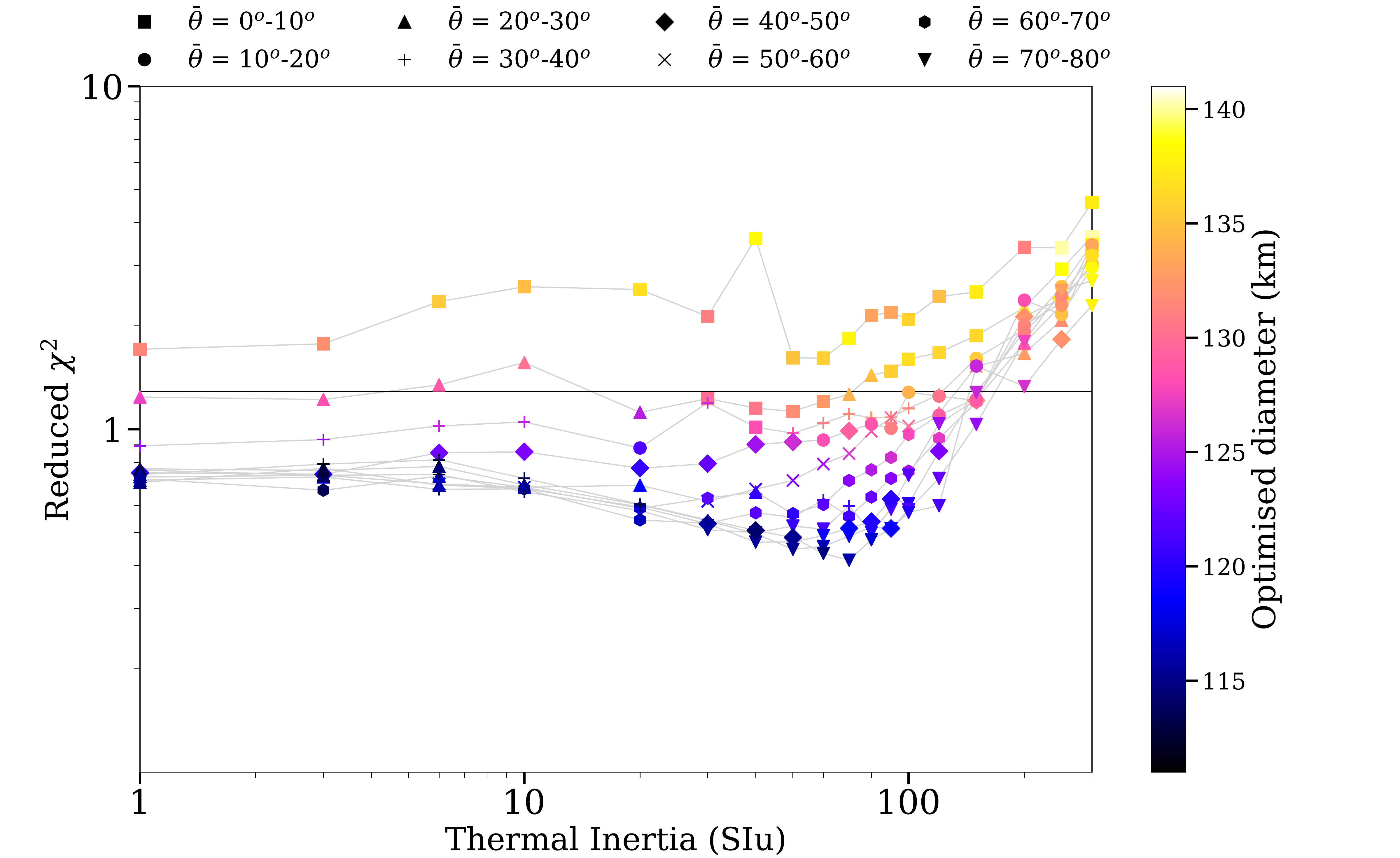}  
        \caption{(279) Thule: reduced $\chi^2$ vs. $\Gamma$ for pole 1.}
        \label{fig:ti_279}
    \end{minipage}

\end{figure}

\begin{figure}[h!]
    \centering
    \begin{minipage}{0.5\textwidth}
        \centering
        \includegraphics[width=\linewidth]{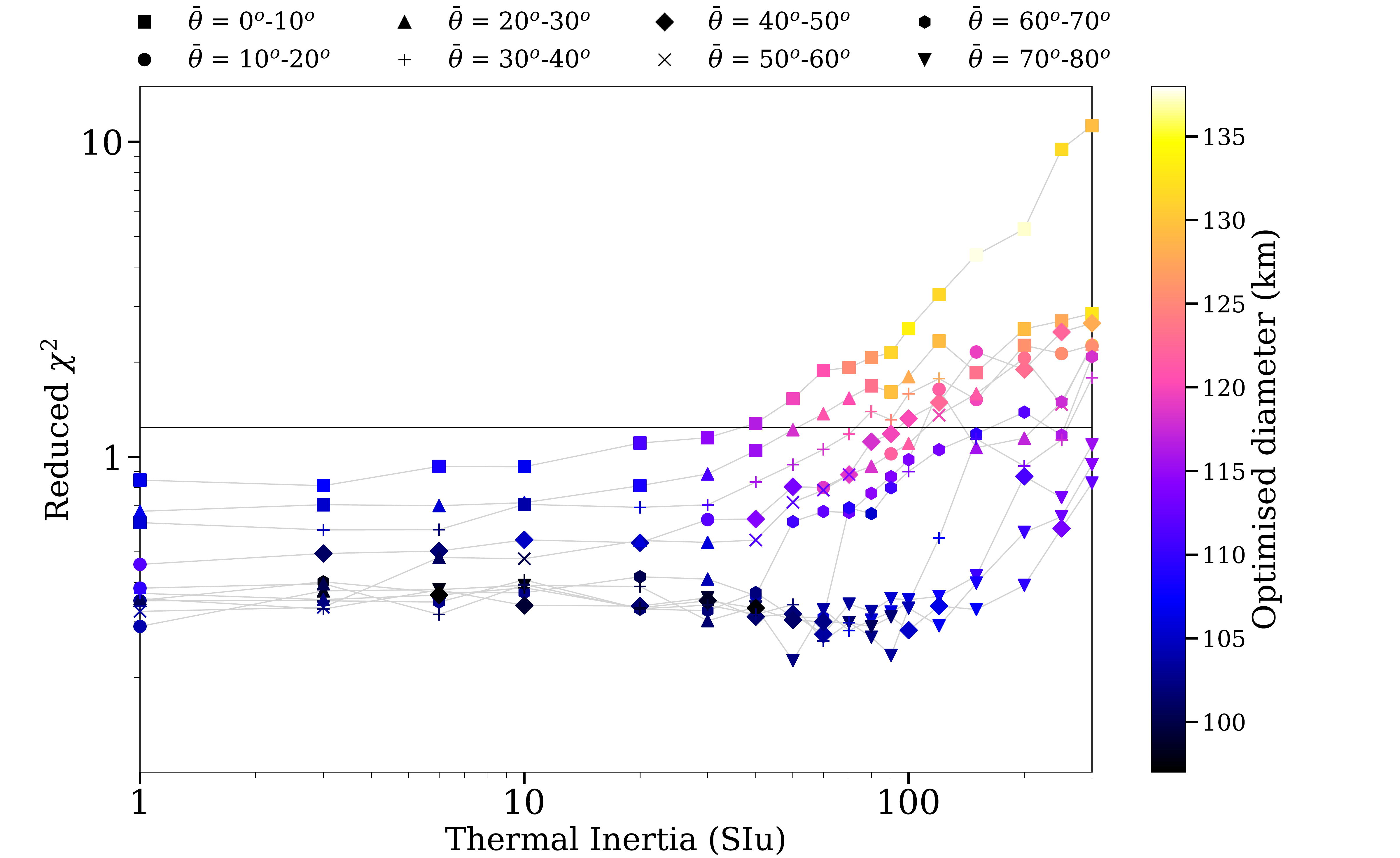}  
        \caption{(357) Ninina: reduced $\chi^2$ vs. $\Gamma$ for pole 1.}
        \label{fig:ti_357}
    \end{minipage}\hfill
    \begin{minipage}{0.5\textwidth}
        \centering
        \includegraphics[width=\linewidth]{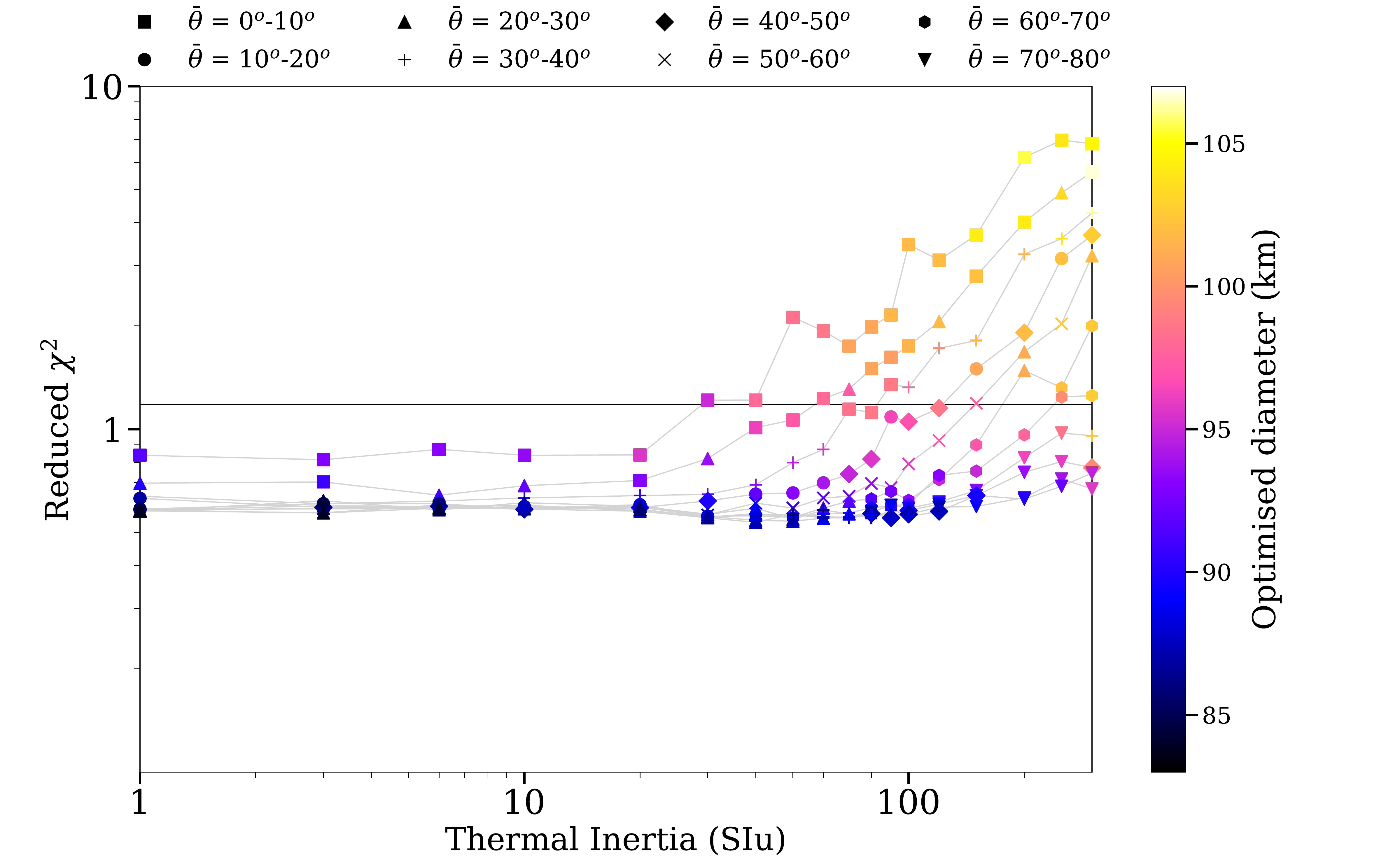}  
        \caption{(366) Vincentina: reduced $\chi^2$ vs. $\Gamma$ for pole 1.}
        \label{fig:ti_366}
    \end{minipage}

\end{figure}

\begin{figure}
    \centering
    \begin{minipage}{0.5\textwidth}
        \centering
        \includegraphics[width=\linewidth]{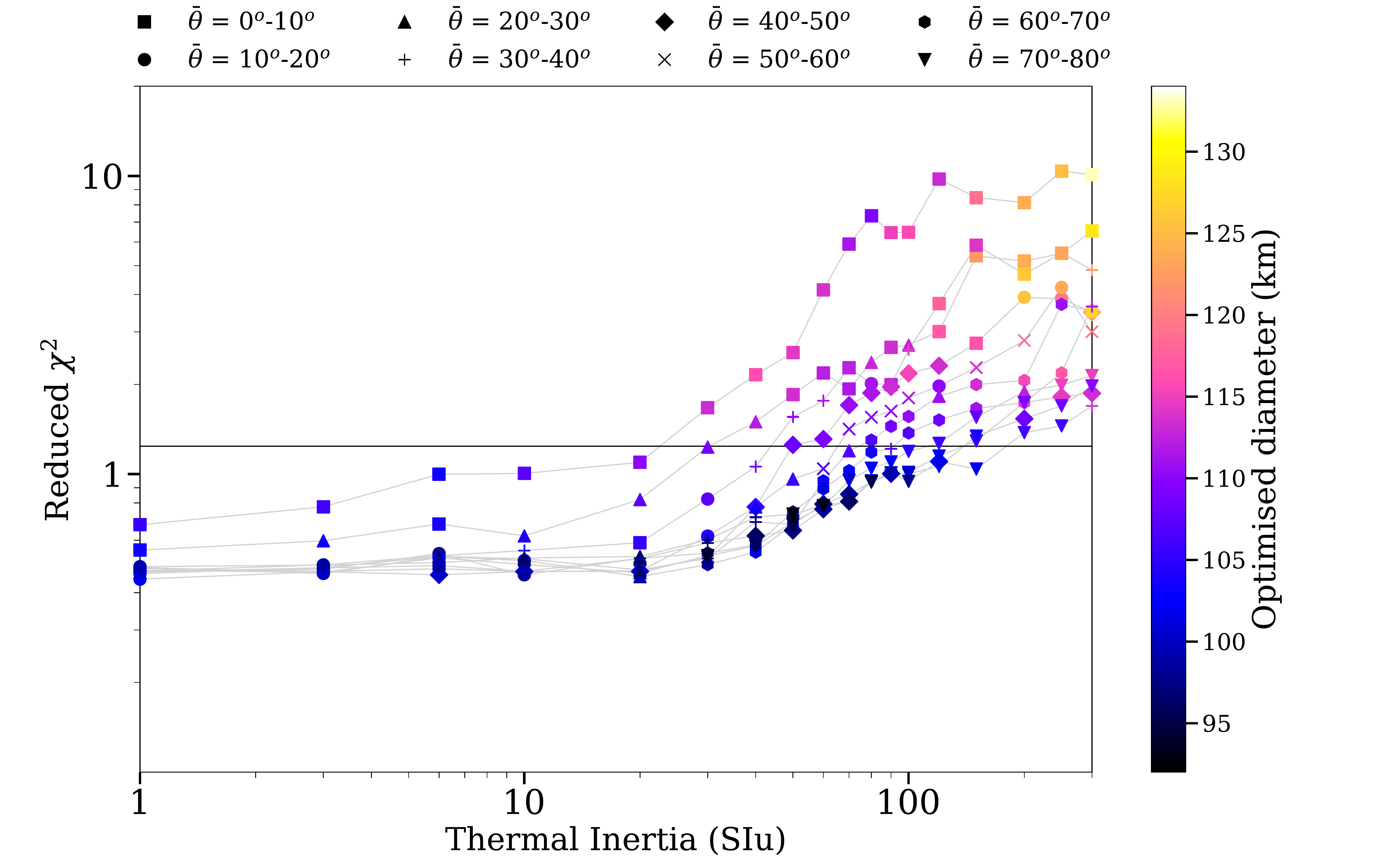} 
        \caption{(373) Melusina: reduced $\chi^2$ vs. $\Gamma$ for pole 2.}
        \label{fig:ti_373}
    \end{minipage}\hfill
    \begin{minipage}{0.5\textwidth}
        \centering
        \includegraphics[width=\linewidth]{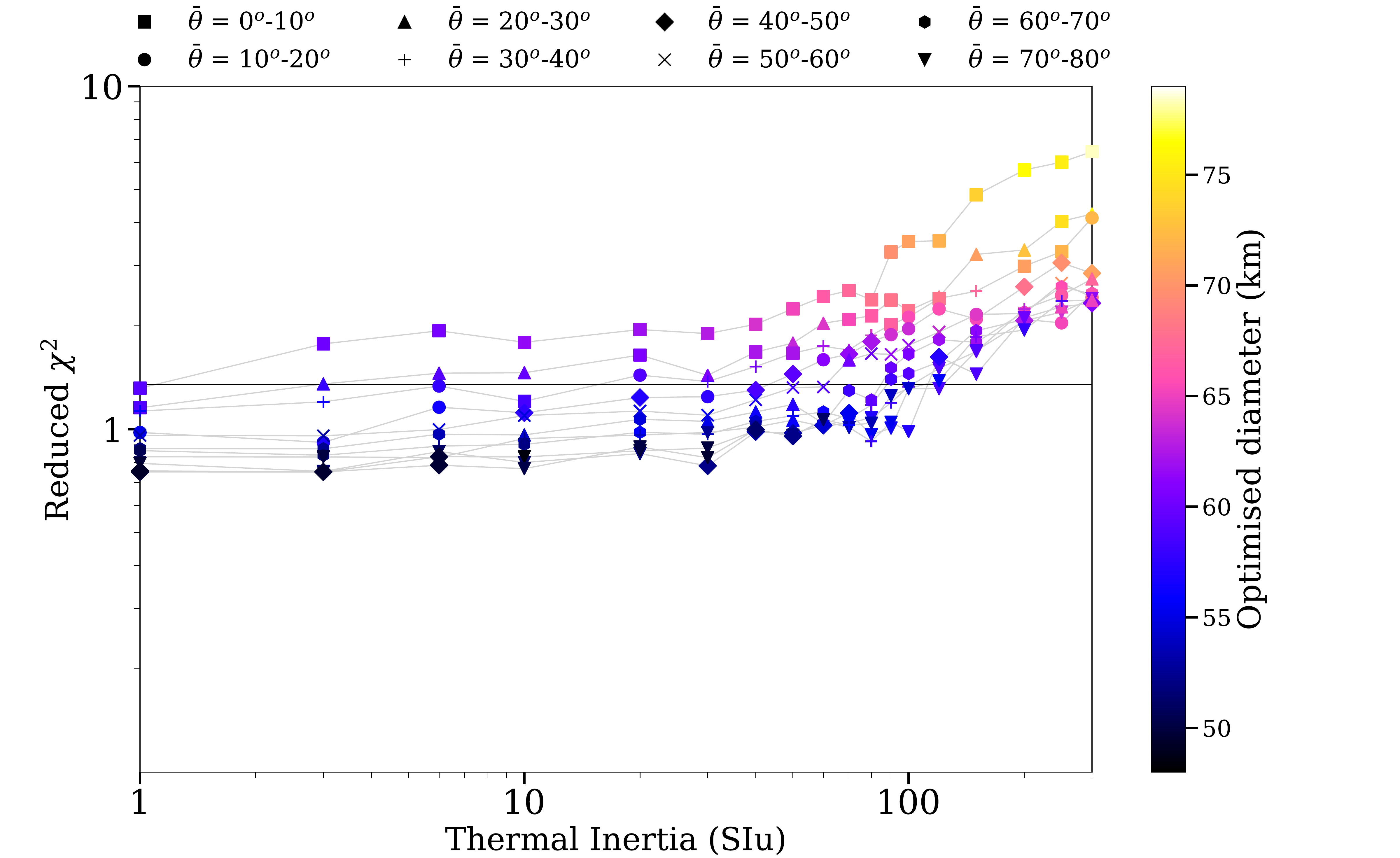}  
        \caption{(395) Delia: reduced $\chi^2$ vs. $\Gamma$ for pole 1.}
        \label{fig:ti_395}
    \end{minipage}
    
\end{figure}

\begin{figure}
    \centering
    \begin{minipage}{0.5\textwidth}
        \centering
        \includegraphics[width=\linewidth]{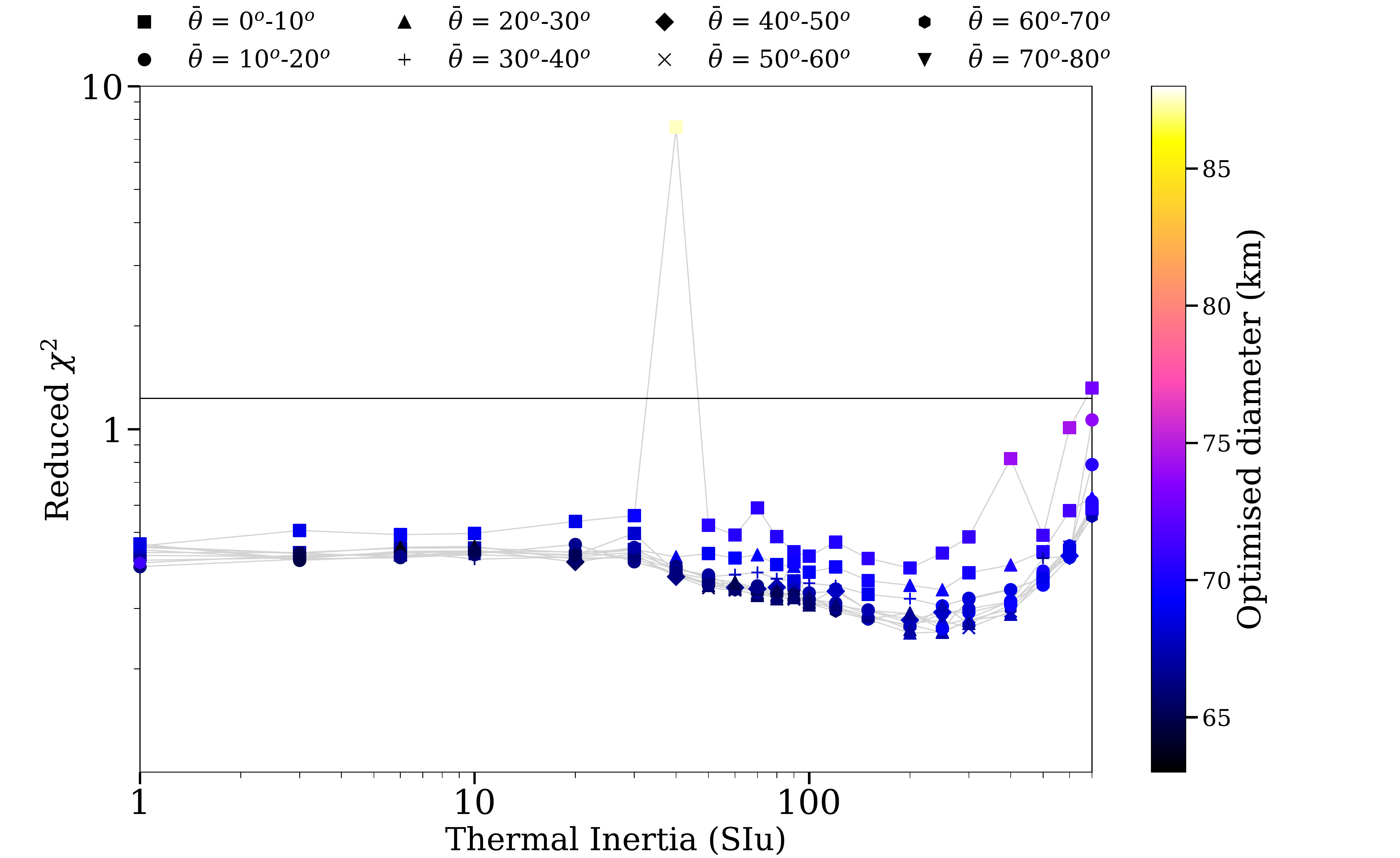} 
        \caption{(429) Lotis: reduced $\chi^2$ vs. $\Gamma$ for pole 1.}
        \label{fig:ti_429}
    \end{minipage}\hfill
    \begin{minipage}{0.5\textwidth}
        \centering
        \includegraphics[width=\linewidth]{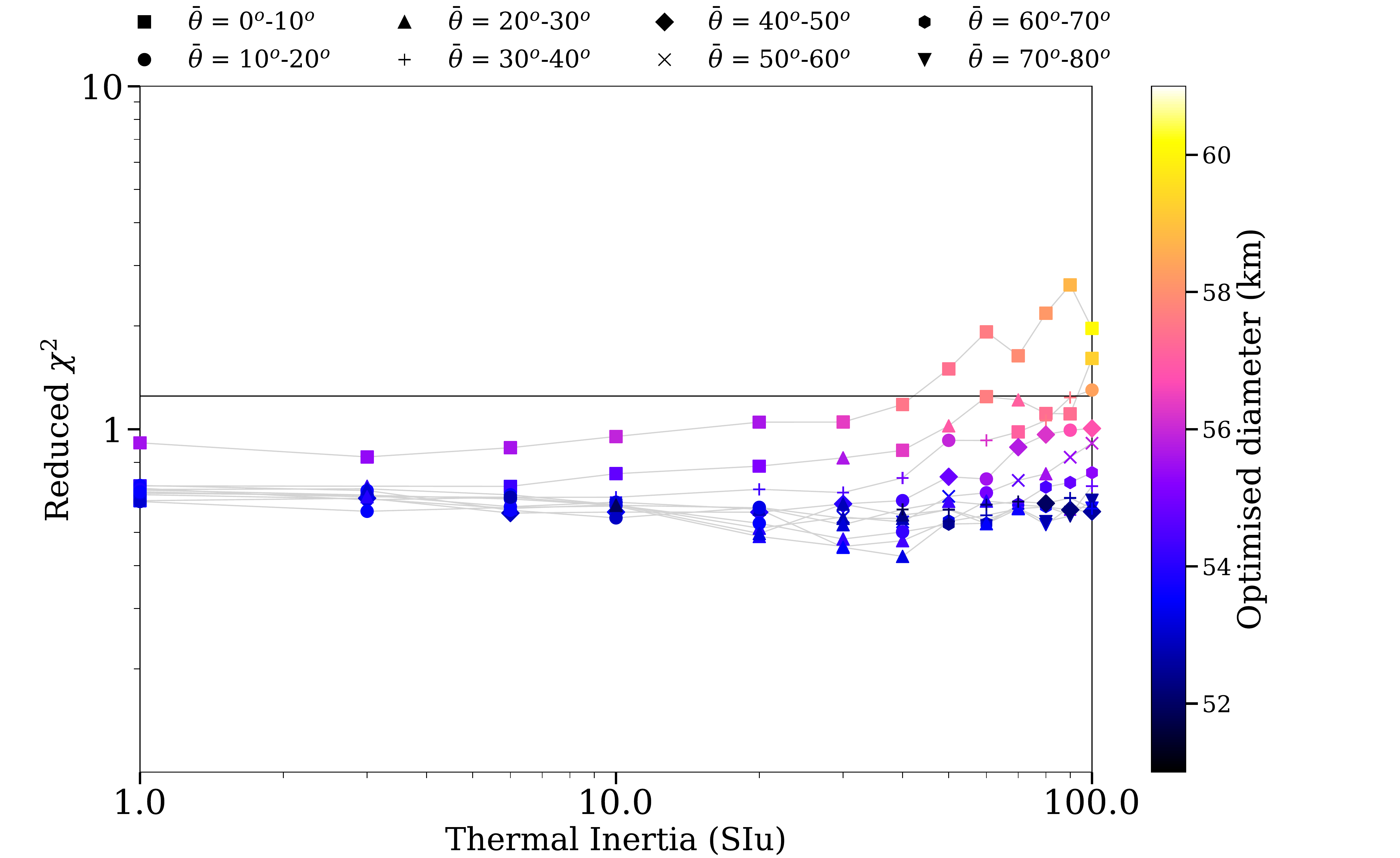} 
        \caption{(527) Euryanthe: reduced $\chi^2$ vs. $\Gamma$ for pole 2.}
        \label{fig:ti_527}
    \end{minipage}
    
\end{figure}

\begin{figure}
    \centering
    \begin{minipage}{0.5\textwidth}
        \centering
        \includegraphics[width=\linewidth]{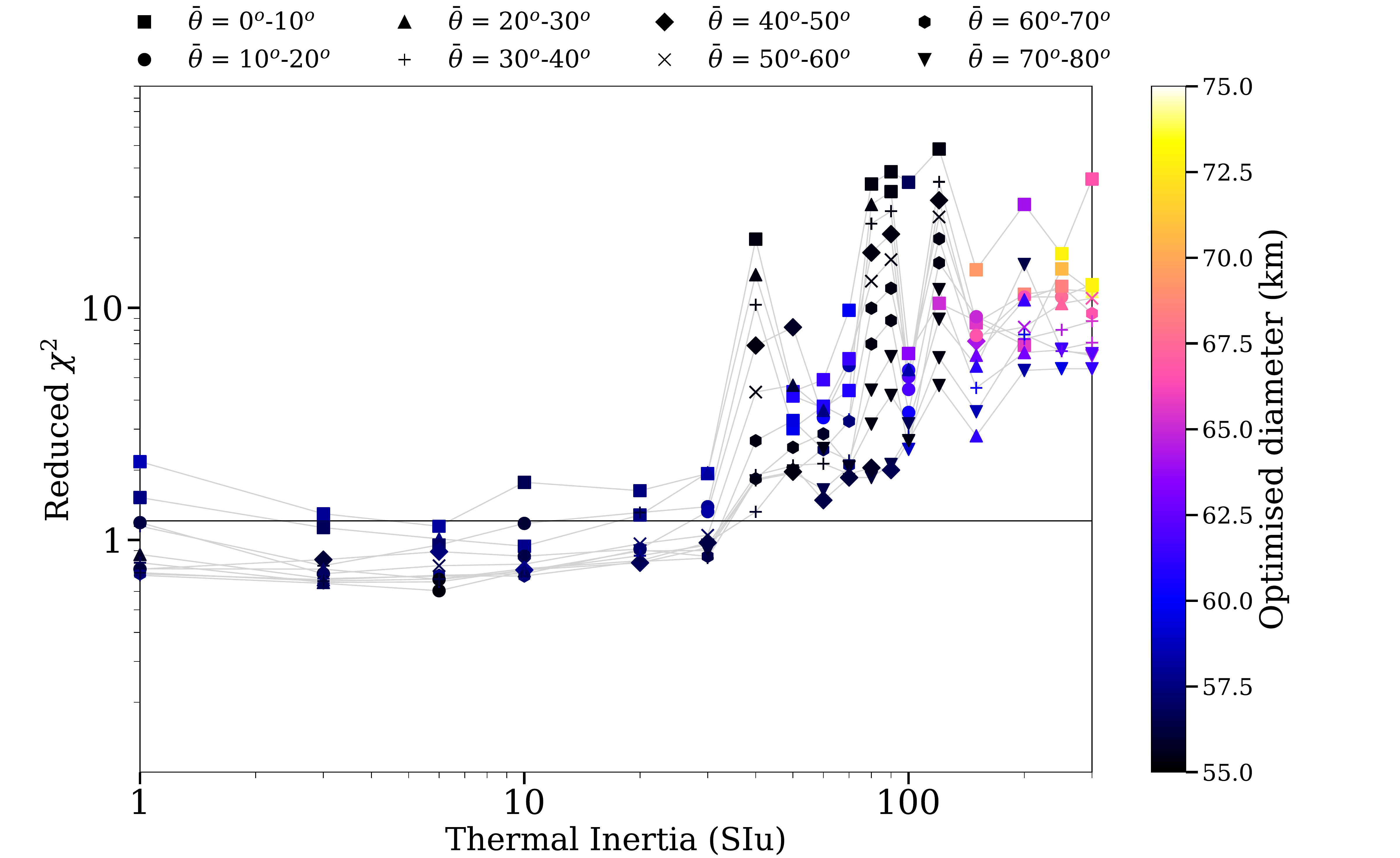} 
        \caption{(541) Deborah: reduced $\chi^2$ vs. $\Gamma$ for pole 2.}
        \label{fig:ti_541}
    \end{minipage}\hfill
    \begin{minipage}{0.5\textwidth}
        \centering
        \includegraphics[width=\linewidth]{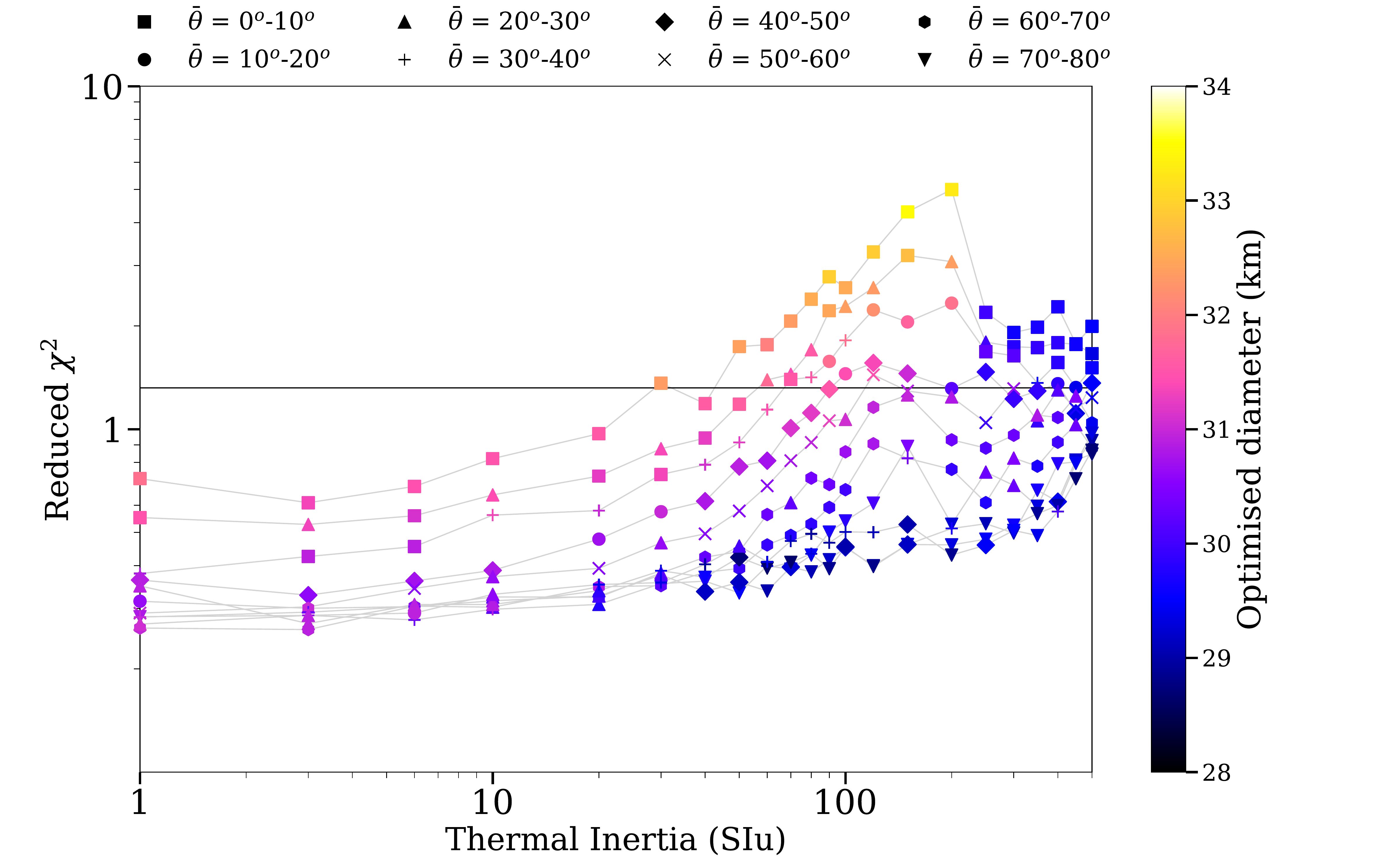} 
        \caption{(672) Astarte: reduced $\chi^2$ vs. $\Gamma$ for pole 1.}
        \label{fig:ti_672}
    \end{minipage}

\end{figure}

\begin{figure}
    \centering
    \begin{minipage}{0.5\textwidth}
        \centering
        \includegraphics[width=\linewidth]{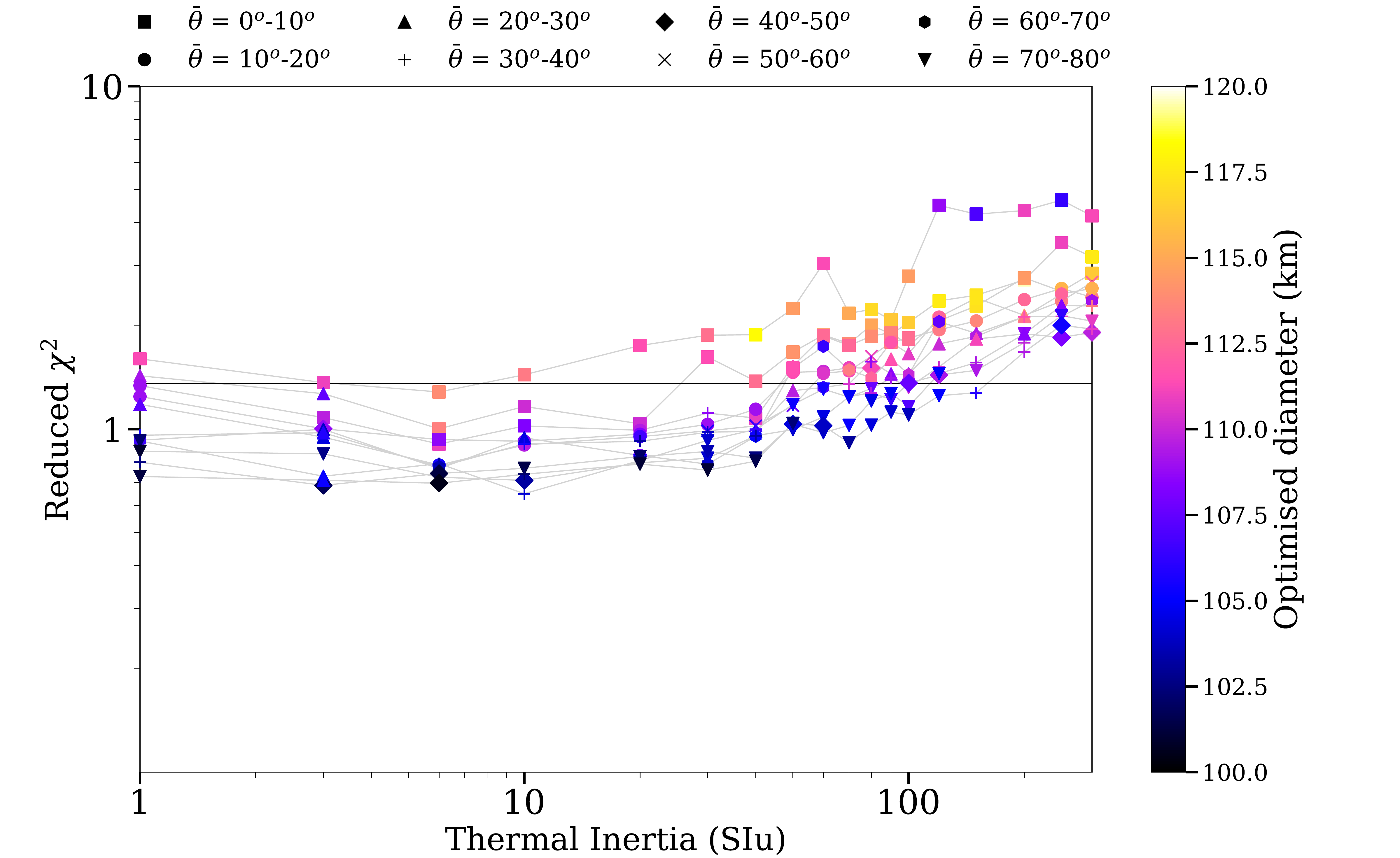}  
        \caption{(814) Tauris: reduced $\chi^2$ vs. $\Gamma$ for pole 1.}
        \label{fig:ti_814}
    \end{minipage}\hfill
    \begin{minipage}{0.5\textwidth}
        \centering
        \includegraphics[width=\linewidth]{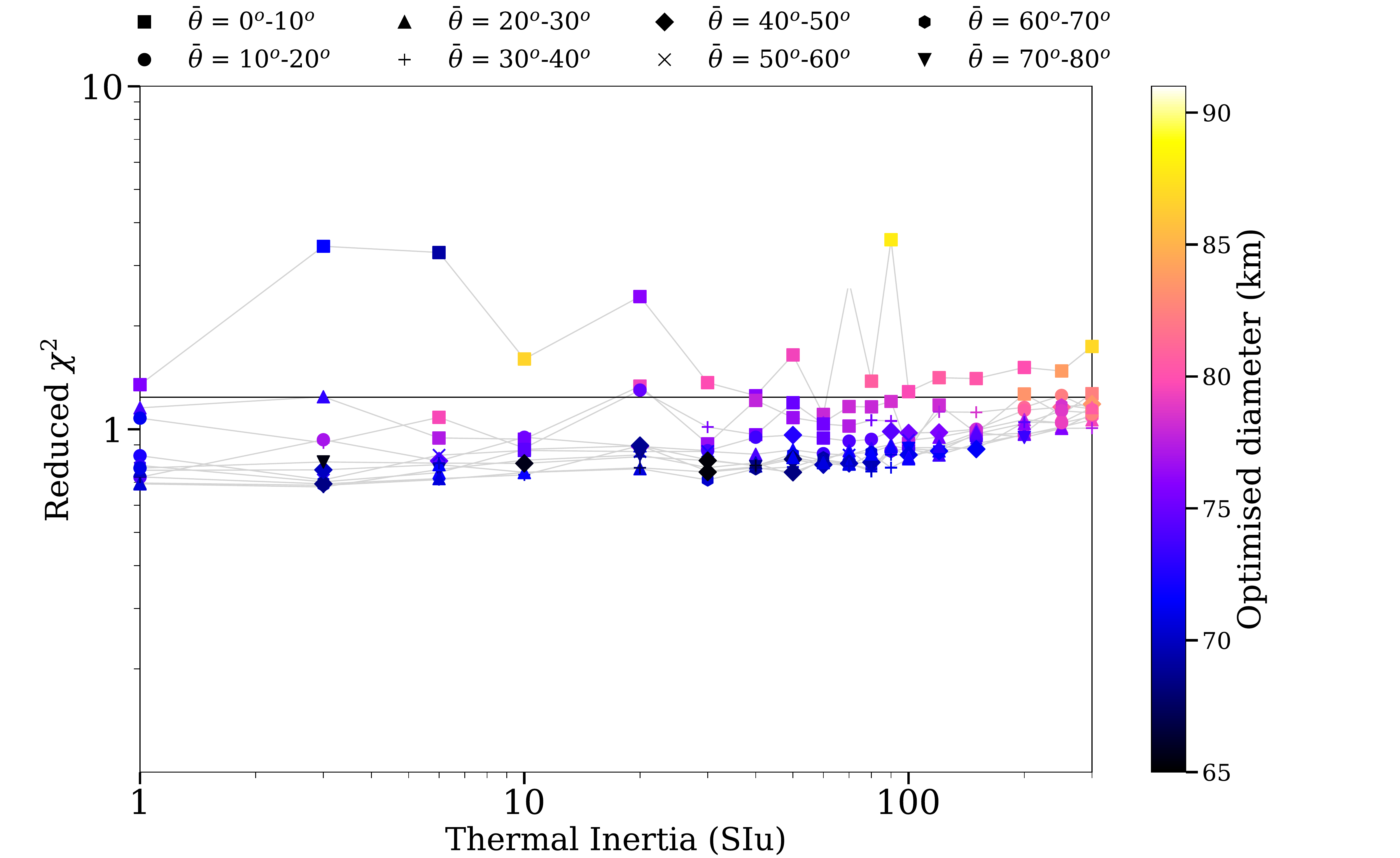}
        \caption{(859) Bouzareah: reduced $\chi^2$ vs. $\Gamma$ for pole 1.}
        \label{fig:ti_859}
    \end{minipage}

\end{figure}

\begin{figure}
    \centering
    \begin{minipage}{0.5\textwidth}
        \centering
        \includegraphics[width=\linewidth]{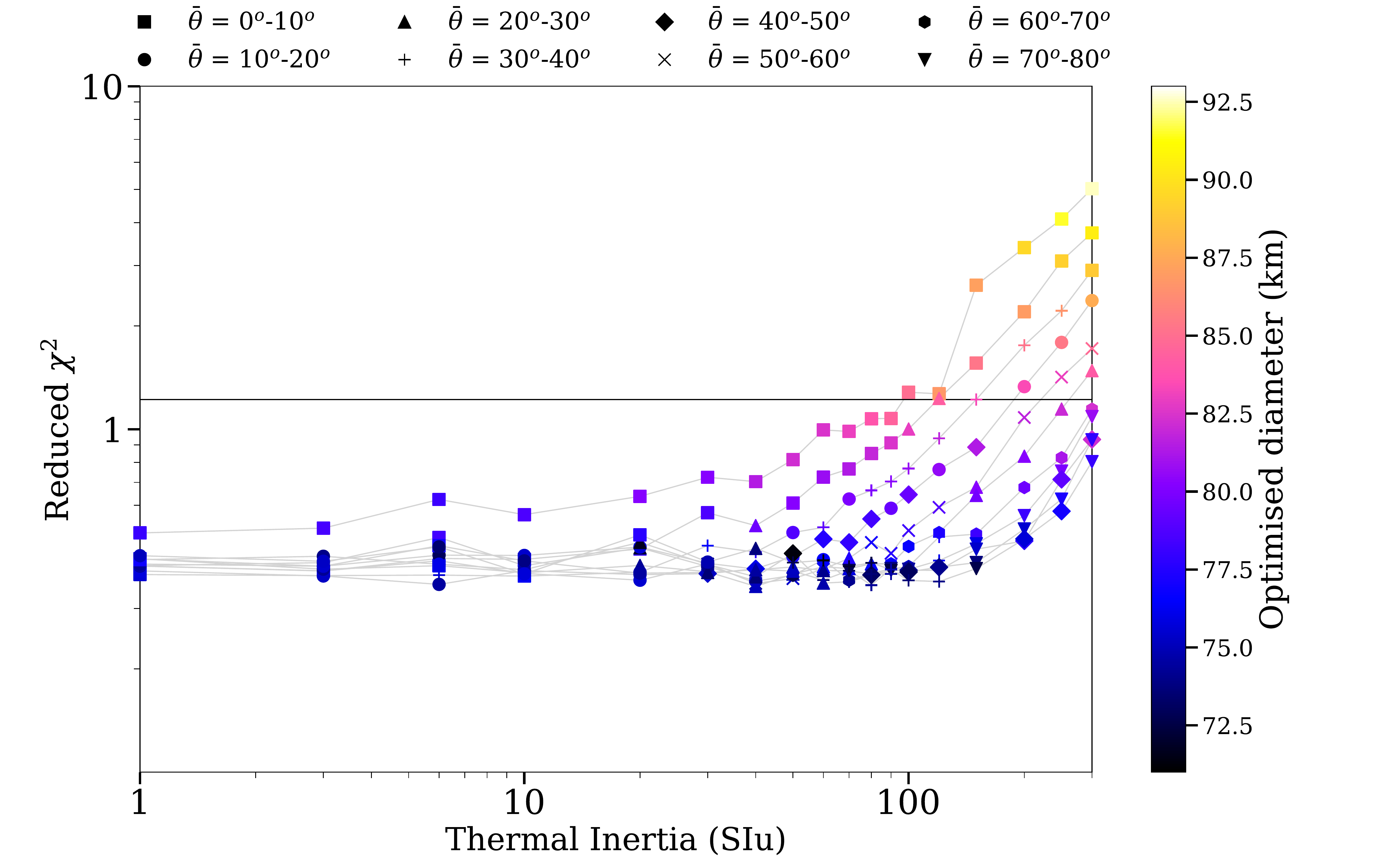} 
        \caption{(907) Rhoda: reduced $\chi^2$ vs. $\Gamma$ for pole 1.}
        \label{fig:ti_907}
    \end{minipage}\hfill
    \begin{minipage}{0.5\textwidth}
        \centering
        \includegraphics[width=\linewidth]{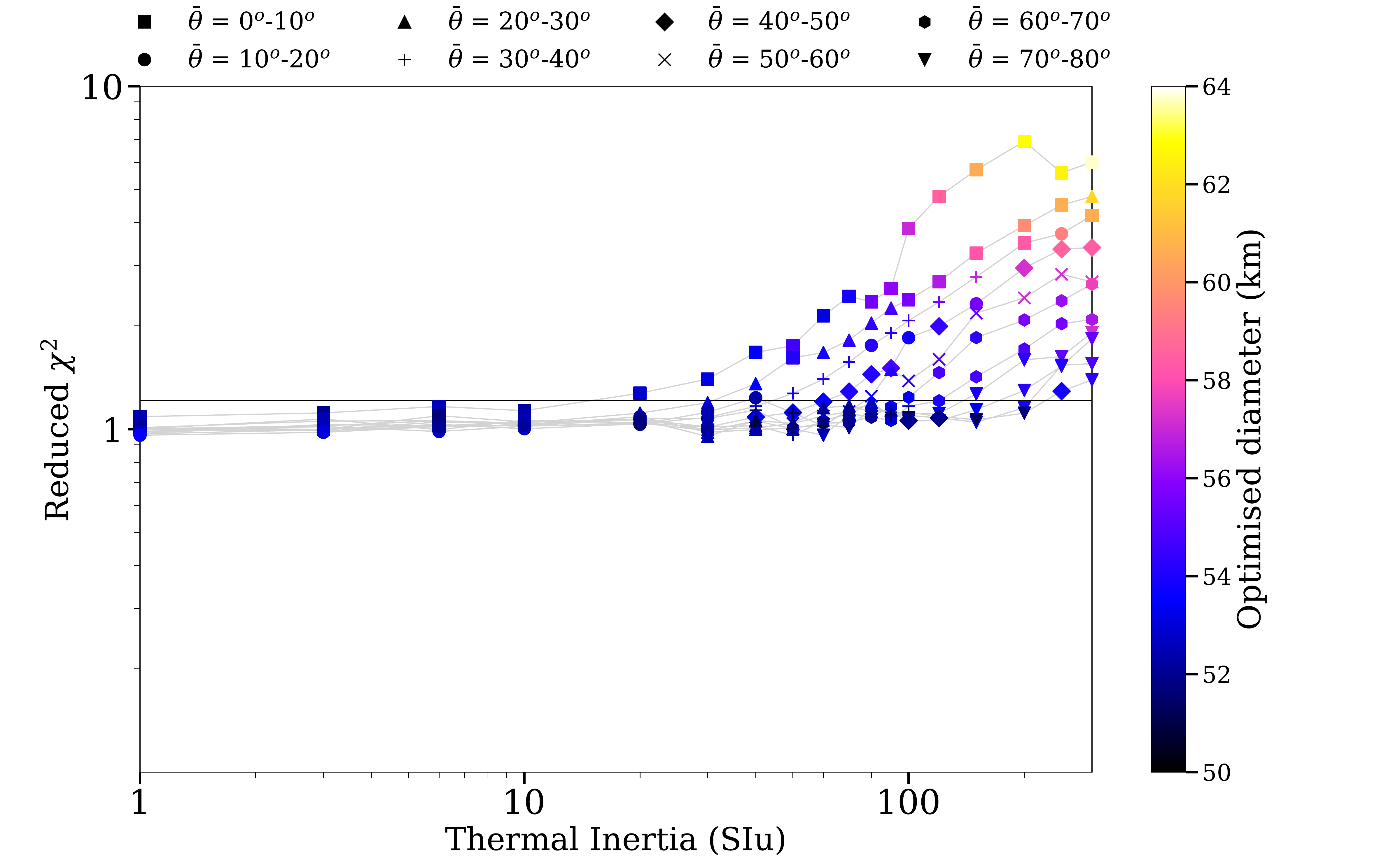} 
        \caption{(931) Whittemora: reduced $\chi^2$ vs. $\Gamma$ for pole 2.}
        \label{fig:ti_931}
    \end{minipage}

\end{figure}

\begin{figure}
    \centering
    \begin{minipage}{0.5\textwidth}
        \centering
        \includegraphics[width=\linewidth]{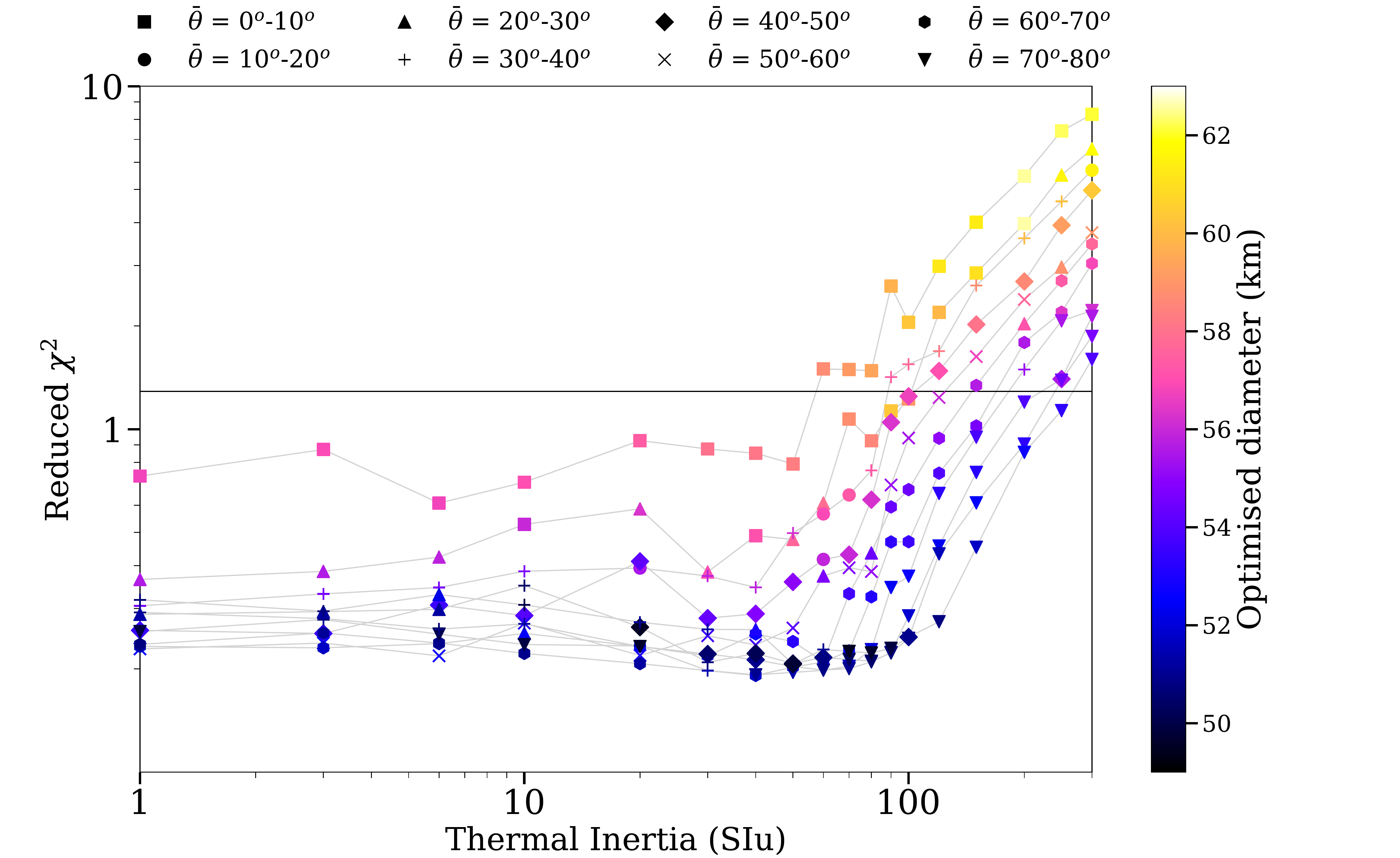}
        \caption{(1062) Ljuba: reduced $\chi^2$ vs. $\Gamma$ for pole 2.}
        \label{fig:ti_1062}
    \end{minipage}\hfill
\end{figure}

\clearpage
\newpage

\section{Example fits to light curves in the visible range}

Model fit to light curves in the visible range obtained from Kepler and TESS spacecrafts (Figures \ref{215lc_fit} - \ref{1062lc_fit}) for the preferred pole solution (see Table~\ref{tab:occ_diam}). When no pole solution was preferred, the fit for pole 1 is shown. Fit to the whole light curve dataset can be viewed in the DAMIT database.
 \begin{table*}[h!]
    \centering

\begin{tabularx}{\linewidth}{XX}
 \includegraphics[width=0.5\textwidth]{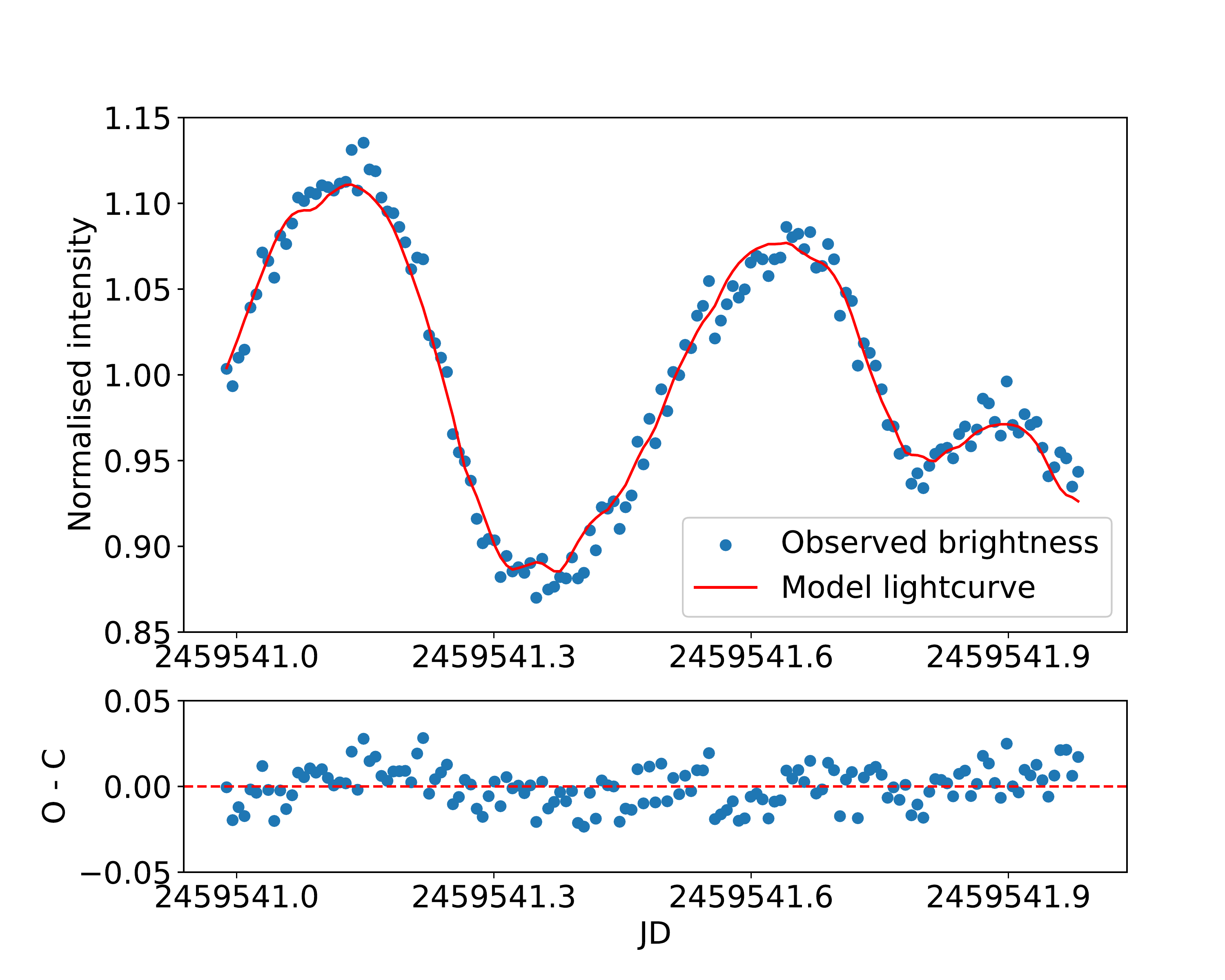}
\captionof{figure}{(215) Oenone model light curve fit (red line) to visible data from TESS spacecraft (blue points)}
\label{215lc_fit}
&
 \includegraphics[width=0.5\textwidth]{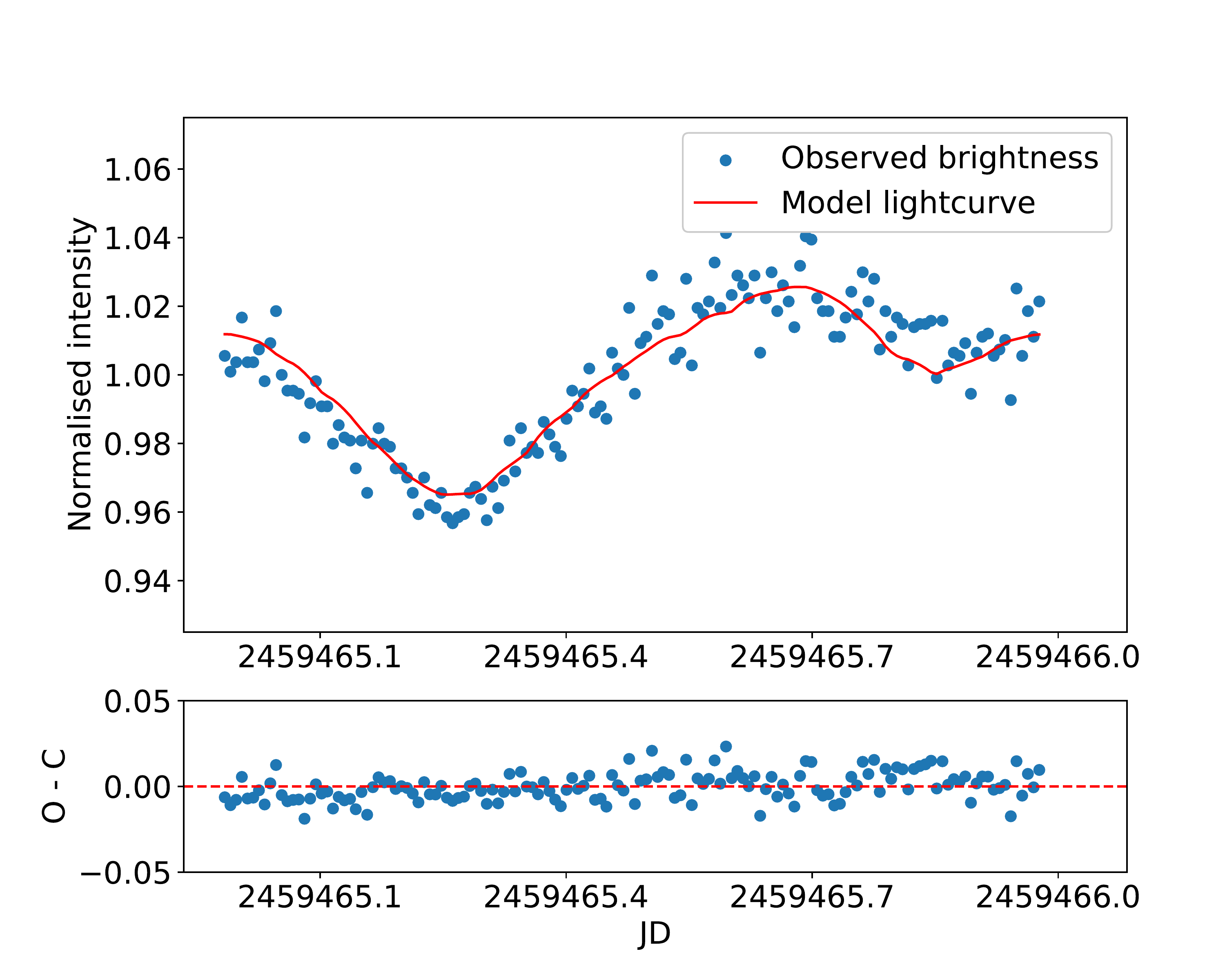}
\captionof{figure}{(279) Thule model fit to TESS data}
\label{279lc_fit}
\\
 \includegraphics[width=0.5\textwidth]{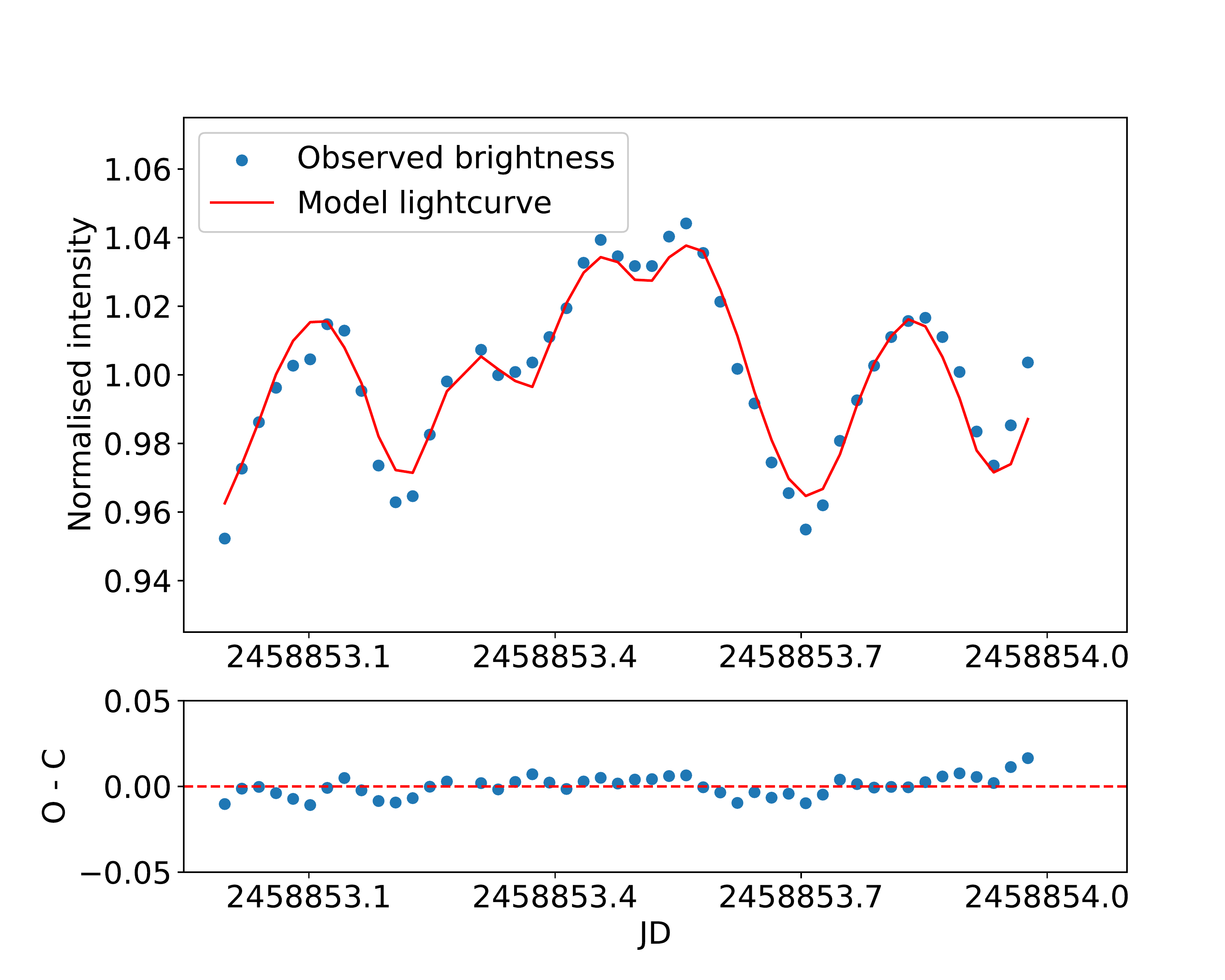}
\captionof{figure}{(366) Vincentina model fit to TESS data}
\label{366lc_fit}
&
 \includegraphics[width=0.5\textwidth]{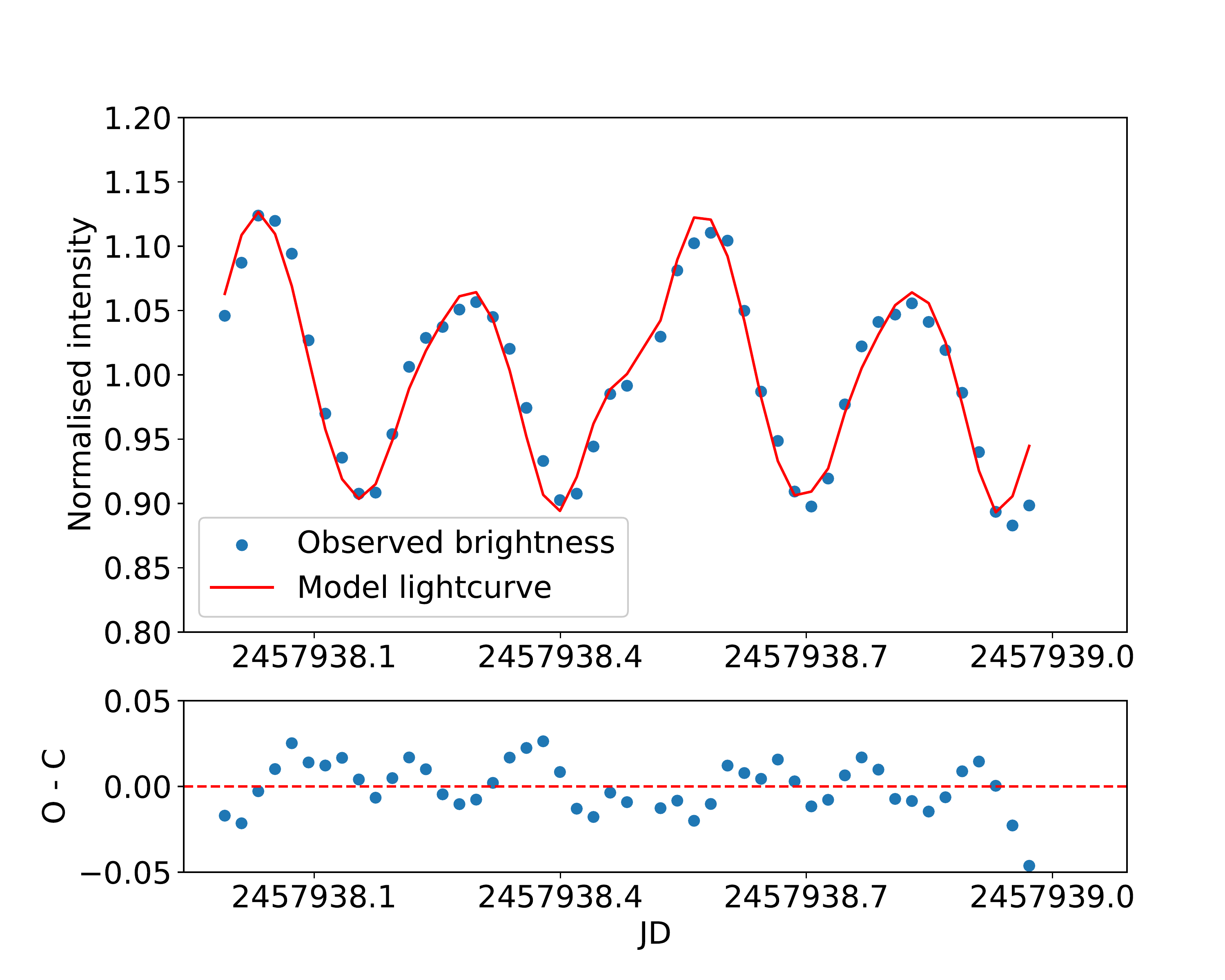}
\captionof{figure}{(373) Melusina model fit to Kepler data}
\label{373lc_fit}
\\
\end{tabularx}
    \end{table*}

\begin{table*}[h!]
    \centering

\begin{tabularx}{\linewidth}{XX}
 \includegraphics[width=0.5\textwidth]{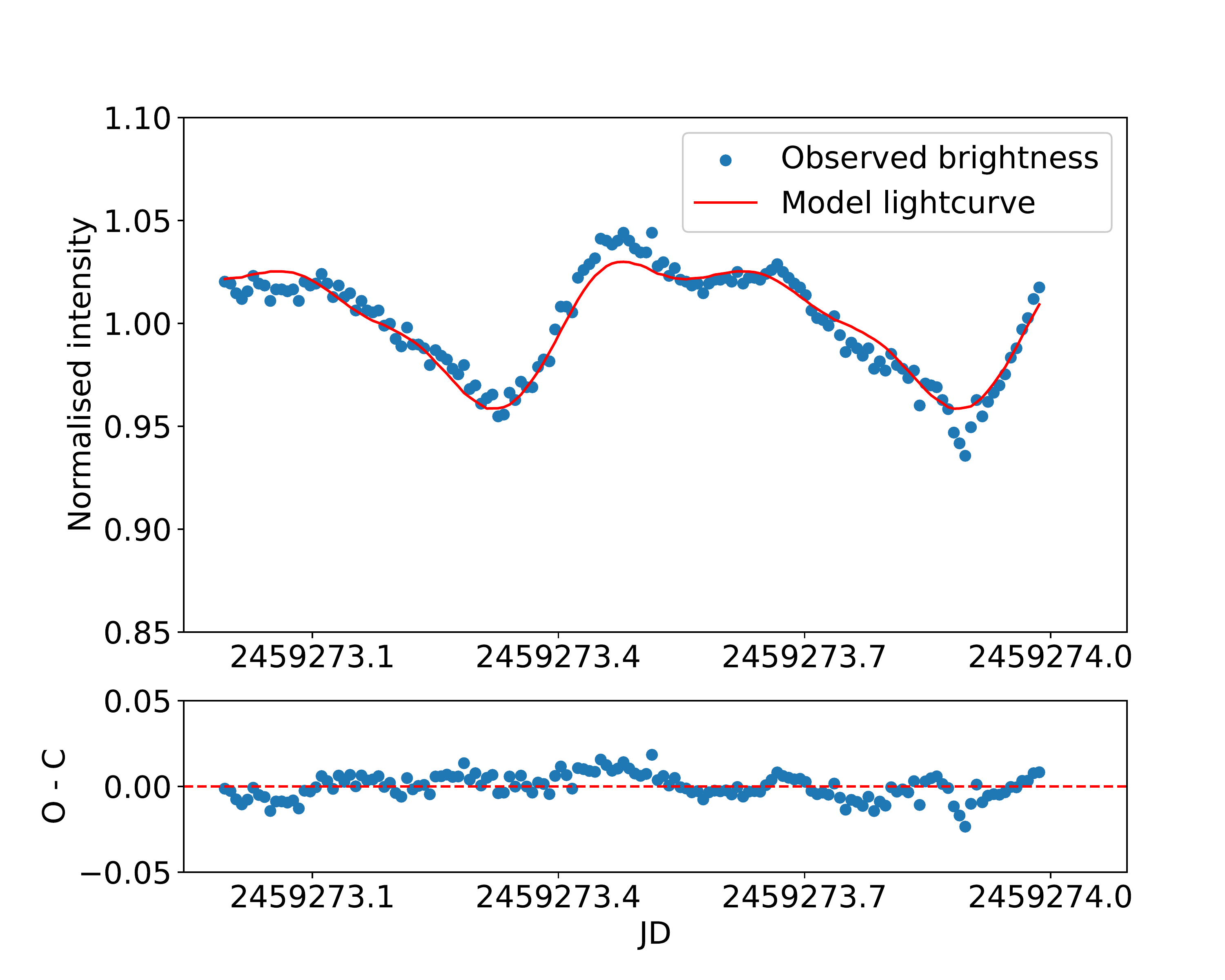}
\captionof{figure}{(429) Lotis model fit to TESS data}
\label{429lc_fit}
&
 \includegraphics[width=0.5\textwidth]{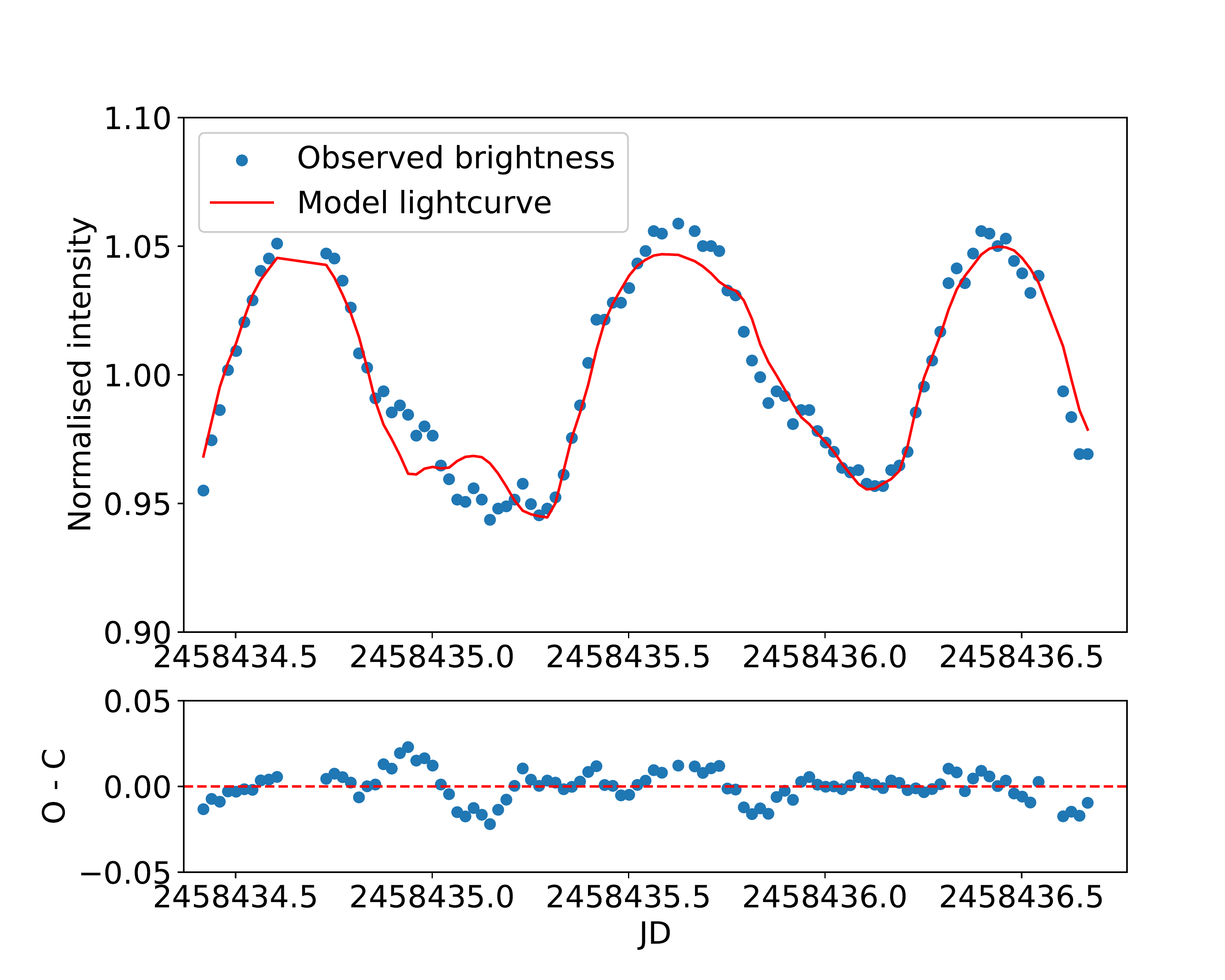}
\captionof{figure}{(527) Euryanthe model fit to TESS data}
\label{527lc_fit}
\\
 \includegraphics[width=0.5\textwidth]{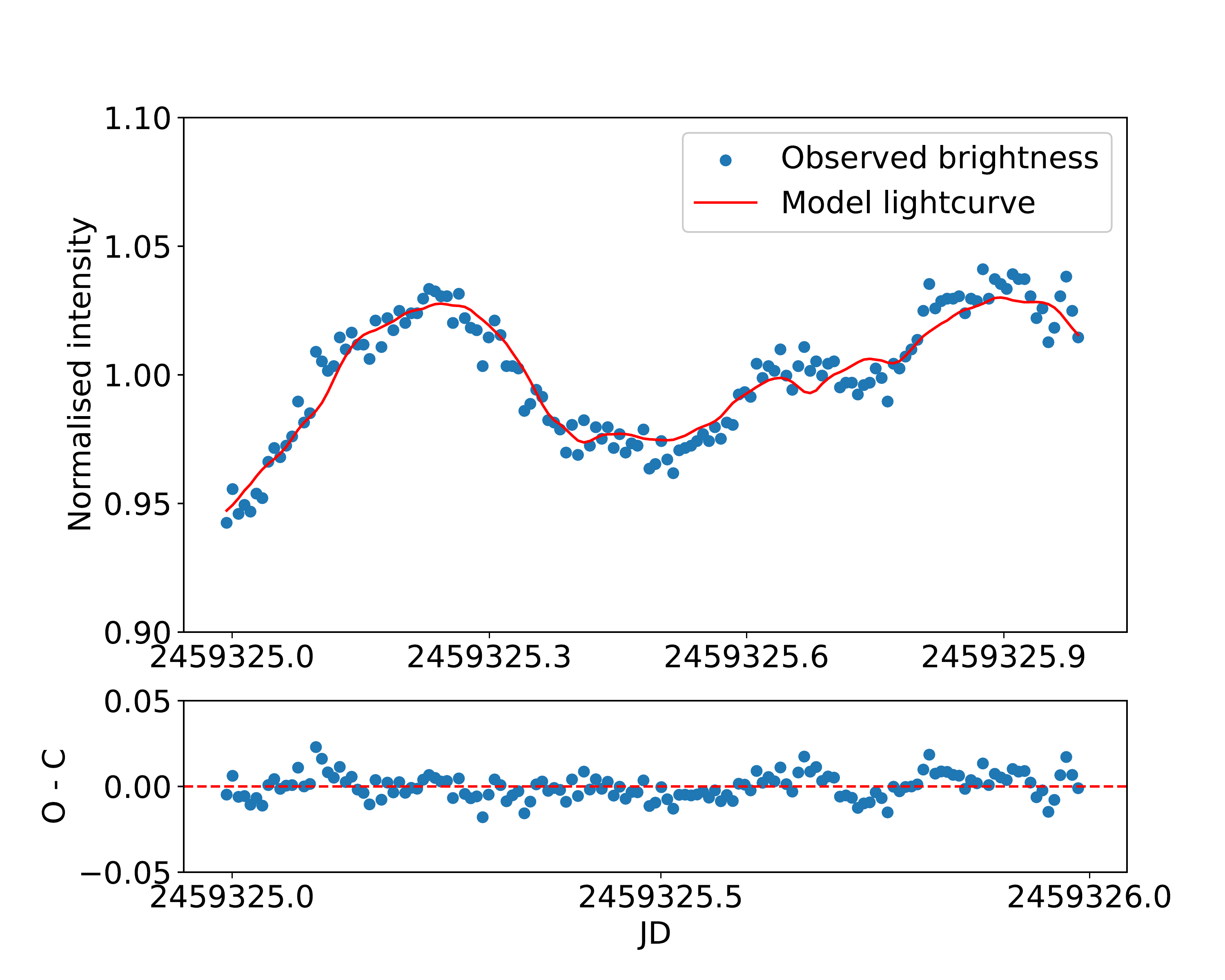}
\captionof{figure}{(541) Deborah model fit to TESS data}
\label{541lc_fit}
&
 \includegraphics[width=0.5\textwidth]{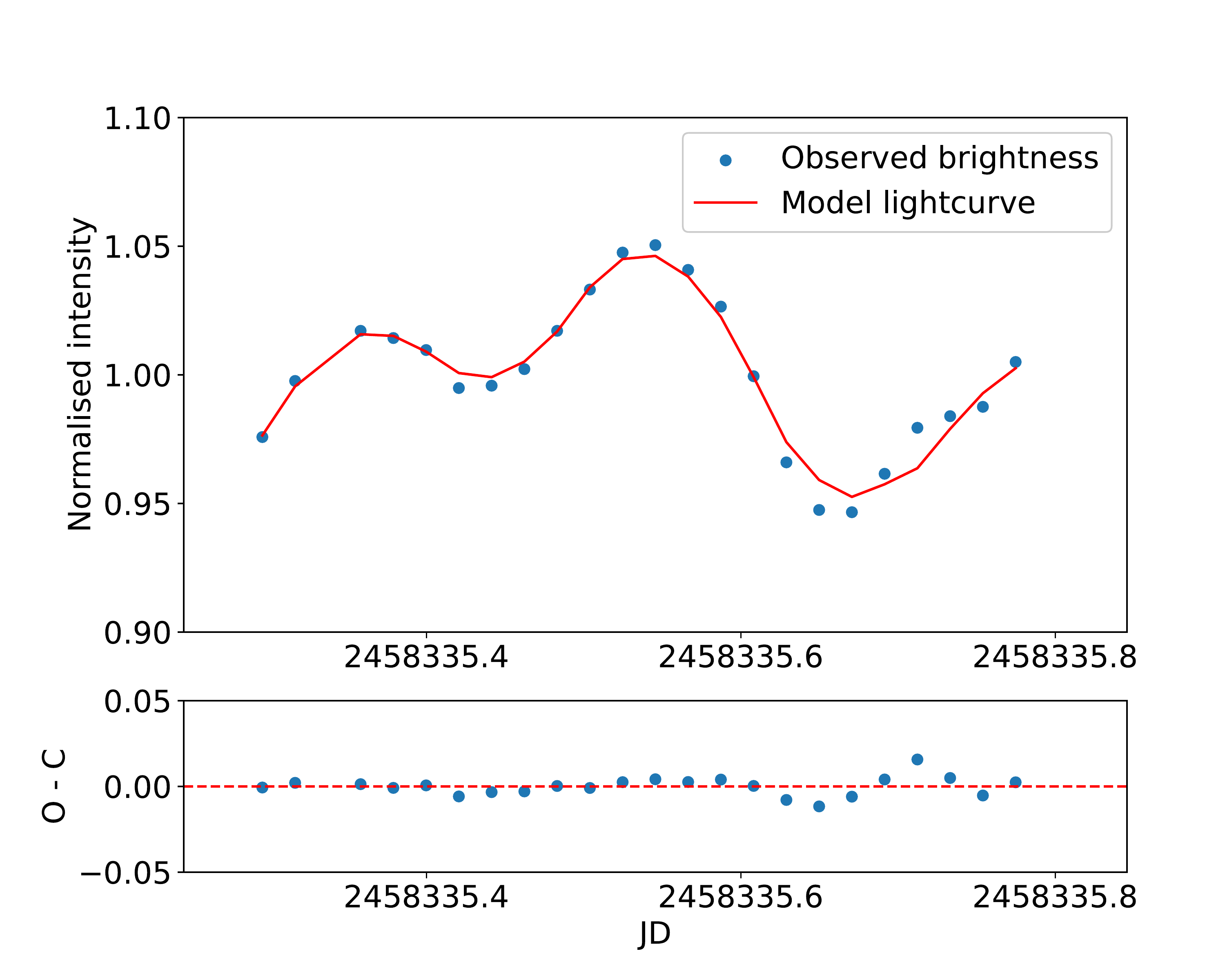}
\captionof{figure}{(672) Astarte model fit to TESS data}
\label{672lc_fit}
\\
 \includegraphics[width=0.5\textwidth]{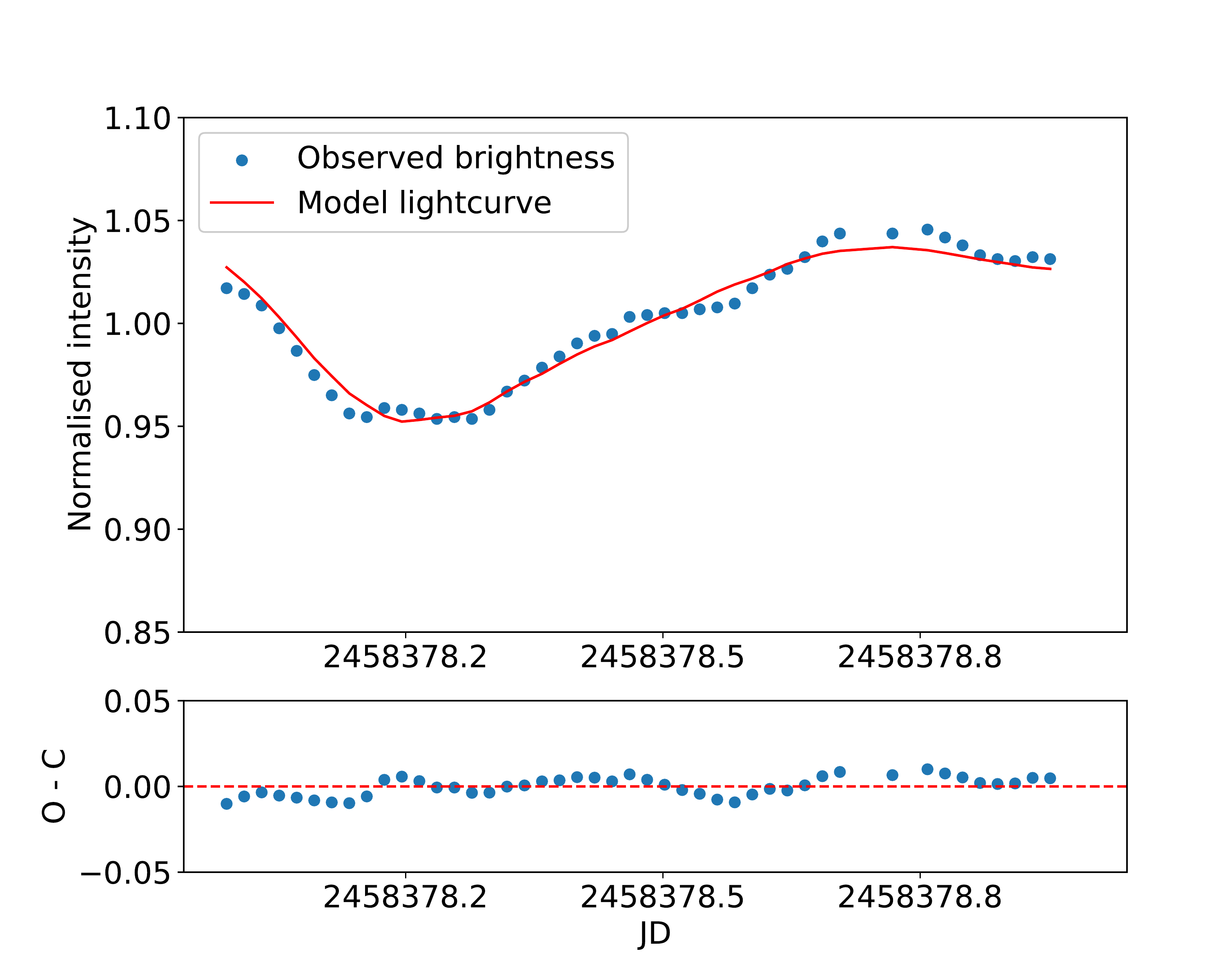}
\captionof{figure}{(1062) Ljuba model fit to Kepler data}
\label{1062lc_fit_Kepler}
&
 \includegraphics[width=0.5\textwidth]{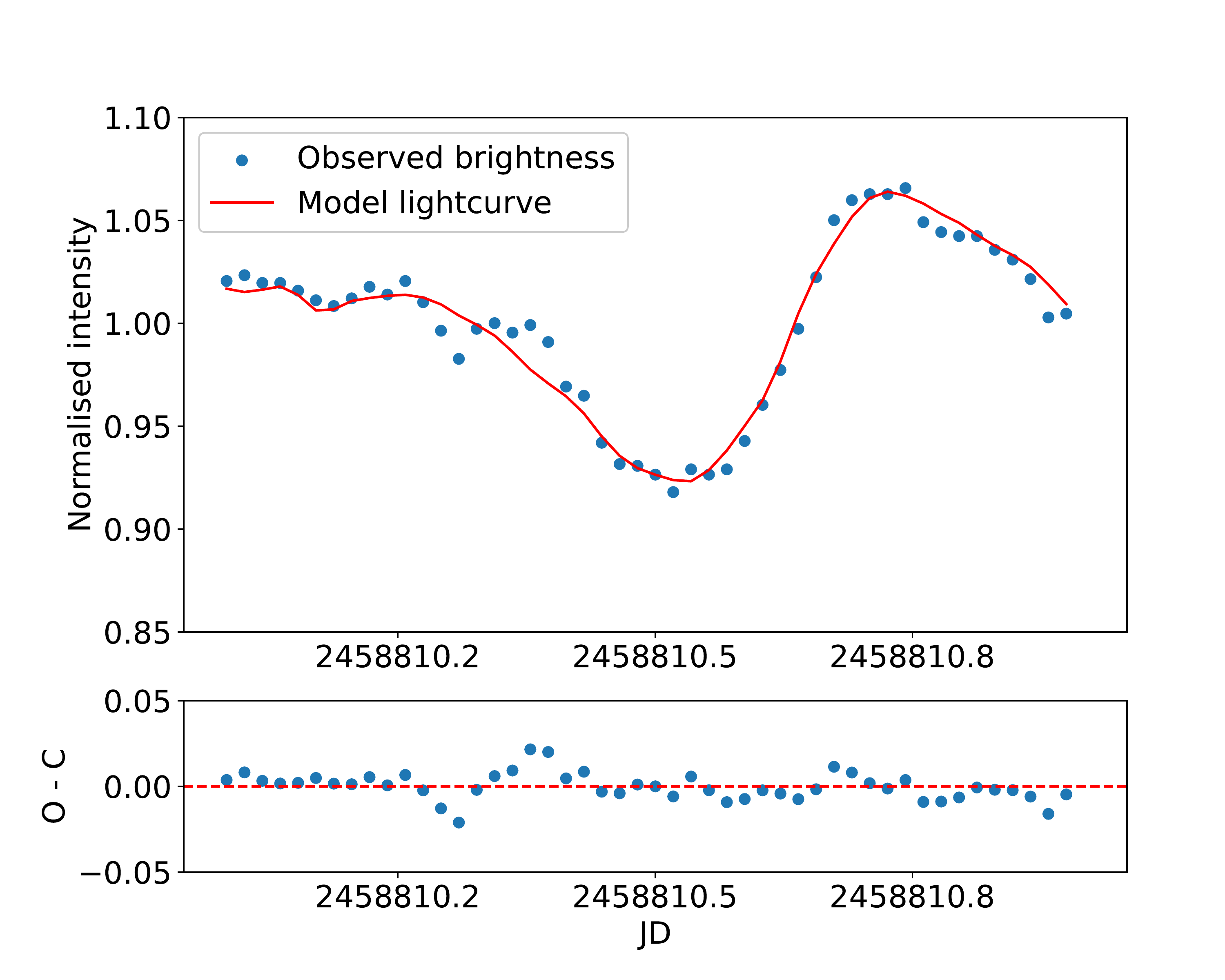}
\captionof{figure}{(1062) Ljuba model fit to TESS data}
\label{1062lc_fit}
\\
\end{tabularx}
    \end{table*}

\clearpage
\newpage

\section{Model fits to thermal light curves}

Model fit to non-saturated WISE thermal light curves (Figures \ref{215irW3} - \ref{1062irW4}) for the preferred pole solution (see Table~\ref{tab:occ_diam}). When no pole solution was preferred, the fit for pole 1 is shown.

    \begin{table*}[h!]
    \centering
\vspace{0.5cm}
\begin{tabularx}{\linewidth}{XXX}
 \includegraphics[width=0.31\textwidth]{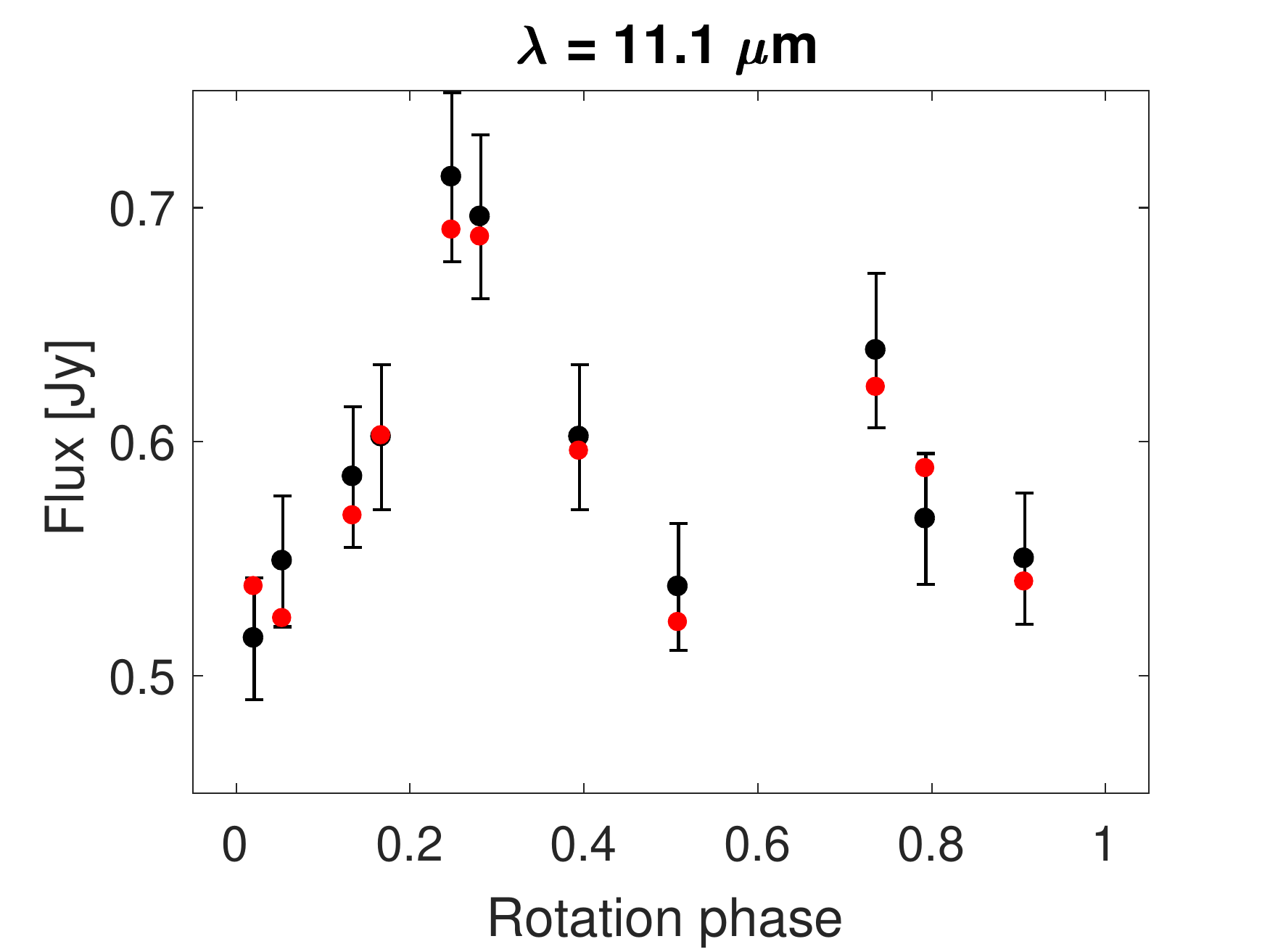}
\captionof{figure}{(215) Oenone model flux (red points) compared with data point (black) from the W3 band from WISE spacecraft.}
\label{215irW3}
&
 \includegraphics[width=0.31\textwidth]{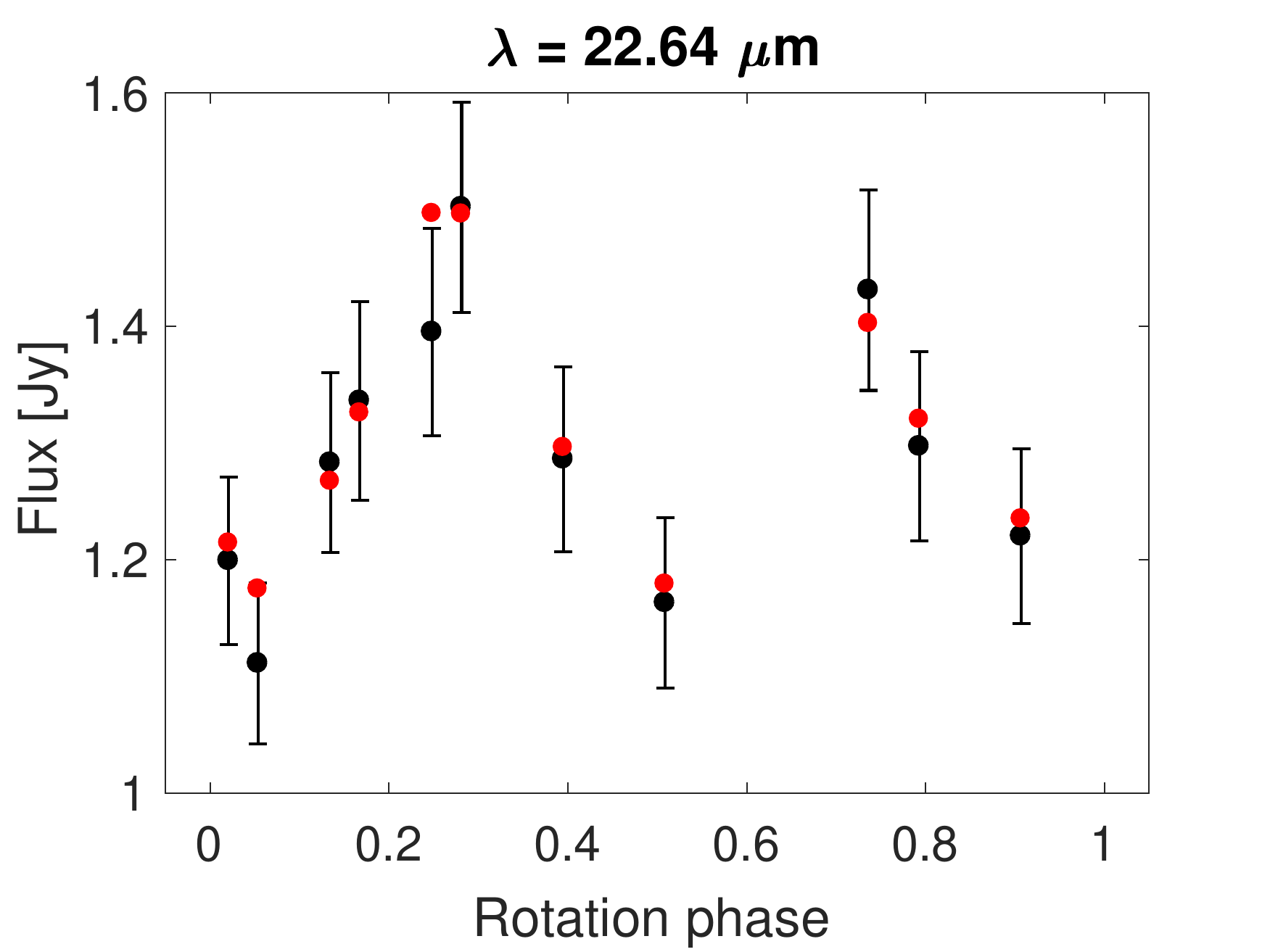}
\captionof{figure}{(215) Oenone, W4 band.}
\label{215irW4}
&
 \includegraphics[width=0.31\textwidth]{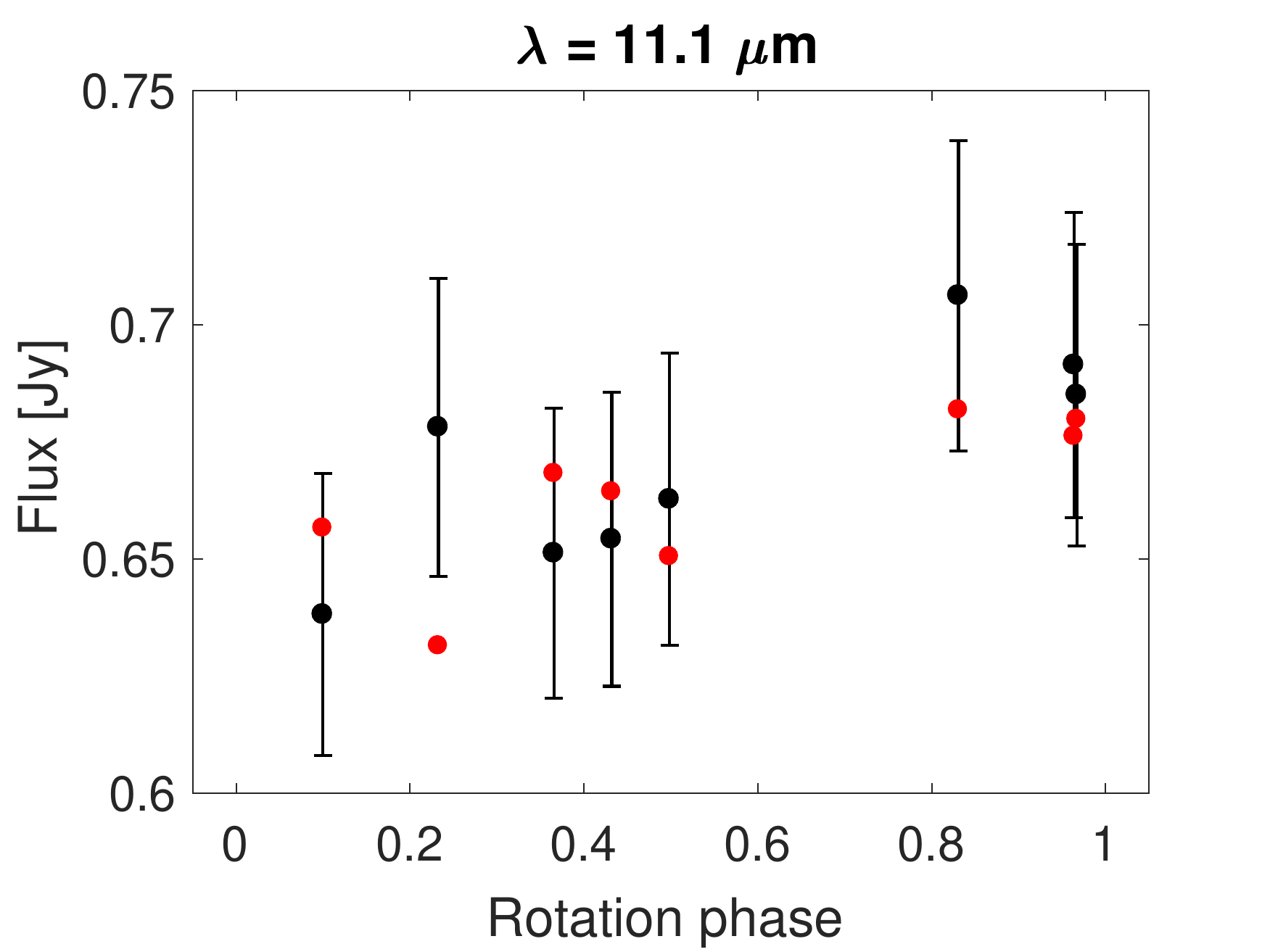}
\captionof{figure}{(279) Thule{\color{highlight}, W3 band.}}
\label{279irW3}
\\
 \includegraphics[width=0.31\textwidth]{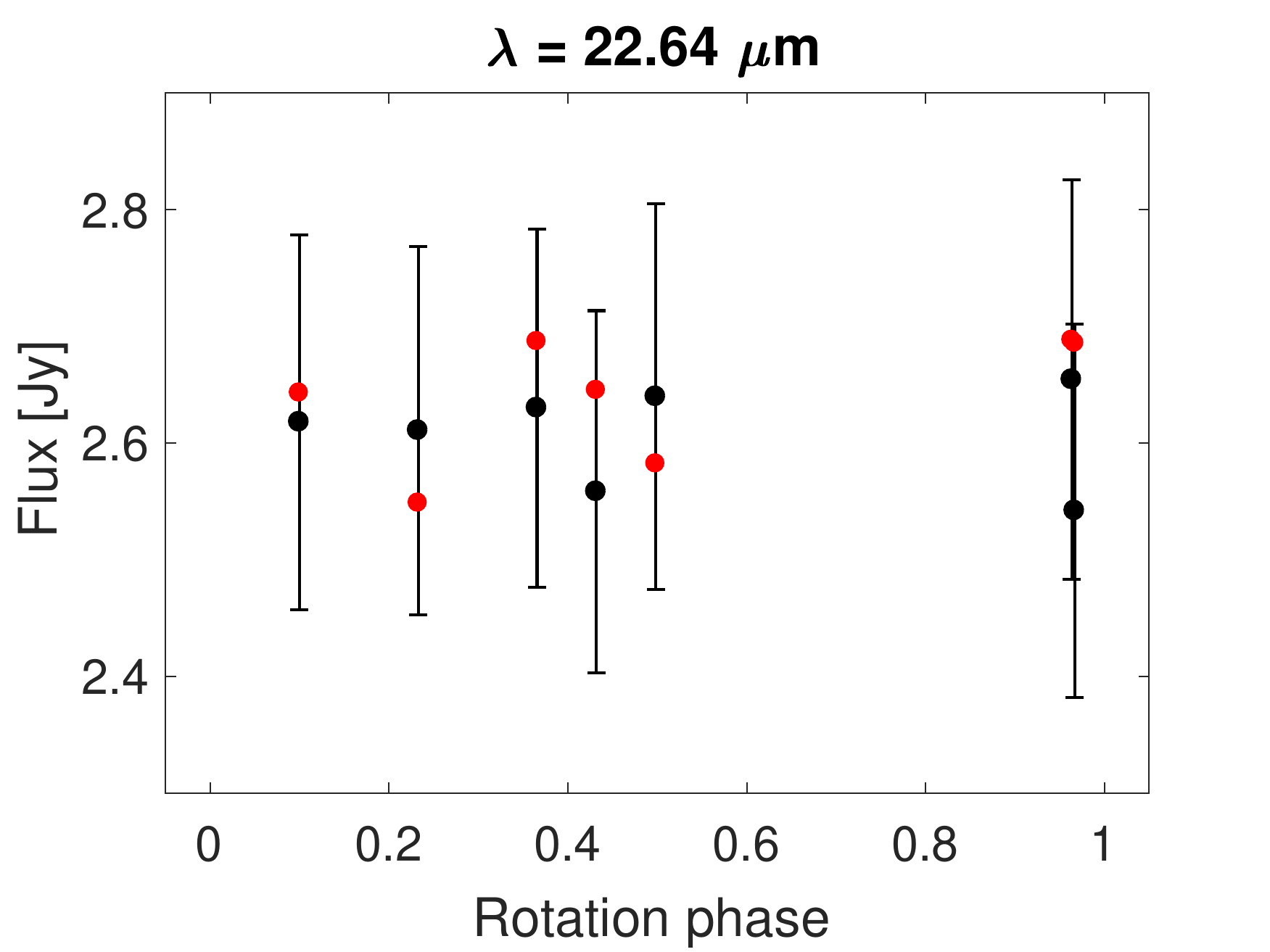}
\captionof{figure}{(279) Thule{\color{highlight}, W4 band.}}
\label{279irW4}
&
 \includegraphics[width=0.31\textwidth]{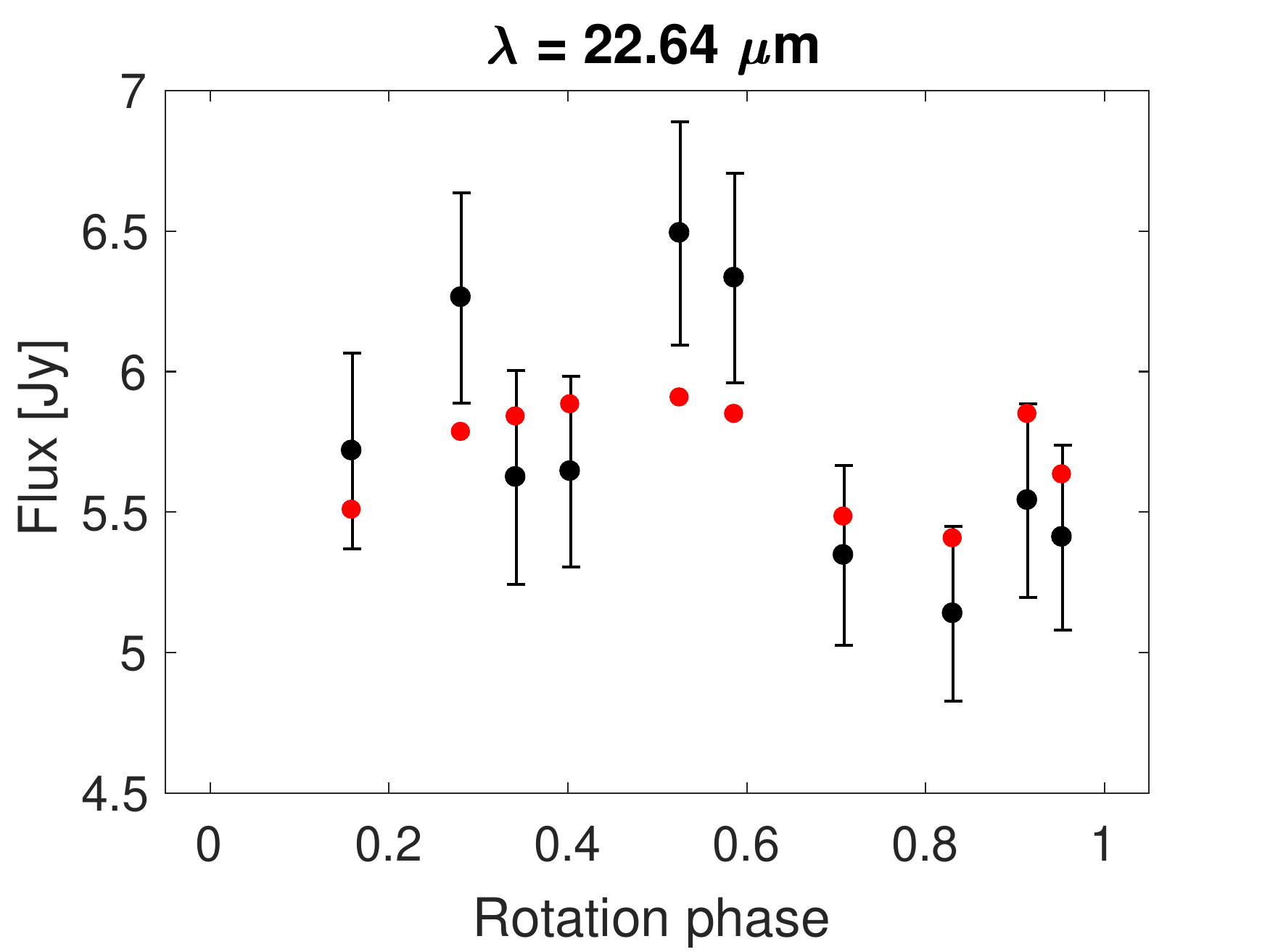}
\captionof{figure}{(357) Ninina{\color{highlight}, W4 band.}}
\label{357irW4}
&
 \includegraphics[width=0.31\textwidth]{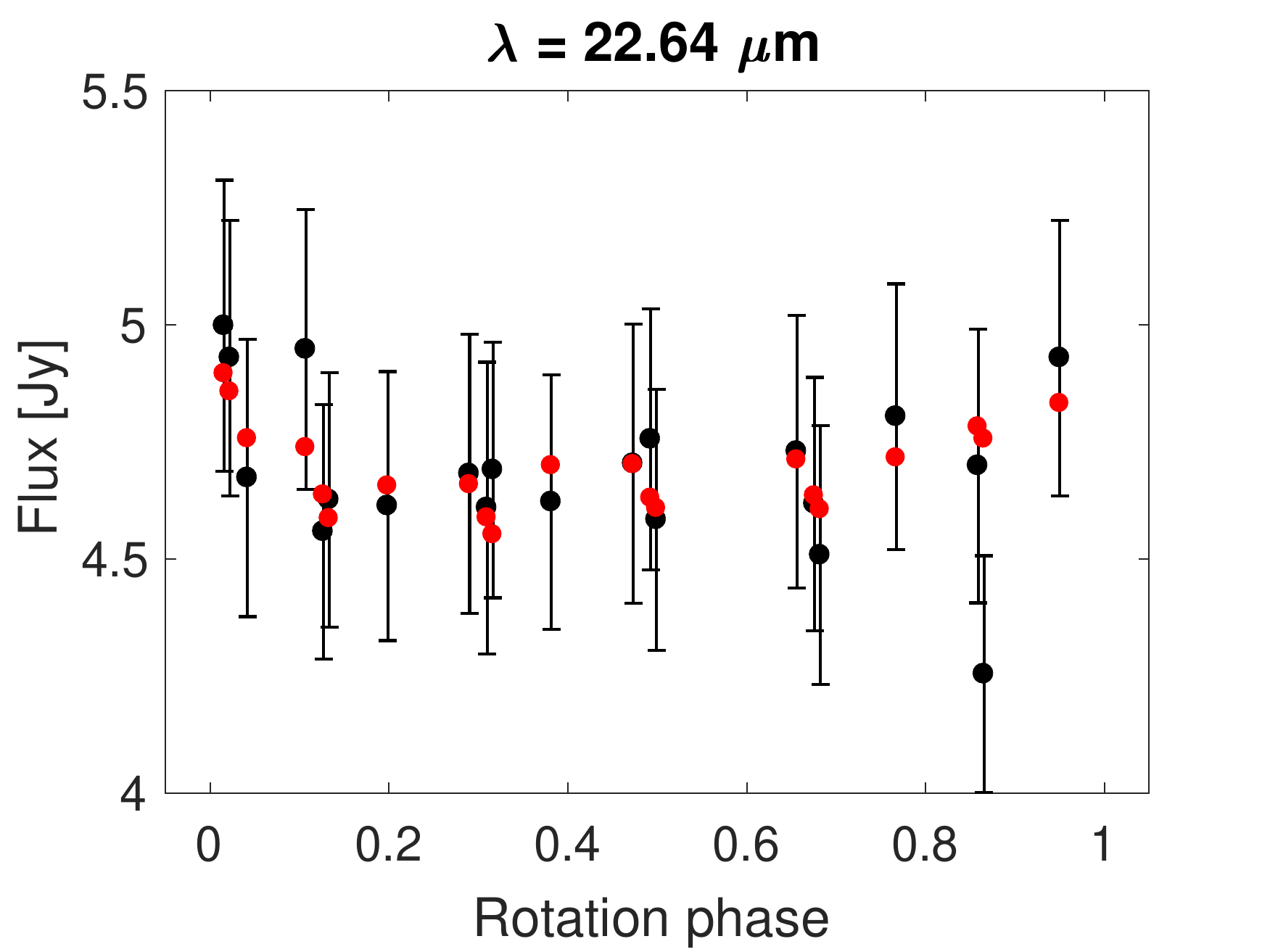}
\captionof{figure}{(366) Vincentina{\color{highlight}, W4 band.}}
\label{366irW4}
\\
 \includegraphics[width=0.31\textwidth]{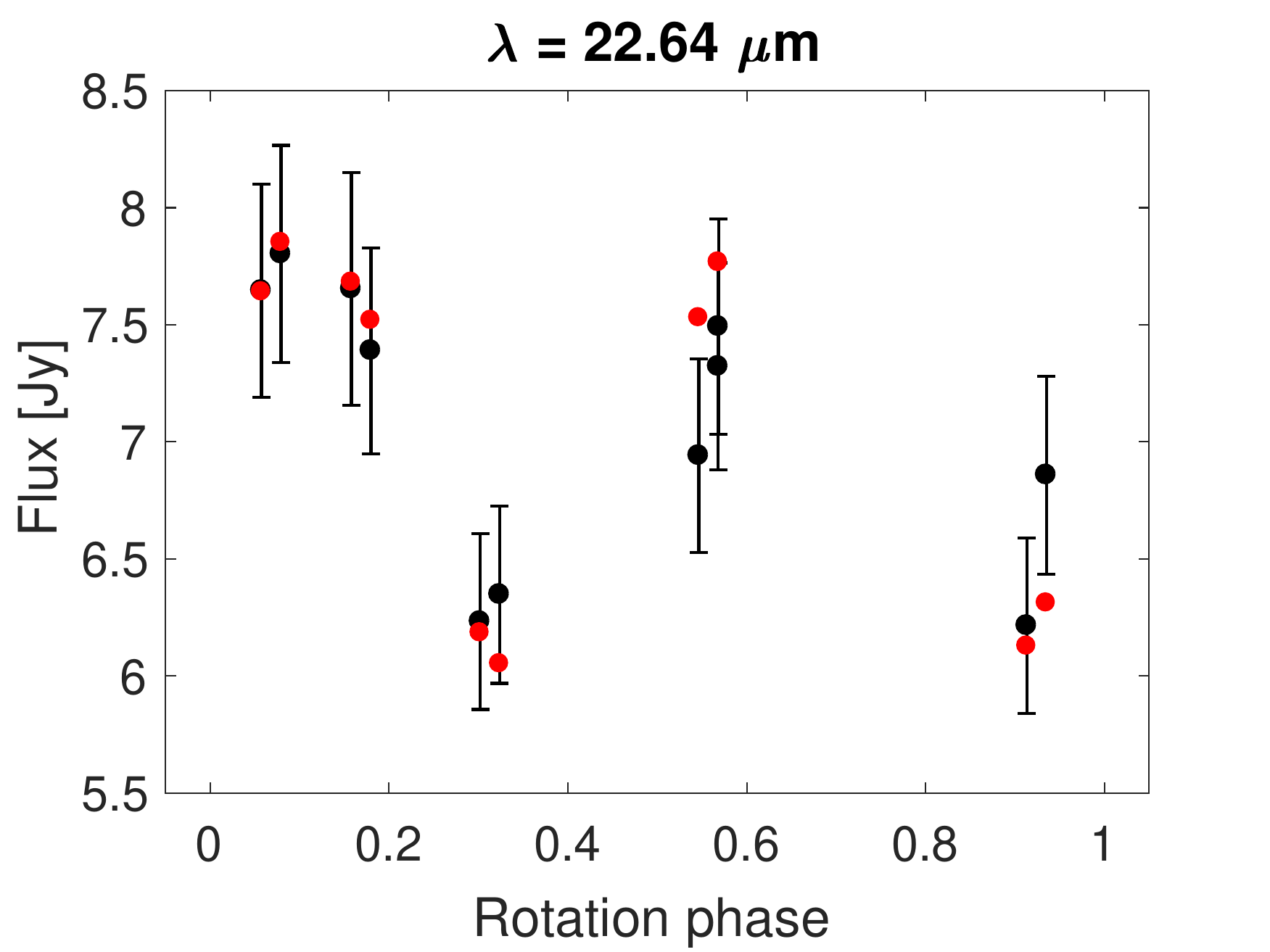}
\captionof{figure}{(373) Melusina{\color{highlight}, W4 band.}}
\label{373irW4}
&
 \includegraphics[width=0.31\textwidth]{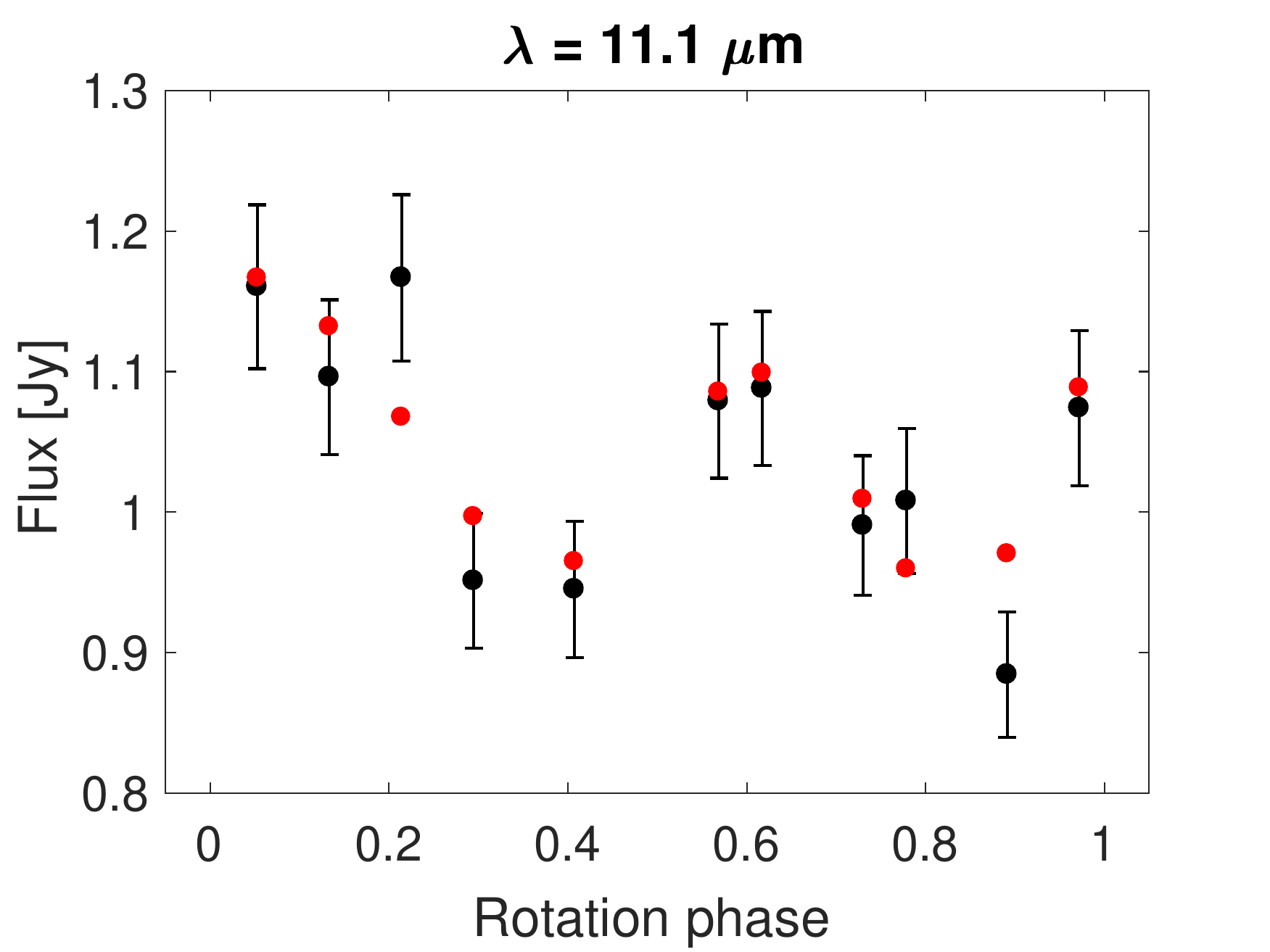}
\captionof{figure}{(395) Delia{\color{highlight}, W3 band.}}
\label{395irW3}
&
 \includegraphics[width=0.31\textwidth]{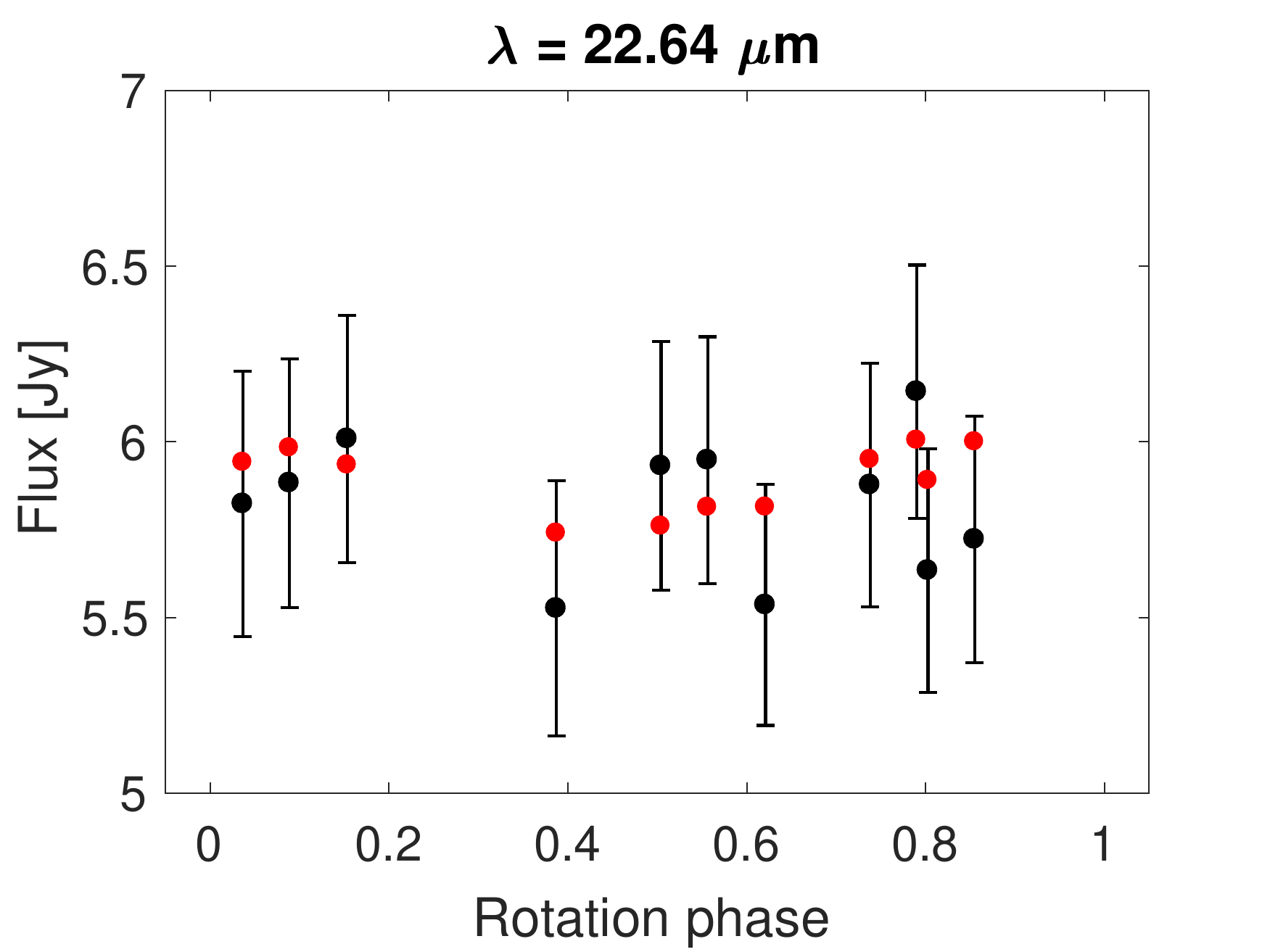}
\captionof{figure}{(429) Lotis{\color{highlight}, W4 band.}}
\label{429irW4}
\\
 \includegraphics[width=0.31\textwidth]{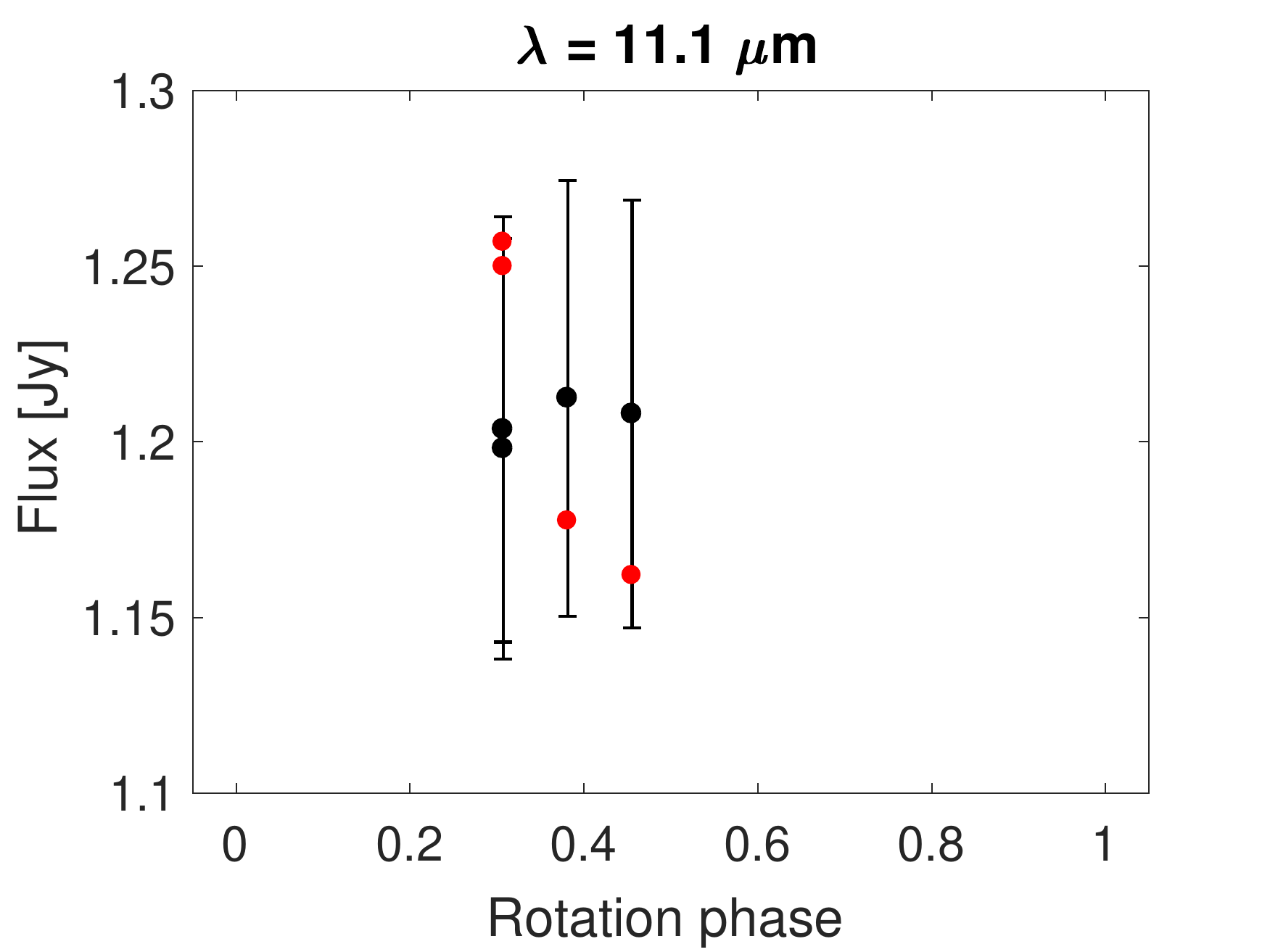}
\captionof{figure}{(527) Euryanthe{\color{highlight}, W3 band.}}
\label{527irW3}
&
 \includegraphics[width=0.31\textwidth]{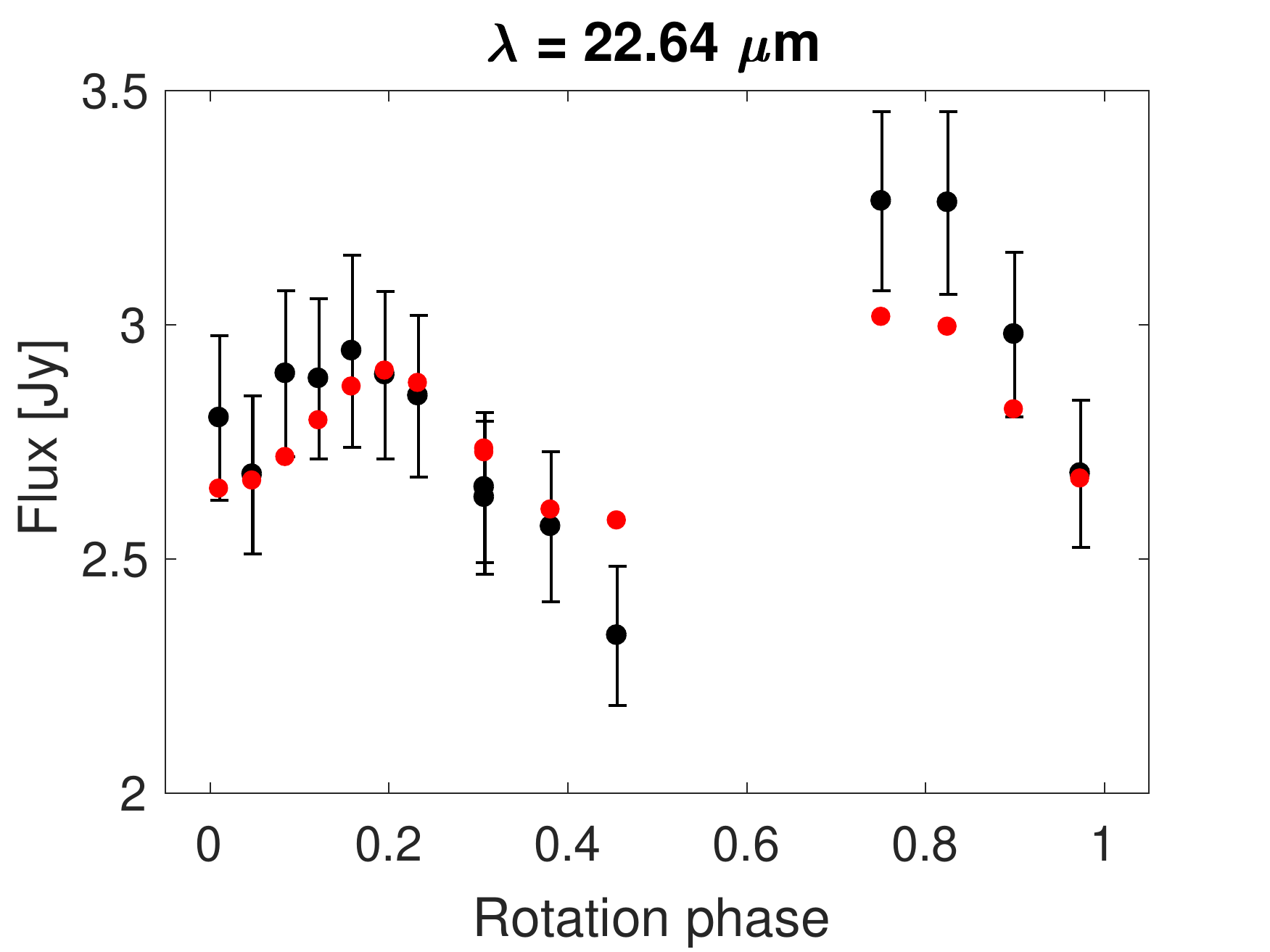}
\captionof{figure}{(527) Euryanthe{\color{highlight}, W4 band.}}
\label{527irW4}
&
 \includegraphics[width=0.31\textwidth]{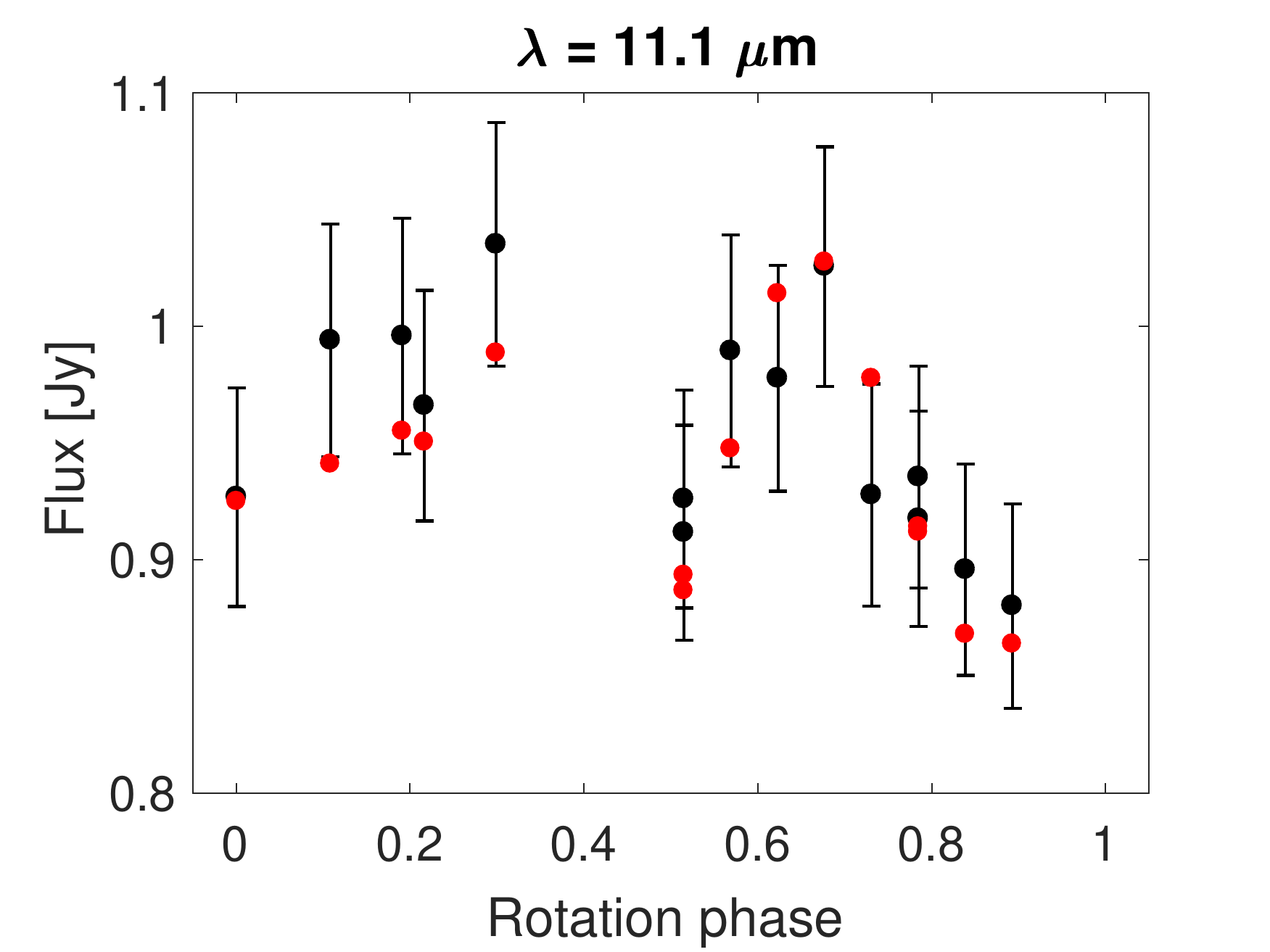}
\captionof{figure}{(541) Deborah{\color{highlight}, W3 band.}}
\label{541irW3}
\\
\end{tabularx}
    \end{table*}

\clearpage
\newpage

\begin{table*}[h!]
    \centering
\vspace{0.5cm}
\begin{tabularx}{\linewidth}{XXX}
 \includegraphics[width=0.33\textwidth]{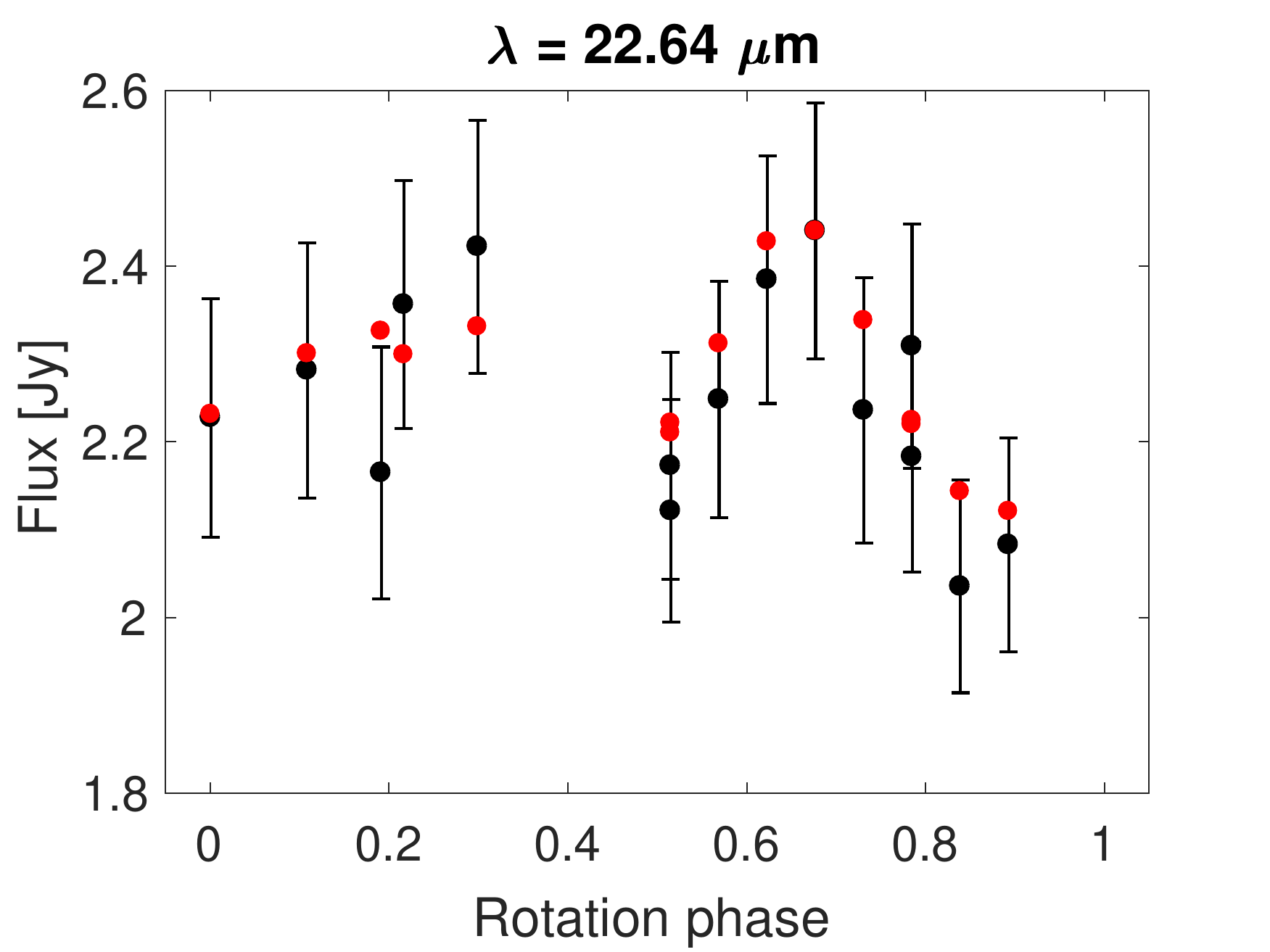}
\captionof{figure}{(541) Deborah{\color{highlight}, W4 band.}}
\label{541irW4}
&
 \includegraphics[width=0.33\textwidth]{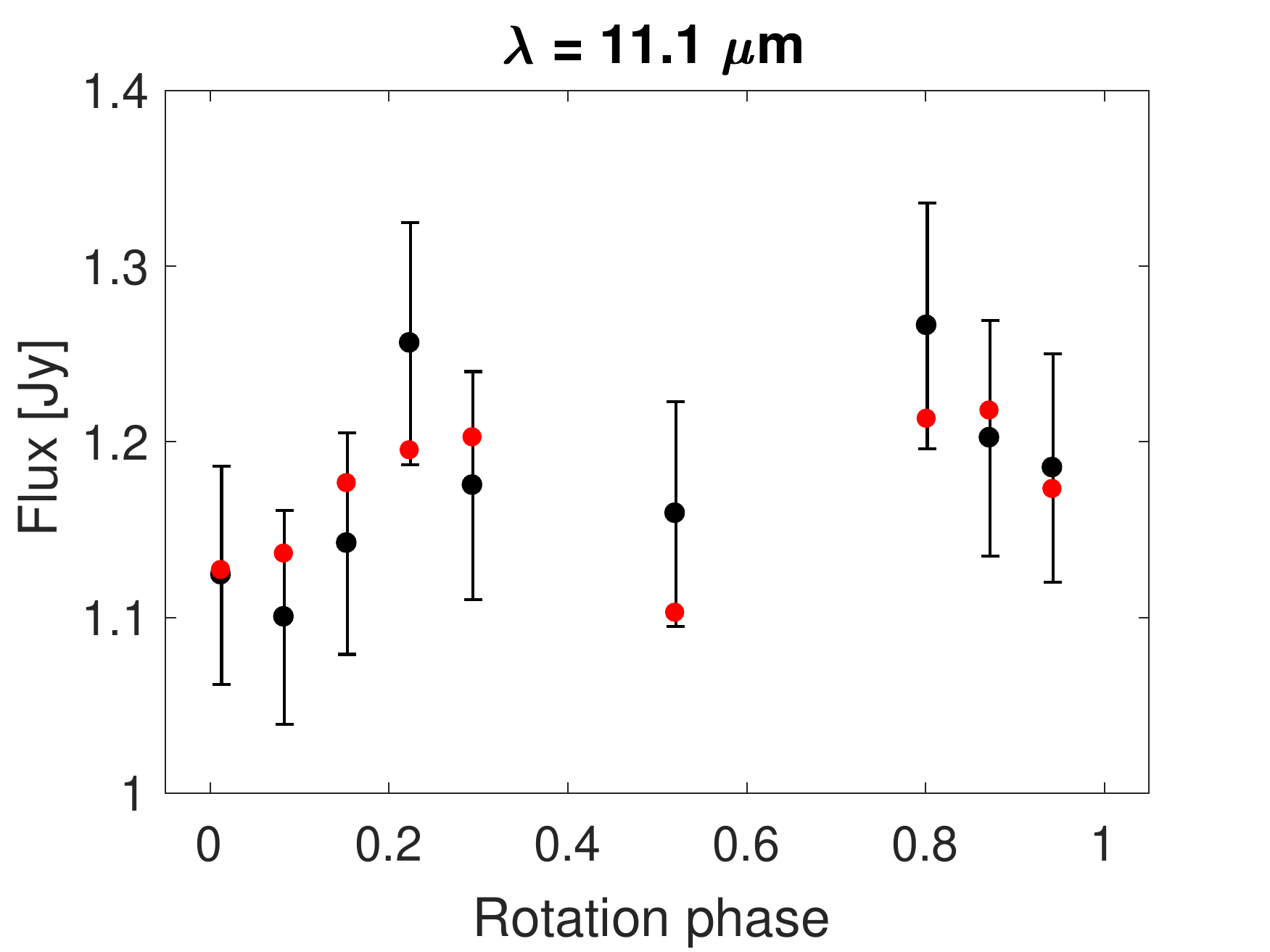}
\captionof{figure}{(672) Astarte{\color{highlight}, W3 band.}}
\label{814irW3}
&
 \includegraphics[width=0.33\textwidth]{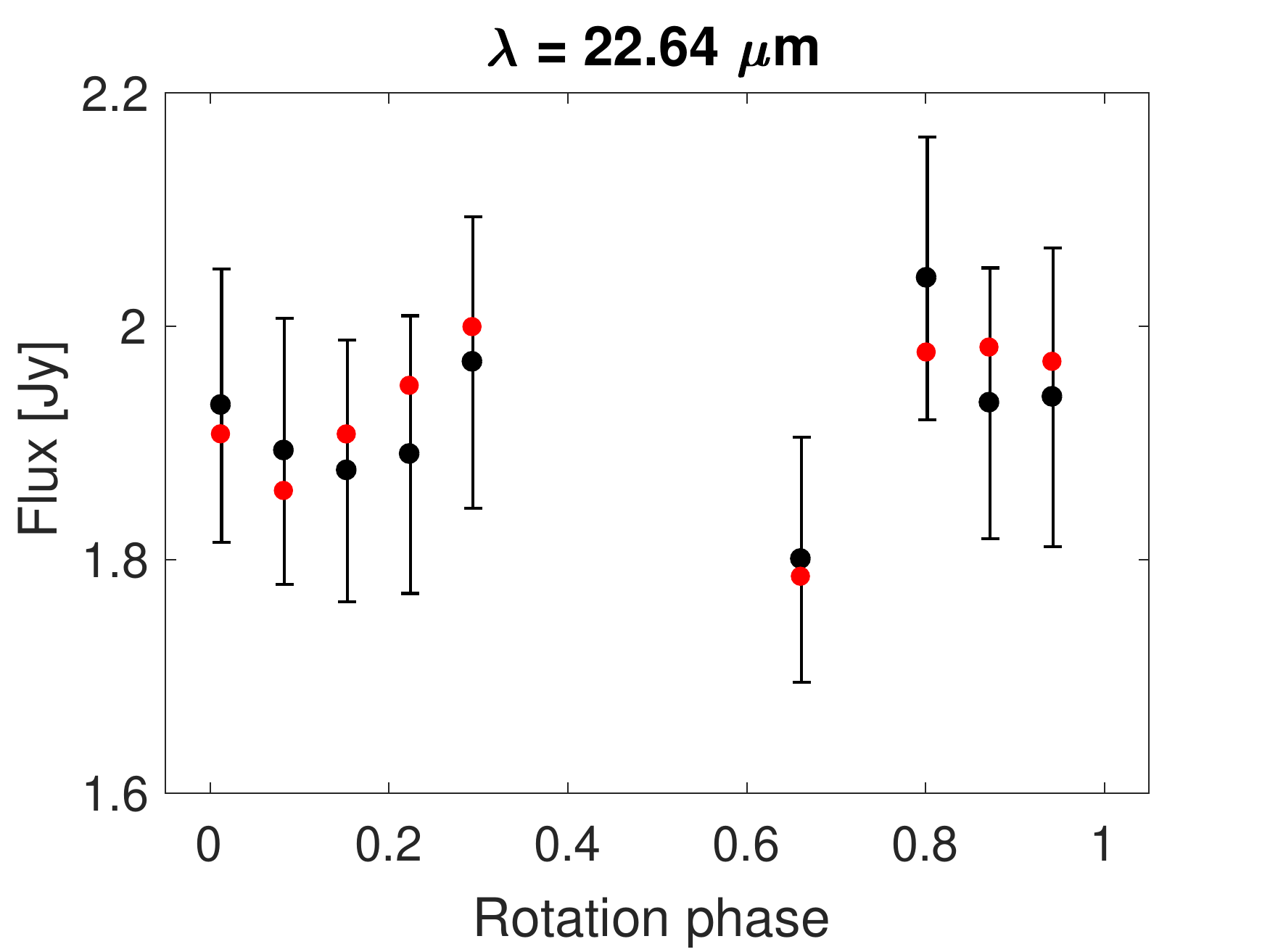}
\captionof{figure}{(672) Astarte{\color{highlight}, W4 band.}}
\label{672irW4}
\\
 \includegraphics[width=0.33\textwidth]{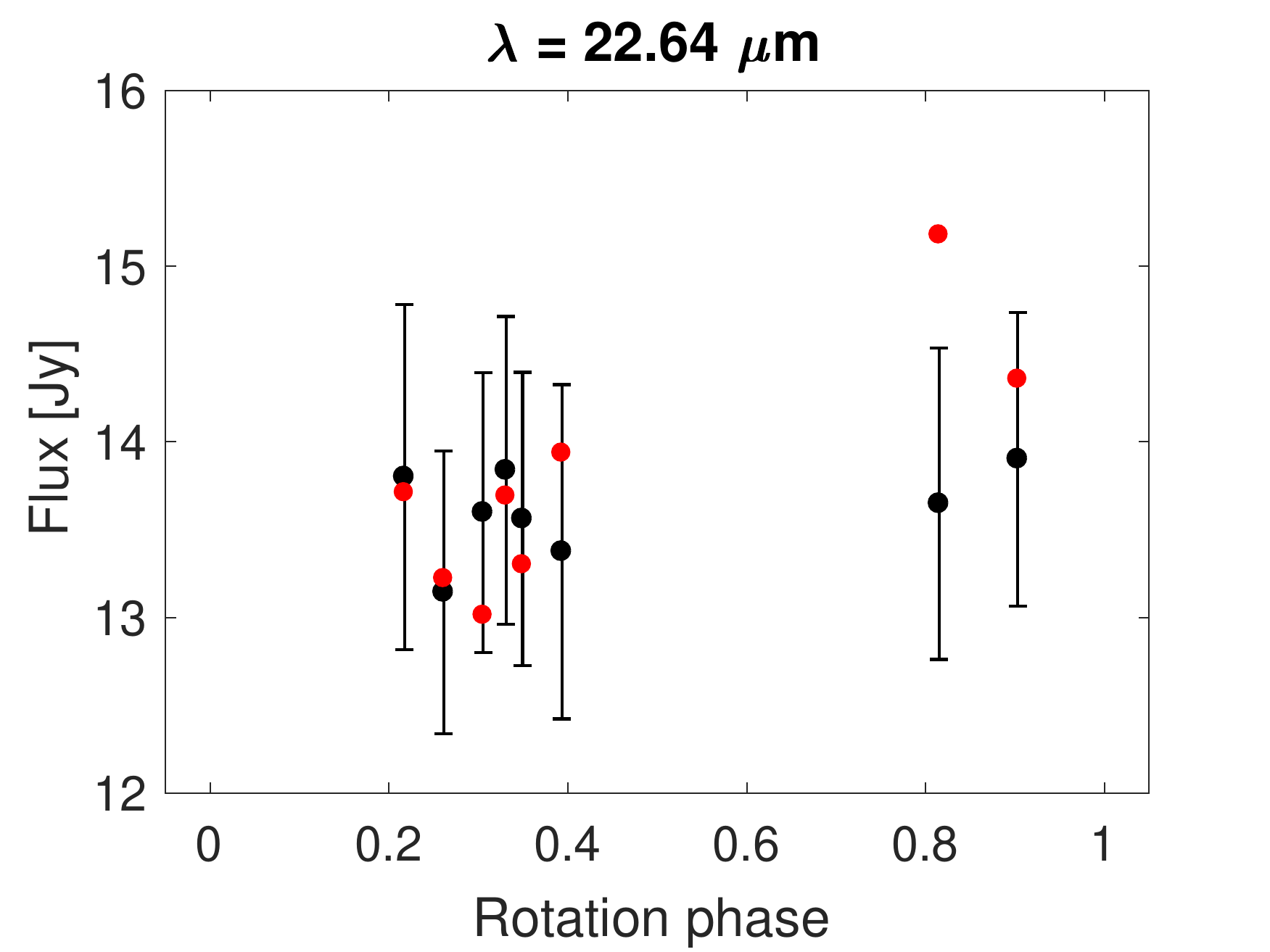}
\captionof{figure}{(814) Tauris{\color{highlight}, W4 band.}}
\label{814irW4}
&
 \includegraphics[width=0.33\textwidth]{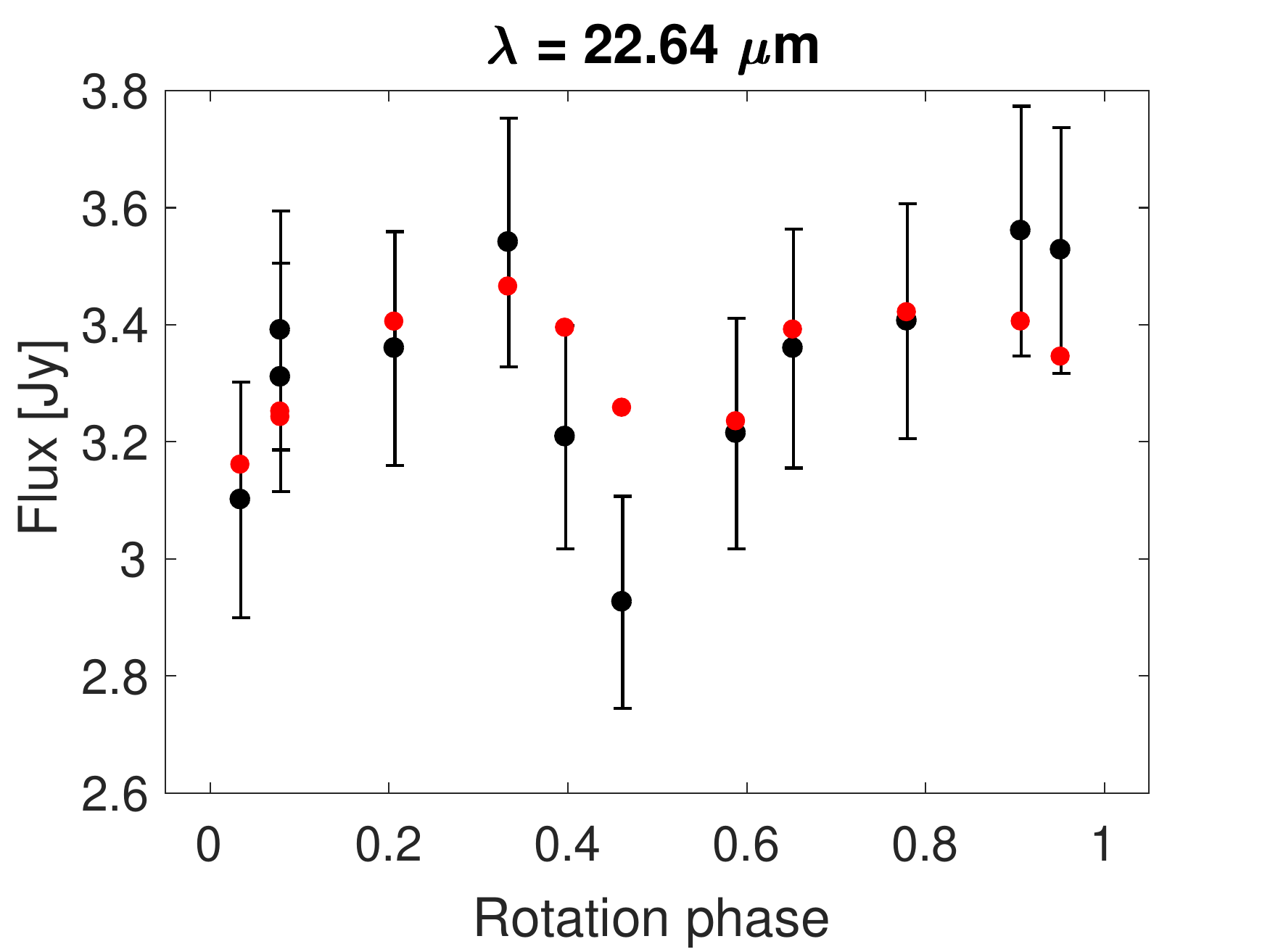}
\captionof{figure}{(859) Bouzareah{\color{highlight}, W4 band.}}
\label{859irW4}
&
 \includegraphics[width=0.33\textwidth]{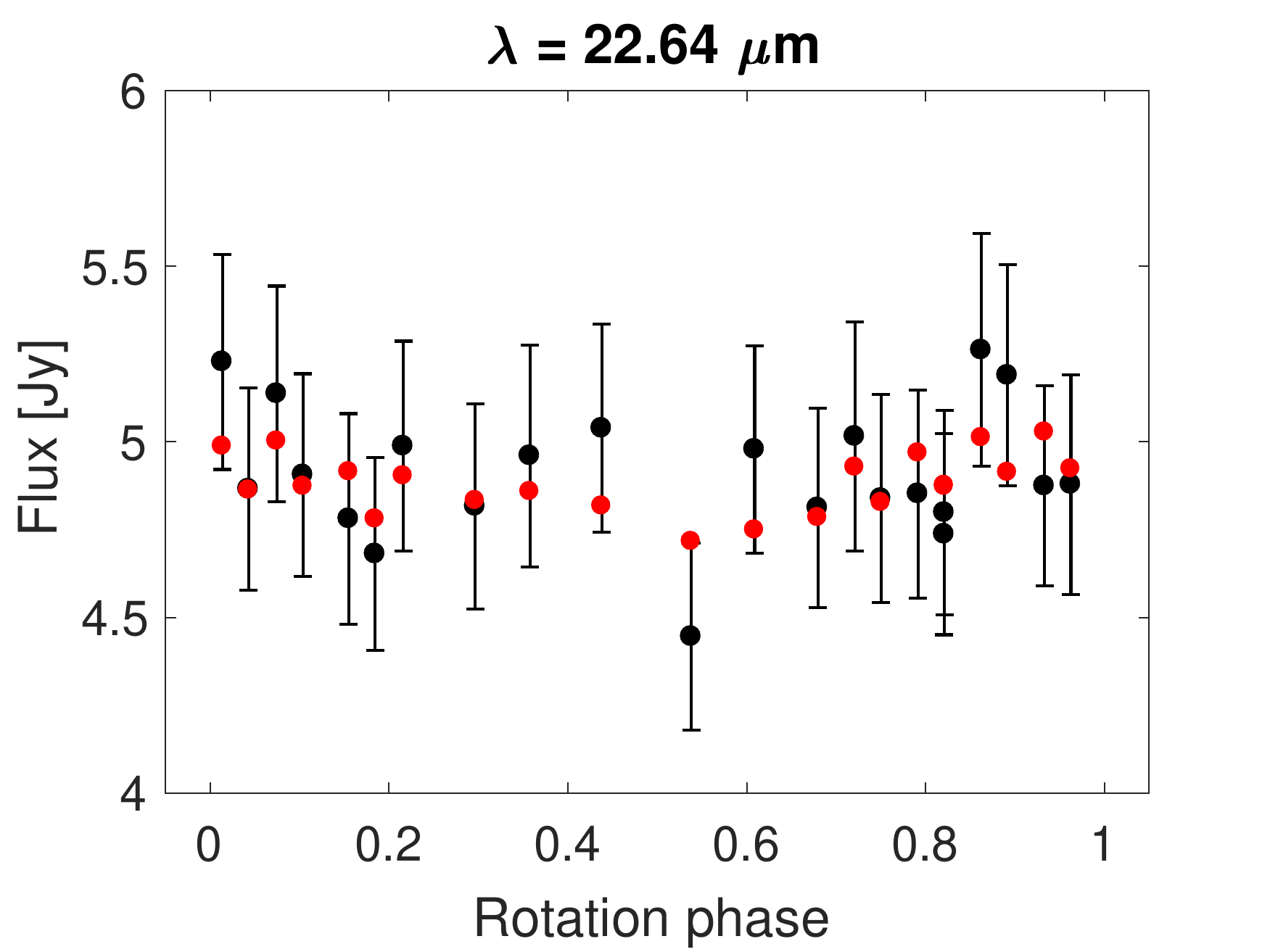}
\captionof{figure}{(907) Rhoda{\color{highlight}, W4 band.}}
\label{907irW4}
\\
 \includegraphics[width=0.33\textwidth]{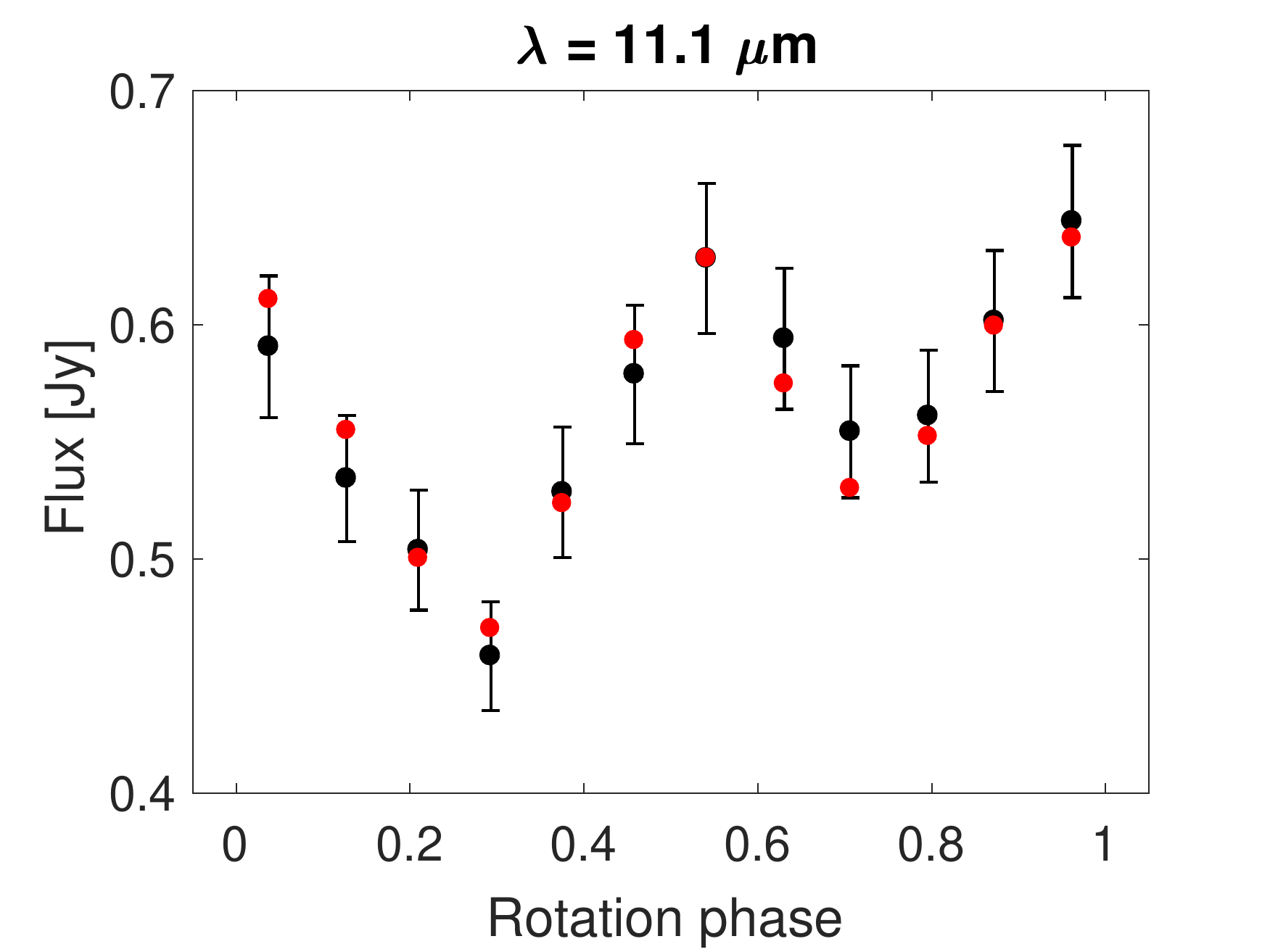}
\captionof{figure}{(931) Whittemora {\color{highlight}, W3 band.}}
\label{931irW3}
&
 \includegraphics[width=0.33\textwidth]{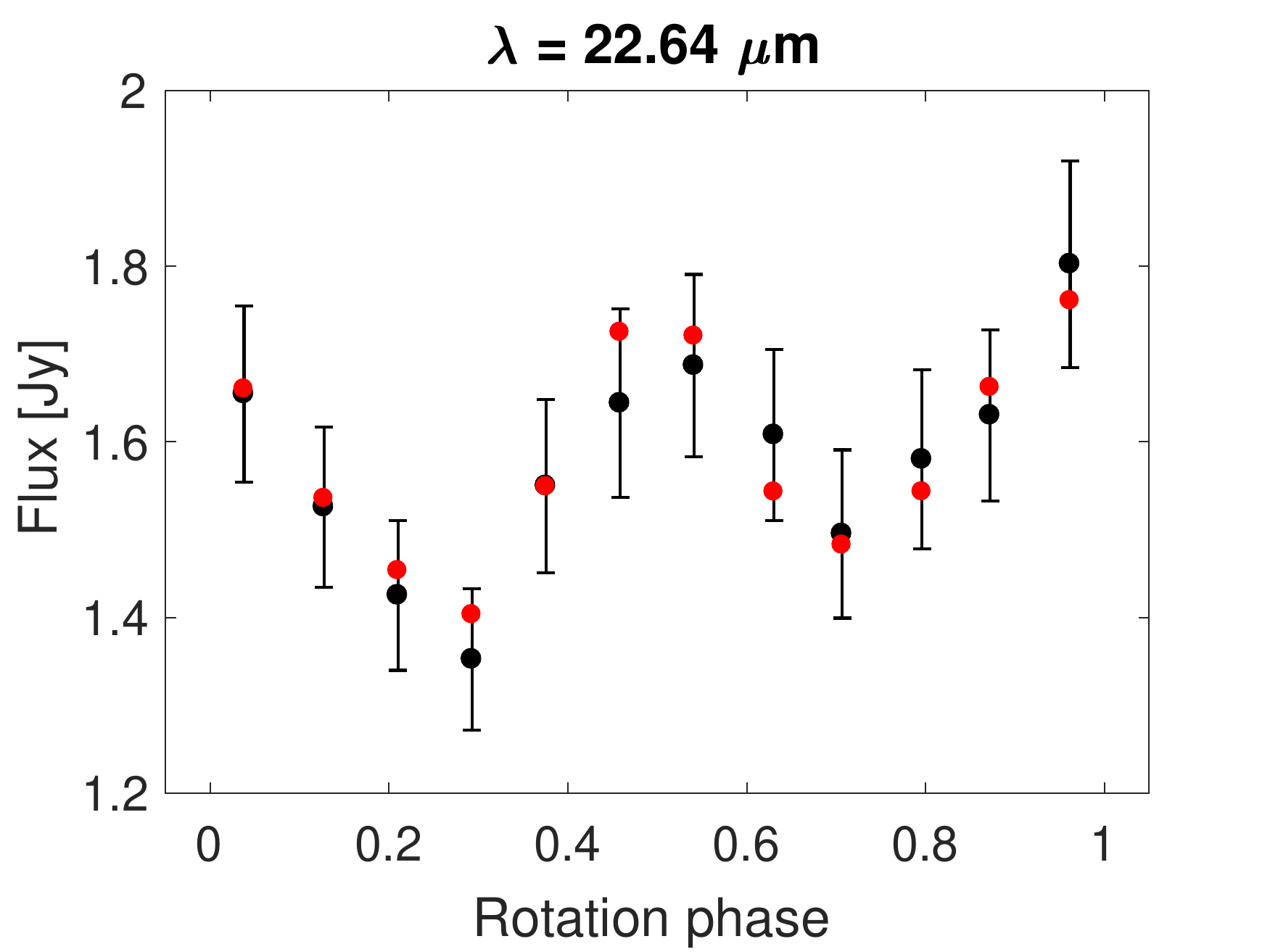}
\captionof{figure}{(931) Whittemora{\color{highlight}, W4 band.}}
\label{931irW4}
&
 \includegraphics[width=0.33\textwidth]{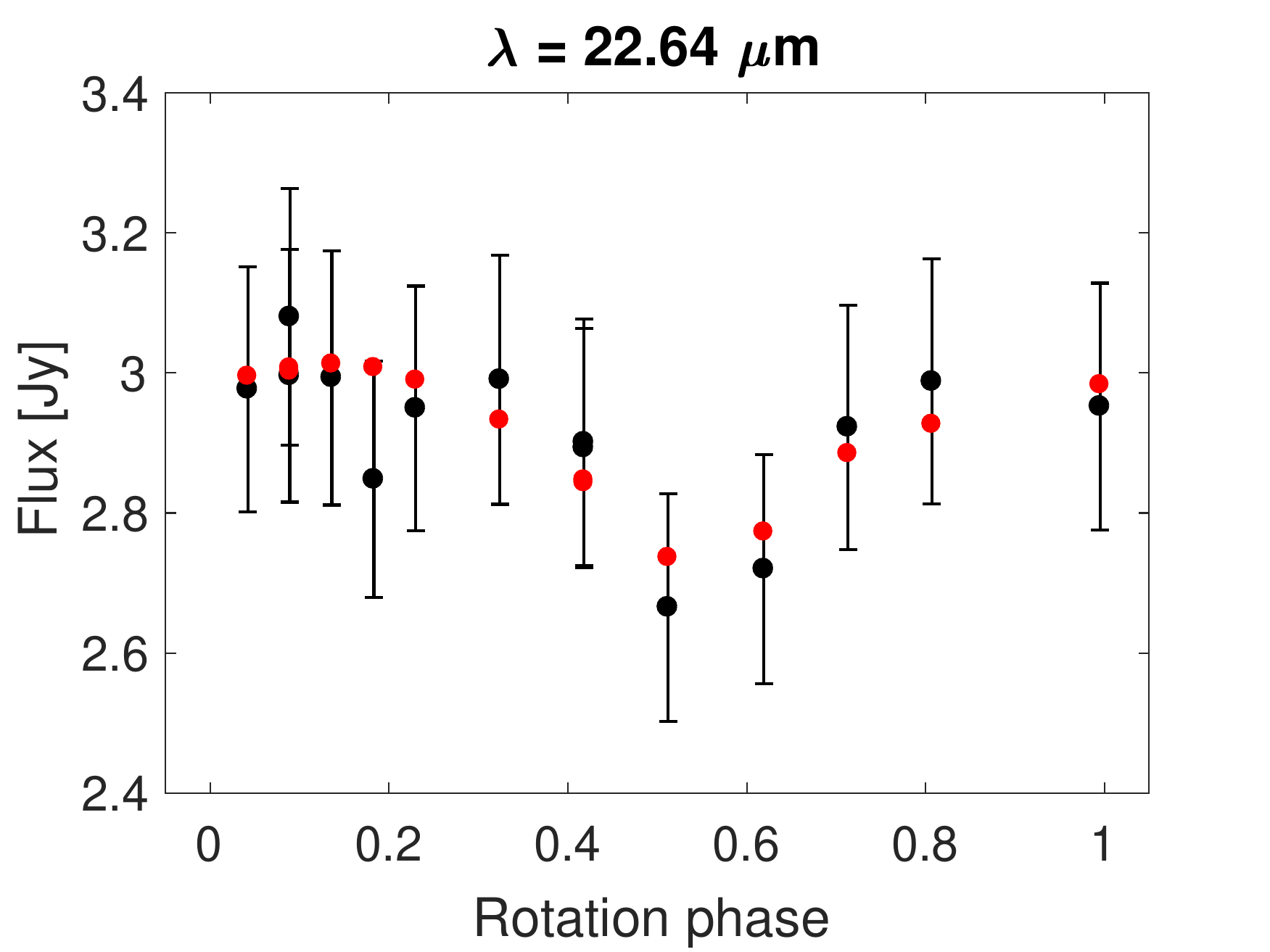}
\captionof{figure}{(1062) Ljuba{\color{highlight}, W4 band.}}
\label{1062irW4}

\end{tabularx}
    \end{table*}

\clearpage
\newpage

\section{Occultation fits}
\begin{figure*}[h!]
\centering
\includegraphics[width=0.33\textwidth]{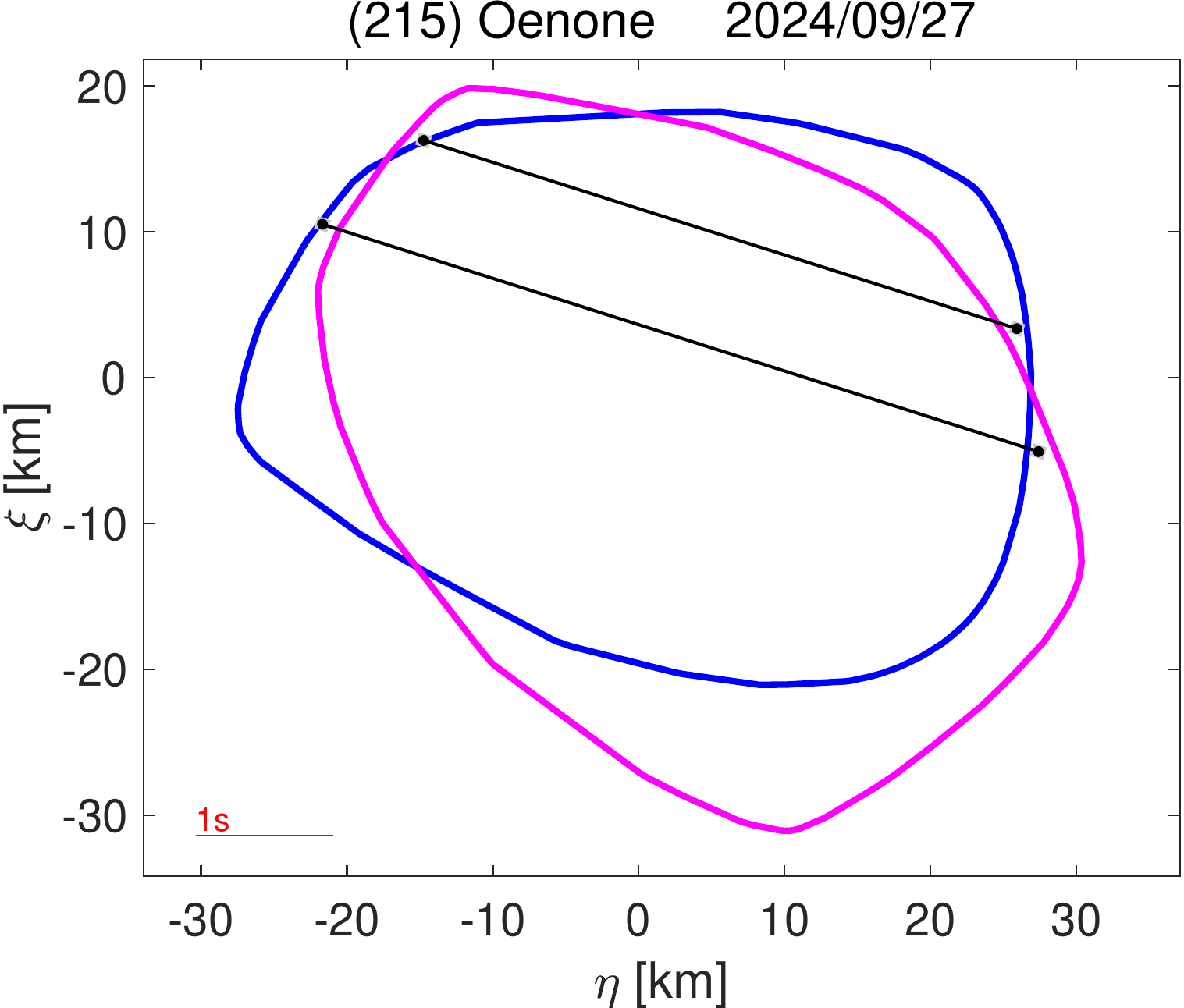}
\caption{Occultation fits for asteroid (215) Oenone}
\label{215_occ}
\end{figure*}

\begin{figure*}[h!]
\includegraphics[width=0.33\textwidth]{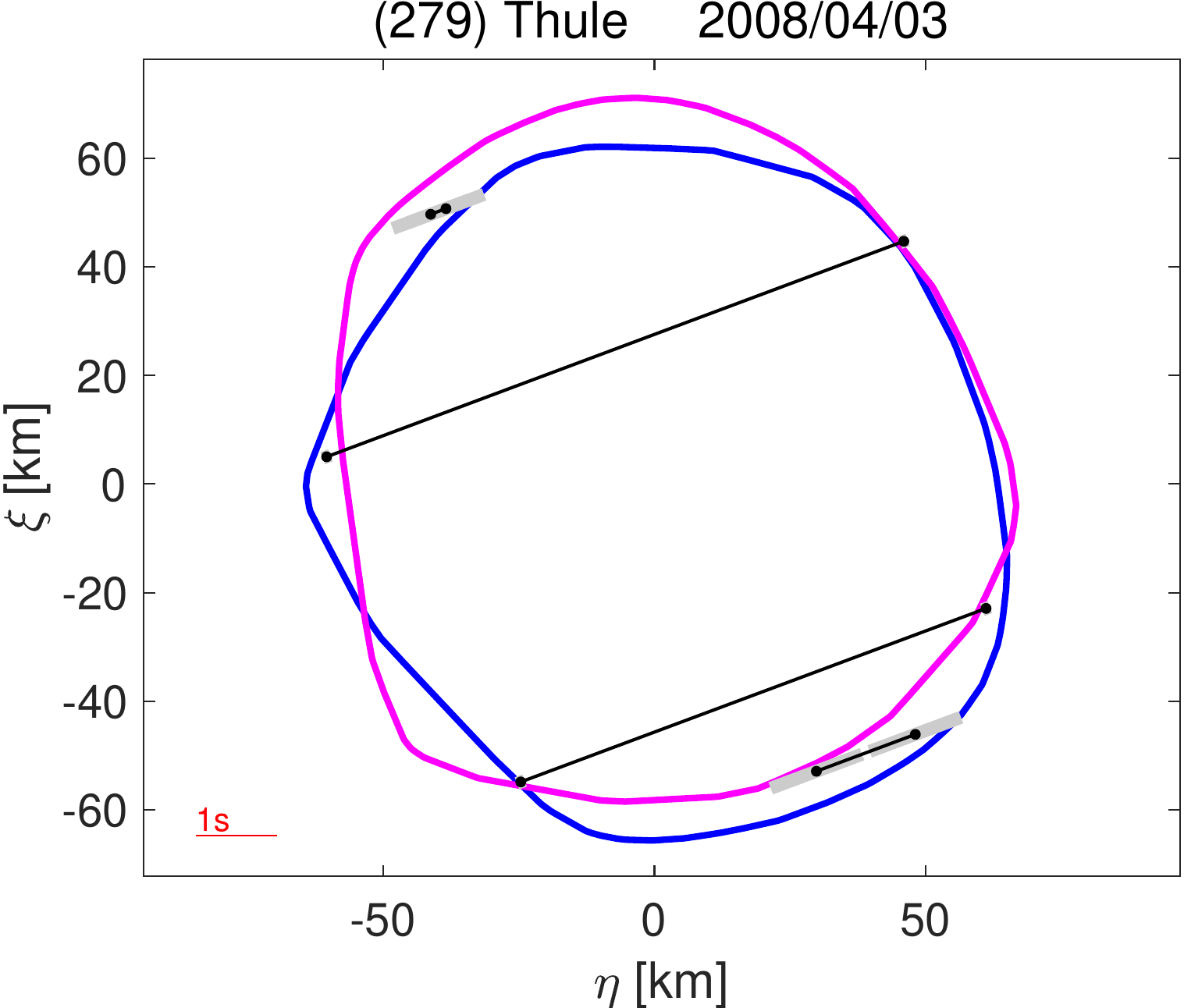}
\includegraphics[width=0.33\textwidth]{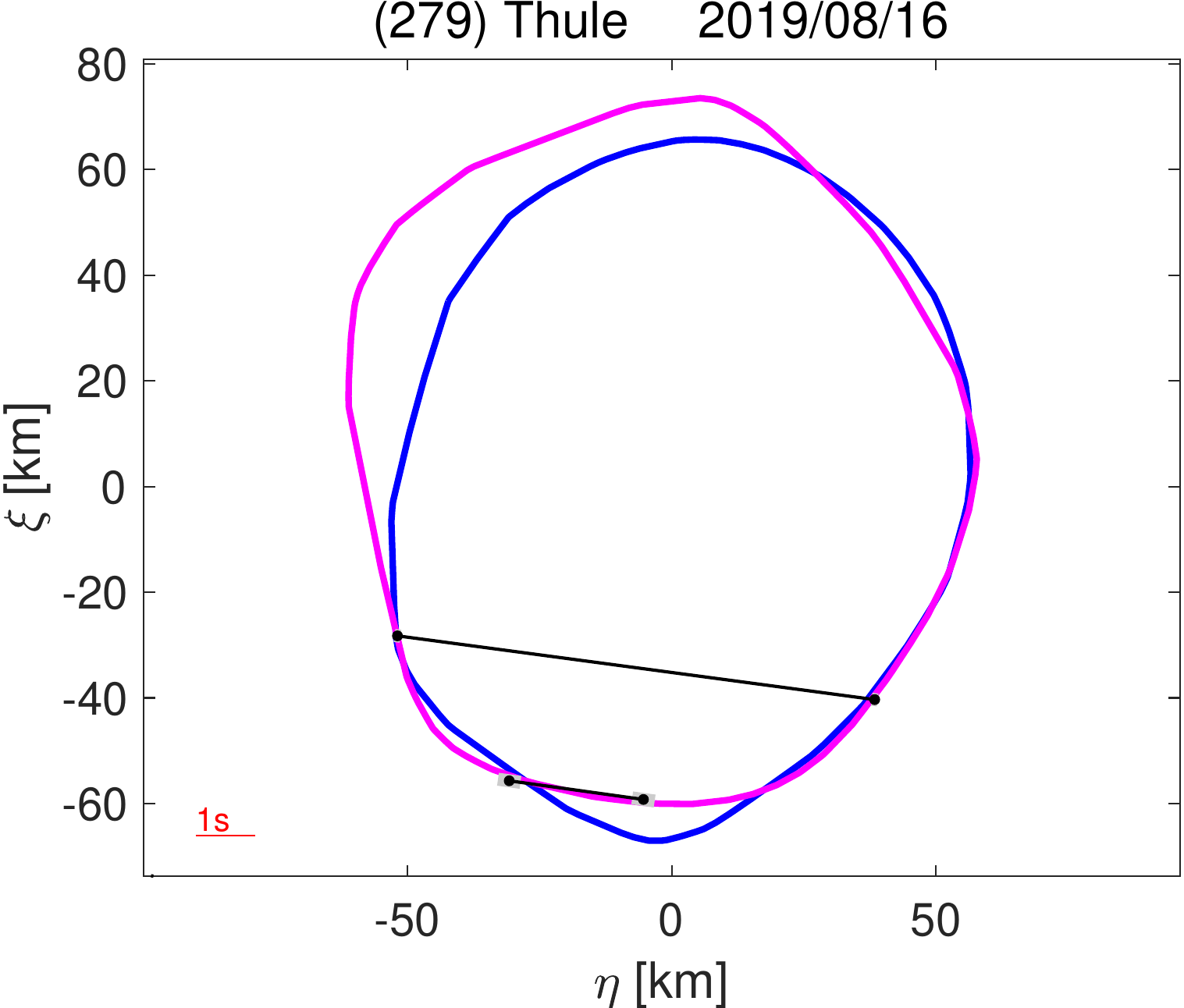}
\includegraphics[width=0.33\textwidth]{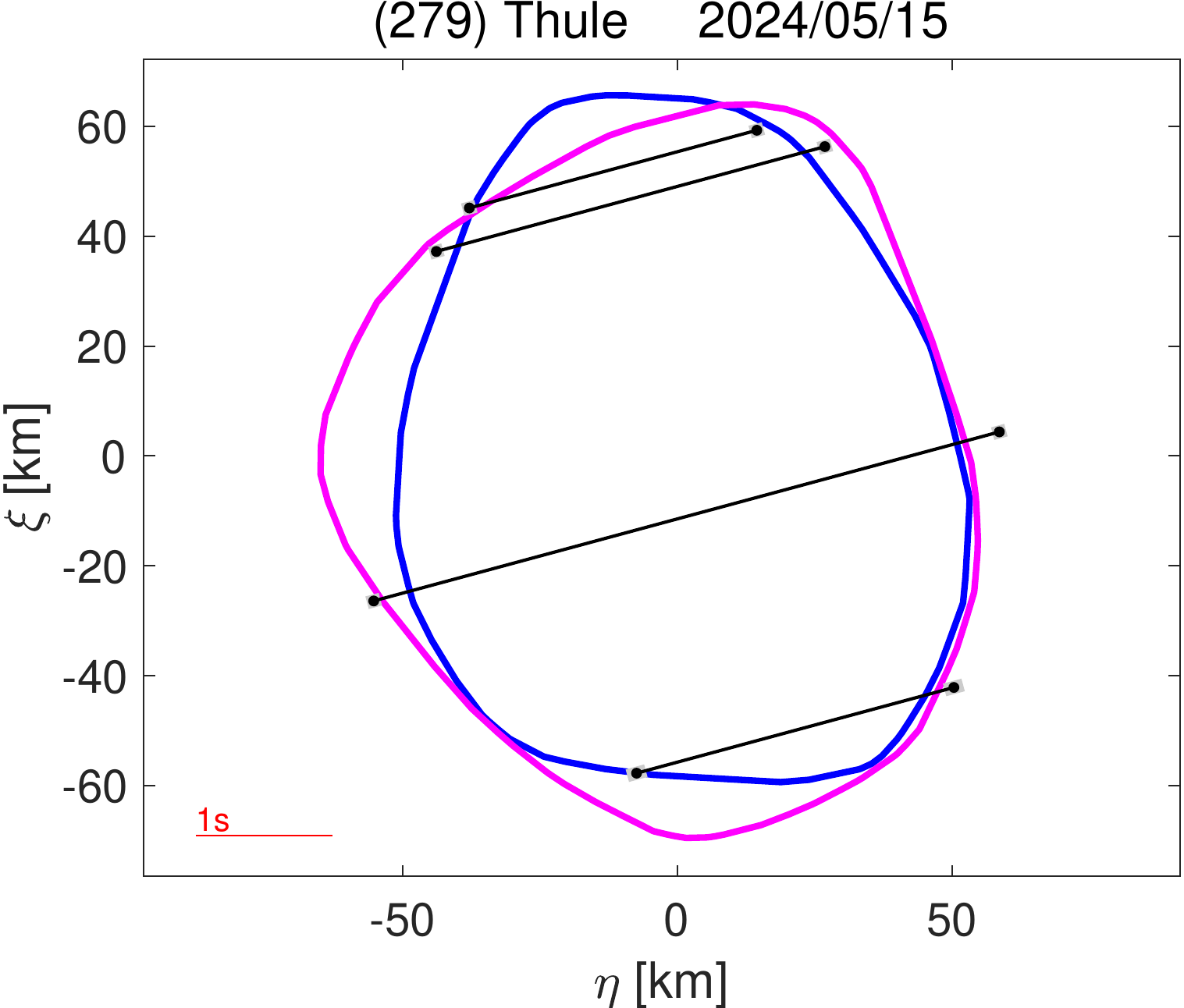}
\caption{Occultation fits for asteroid (279) Thule}
\label{279_occ}
\end{figure*}

\begin{figure*}
\centering
\includegraphics[width=0.33\textwidth]{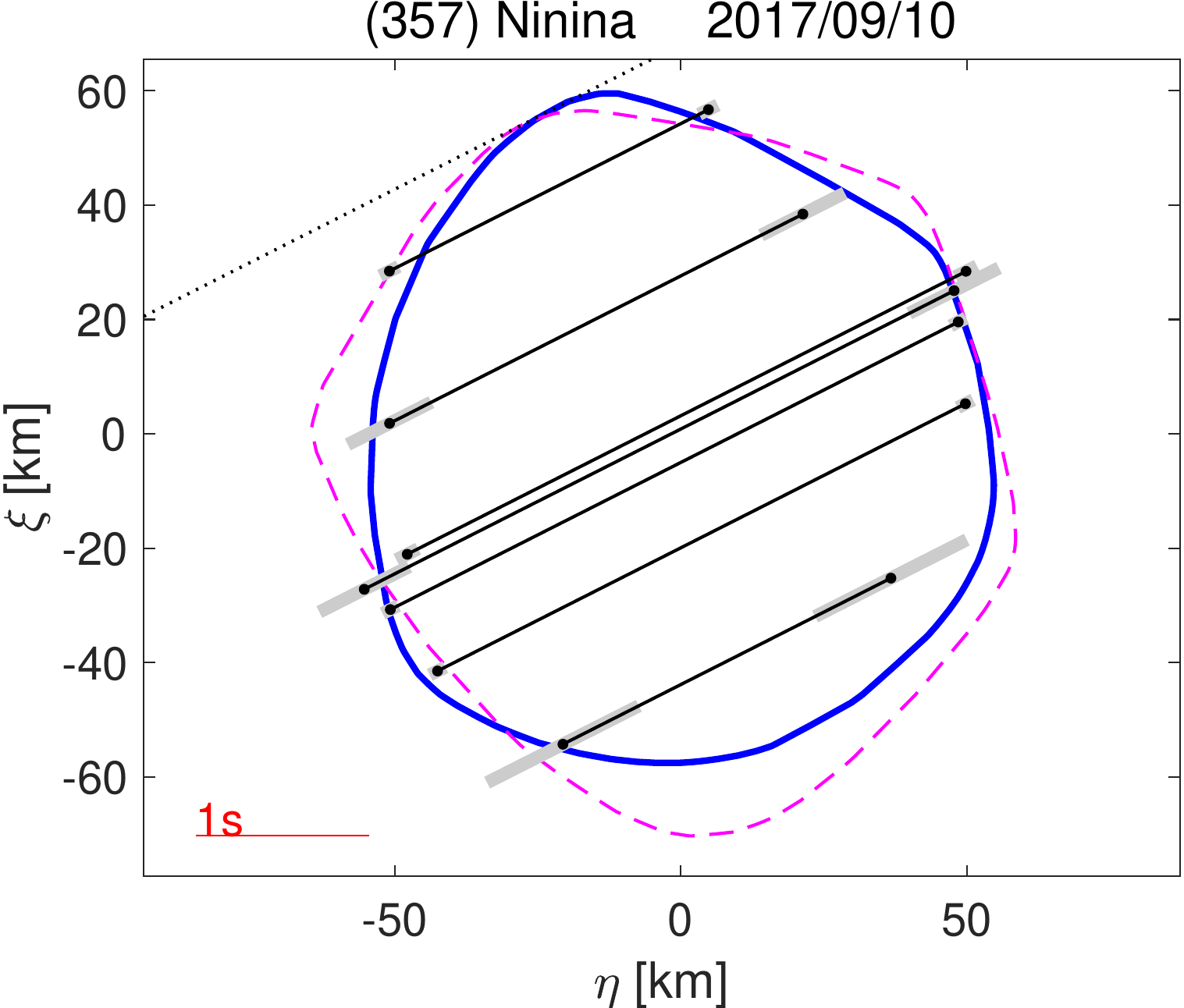}
\includegraphics[width=0.33\textwidth]{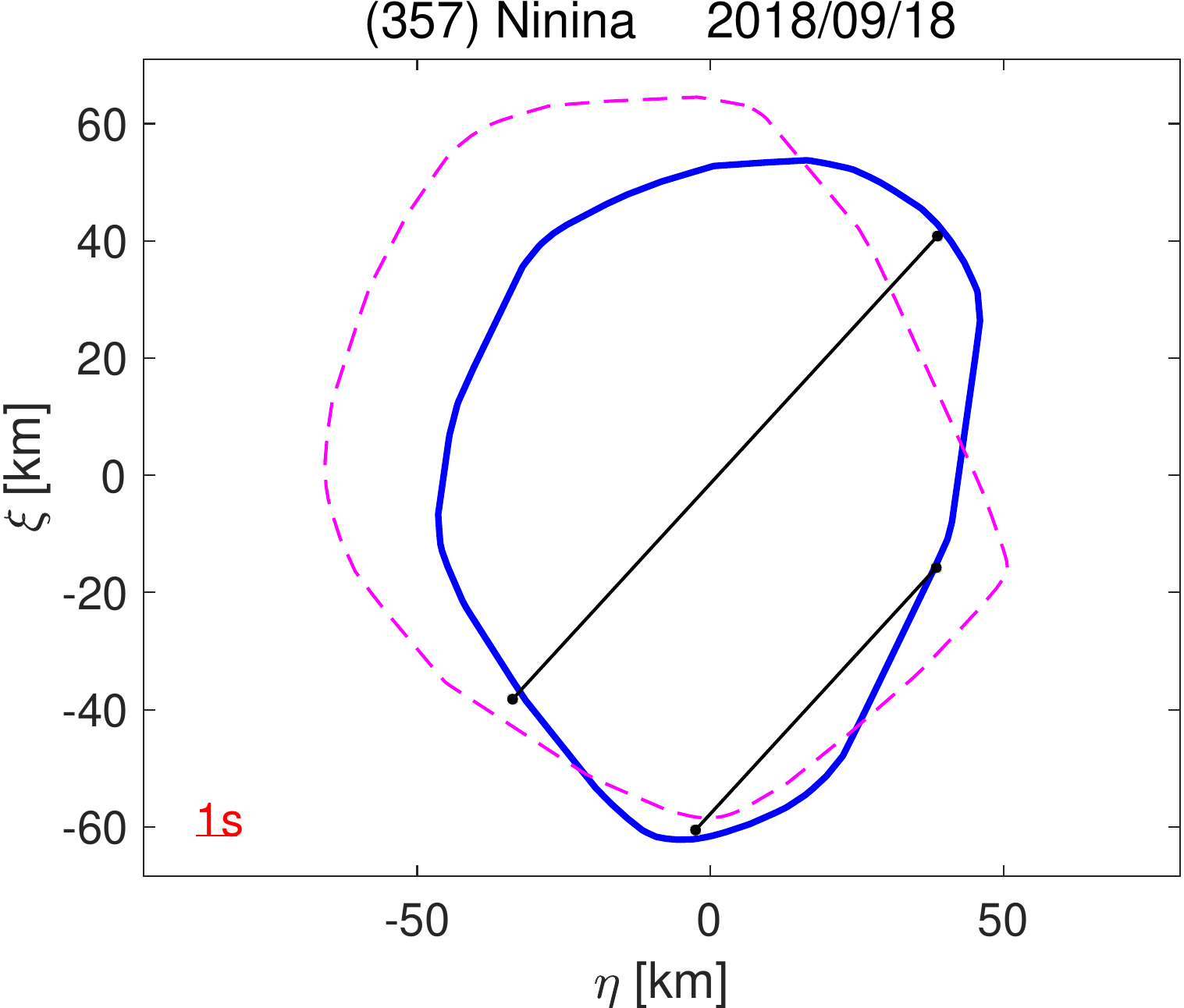}

\includegraphics[width=0.33\textwidth]{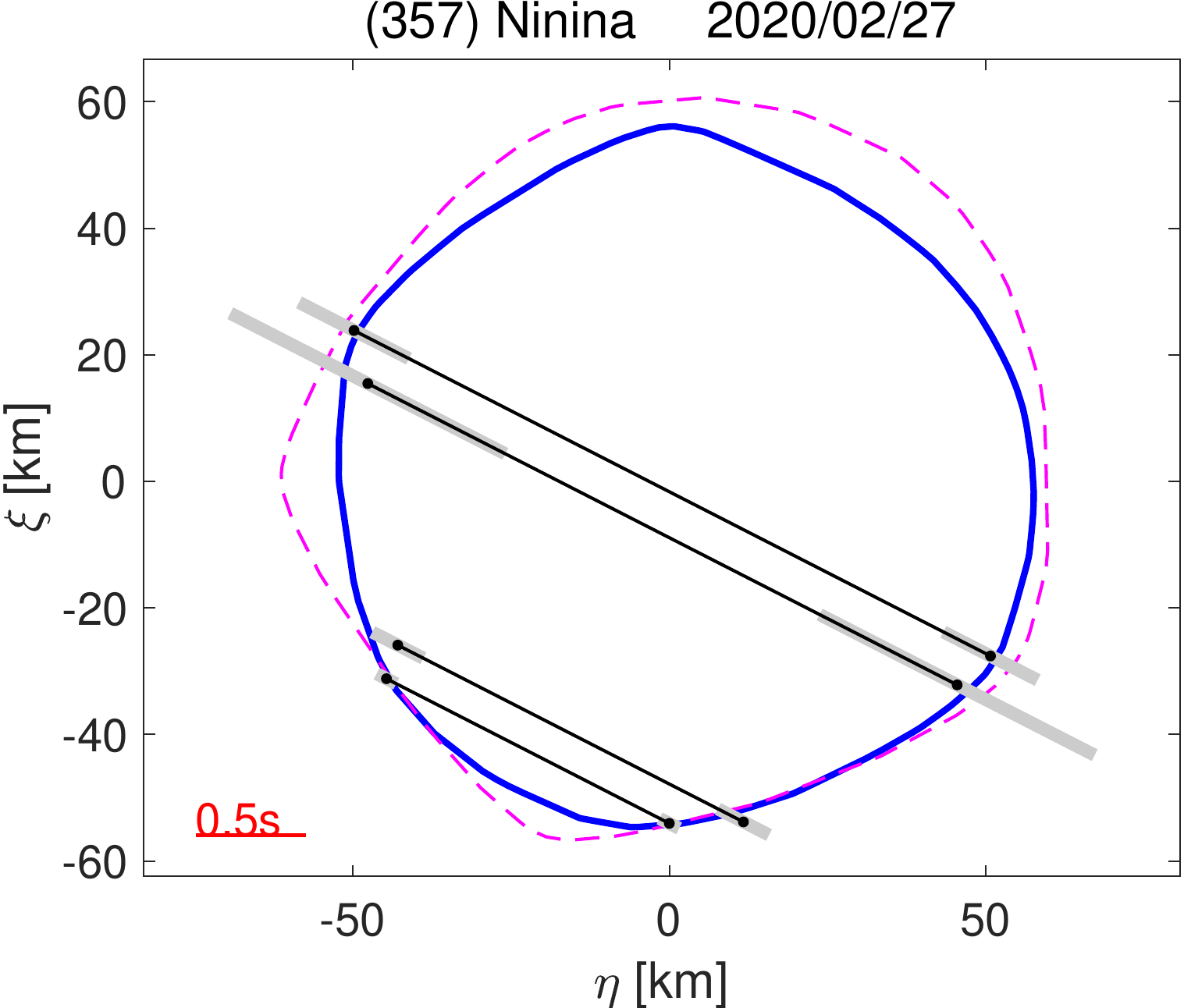}
\includegraphics[width=0.33\textwidth]{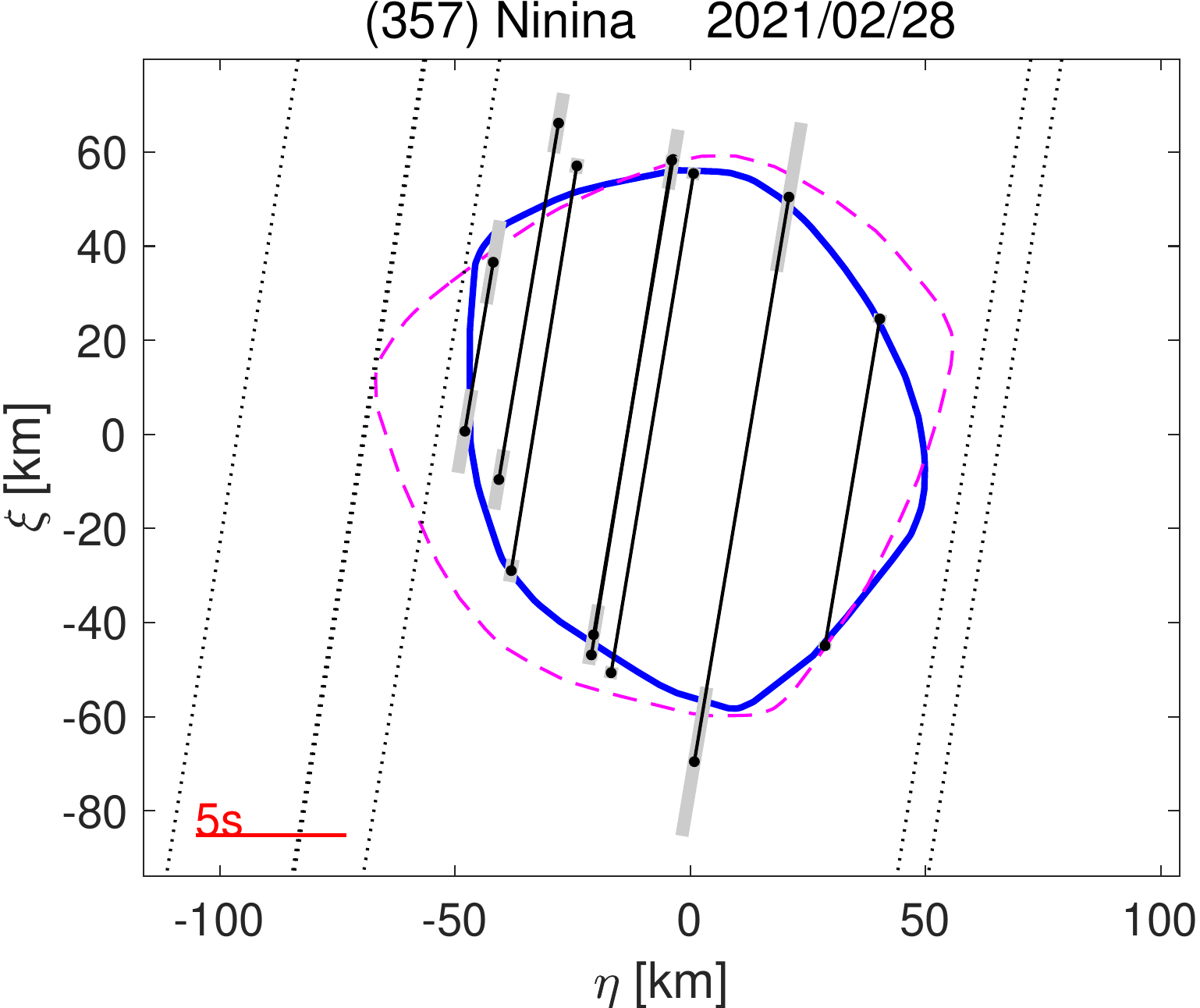}
\caption{Occultation fits for asteroid (357) Ninina. The pole 1 shape model is preferred.}
\label{357_occ}
\end{figure*}

\begin{figure*}
\includegraphics[width=0.33\textwidth]{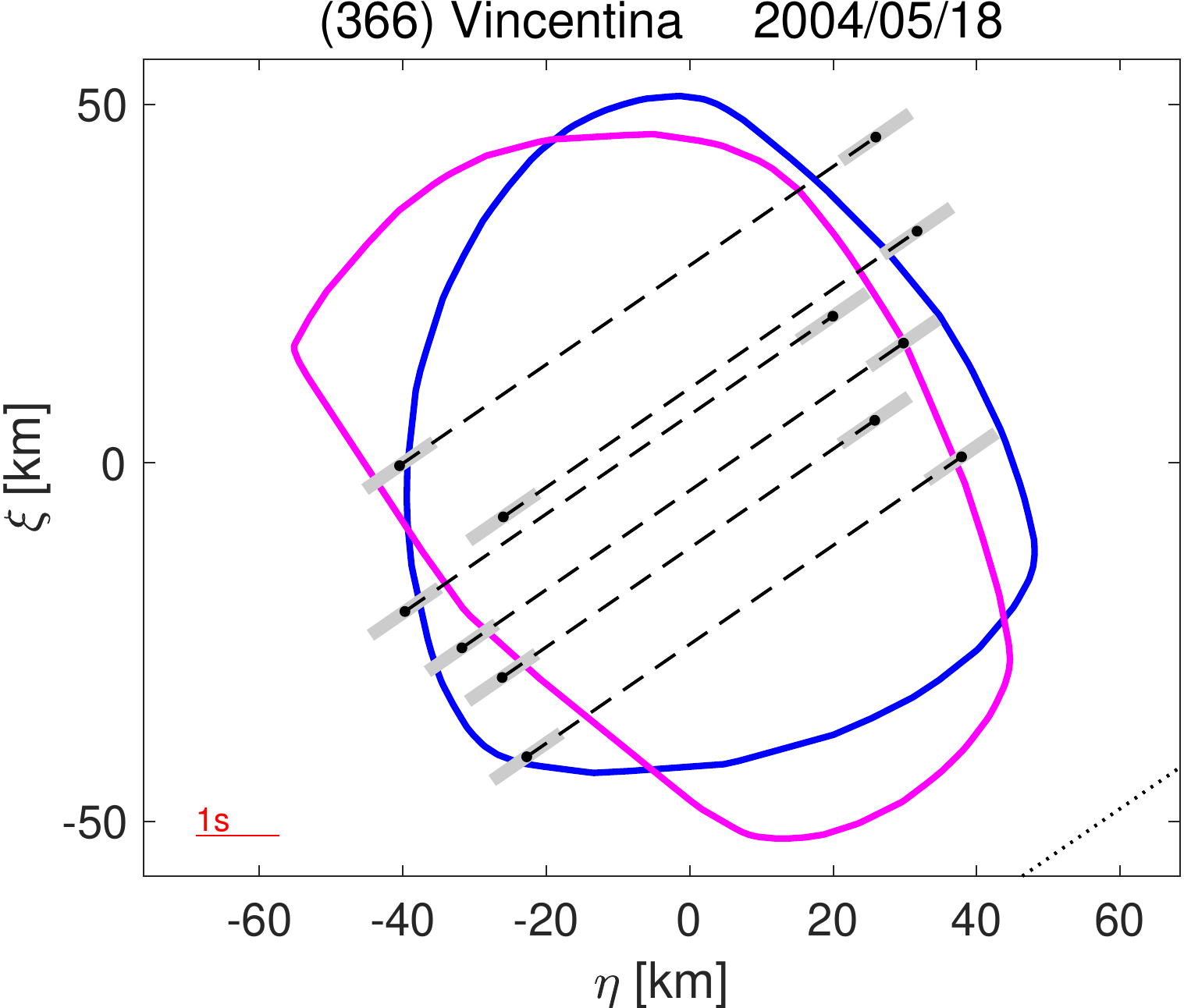}
\includegraphics[width=0.33\textwidth]{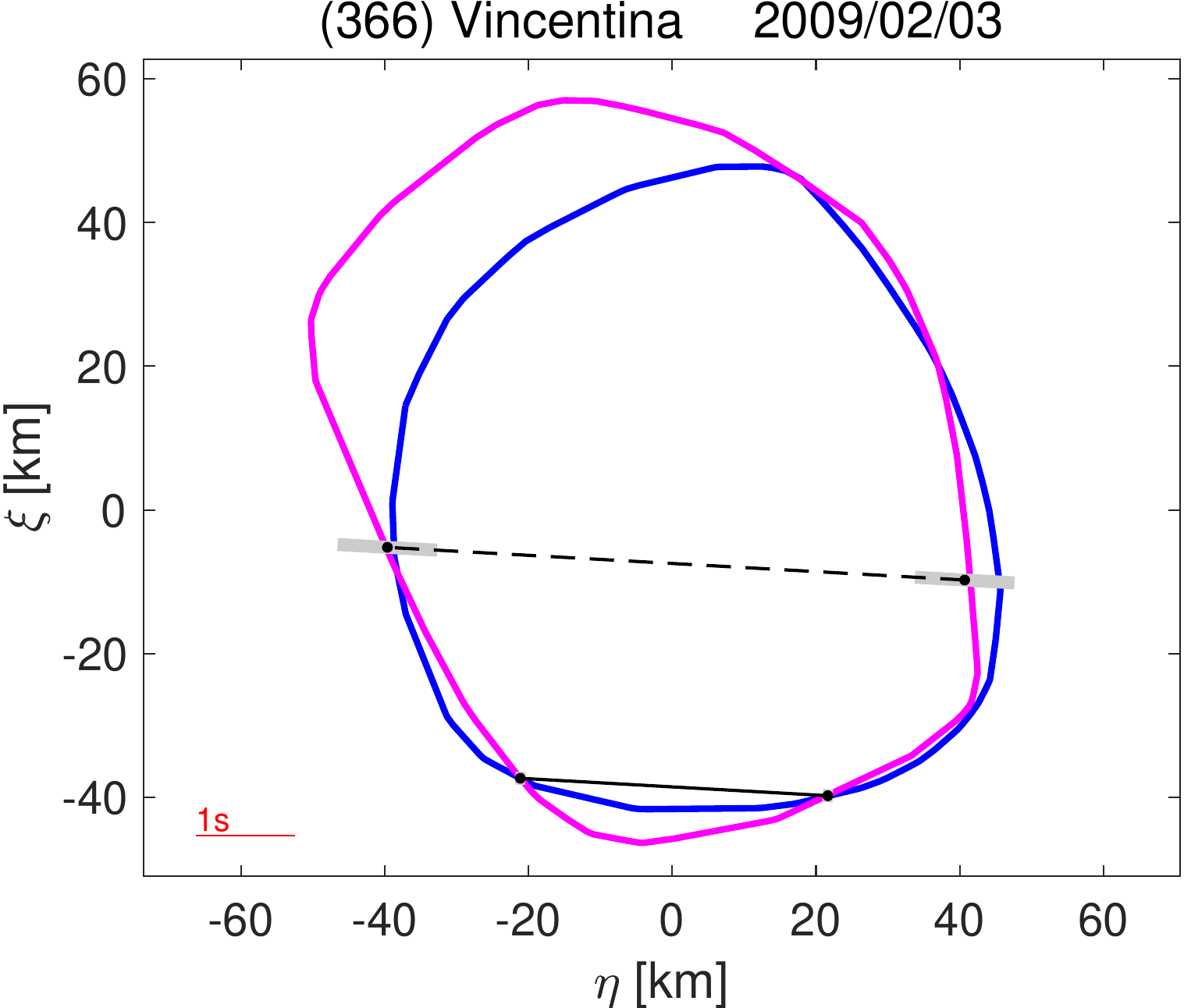}
\includegraphics[width=0.33\textwidth]{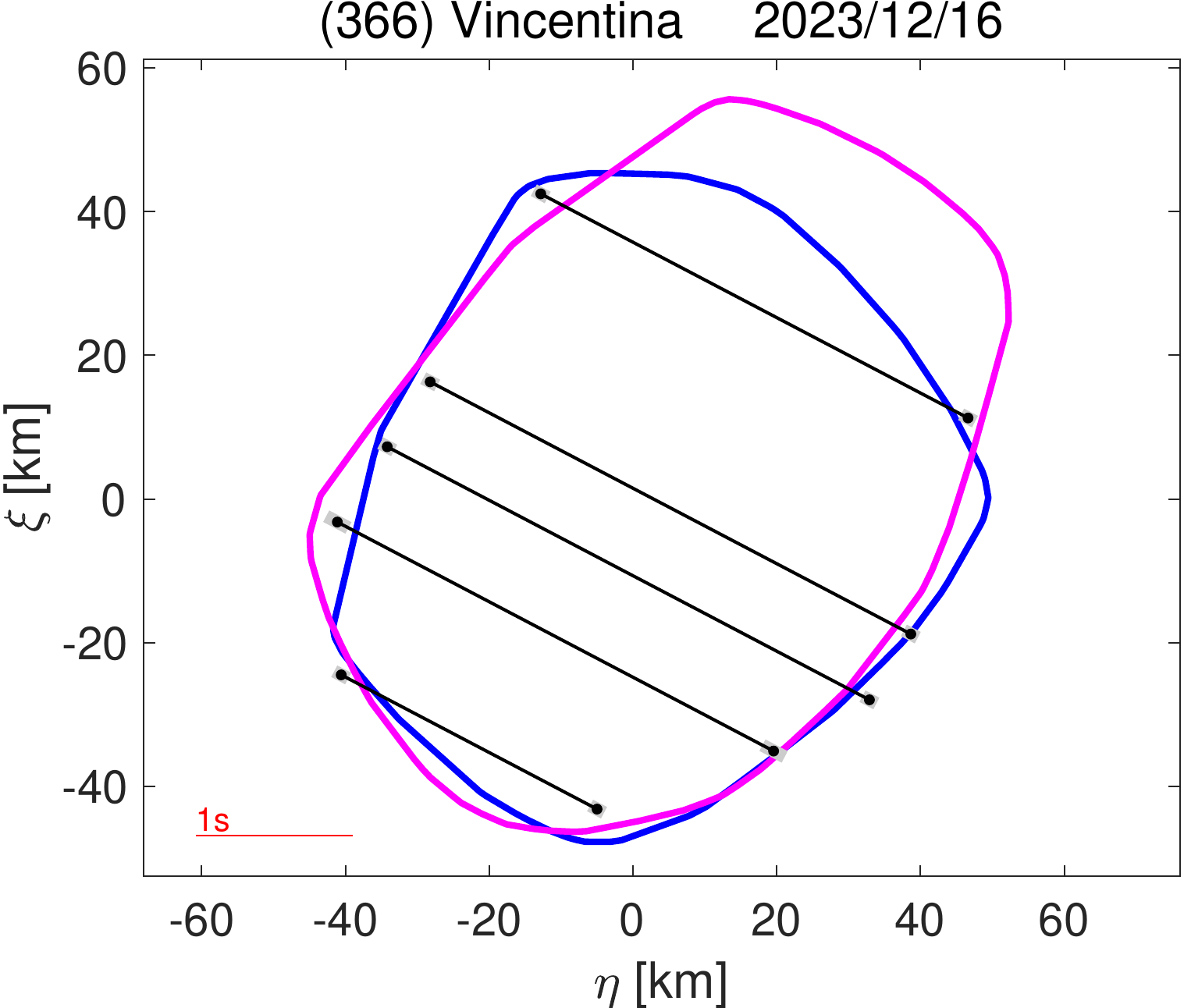}
\caption{Occultation fits for asteroid (366) Vincentina}
\label{366_occ}
\end{figure*}

\begin{figure*}
\includegraphics[width=0.33\textwidth]{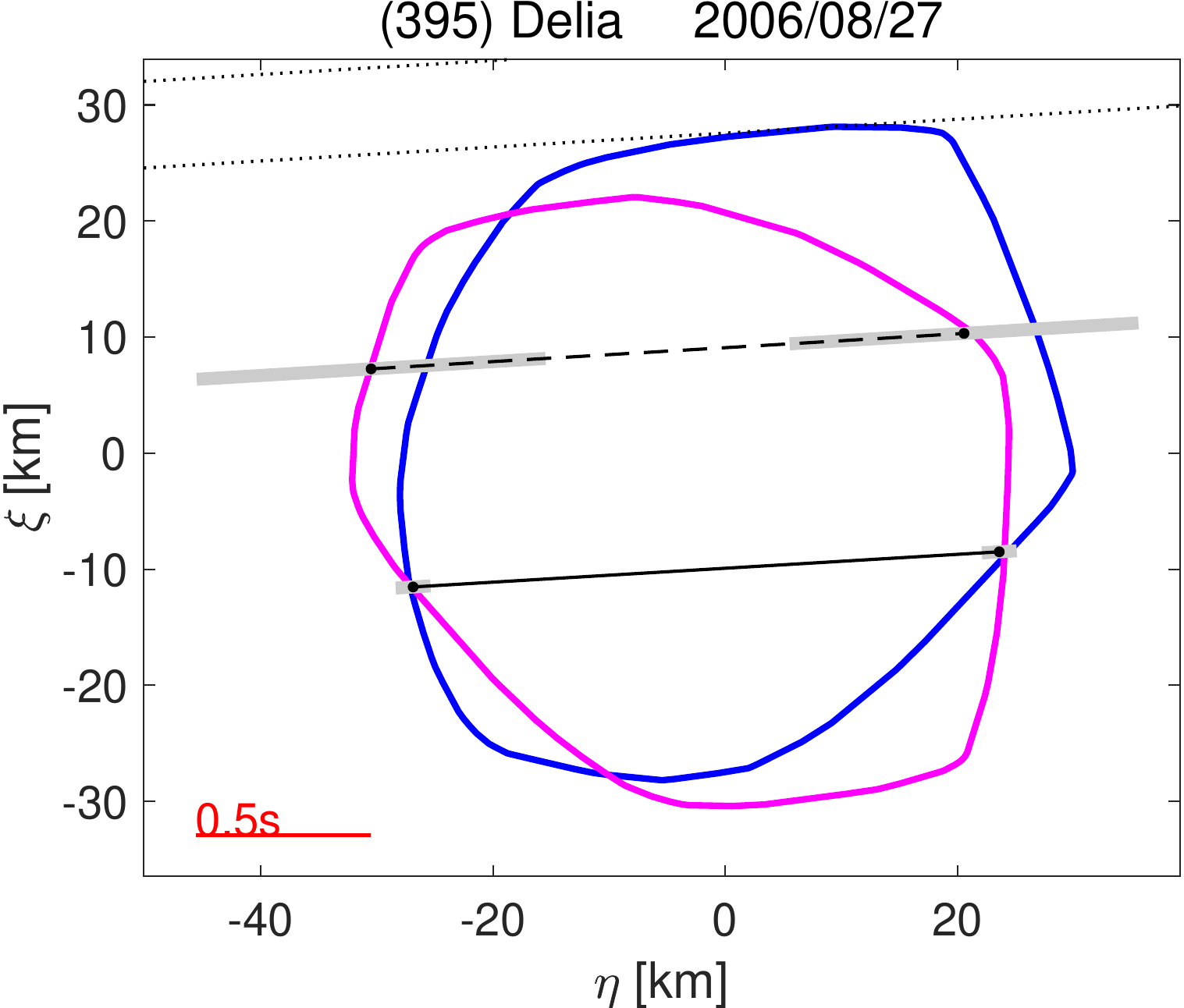}
\includegraphics[width=0.33\textwidth]{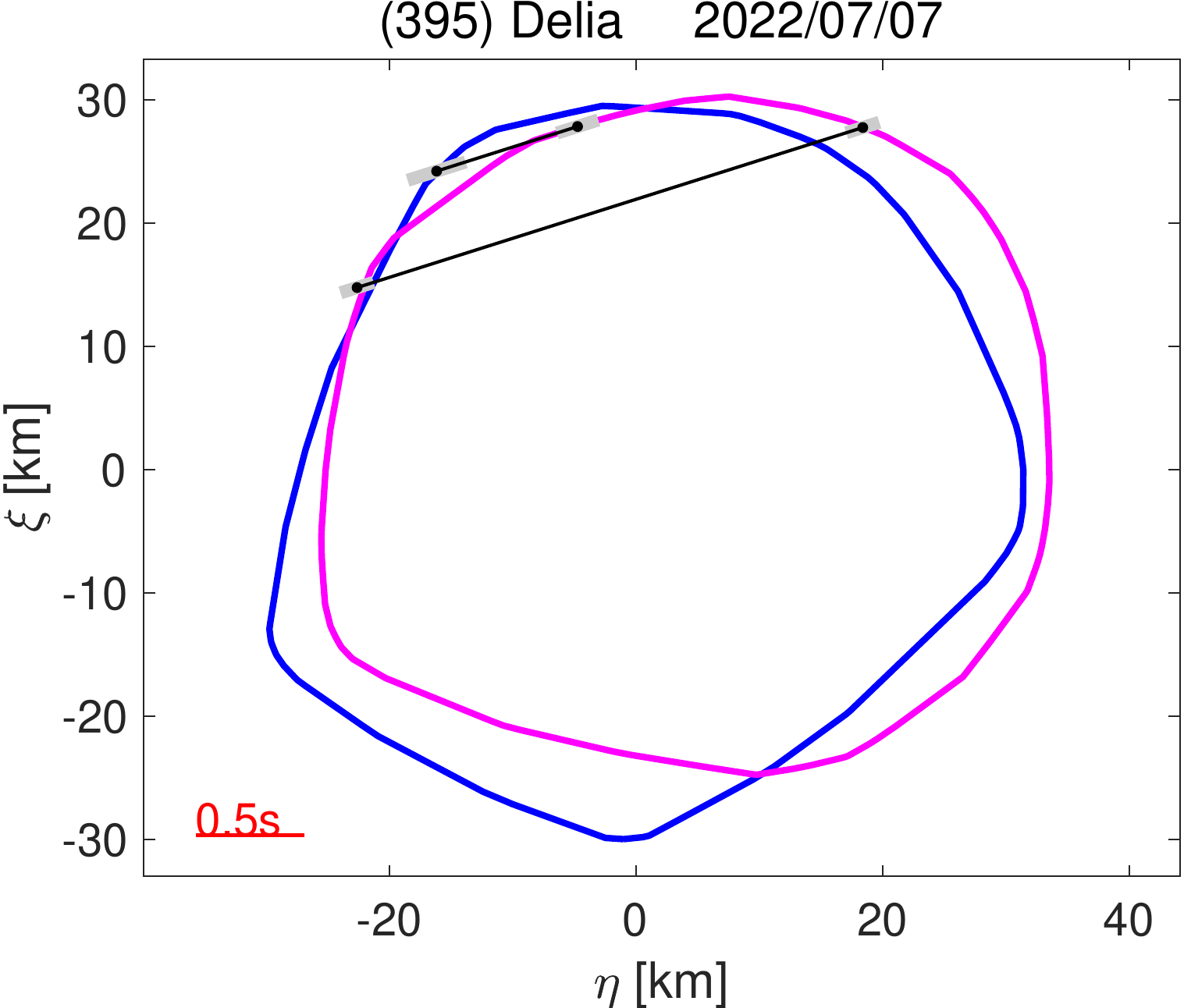}
\includegraphics[width=0.33\textwidth]{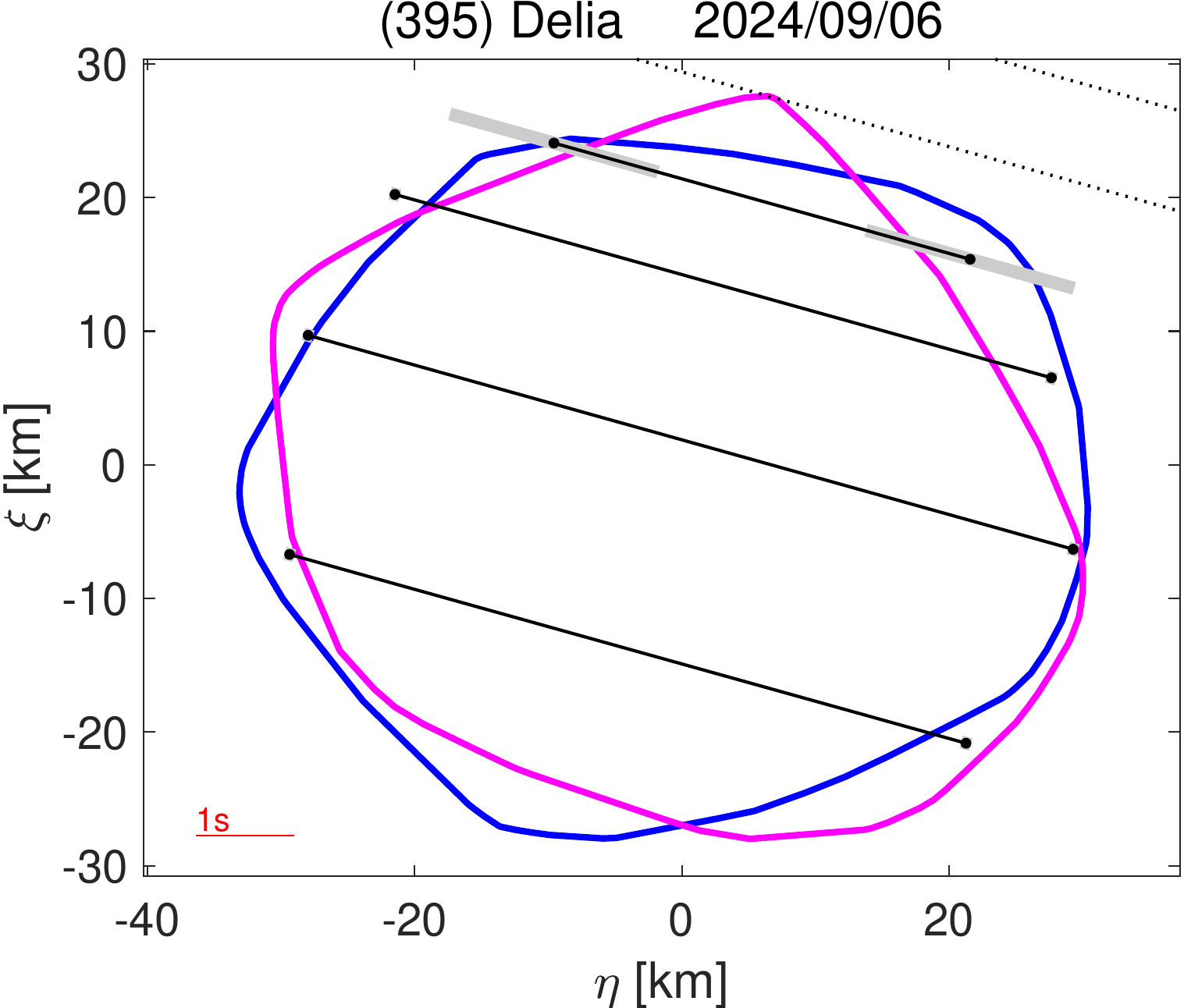}
\caption{Occultation fits for asteroid (395) Delia}
\label{395_occ}
\end{figure*}

\begin{figure*}
\centering
\includegraphics[width=0.33\textwidth]{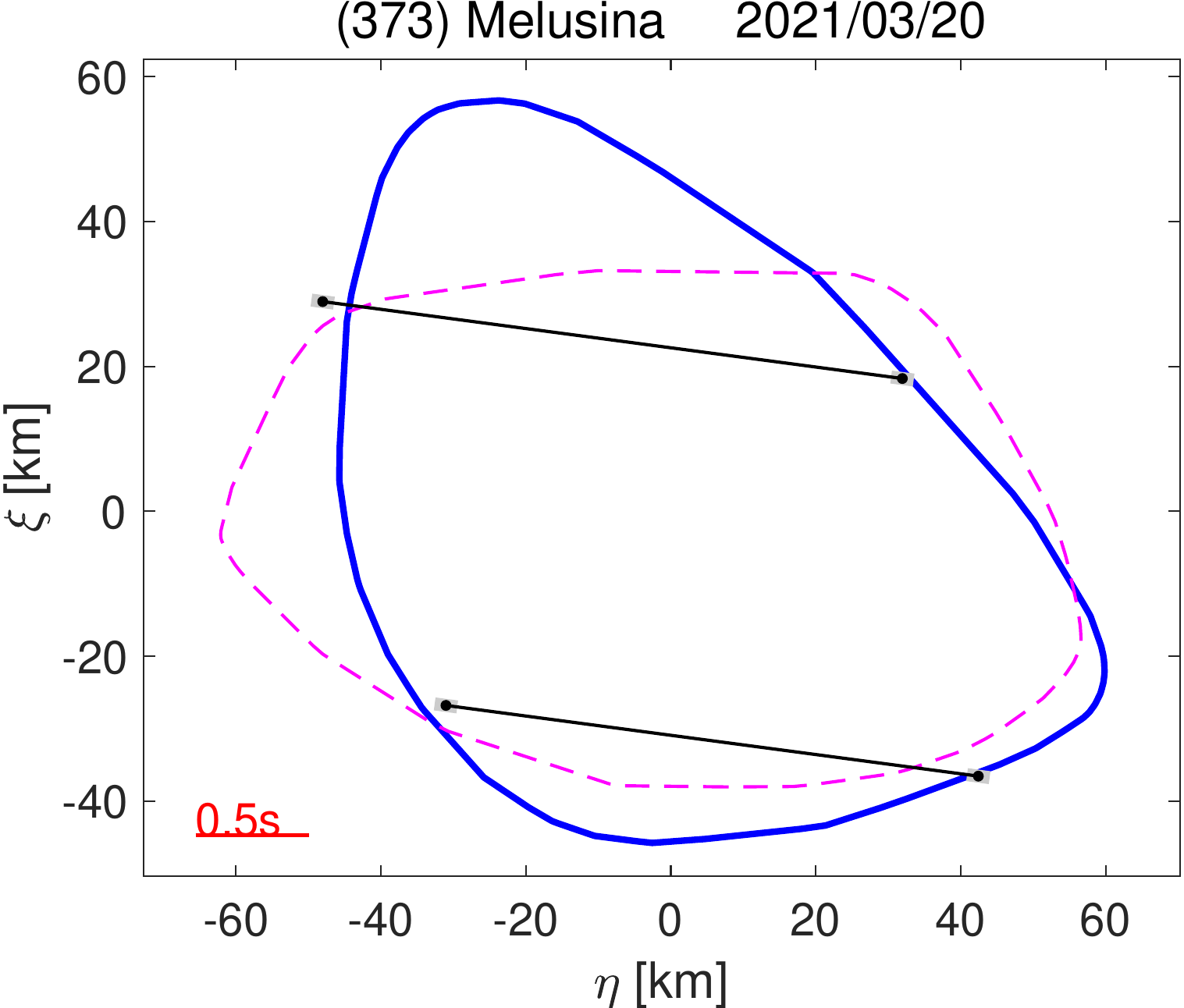}
\includegraphics[width=0.33\textwidth]{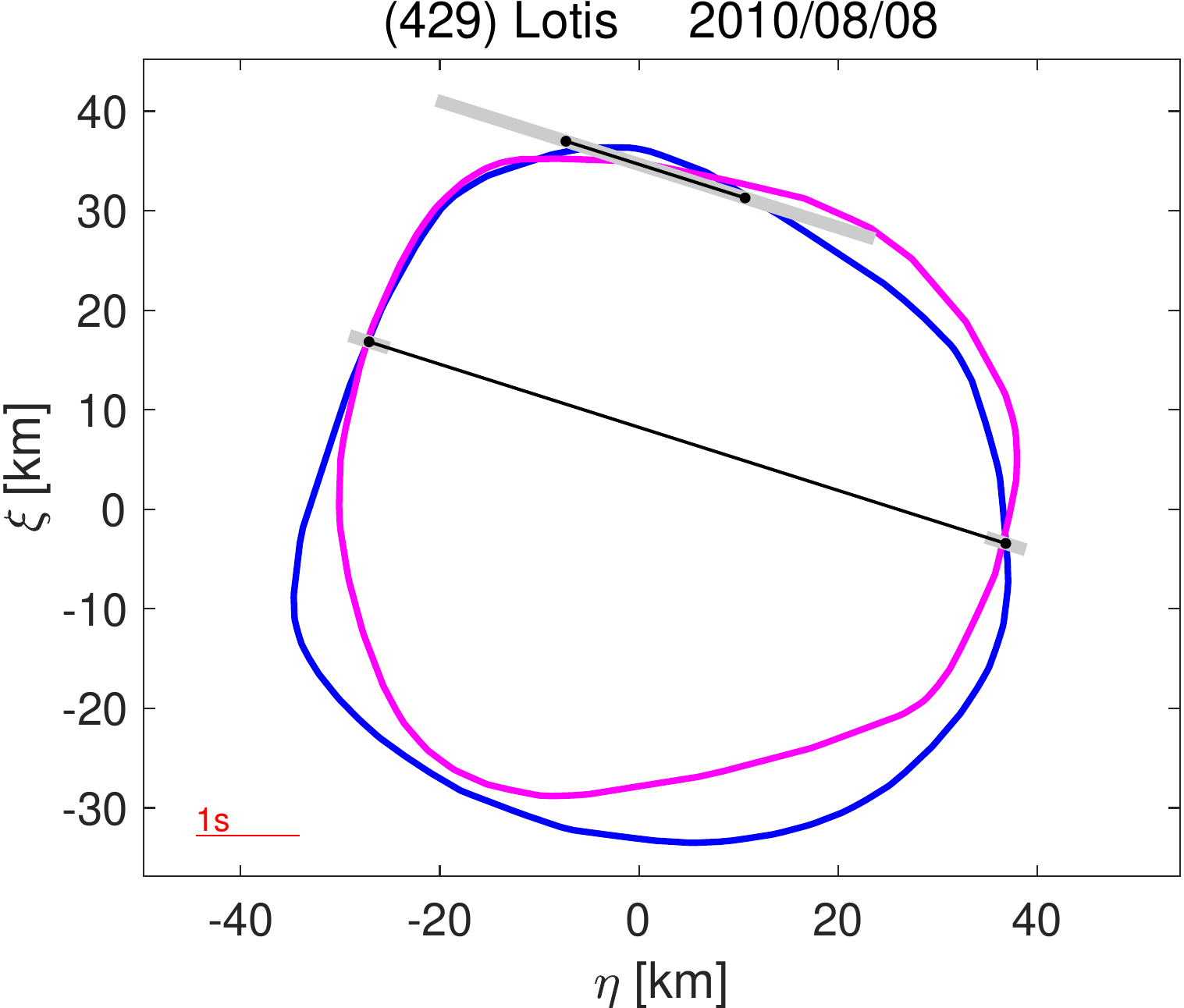}
\caption{Occultation fits for asteroids (373) Melusina and (429) Lotis. The preferred pole 
for Melusina is pole 2 (blue contour). For Lotis the slightly preferred solution (pole 2) 
is also marked in blue.}
\label{373_occ}
\end{figure*}

\begin{figure*}
\centering
\includegraphics[width=0.33\textwidth]{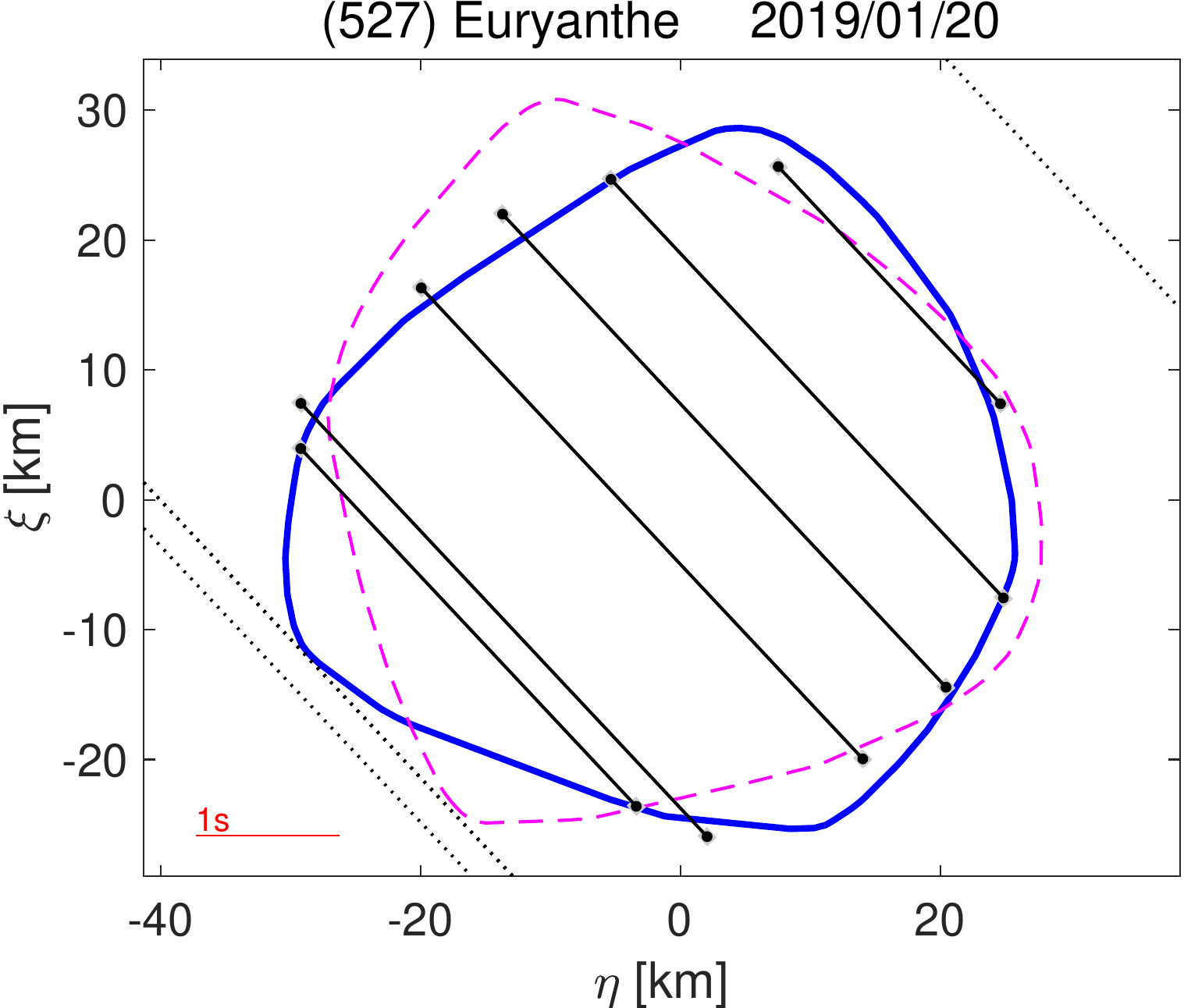}
\includegraphics[width=0.33\textwidth]{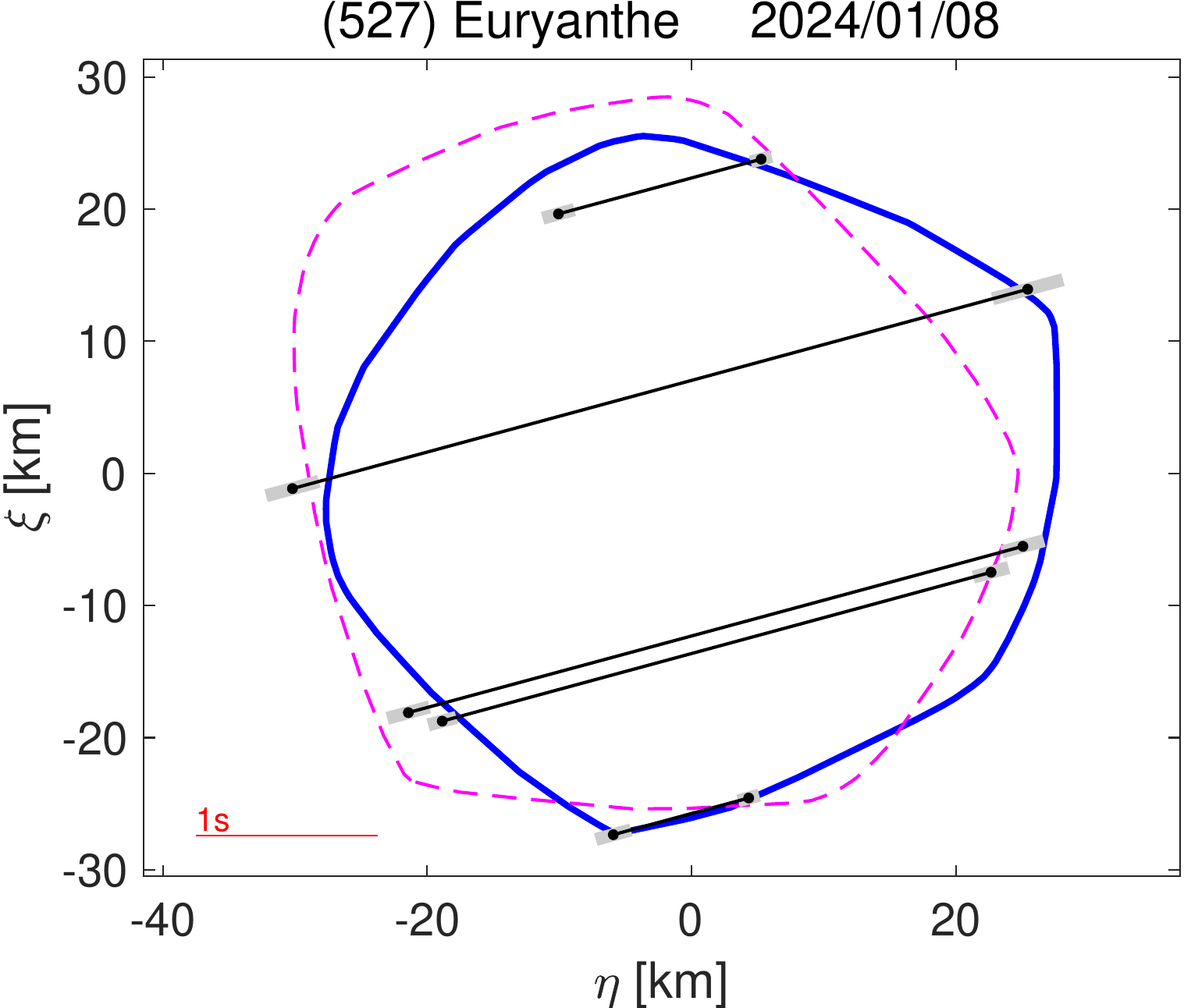}
\caption{Occultation fits for asteroid (527) Euryanthe. Pole 2 (blue contour) is preferred.}
\label{527_occ}
\end{figure*}

\begin{figure*}
\centering
\includegraphics[width=0.33\textwidth]{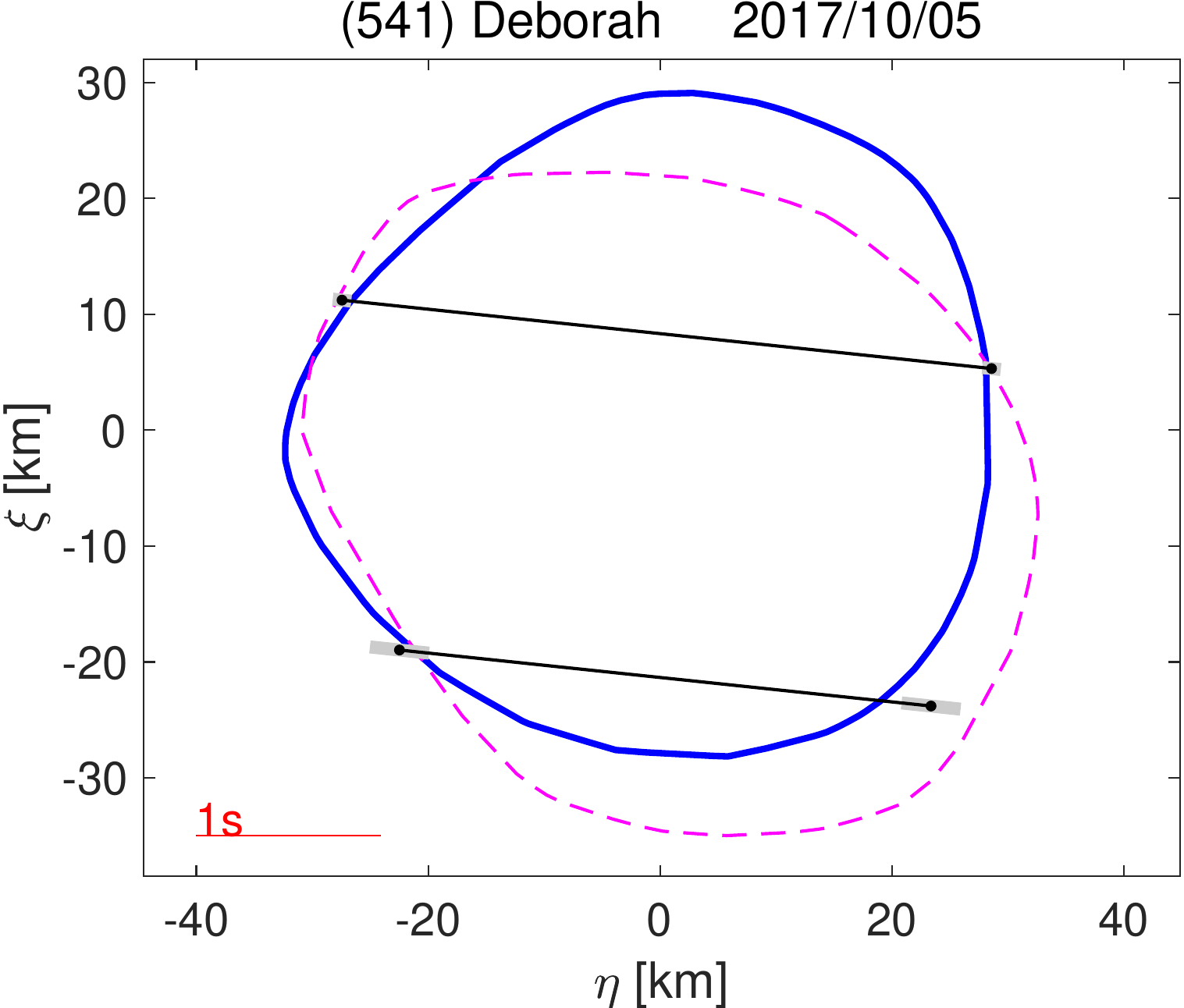}
\includegraphics[width=0.33\textwidth]{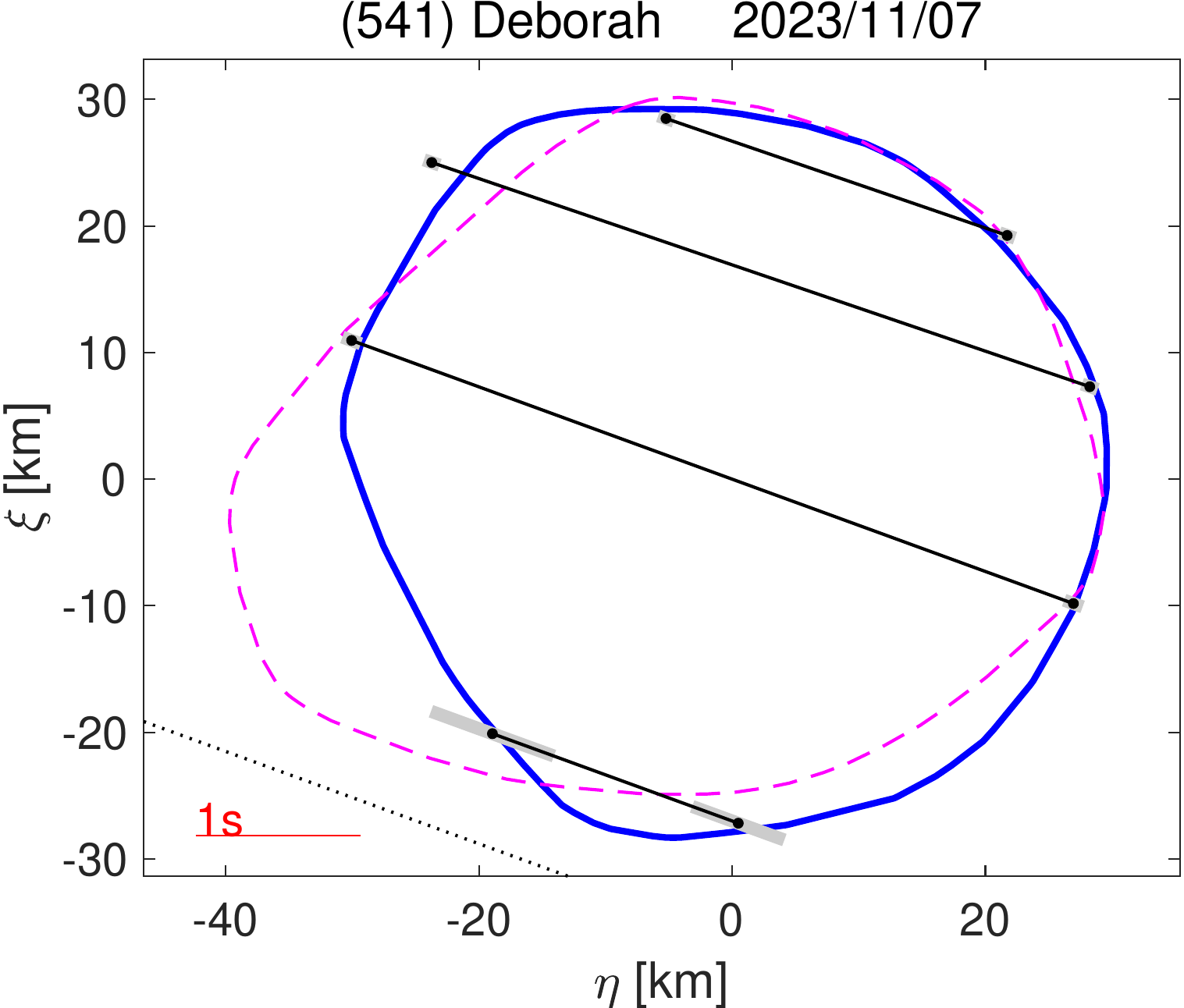}
\includegraphics[width=0.33\textwidth]{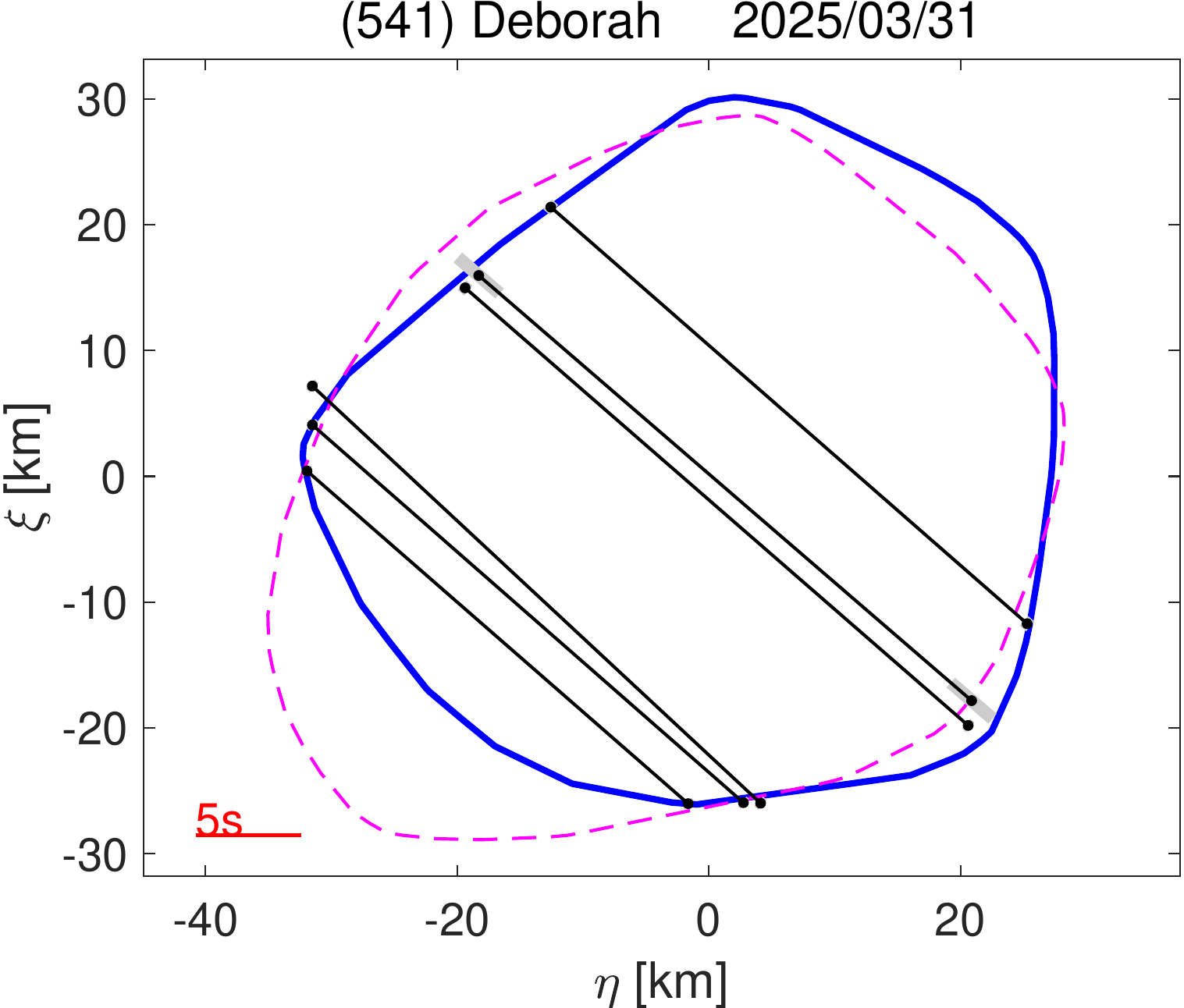}
\caption{Occultation fits for asteroid (541) Deborah. Pole 2 (blue contour) is preferred.}
\label{541_occ}
\end{figure*}

\begin{figure*}
\centering
\includegraphics[width=0.33\textwidth]{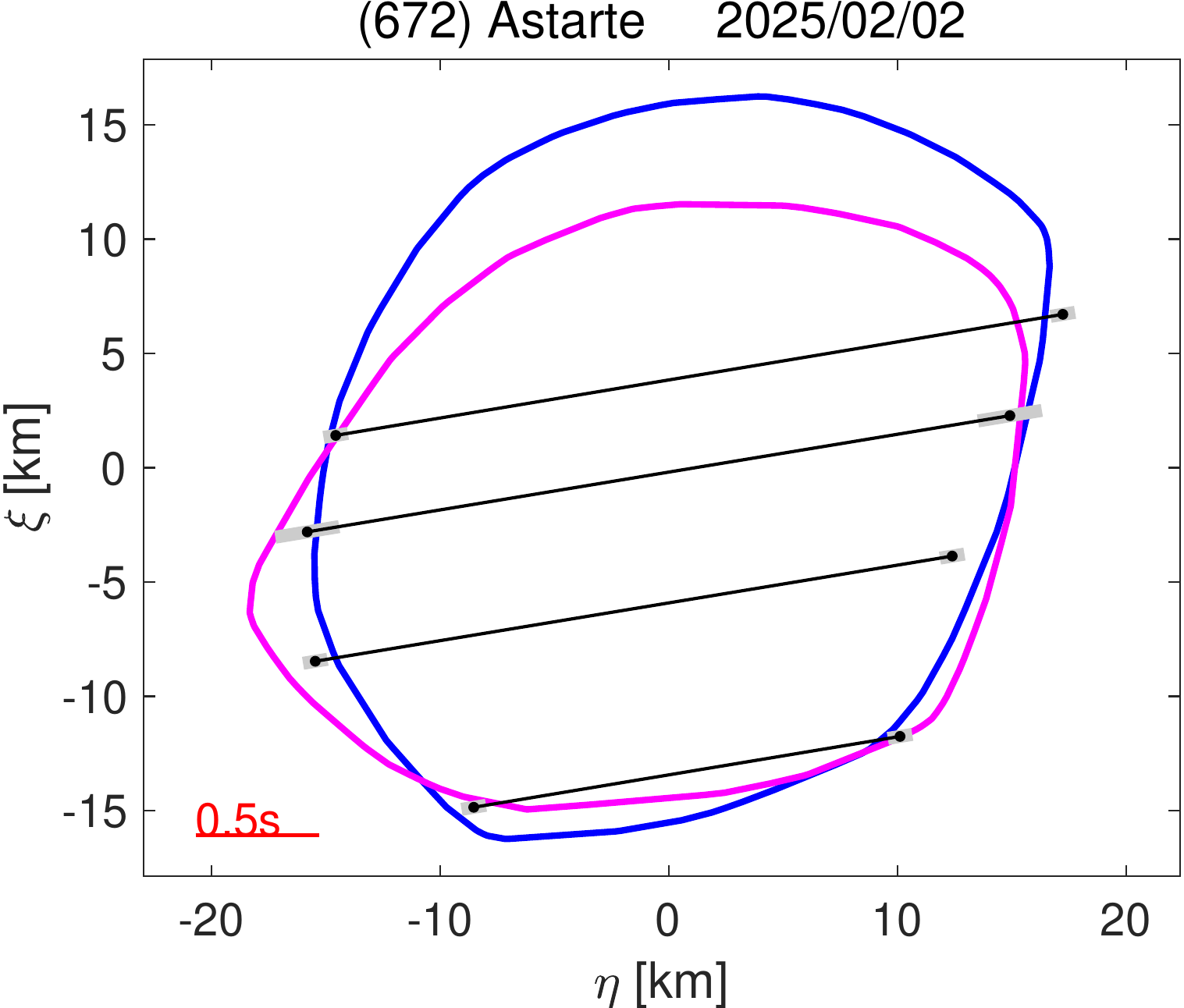}
\caption{Occultation fits for asteroid (672) Astarte. Blue contour is for pole 2.}
\label{672_occ}
\end{figure*}

\begin{figure*}
\includegraphics[width=0.33\textwidth]{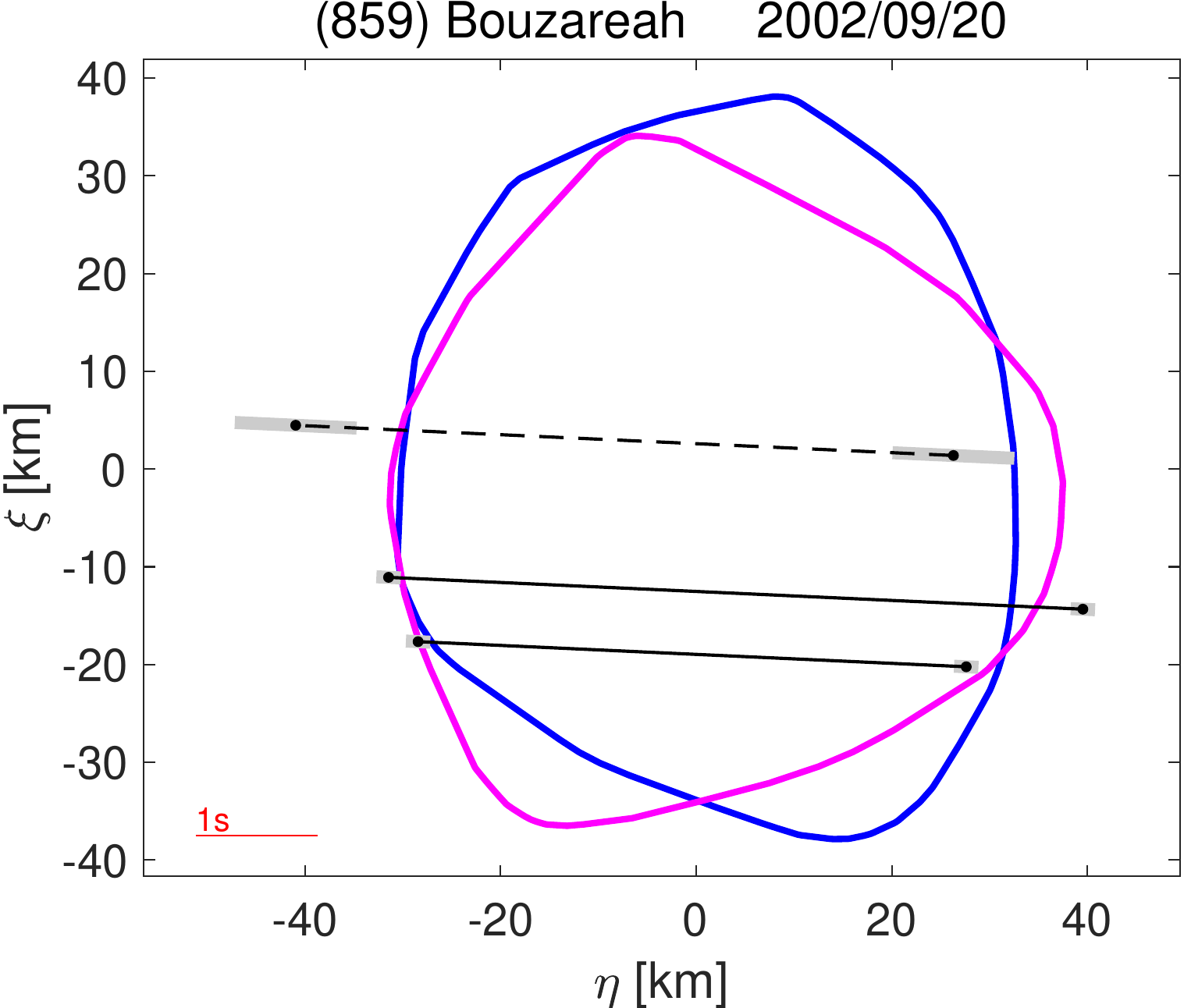}
\includegraphics[width=0.33\textwidth]{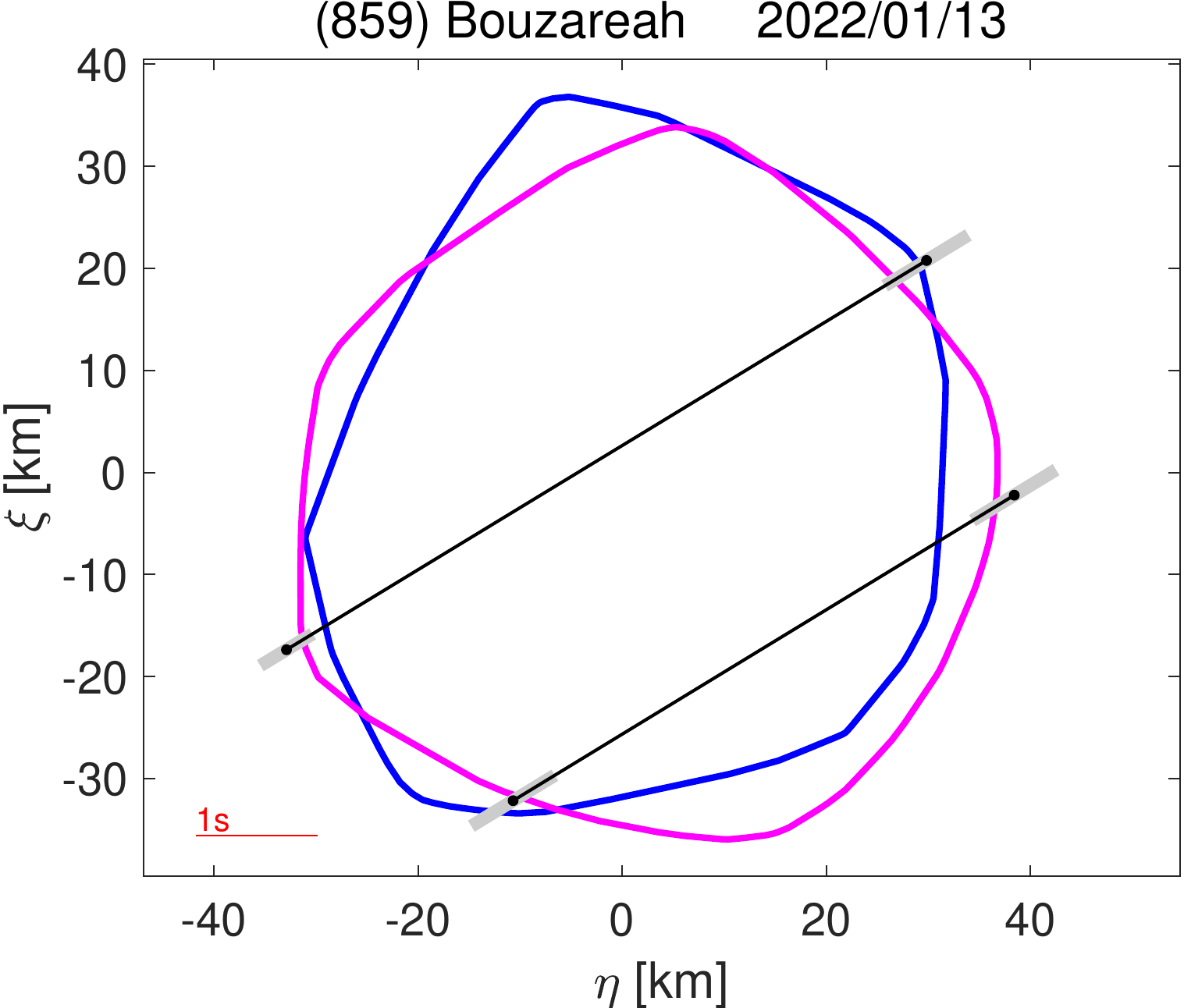}
\includegraphics[width=0.33\textwidth]{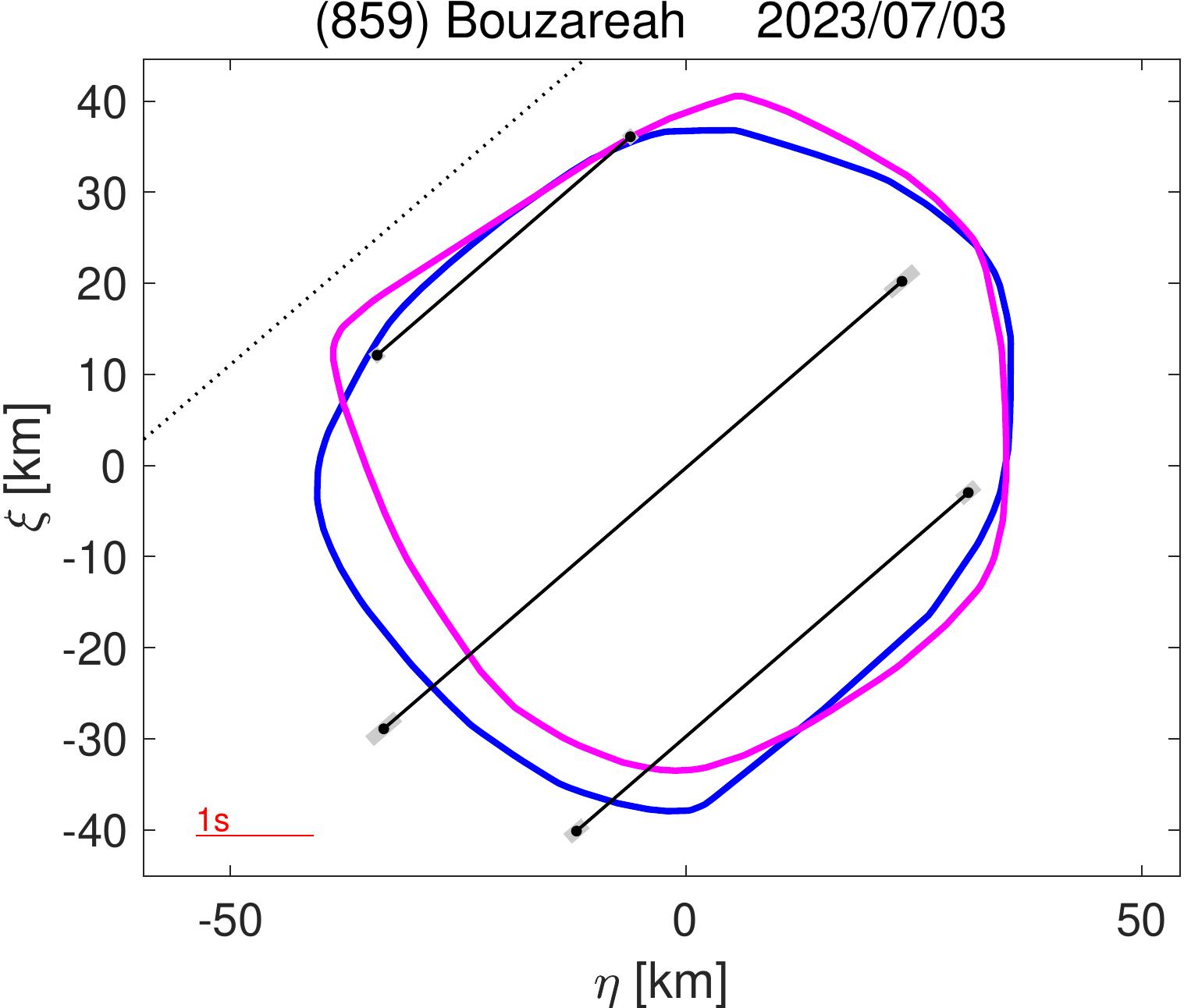}
\caption{Occultation fits for asteroid (859) Bouzareah}
\label{859_occ}
\end{figure*}

\begin{figure*}
\centering
\includegraphics[width=0.33\textwidth]{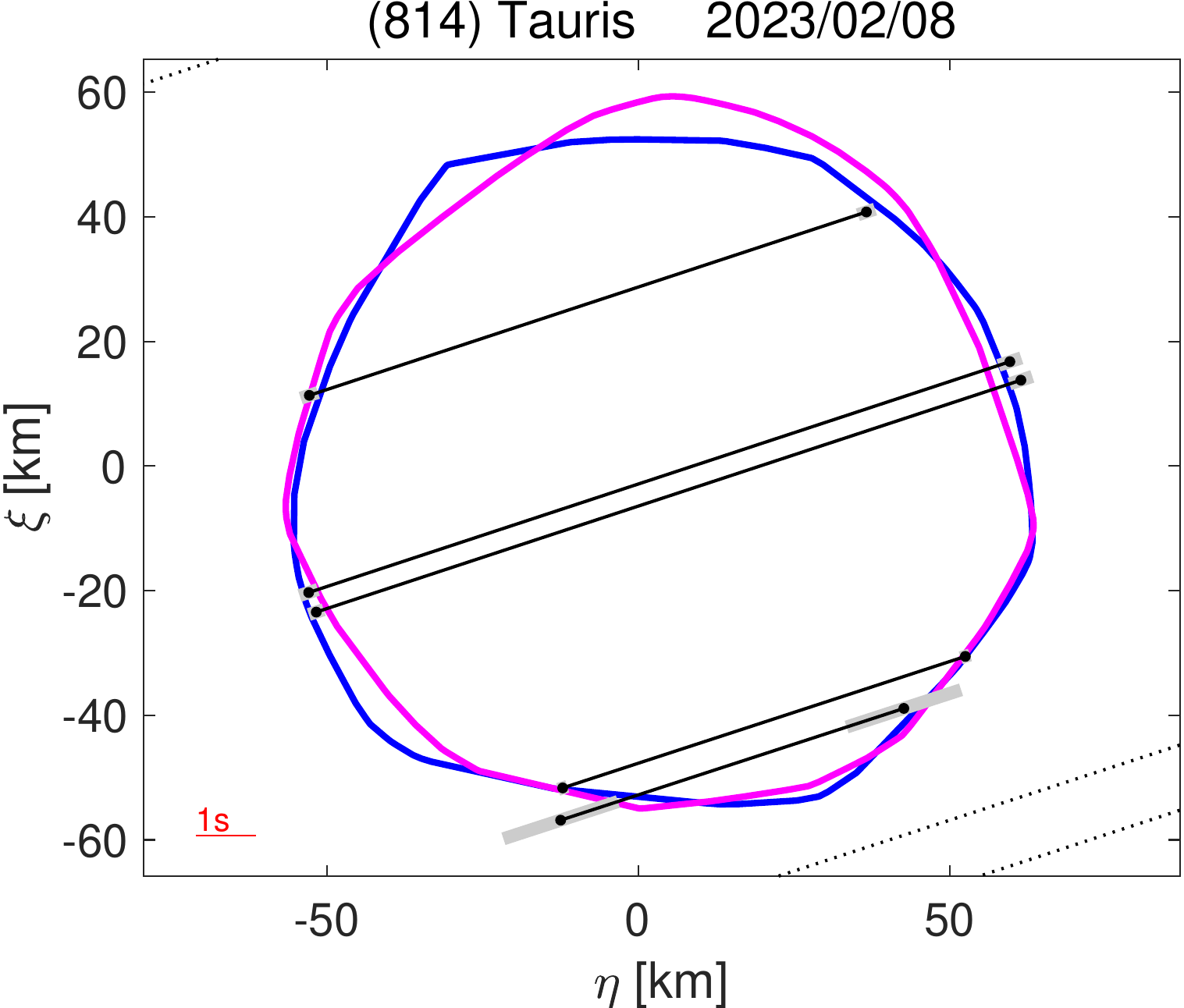}
\includegraphics[width=0.33\textwidth]{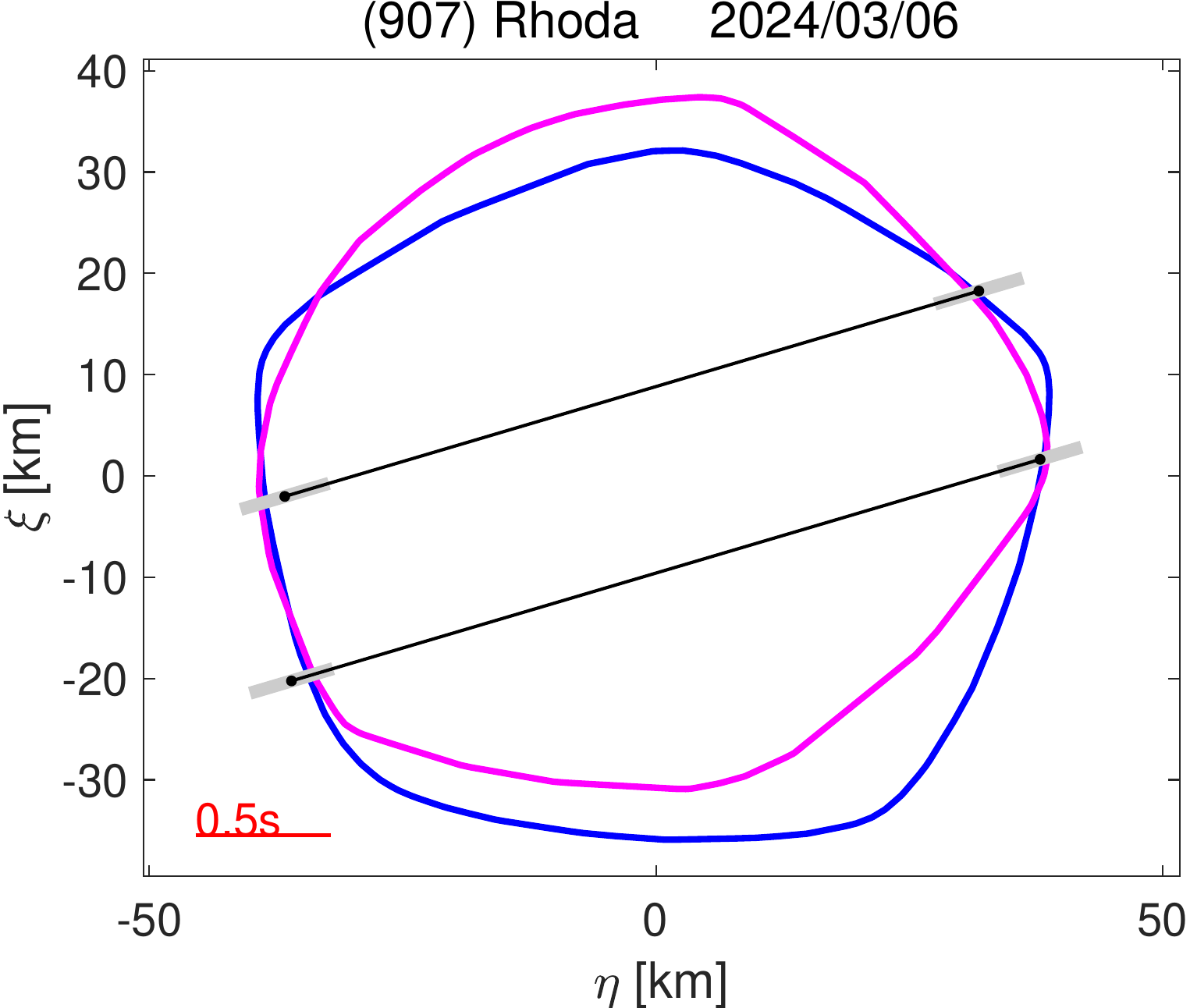}
\caption{Occultation fits for asteroids (814) Tauris and (907) Rhoda. The blue contour is for pole 2 in the case of Rhoda.}
\label{814_occ}
\end{figure*}

\begin{figure*}
\includegraphics[width=0.33\textwidth]{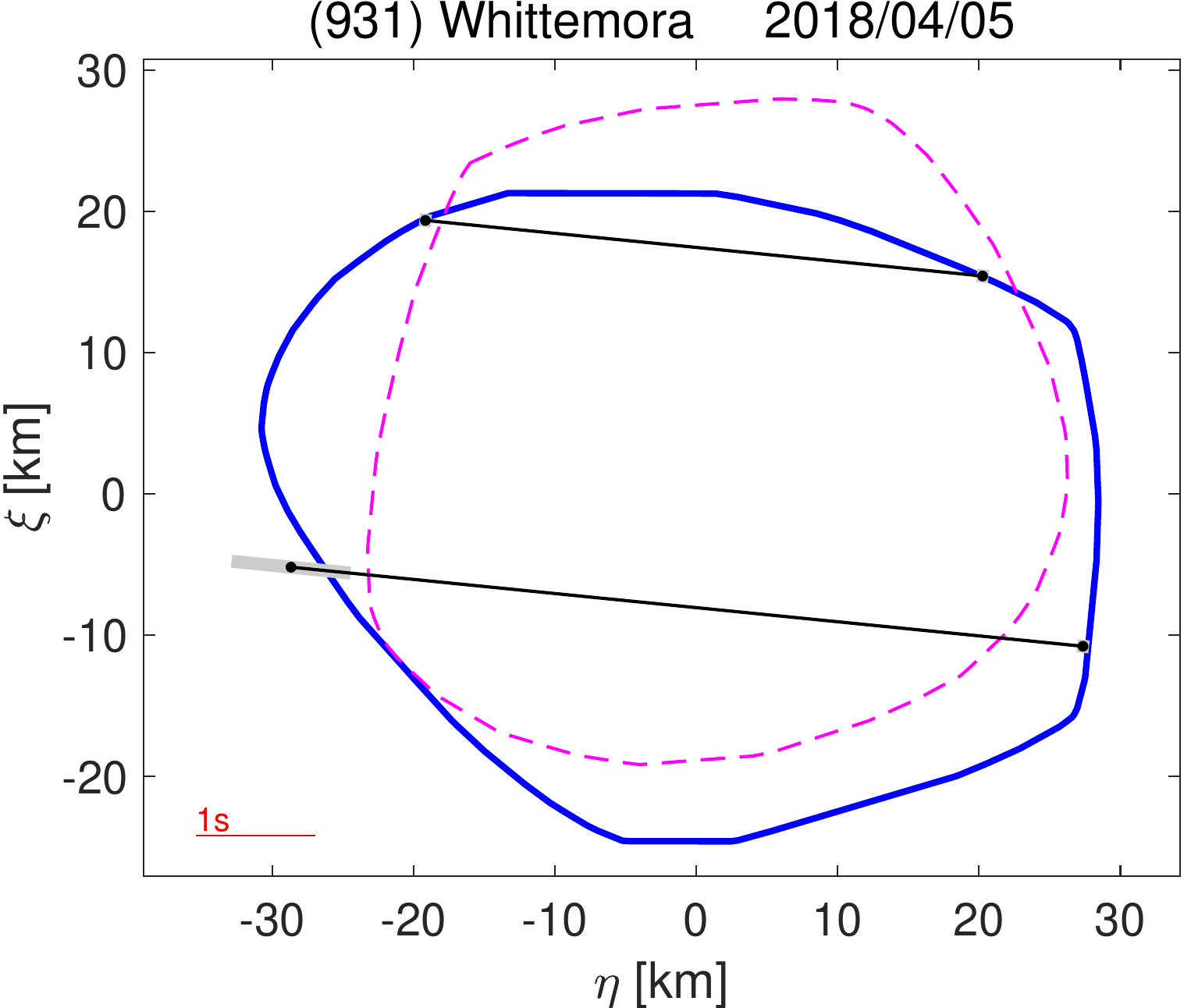}
\includegraphics[width=0.33\textwidth]{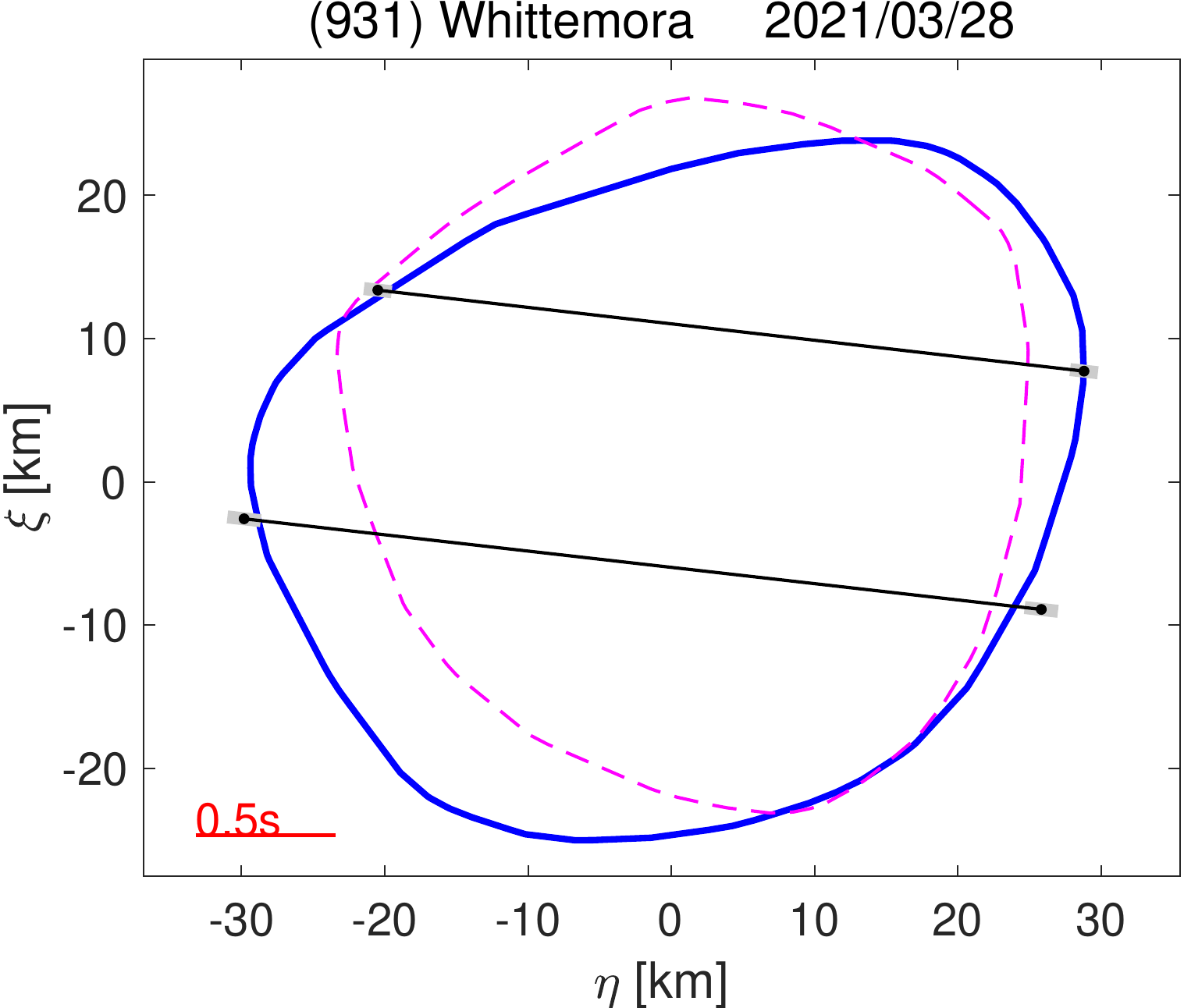}
\includegraphics[width=0.33\textwidth]{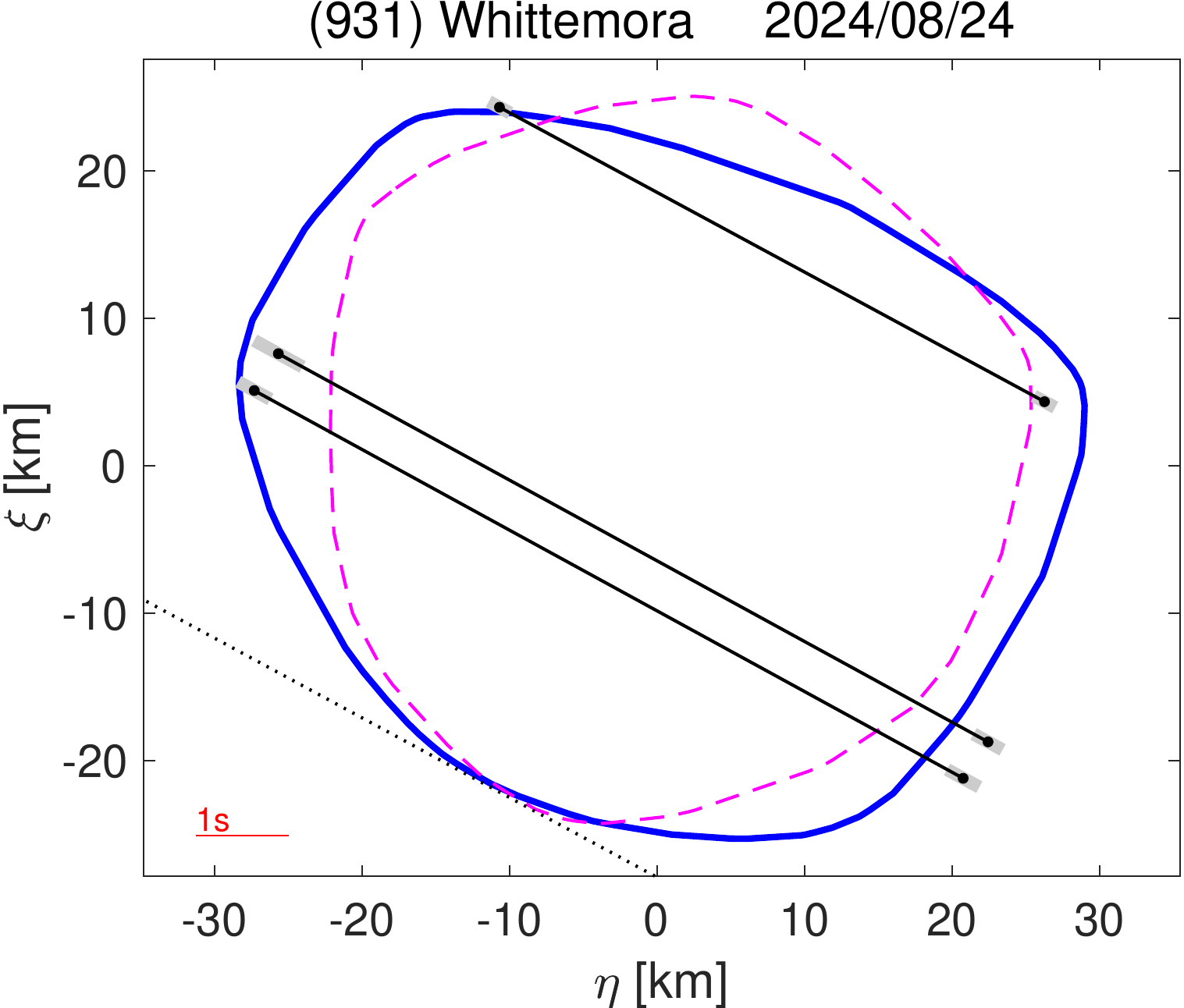}
\caption{Occultation fits for asteroid (931) Whittemora. Pole 2 is preferred (blue contour).}
\label{931_occ}
\end{figure*}

\begin{figure*}
\includegraphics[width=0.33\textwidth]{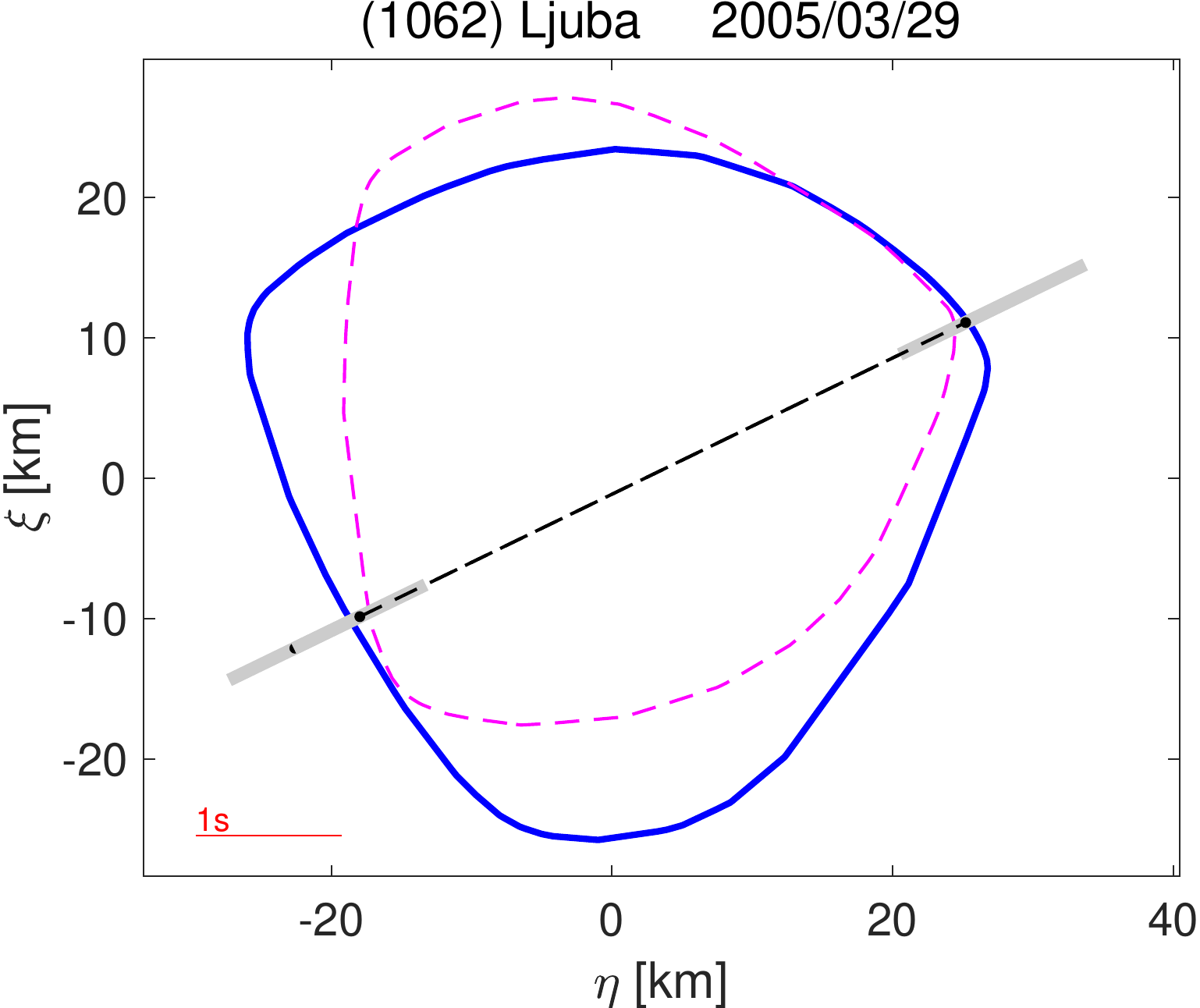}
\includegraphics[width=0.33\textwidth]{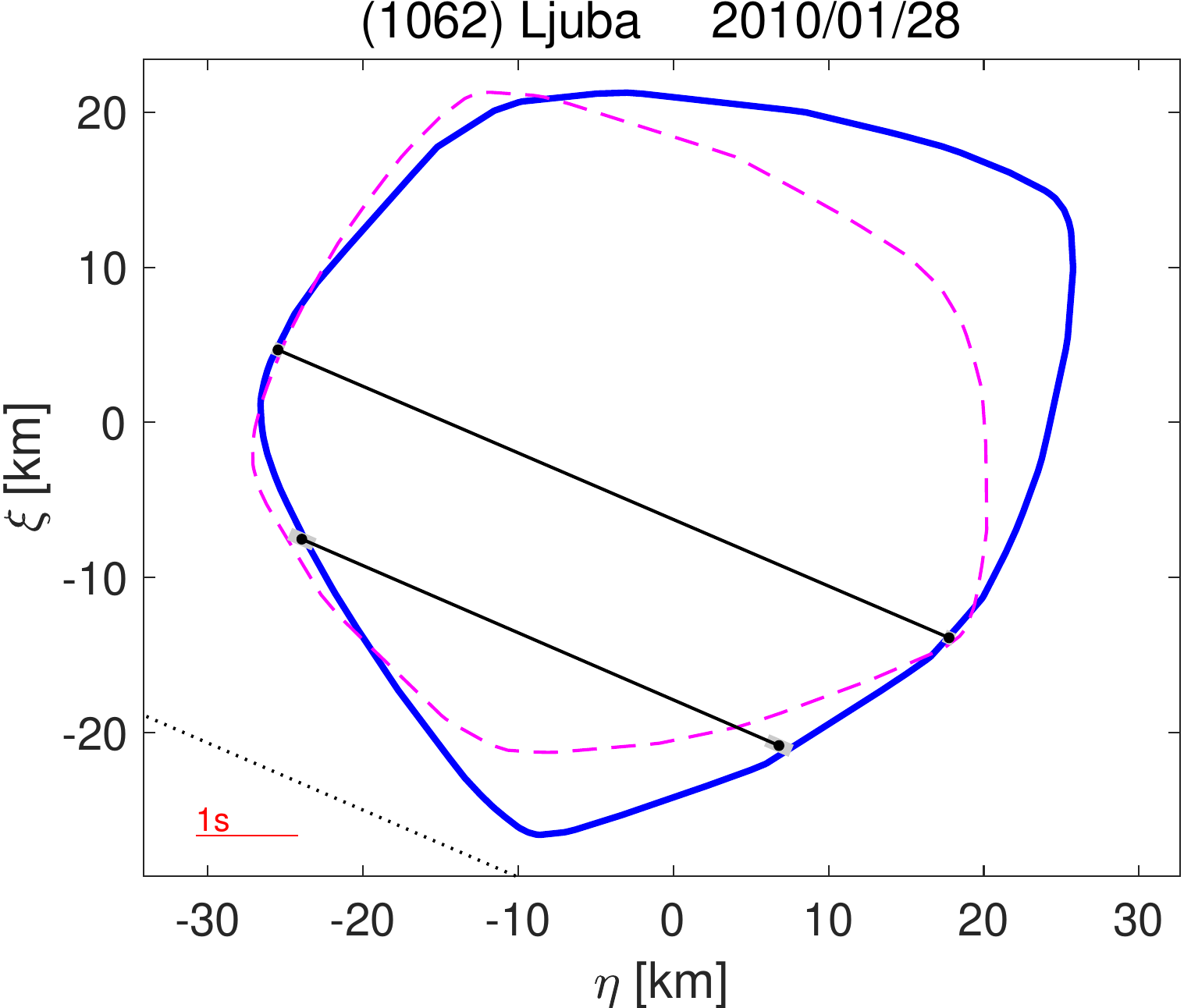}
\includegraphics[width=0.33\textwidth]{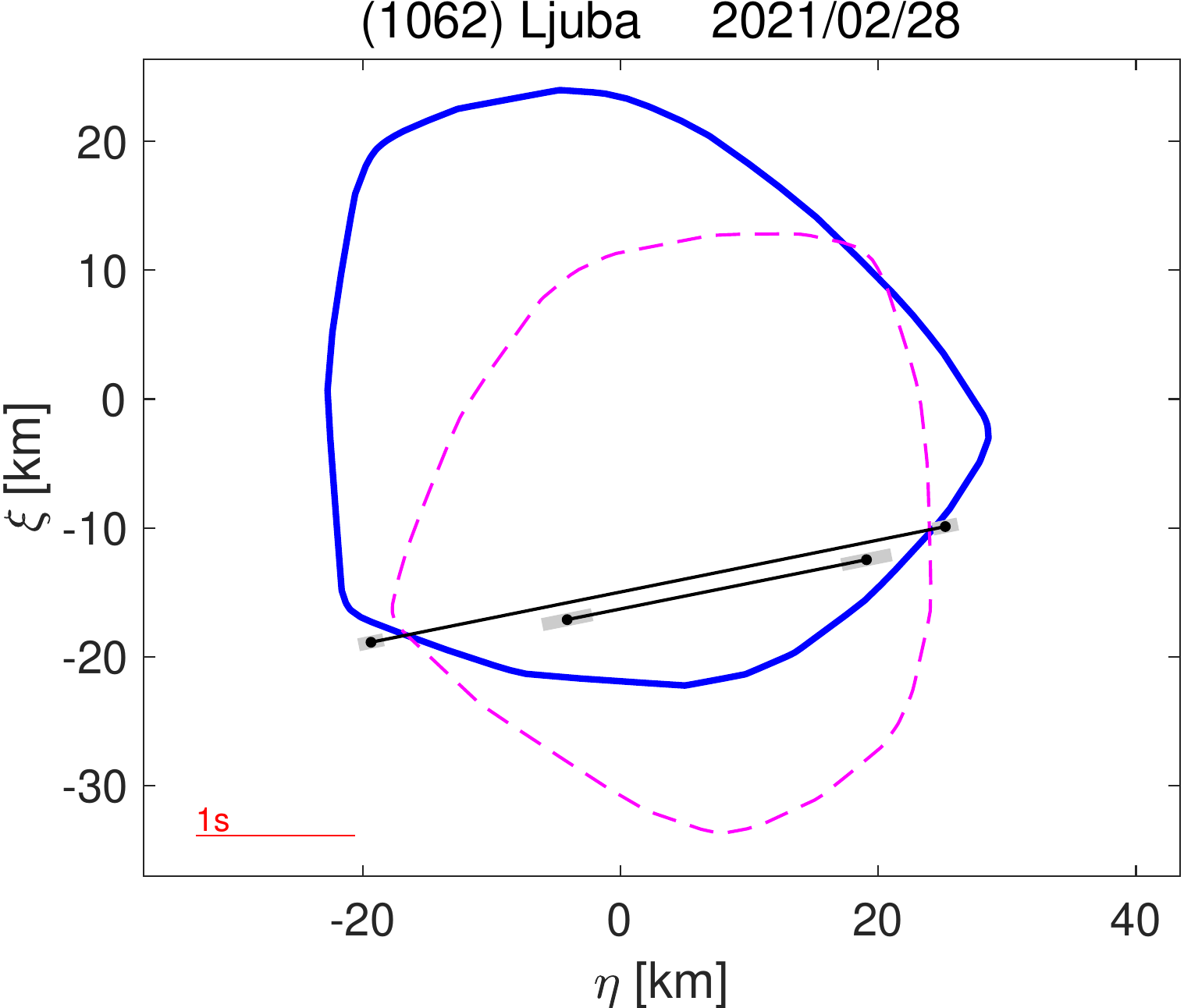}
\caption{Occultation fits for asteroid (1062) Ljuba. Pole 2 is preferred (blue contour).}
\label{1062_occ}
\end{figure*}

Instantaneous silhouettes of asteroid models derived here with the light curve inversion, fitted on the fundamental plane to the chords from stellar occultations (Figures \ref{215_occ} to \ref{1062_occ}). The scale is in kilometres. Blue contour by default is for pole 1 shape solution, and magenta for pole 2, unless one of them is preferred -- then this one one is shown with the solid contour, and less preferred in dashed contour. Continuous black lines join photoelectrically determined occultation timings, creating more reliable chords than those from visual observations marked with dashed black lines. The dotted lines are for negative (non-detection) observations. The grey segments show the chord uncertainties from the occultation timings, while the red segment shows the distance covered by the asteroid shadow in a given time. 

\clearpage

\end{appendix}
\twocolumn

\end{document}